\documentclass{aa} 
\usepackage[breaklinks, colorlinks, citecolor=blue, linkcolor=blue]{hyperref}
\usepackage{amsmath}
\usepackage{mathtools}
\usepackage{graphicx}
\usepackage{placeins}
\usepackage{txfonts}
\usepackage{float}
\usepackage{lscape}

\makeatletter
\DeclareRobustCommand{\ion}[2]{%
  \text{#1\,\check@mathfonts\fontsize\sf@size\z@\selectfont #2}%
}
\makeatother

\begin{document}

   \title{First statistical constraints on galactic scale outflows properties traced by their extended Mg II emission with MUSE}

   \author{
   Ismael Pessa\inst{\ref{AIP}} \and 
   Lutz Wisotzki\inst{\ref{AIP}} \and 
   Tanya Urrutia \inst{\ref{AIP}} \and 
   Nicolas F. Bouch{\'e} \inst{\ref{Lyon}} \and 
   Floriane Leclercq\inst{\ref{Lyon}} \and
   Ramona Augustin \inst{\ref{AIP}} \and
   Yucheng Guo\inst{\ref{Arizona}} \and
   Daria Kozlova \inst{\ref{AIP}} \and
   Haruka Kusakabe \inst{\ref{UTokyo}} \and 
   John Pharo \inst{\ref{AIP}}
   }
   \institute{   
   Leibniz-Institut for Astrophysik Potsdam (AIP), An der Sternwarte 16, 14482 Potsdam, Germany\label{AIP}\\
        \email{ipessa@aip.de}
   \and
   Univ of Lyon1, Ens de Lyon, CNRS, Centre de Recherche Astrophysique de Lyon (CRAL) UMR5574, F-69230 Saint- Genis-Laval, France\label{Lyon}
   \and
   NSF’s National Optical-Infrared Astronomy Research Laboratory,
950 N. Cherry Ave., Tucson, AZ 85719, USA\label{Arizona}
   \and Department of General Systems Studies, Graduate School of Arts and Sciences, The University of Tokyo, 3-8-1 Komaba, Meguro-ku, Tokyo, 153-8902, Japan\label{UTokyo}
    }
   \date{Received 28 October 2025; Accepted 10 February 2026}

  \abstract
  {Galaxies evolve within vast, gaseous halos that fuel star formation and carry signatures of feedback-driven outflows. Deep integral field data have enabled the study of \ion{Mg}{II} $\lambda\lambda$2796, 2803 halos, which trace galaxy-scale outflows in emission, and while individual detections of \ion{Mg}{II} halos have revealed extended circumgalactic structures, their faintness has limited studies to single-object analyses. Here, we present the first statistical study of \ion{Mg}{II}-emitting halos using deep MUSE observations of 47 star-forming galaxies at $0.7<z<2.0$. Building on our previous work, where we developed and applied an outflow modeling framework to a single \ion{Mg}{II} halo, we now extend this approach to a larger sample, enabling robust population-level insights on the properties of circumgalactic outflows traced by their extended \ion{Mg}{II} emission for the first time. We detect extended \ion{Mg}{II} emission out to tens of kiloparsecs and model the outflows as an ensemble of radially accelerating shells. Galaxies with \ion{Mg}{II} outflows tend to have higher star formation rates (SFRs), specific star formation rates (sSFRs), and younger stellar populations—consistent with star-formation-driven winds. The observations are consistent with winds that accelerate linearly with radius ($v\propto r$) from launching velocities of $v\sim 60$ km s$^{-1}$ up to maximum velocities that correlate with galaxies’ stellar mass, and are of about $\sim490$ km s$^{-1}$. Their inner regions are highly opaque ($\log \tau \sim2.6$), and we also identify a tentative trend between stellar mass and central optical depth. The opening angle of the outflow shows some dependency on the host-galaxy stellar mass, with less massive galaxies showing primarily wide opening angles (i.e., nearly isotropic outflows), and more massive galaxies showing a broader range of values, with both wide and narrow opening angles. The distribution of the spatial extent of \ion{Mg}{II} halos exhibits a clear peak at half-light radius (HLR) of $\sim5$ kpc, with an extended tail of larger HLR values, up to $\sim20$ kpc. Compact halo sizes (HLR $< 8$ kpc) correlate with stellar mass, but extended halos do not, which could suggest a difference in the powering mechanism between compact and extended halos. This work opens a new window into the structure and dynamics of the circumgalactic medium and its role in galaxy evolution, by constraining for the first time the properties of circumgalactic outflows with a statistically significant sample of galaxies with an extended \ion{Mg}{II} emission halo.}

\titlerunning{}
\authorrunning{I. Pessa}
\maketitle

\section{Introduction}

Observing the galactic halos in emission poses a major challenge due to the low densities of the gas within \citep[on the order of n$_{\rm H}\sim10^{-4}$ cm$^{-3}$, see, e.g.,][]{Werk2014}. At high redshifts ($z>3$), long exposure time integral field data have enabled the detection and characterization of individual galactic halos probed by their Ly$\alpha$ emission \citep[e.g.,][]{Wisotzki:2016hw, Leclercq:2017cp, Kusakabe2022, Erb2023}. These Ly$\alpha$ halos are typically ten times larger in scale length than the stellar component of their host galaxies \citep{Leclercq:2017cp, Wisotzki:2016hw}, and several studies indicate that they originate from the resonant scattering of photons produced by the central galaxy with neutral hydrogen atoms in the circumgalactic medium \citep[CGM, e.g.,][]{Kusakabe2019, Song2020, Byrohl2020}. For this reason, the complex spectral profiles and morphologies observed in the Ly$\alpha$ halos carry crucial information about outflows and inflows (i.e., baryon cycle) taking place in these galaxies \citep{Erb:2018dw, Remolina2019,Mitchell2021}. 

However, given that Ly$\alpha$ is in the rest-frame UV spectrum of galaxies, it is inaccessible for lower-redshift galaxies ($z < 3$) from ground-based observing facilities. The \ion{Mg}{II} $\lambda\lambda 2796$, $2803$ doublet provides an alternative resonant line that, like Ly$\alpha$, allows photons to scatter off ions in the CGM, but is accessible at lower redshifts. It is one of the most abundant metal ions in the Universe, and has an ionization potential slightly lower than that of hydrogen, making it a good tracer of the cool ($T \sim 10^{4}$ K) phase of the CGM. Unlike Ly$\alpha$, which originates primarily from recombination in \ion{H}{II} regions powered by young, massive stars, \ion{Mg}{II} halos are predominantly produced through the scattering of stellar continuum photons \citep[e.g.,][]{Prochaska:2011eqa}. Another alternative tracers of the cool gas in the CGM used in previous works are provided by the fluorescent \ion{Fe}{II*} \citep[e.g.,][]{Finley:2017eg, Finley:2017gc} or \ion{Si}{II*} \citep{Kusakabe2024} emission lines. However, both of these lines are weaker, and lie at bluer wavelengths than \ion{Mg}{II}.

The first detection of extended \ion{Mg}{II} emission around a galaxy was reported by \citet{Rubin:2011bm}, using slit-spectroscopy. They find an excess of \ion{Mg}{II} emission with respect to the stellar continuum around a galaxy at $z=0.69$ out to distances of $\sim10$ kpc from the central galaxy. This emission is interpreted as ions present in outflows scattering with photons produced by the central galaxy. The same object was later observed using KCWI integral field spectroscopy \citep{Burchett2021} that enabled to study the extended \ion{Mg}{II} emission in a spatially resolved manner and compare it with the radiative transfer models from \citet{Prochaska:2011eqa} to infer the physical properties of the outflow, such as orientation and velocity profile, and find that the data is consistent with a isotropic and radially decelerating wind.

Another example of such spatially resolved studies is \citet{Zabl2021}, where the authors use the unique capabilities, in terms of angular resolution and sensitivity, of the Multi-Unit Spectroscopic Explore \citep[MUSE,][]{Bacon:2010jn} to study the extended \ion{Mg}{II} emission up to distances of $\sim25$ kpc from a galaxy at $z=0.7$, whose CGM is also probed in absorption by a QSO sightline at an impact parameter of $\sim39$ kpc. They find their data to be consistent with the expectation from a simple toy model of a biconical outflow, with a constant slow velocity (130 km s$^{-1}$).  \citet{Leclercq2022} also uses MUSE data from the MUSE eXtremely Deep Field \citep[MXDF,][]{Bacon:2017hn, Bacon2023} to report the discovery of extended \ion{Mg}{II} emission emerging from the intragroup medium of a group of five star-forming galaxies at $z = 1.31$, and concludes that the intragroup medium is enriched by both outflows and tidal stripping from galaxy interactions. More recently, \citet{Pessa2024} have reported the discovery of an extended \ion{Mg}{II} halo around a star-forming galaxy at $z\sim0.7$ in the MUSE Hubble Ultra Deep Field mosaic \citep{Bacon:2017hn, Bacon2023} with \ion{Mg}{II} emission present up to distances of 30-40 kpc from the central galaxy, and found that the observed emission is consistent with radially accelerating winds. 

Due to the intrinsic difficulty of studying the CGM in emission, primarily because of its low density, the detections of extended \ion{Mg}{II} emission in individual galaxies listed previously have relied on very deep MUSE observations ($10 - 140$ hrs). These studies have revealed detailed morphologies and kinematic structures in single objects, highlighting the power of integral field spectroscopy for resolving the CGM of individual galaxies.

In addition to these individual studies, stacked analyses have enabled the detection of extended \ion{Mg}{II} halos in a statistical sense by enhancing the low surface brightness signal \citet{Guo2023b,Dutta2023}. \citet{Guo2023b} stacked MUSE data cube segments for 172 galaxies from the MUSE Hubble Ultra Deep Field surveys and identified a significant enhancement of \ion{Mg}{II} emission along the minor axis, consistent with bipolar outflows, extending beyond 10 kpc. This finding contrasts with the results from \citet{Dutta2023}, where the authors stack the \ion{Mg}{II} emission of nearly 40 galaxies from the MUSE Ultra Deep Field survey \citep[MUDF][]{Fossati2019}, and do not find a significant difference between the major and minor axes. However, stacking inherently averages over diverse morphologies and environments, losing the ability to study the specific characteristics of individual systems.

Both deep observations of individual galaxies and stacked analyses have played crucial roles in advancing our understanding of extended \ion{Mg}{II} emission. Individual detections have revealed the detailed morphologies, orientations, and kinematic features of \ion{Mg}{II} halos, offering insight into the physical mechanisms at play in specific systems. In contrast, stacked studies have demonstrated the ubiquity of such emission on average, confirming the presence of diffuse \ion{Mg}{II} halos across galaxy populations,

However, despite these important contributions, neither approach alone provides a statistical baseline to study the properties of the \ion{Mg}{II} halos on a population level, and how their properties depend on the host-galaxy properties. 

Transitioning to sample-level analyses (i.e., studying in a comparable manner the properties of a collection of objects, rather than focusing on the properties of a single galaxy, often chosen due to its special properties, rather than being a representative case) is thus paramount to understanding the complex interplay between galaxies and their CGM. In this paper, we use deep MUSE data to build, for the first time, a sample of galaxies with extended \ion{Mg}{II} emission, and study the properties of \ion{Mg}{II} halos on a population level, going beyond single-object discoveries. We use the outflow model introduced in \citet{Pessa2024} to infer the physical properties of the galactic winds shaping these \ion{Mg}{II} halos, and we explore correlations between host galaxies and outflow properties. The paper is structured as follows. In Sec.~\ref{sec:obs}, we present our data, and in Sec.~\ref{sec:sample_selection}, we describe our sample selection and the general properties of our sample galaxies. In Sec.~\ref{sec:reconstruction} we present the size distribution of the \ion{Mg}{II} halos in our sample, and in Sec.~\ref{sec:fitting_results}, we show the results of fitting our outflow model to the sample galaxies. In Sec.~\ref{sec:discussion}, we discuss the implications of our results and compare the physical properties of our sample galaxies with those inferred for their galactic winds. Finally, in Sec.~\ref{sec:summary}, we summarize our findings and conclusions. In Appendix~\ref{sec:methods}, we describe some of the relevant aspects of the methodology employed for our analyses, including a description of the outflow model used in Appendix~\ref{sec:model}.  We used a flat concordance cosmological model with $H_0 = 70$ km s$^{-1}$ Mpc$^{-1}$ and $\Omega_m = 0.3$. All quoted transverse distances are in physical units, not co-moving.

\section{Observational data}
\label{sec:obs}
\subsection{MUSCATEL survey}
\label{sec:muscatel}
The MUSE Cosmic Assembly survey Targeting Extragalactic Legacy fields (MUSCATEL, PIs: Lutz Wisotzki, Roland Bacon, Thierry Contini) is based on observations with the MUSE instrument on the ESO-VLT \citep{Bacon:2010jn}. MUSE is a panoramic integral-field spectrograph with a field of view (FoV) of $1'\times 1'$ (in Wide Field Mode, used for these observations) at a spatial sampling of $0\farcs2\times 0\farcs2$. The spectral range is 4700~\AA--9350~\AA, with a spectral sampling of 1.25~\AA\ and a wavelength-dependent velocity resolution between $\sim$190 and $\sim$80~km~s$^{-1}$ (full width at half maximum; FWHM) at the blue and red ends of the spectrum, respectively. 

 MUSCATEL consists of MUSE observations of parallel fields of the Hubble Frontier Fields \citep[HFF,][]{Pagul2021}, for 4 clusters accessible to the VLT (A2744, M0416, AS1063, A370).  We employed a wedding cake approach for the observing strategy, with three different levels of exposure time. First, a large shallow field composed of a 3\arcmin$\times$3\arcmin mosaic of 100 min exposure time following the Hubble Space Telescope (HST) Advanced Camera for Surveys (ACS) footprint of the parallel fields. Then, the medium field lies on top of the shallow field, consisting of a 2’x2’ mosaic of 5 hours depth, following the HST Wide Field Camera 3 (WFC3) footprint. Finally, the deep field consists of a single 1’x1’ MUSE pointing on top of the medium field, with 25 hours of observing time and a suitable star for the PSF characterization\footnote{For A370, only one `deep field' with a total exposure time of 10 hours was observed.}.

Some of the science goals of the MUSCATEL survey include increasing the number of deep fields, reducing cosmic variance, and improving statistics for studies that require deep data (e.g., clustering). All the observations have been carried out using the VLT adaptive optics facility \citep[AO,][]{Leibundgut2017}, with the exception of one of the shallow field pointings from the AS1063 clusters, and a small number of additional individual exposures. The separation from the parallel fields to the center of the clusters is large ($6$ arcmin), meaning that the parallel fields are not affected by gravitational lensing from cluster galaxies. Furthermore, excluding redshifts consistent with the cluster, the redshift distribution of galaxies in the field is consistent with those from other deep fields.

We have proceeded with the data reduction in a similar way as \citet{Urrutia2019}, using the standard MUSE data reduction software \citep{Weilbacher2014, Weilbacher2020}, but with additional post-processing steps, for instance, superflat correction as described in \citet{Bacon2023}. We have then used the LSDCat software \citep[Line Source Detection and Cataloguing,][]{Herenz:2017er} to search for emission line peaks in the MUSCATEL reduced data cubes. We have imposed a S/N threshold in the emission line of 5 to get rid of spurious sources\footnote{For the shallow fields, we used a S/N threshold of 6.}, and created a catalog of emission-line-selected galaxies, for which we have carried out a careful supervised redshift determination. The final catalog contains a total of 3245 galaxies with a measured redshift.

The details of the redshift determination, the construction of the galaxy catalog, as well as other critical aspects of the MUSCATEL survey, such as an exhaustive description of the data and data reduction will be reported in a future MUSCATEL dedicated survey paper.

\subsection{Photometric data}
\label{sec:photometry}
In addition to the MUSE observations, there is extensive multiwavelength photometric archival data available for the fields targeted by the MUSCATEL survey. In particular, deep optical and near-infrared images were obtained with the Hubble Space Telescope \citep[HST,][]{Shipley2018, Pagul2021}. The optical bands available include the F435W, F606W, and F814W bands from the HST/ACS. The NIR bands include the F105W, F125W, F140W, and F160W bands from the HST/WFC3. There is also available deep James Webb Space Telescope (JWST) data for these fields \citep[e.g,][]{Bezanson2024, Rihtar2025}. However, these data are currently still undergoing postprocessing; thus, we do not use them in this work.

For each galaxy in the catalog of emission-line-selected galaxies, we have identified its counterpart in the HST images and used the photometric data to perform a fitting of its spectral energy distribution (SED). We estimated the stellar mass, star formation rate, and age of each galaxy by fitting the broad-band spectral energy distribution with the FAST code \citep{Kriek:2009cs}, employing all available HST photometric bands, and the models from the \citet{Bruzual2003} stellar library, assuming a Chabrier IMF \citep{Chabrier2003} and an exponentially declining star formation history. We use four possible stellar metallicities to fit the data, namely Z = [0.004, 0.008, 0.02, 0.05]. The uncertainties in stellar mass, SFR, and age were estimated by performing 300 Monte Carlo iterations, and are on the order of $0.2$, $0.3$, and $0.2$ dex, respectively. The SED fitting also yielded a global estimate of the dust extinction acting on the starlight, assuming a \citet{Calzetti:2000iy} extinction curve. The stellar masses, SFRs, and ages computed via SED fitting for each galaxy are provided in Table~\ref{tab:sample_table}.

The counterpart identification is carried out by cross-matching the location of the galaxies identified in the MUSCATEL dataset with the objects in the HST catalogs. A subsequent visual inspection ensured the consistency of the identifications. Since the galaxies in our sample are relatively continuum bright\footnote{Median F814W mag of $\sim23.4$, with the faintest having a F814W mag of $\sim25.5$, significantly brighter than the detection threshold of the HFF HST data of $\sim 28$ mag, see \citet{Shipley2018}.}, there is little ambiguity in the counterpart identification. For only one of our sample galaxies, we were not able to unambiguously determine its HST counterpart, due to the crowded field and irregular morphology of the galaxies near the MUSE detection (see caption of Table~\ref{tab:sample_table}).

\FloatBarrier

\section{Galaxy sample}
\label{sec:sample_selection}
\subsection{Sample selection}
Our goal is to build a sample of galaxies where we can robustly detect extended \ion{Mg}{II} outflow emission and measure its properties. However, designing a selection criterion for this purpose is not straightforward. A simple flux threshold in the \ion{Mg}{II} line could potentially be misleading, depending on the geometry of the outflow and the size of the galaxy. The spectrum extracted from a fixed-size aperture could be, for instance, strongly dominated by \ion{Mg}{II} in absorption, with the emission present at larger impact parameters. Another possibility is that the emission and absorption components partially or totally cancel out each other, biasing the strength of the \ion{Mg}{II} emission in the aperture-integrated spectrum. On the other hand, a minimum S/N threshold in the stellar continuum emission might also be suboptimal since it would not be informative about the relative strength of the \ion{Mg}{II} emission with respect to the continuum. Additionally, it would also imply indirectly imposing a stellar mass limit in our sample, and the work from \citet{Feltre:2018in} shows that there is a correlation between the observed \ion{Mg}{II} spectral profile and galaxy stellar mass, where most massive galaxies more likely show \ion{Mg}{II} in absorption, and less massive galaxies show mostly \ion{Mg}{II} in emission.

Thus, we have opted for the following approach. First, we have kept only those galaxies in the MUSCATEL catalog within the redshift range where the \ion{Mg}{II} doublet is available in the MUSE wavelength range, that is, $0.7 \lesssim z \lesssim 2.3$, and with redshift confidence of $3$ (i.e., a secure redshift determination based on more than one spectral feature). This leaves us with an initial sample of 545 galaxies. Hereafter, we will refer to this sample as our parent sample.

Then, we have visually inspected the aperture-integrated spectrum of each source (in an aperture of angular size $1\farcs0\times 1\farcs0$, approximately $8^2$ kpc$^{2}$ at $z=1$), as well as their continuum-subtracted \ion{Mg}{II} pseudo-narrowband image constructed from a cutout of the MUSE data cube around the \ion{Mg}{II} doublet wavelength ($2790 - 2810\, \AA$ in the rest-frame of each galaxy). We kept those galaxies where we either find evidence of \ion{Mg}{II} emission or absorption in the integrated spectrum, or in the continuum-subtracted pseudo-narrowband image, without imposing a priori any S/N threshold.

We ended up with a sample of 89 galaxies, to which we will refer as our preliminary sample. Next, we examined their spectra in concentric annular apertures ($r < 0\farcs4$, $0\farcs4 < r < 0\farcs8$, $0\farcs8 < r < 1\farcs2$, $1\farcs2 < r < 2\farcs0$, $2\farcs0 < r < 3\farcs5$, which correspond to physical impact parameters of $r < 3.2$ kpc, $3.2$ kpc $< r < 6.4$ kpc, $6.4$ kpc $< r < 9.6$ kpc, $9.6$ kpc $< r < 16.0$ kpc, $16.0$ kpc $< r < 28.0$ kpc at $z=1$), and classified the \ion{Mg}{II} profiles of the sample galaxies as P-Cygni (Pcyg), absorption-only (Abs), and emission-only (Ems), depending on whether they show clear \ion{Mg}{II} absorption and/or emission along these radial apertures. For instance, for some galaxies, their integrated spectrum in a circular aperture could be dominated by \ion{Mg}{II} in emission, but if they exhibit \ion{Mg}{II} in absorption in an inner, smaller aperture, they were classified as a P-Cygni profile. Inspecting the spectra of galaxies in radial apertures is, thus, useful for disentangling strong central absorption from \ion{Mg}{II} emission at larger radii. The profile classification has been done in a qualitative manner, based on the presence of significant absorption and/or emission in the spectra extracted from annular apertures.

\begin{figure*}
\includegraphics[width=\textwidth]{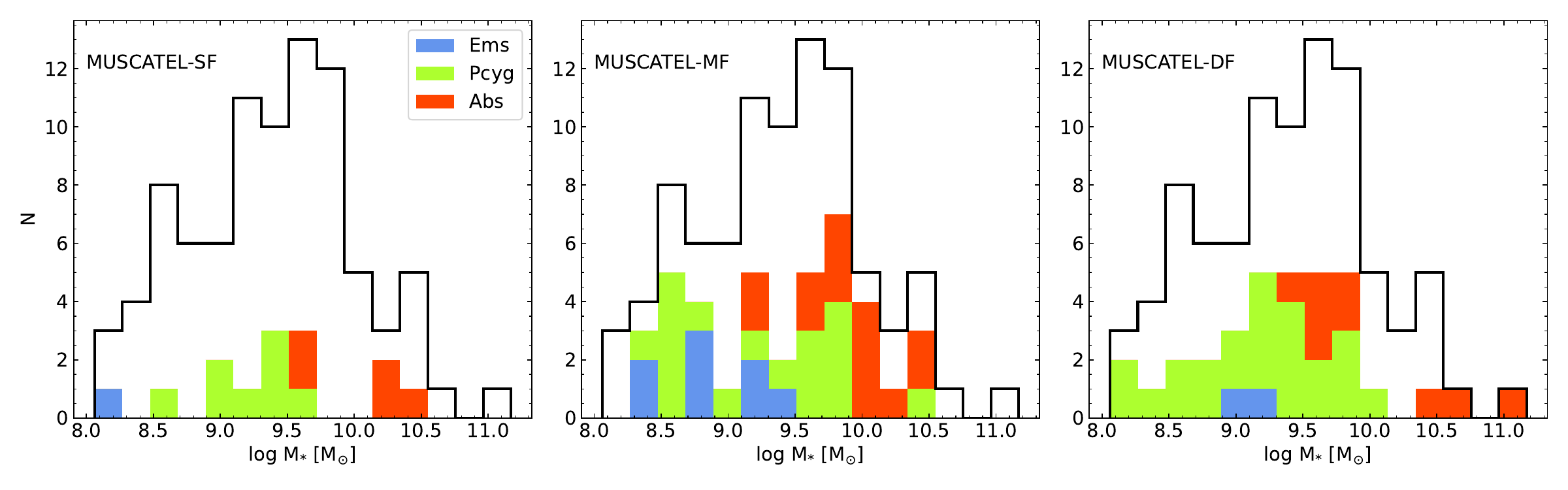}
\caption{Stellar mass distribution of our preliminary sample of galaxies. The sample consists of 89 galaxies drawn from the MUSCATEL survey that exhibit \ion{Mg}{II} either in emission, absorption, or a P-Cygni profile, determined after a careful inspection of their integrated spectra and their continuum-subtracted \ion{Mg}{II} pseudo-narrowband image. A direct S/N threshold has not been applied. Each stacked histogram shows the distribution of the galaxies that belong to each one of the depth levels of the MUSCATEL survey (shallow field, medium field, and deep field, see Sec.~\ref{sec:obs}), indicated in the top-left part of each panel. The black line shows the distribution of the full preliminary sample in all the panels. Stellar masses are calculated via SED fitting, as described in Sec.~\ref{sec:photometry}. The stacked histograms are color-coded by the different spectral profiles shown in \ion{Mg}{II}, as indicated in the legend.}
\label{fig:stellar_mass_dist}
\end{figure*}

Figure~\ref{fig:stellar_mass_dist} shows the stellar mass distribution (computed from the SED fitting, see Sec.~\ref{sec:photometry} for details) of the galaxies in our preliminary sample, separated by MUSCATEL depth and profile type. Our sample is consistent with the findings from \citet{Feltre:2018in}, where galaxies that show \ion{Mg}{II} in absorption are generally on the high-mass side of the distribution, galaxies that show \ion{Mg}{II} in emission lie generally on the lower-mass side, whereas galaxies that exhibit a P-Cygni profile lie somewhere in between (albeit the stellar mass range of our sample galaxies is narrower than that of \citet{Feltre:2018in}, thus, the correlation of spectral profile and stellar mass is less evident here).

From the preliminary sample, we then opted to keep only those galaxies that show a P-Cygni profile in \ion{Mg}{II}, since this feature is interpreted as a signature of resonant scattering in an optically thick outflow, with absorption in the approaching (blueshifted) regions of the outflow and a redshifted emission component due to backscattering in the receding material \citep[e.g.,][]{Castor1970, Castor1979, Scuderi1992}. This clear outflow signature makes them more suitable for studying the physical properties of galactic winds. Out of the 89 galaxies in the preliminary sample, 50 show a clear P-Cygni profile. However, we stress that while the non-P-Cygni galaxies will be excluded from the analyses carried out in this paper, we will include them in our sample in a future paper, where we will investigate the properties of this broader halo population.

\subsection{\ion{Mg}{II} significance}
We then used our outflow modeling scheme to model the MUSE data cube cutout of each one of the P-Cygni galaxies. Our outflow model is described in detail in \citet{Pessa2024}. In Appendix~\ref{sec:model} we include a summarized description of the model. At this stage, we used our model only to reproduce the spectral profile of the \ion{Mg}{II} absorption. By modeling the absorption of the photons produced by the central source, we are able to correct for it (i.e., add the flux from the absorbed photons back into the observed spectra), and then produce a \ion{Mg}{II} emission-only data cube for each galaxy. We disentangle the \ion{Mg}{II} emission from the absorption in order to quantify the significance of the \ion{Mg}{II} emission.

We stress that the model is only used to reproduce the absorption profile at this point. The actual best-fitting parameters, or the performance of the model at reproducing the extended emission, are not relevant here. In principle, we could use different methods to perform this correction \citep[e.g., optical depth and covering fractions that vary as a function of projected velocity, see][]{Xu2022}. Still, for internal consistency, we opted to use the same model that we later use to infer the properties of the galactic winds.

Next, we produced a continuum cube for each galaxy and removed the continuum from the \ion{Mg}{II} emission-only (already absorption corrected) data cube. The continuum cubes were constructed using a running median across the spectral dimension of the original cube, within a window of 200 wavelength channels. 

We then collapsed the continuum-subtracted \ion{Mg}{II} emission-only data cubes along the wavelength axis (using a mean) to produce an emission-only continuum-subtracted \ion{Mg}{II} pseudo-narrowband image, for a wavelength range of approximately $2790\AA < \lambda < 2809\,\AA$ rest-frame (this range can be slightly different for some galaxies if they exhibit, for instance, a broader spectral profile, or contamination due to a nearby sky emission line, as detailed in Appendix~\ref{sec:fitting_desc}). We used this pseudo-narrowband image to compute the S/N of the \ion{Mg}{II} emission for each galaxy. We computed the S/N of each galaxy in two different ways. First, we simply integrated the \ion{Mg}{II} flux across the full modeled region (a circular aperture of $3\farcs5$ radius, see Appendix~\ref{sec:fitting_desc}), and divided it by the square root of the integrated variance across the same region. Secondly, we did the same calculation, but for the different annular apertures defined earlier at the beginning of this section separately, and we kept the highest S/N among the different apertures (excluding the central one, in order to minimize the contribution from nebular emission and focus on the extended component). We finally opted to use the highest of the two different values (integrated in the whole modeled region vs. integrated in an annular aperture). The motivation behind this criterion is that while some halos are more spatially extended, and integrating over larger apertures maximizes the significance of the \ion{Mg}{II} emission, others are more compact and radially concentrated. For the latter, integrating over large apertures leads to a dilution of the total signal. We did not aim at discarding some halo morphologies with our selection criterion; thus, we used this flexible definition to characterize the significance of the \ion{Mg}{II} emission.

\begin{figure*}
\includegraphics[width=\textwidth]{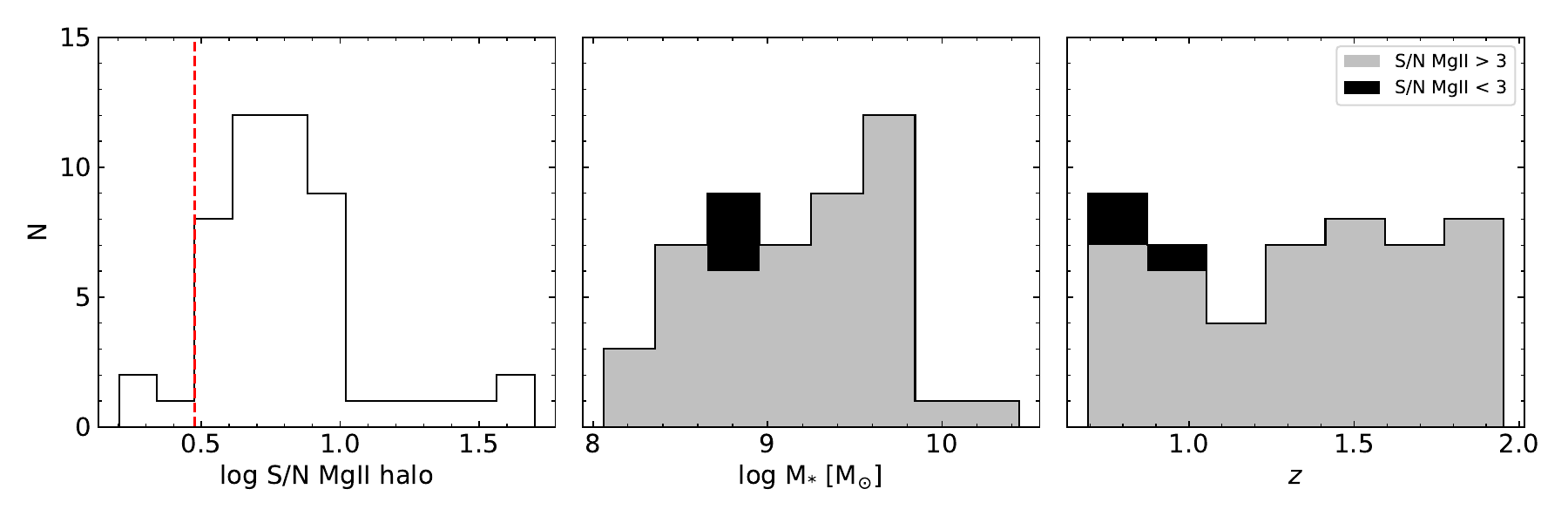}
\caption{Significance of the extended \ion{Mg}{II} emission, stellar mass, and redshift distribution for the galaxies in our preliminary sample that exhibit a P-Cygni profile in \ion{Mg}{II}. \textit{Left}: Significance of the extended \ion{Mg}{II} emission, in terms of its spatially integrated S/N, for the 50 galaxies in our preliminary sample that exhibit a P-Cygni profile in \ion{Mg}{II}. We further removed from our preliminary sample those galaxies where the integrated S/N of the modeled emission is lower than $3\,\sigma$ (indicated with a vertical dashed red line). \textit{Middle}: Stacked histogram that shows the stellar mass distribution of galaxies in the preliminary sample with a P-Cygni profile in \ion{Mg}{II}. The gray histogram shows the stellar mass distribution of those galaxies where the significance of the \ion{Mg}{II} emission is above the $3\,\sigma$ threshold. The black histogram shows the stellar mass distribution of galaxies below this significance threshold. \textit{Right}: Same as middle panel, for the redshift distribution of the sample galaxies.}
\label{fig:SN_halos}
\end{figure*}

After calculating the S/N of each galaxy, we kept in our sample only those galaxies with a mean S/N (wavelength averaged) higher than 3. The left panel of Fig.~\ref{fig:SN_halos} shows the distribution of the S/N values for our sample galaxies. Out of the 50 galaxies that show a P-Cygni profile, 47 present a S/N above the minimum significance threshold of $3\,\sigma$, and we deem these galaxies to be robust detections of extended \ion{Mg}{II} halos tracing galactic-scale outflows. These 47 galaxies form our final sample. 22 of these galaxies are drawn from the MUSCATEL deep field, 17 from the medium field, and 8 from the shallow field (see Sec.~\ref{sec:obs}). 

While our objective is to search for primarily extended \ion{Mg}{II} halos (hence, we excluded the central aperture from the annular apertures), we do not filter our sample by halo size at this stage, that is, our sample could potentially include galaxies for which the size of the \ion{Mg}{II} emission is consistent with the size of their continuum counterpart. However, even if this is the case, the observed light distribution of \ion{Mg}{II} will always be different from the continuum light distribution, because, by construction, all the galaxies in our final sample exhibit some level of \ion{Mg}{II} absorption in their central region. Furthermore, due to the generally irregular morphology of the \ion{Mg}{II} halos, parameterizing their sizes is not straightforward. In Secs.~\ref{sec:reconstruction} and ~\ref{sec:sizes} we present and discuss the size of the \ion{Mg}{II} emission of our sample galaxies in detail, and compare these sizes with the size of their continuum counterpart.

The middle panel of Fig.~\ref{fig:SN_halos} shows the stellar mass distribution for the galaxies above and below the significance threshold. Galaxies with lower \ion{Mg}{II} S/N are preferentially on the low-mass half of the distribution, with stellar masses of  $\log \mathrm{M}_{*} \sim 8.8 \, M_{\odot}$ (although still close to the median of the distribution of $\log \mathrm{M}_{*} = 9.2 \, M_{\odot}$). The right panel of the figure shows the redshift distribution for galaxies above and below our S/N threshold. The galaxies below our S/N threshold are all at $z<1$. Our final sample presents a relatively flat redshift distribution. The slightly smaller number of galaxies at $z\sim1.1$ is caused by the AO wavelength gap coinciding with the wavelength of the \ion{Mg}{II} doublet at that redshift. 

\subsection{Sample characterization}

\begin{figure}
\includegraphics[width=\columnwidth]{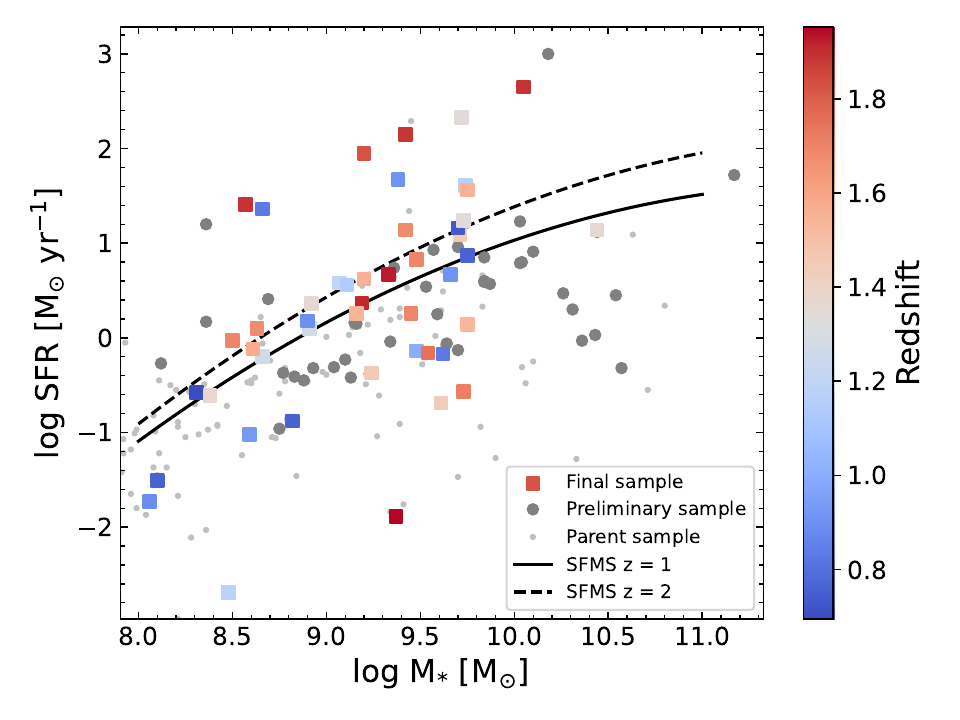}
\caption{Star formation rate as a function of stellar mass, for our sample galaxies. Both quantities are obtained via SED fitting (see Sec.~\ref{sec:photometry}). Our final sample of 47 galaxies that exhibit a P-Cygni profile and a significant detection of modeled extended emission is shown with squares, color-coded by redshift. The preliminary sample is shown as dark gray circles, and the original parent sample is shown as smaller light gray dots. For reference, we show the SFMS as measured by \citet{Popesso2023} for $z=1$ (solid) and $z=2$ (dashed), by collecting different measurements of the SFMS for a wide range of redshifts.}
\label{fig:SFMS_sample}
\end{figure}

Figure~\ref{fig:SFMS_sample} shows the SFR versus the stellar mass of the galaxies in our final sample, as well as the rest of the galaxies in our preliminary and parent samples. Both quantities are measured via SED fitting, see Sec.~\ref{sec:photometry} for details. Most galaxies in our final sample lie relatively close to the star-forming main sequence \citep[SFMS, see, e.g.,][]{Brinchmann2004,Daddi2007,Noeske2007, Popesso2023}. However, there is significant scatter, with some galaxies that exhibit considerably higher and lower SFRs, compared to their expected value, given their stellar mass. This implies that \ion{Mg}{II} halos are not necessarily always associated with starbursting galaxies, as previous works could suggest \citep[e.g., ][]{Zabl2021}. Indeed, only a small fraction of our sample galaxies could be considered as starbursts, significantly above the SFMS. On the contrary, most of them lie closer to or even below the SFMS.

Figure~\ref{fig:sample_1} shows the continuum-subtracted \ion{Mg}{II} pseudo-narrowband image (not corrected by absorption) overlaid onto the HST/F814W images for each galaxy in our final sample, for a square FoV of 40 kpc $\times$ 40 kpc. The \ion{Mg}{II} emission is shown in black, and the stellar light from the HST data is in blue. The continuum-subtracted pseudo-narrowband images were created by collapsing the continuum-subtracted cubes across their wavelength dimension, for the full modeled wavelength range that encloses the complete P-Cygni profile of both \ion{Mg}{II} lines (approximately $2790\AA < \lambda < 2809\AA$ rest-frame. The exact modeled wavelength range used for each galaxy is shown in the figures in Appendix~\ref{sec:model_results_all_galaxies}). For visualization purposes only, the continuum-subtracted \ion{Mg}{II} pseudo-narrowband images have been smoothed with a Gaussian kernel of $1\farcs0$ FWHM. 

In many cases, the halos present strong \ion{Mg}{II} absorption in the central part, and while some galaxies seem to display only \ion{Mg}{II} emission in the pseudo-narrowband, by construction, all of them exhibit some level of absorption (the emission might overcome the absorption when the profile is integrated along the wavelength axis). The figure also shows the observed surface brightness profiles of the \ion{Mg}{II} emission, which, for the galaxies that present strong central absorption, are negative at small galactocentric distances. Hereafter, for simplicity, we will refer to galaxies in the final sample as our sample galaxies. Table~\ref{tab:sample_table} summarizes the general properties of our sample galaxies.

We acknowledge, however, that our sample selection criteria inevitably lead to biasing our sample in the parameter space of galaxy properties. For instance, by keeping only the galaxies that exhibit a P-Cygni profile in the \ion{Mg}{II} line, we are more likely to drop galaxies in the extremes of the mass distribution \citep[see, e.g.,][and Fig.~\ref{fig:stellar_mass_dist}]{Feltre:2018in}. Similarly, by setting a minimum S/N of the \ion{Mg}{II} extended emission, we are also dropping preferentially galaxies on the low-mass side of the distribution (see Fig.~\ref{fig:SN_halos}). Additionally, since at a constant total luminosity, more extended halos are less likely to be detected than more compact ones \citep[see, e.g.,][in the context of Ly$\alpha$ halos]{Pharo2024}, this means that we could be excluding from our sample those more extended halos in less massive galaxies. Also, these biases in the sample ultimately mean that we are confined to limited regions in the parameter space and that our conclusions cannot be extended to, for instance, the low- and high-end of the mass distribution.

\begin{figure*}[!htb]
    \centering
    \begin{minipage}{0.49\textwidth}
        \centering
        \includegraphics[width=0.99\columnwidth]{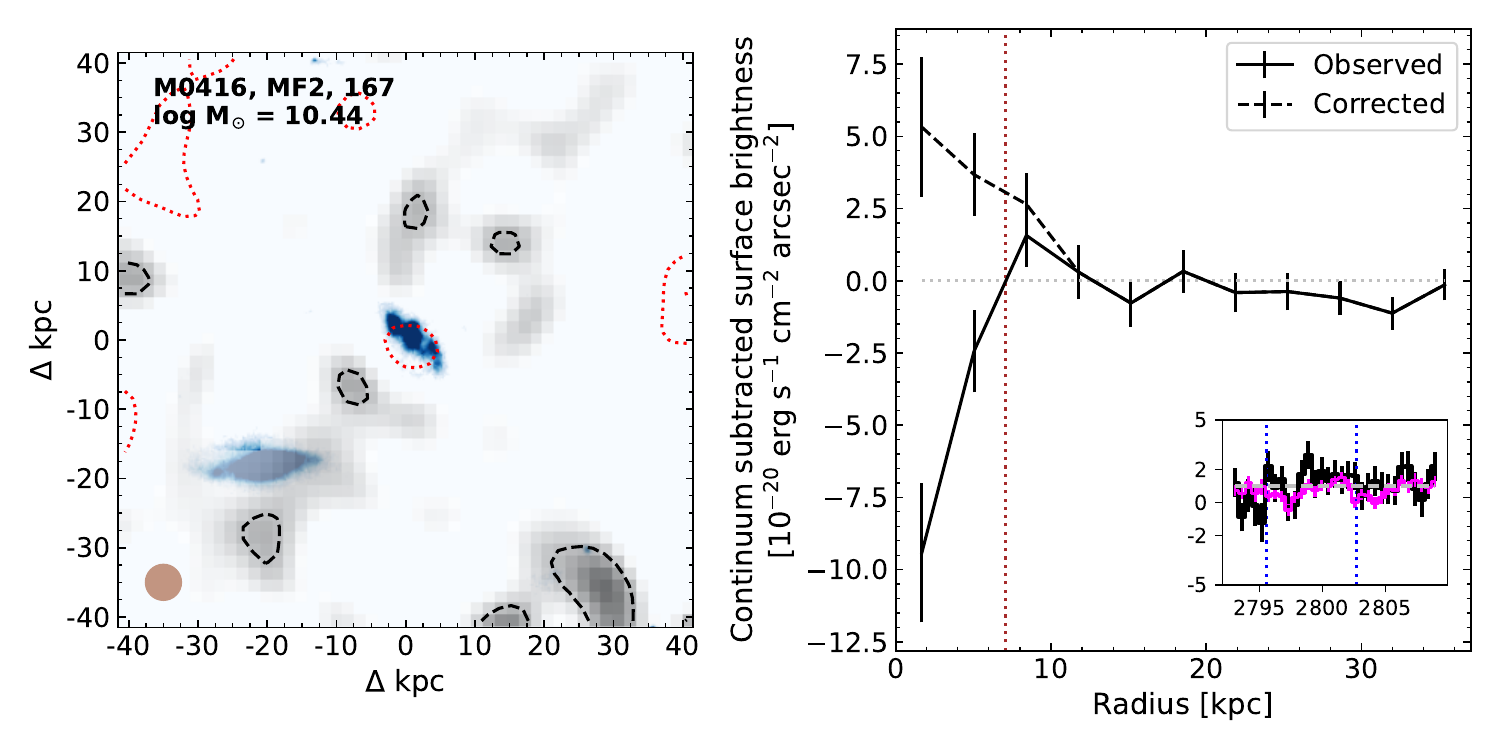}
    \end{minipage}%
    \begin{minipage}{0.49\textwidth}
        \centering
        \includegraphics[width=0.99\columnwidth]{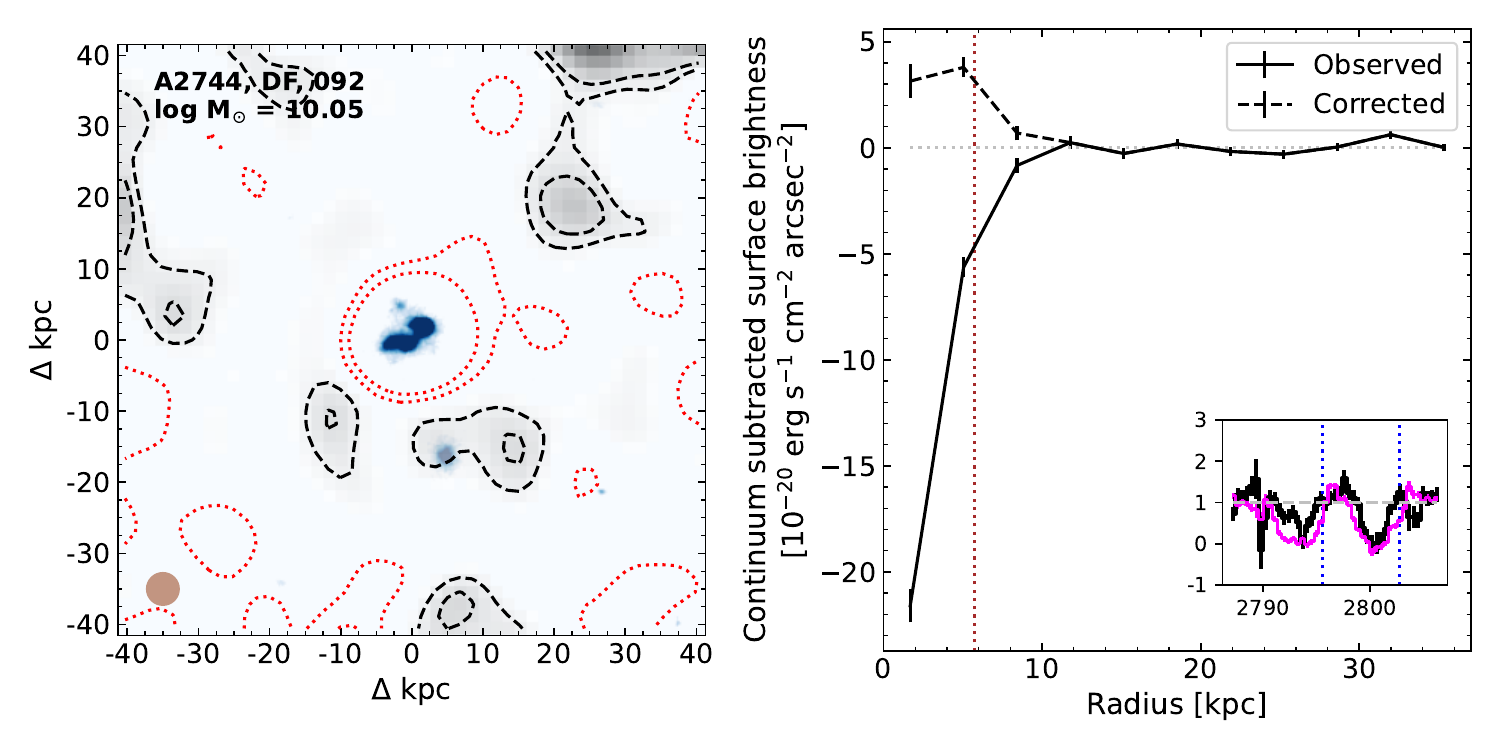}
    \end{minipage}

    \begin{minipage}{0.49\textwidth}
        \centering
        \includegraphics[width=0.99\columnwidth]{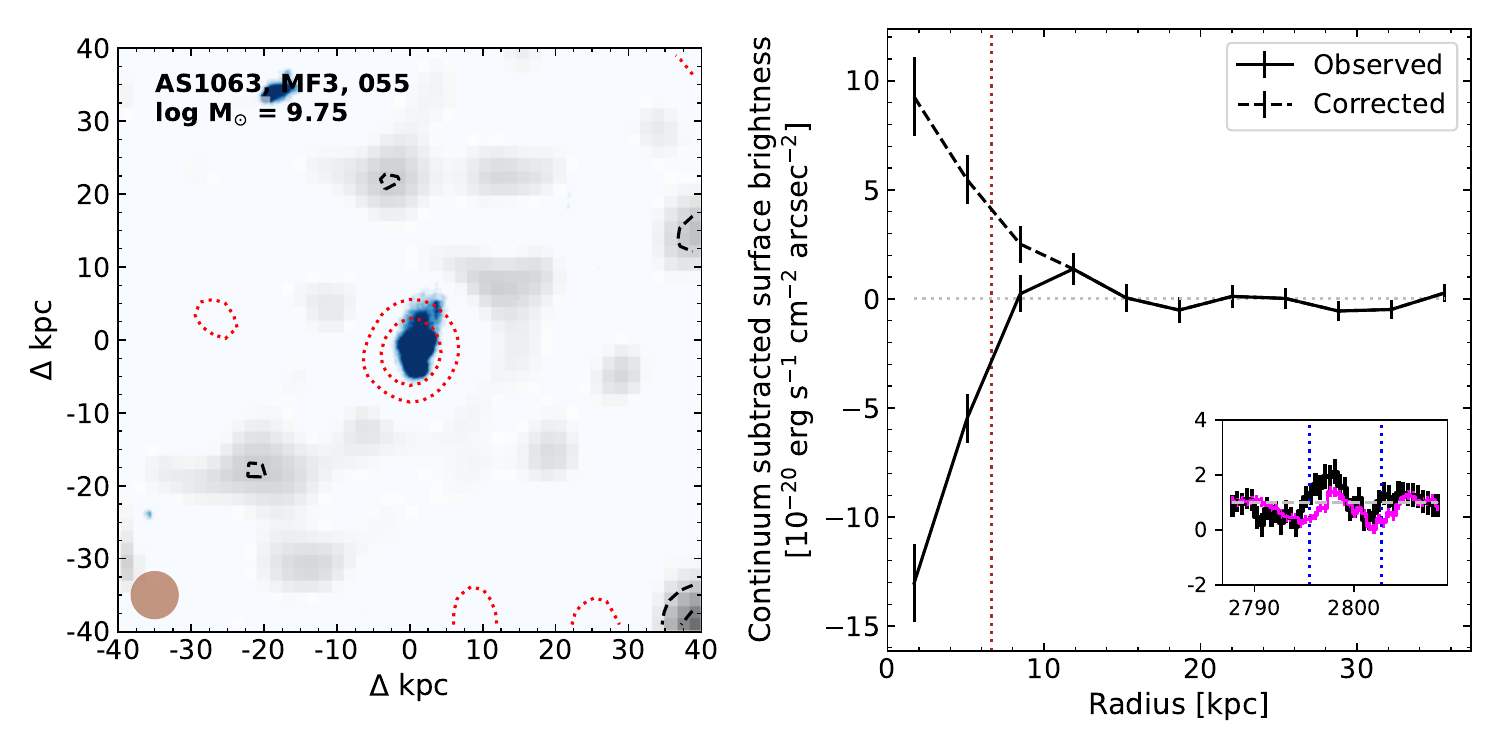}
    \end{minipage}%
    \begin{minipage}{0.49\textwidth}
        \centering
        \includegraphics[width=0.99\columnwidth]{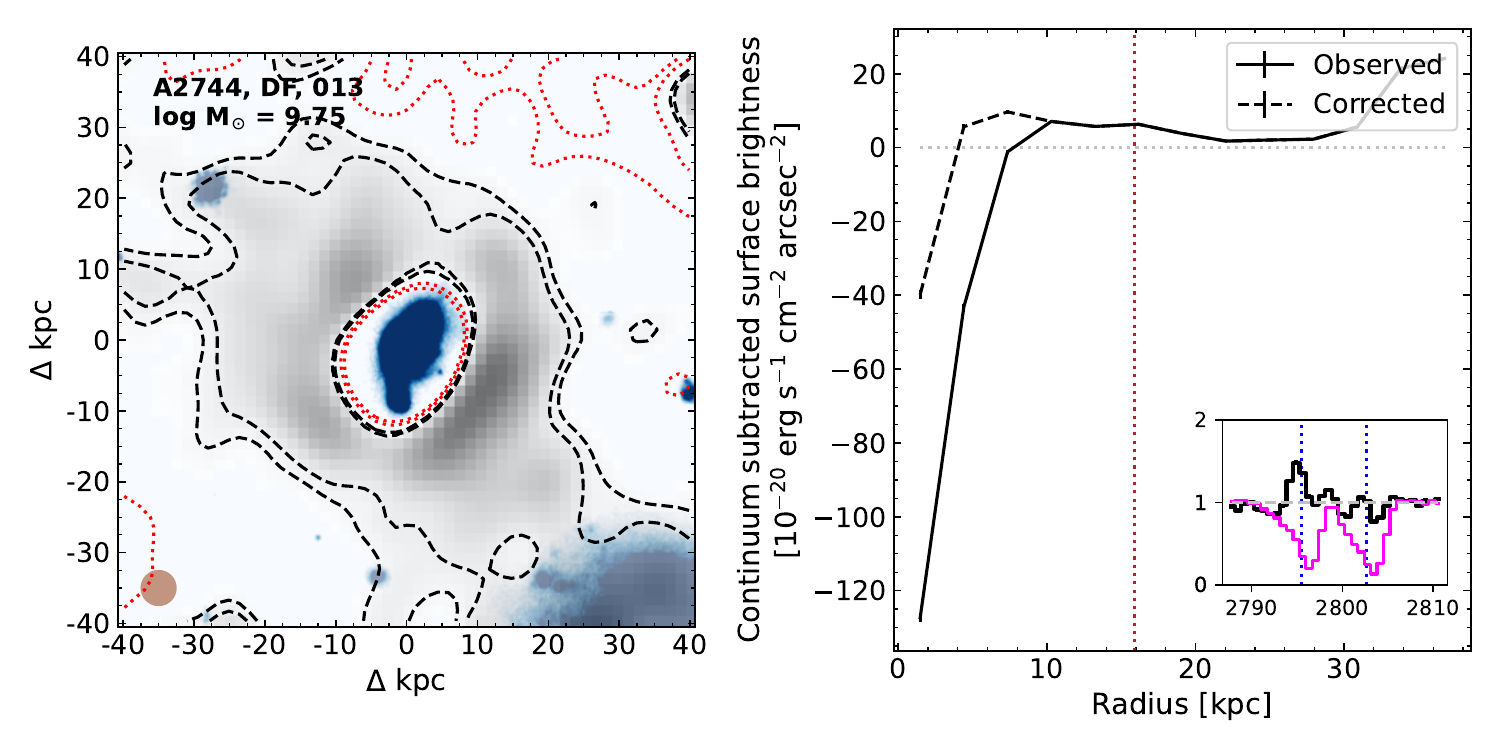}
    \end{minipage}

    \begin{minipage}{.49\textwidth}
        \centering
        \includegraphics[width=0.99\columnwidth]{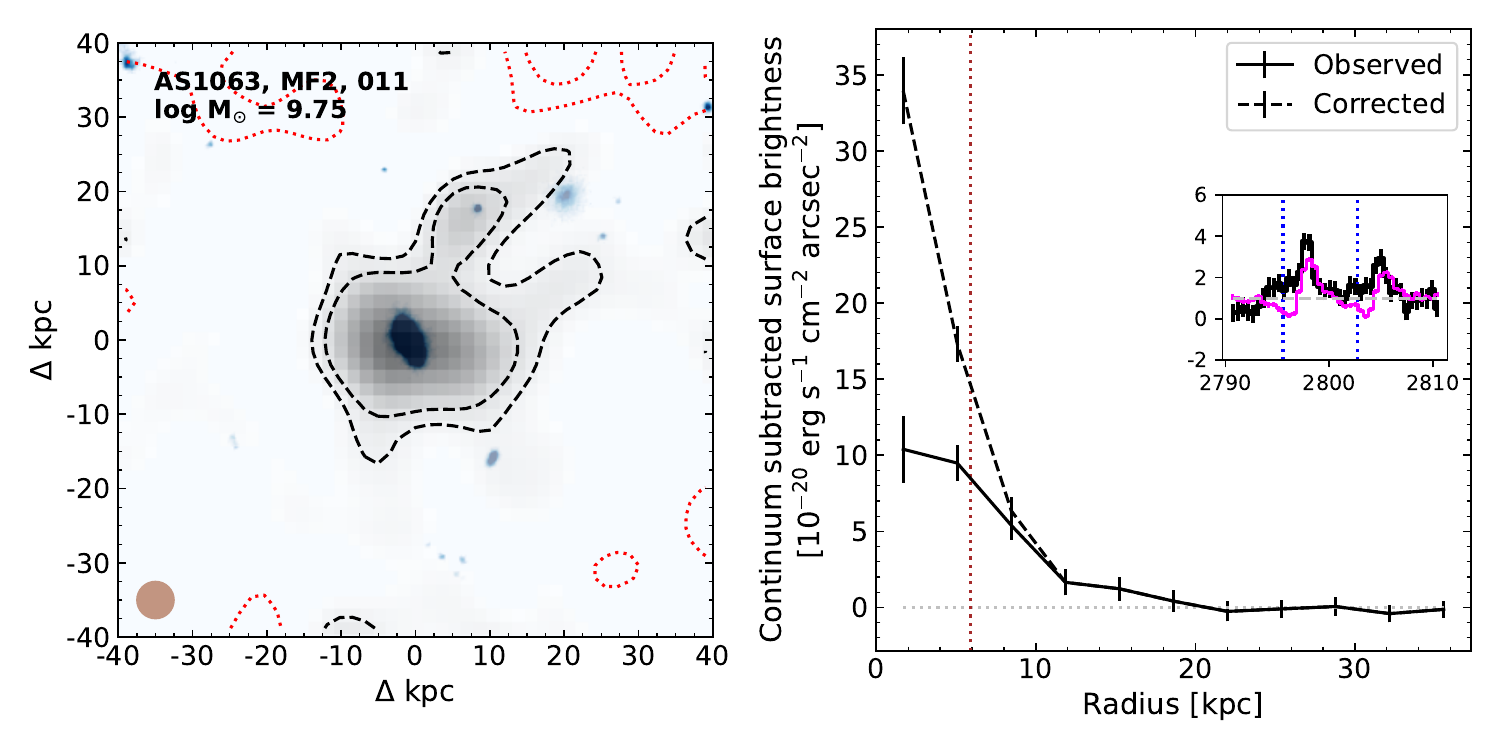}
    \end{minipage}%
    \begin{minipage}{0.49\textwidth}
        \centering
        \includegraphics[width=0.99\columnwidth]{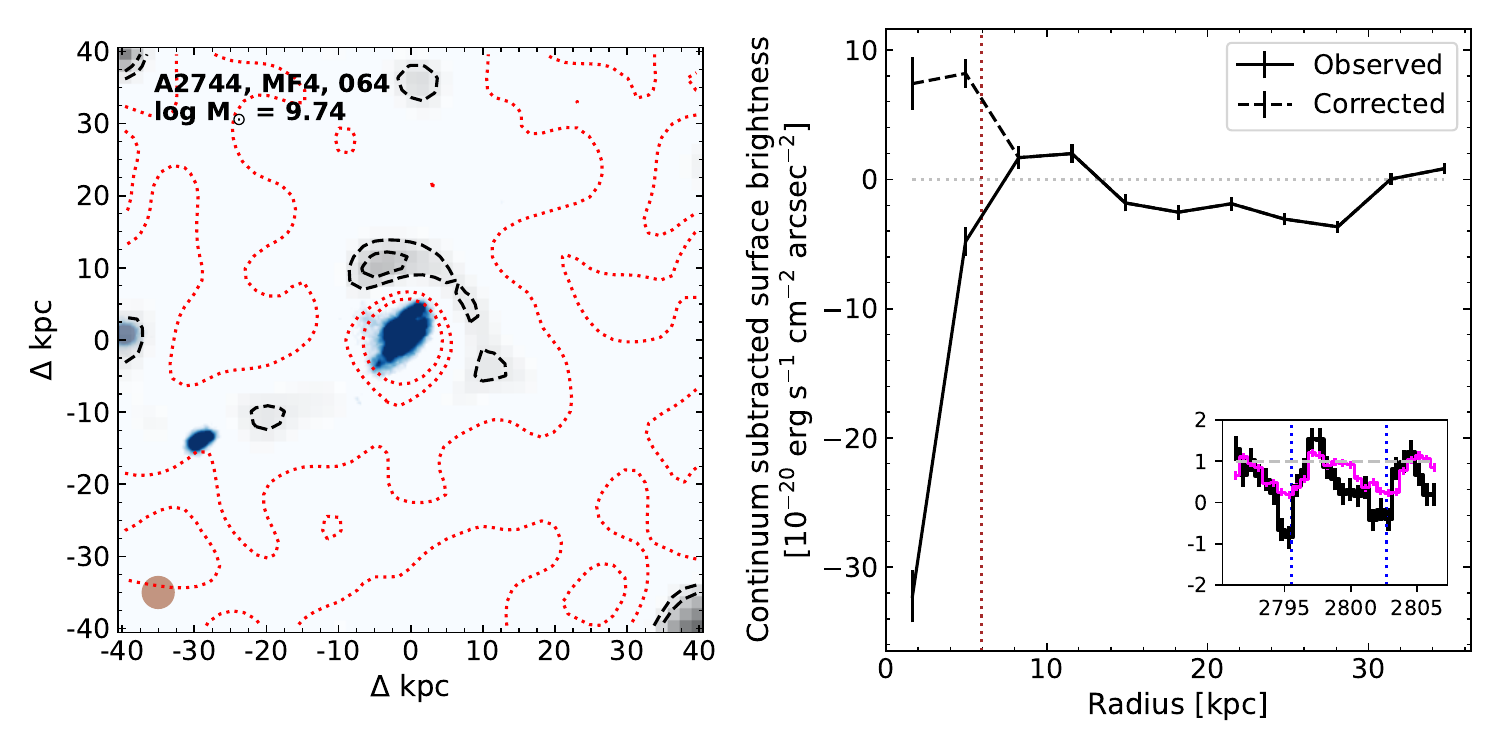}
    \end{minipage}

    \begin{minipage}{.49\textwidth}
        \centering
        \includegraphics[width=0.99\columnwidth]{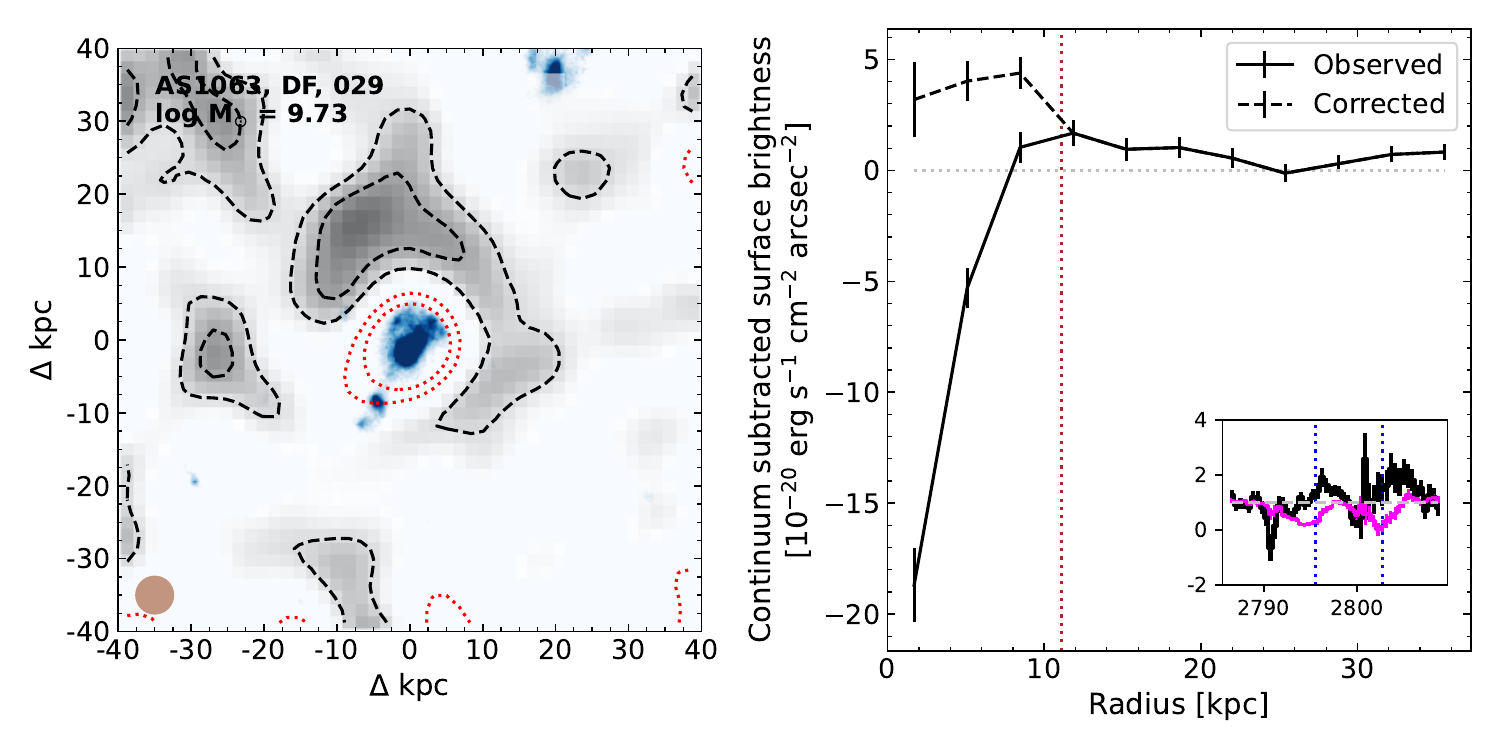}
    \end{minipage}%
    \begin{minipage}{0.49\textwidth}
        \centering
        \includegraphics[width=0.99\columnwidth]{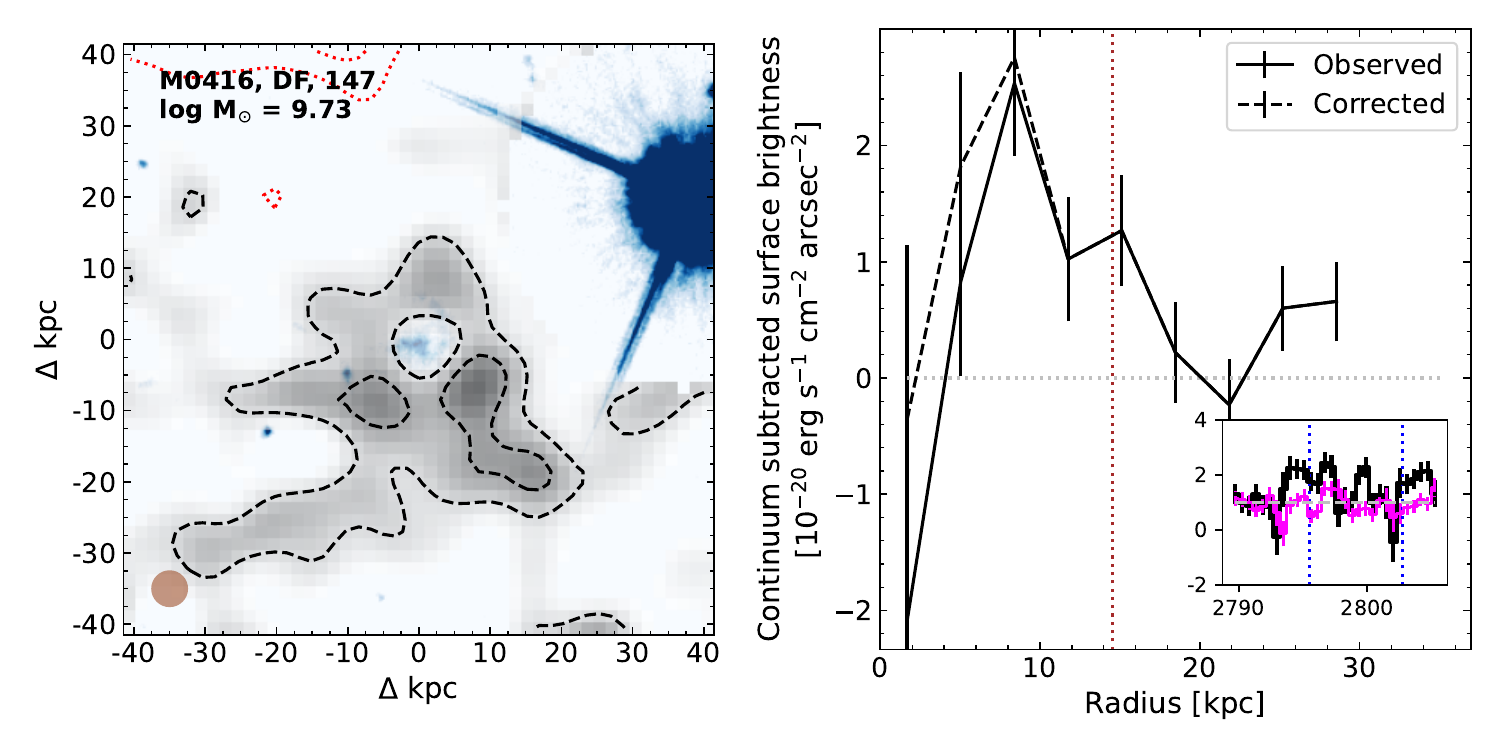}
    \end{minipage}

    \caption{Compilation of continuum-subtracted \ion{Mg}{II} pseudo-narrowband images (gray color scale) and HST F814W images (blue color scale, probing the stellar light) for our final sample of galaxies, that exhibit a P-Cygni profile in \ion{Mg}{II}, as well as a significant detection of extended \ion{Mg}{II} emission. Galaxies are sorted by stellar mass in descending order. The continuum-subtracted \ion{Mg}{II} pseudo-narrowband images are computed by collapsing the MUSE data across the wavelength axis, for wavelengths that enclose the full P-Cygni profile of the \ion{Mg}{II} doublet, for a FoV of 80 x 80 kpc$^{2}$. The brown circle in the bottom left part of the panel shows the size of the MUSE PSF. The black contours show the $1-\sigma$ and $2-\sigma$  detection levels of \ion{Mg}{II} net emission in the pseudo-narrowband images. The red contours show the $2-\sigma$ and $4-\sigma$  levels of \ion{Mg}{II} net absorption in the pseudo-narrowband images. For visualization purposes, the continuum-subtracted \ion{Mg}{II} pseudo-narrowband images have been smoothed using a Gaussian kernel with a full width at half maximum of 1 arcsec. For each galaxy, we also show the radial profile of the \ion{Mg}{II} emission as observed in the data (solid) and after correcting by self-absorption using our best-fitting model (dashed, see Sec.~\ref{sec:reconstruction}). The vertical brown dotted line shows the half-light radius measured for the self-absorption corrected radial profiles. The inset shows the \ion{Mg}{II} spectrum extracted from the MUSE data cube for each galaxy, across the full modeled region (black) and a small central aperture of radius $0\farcs4$.}
    \label{fig:sample_1}
    
\end{figure*}

\addtocounter{figure}{-1}
\begin{figure*}[!htb]
    \centering

    \begin{minipage}{.49\textwidth}
        \centering
        \includegraphics[width=0.99\columnwidth]{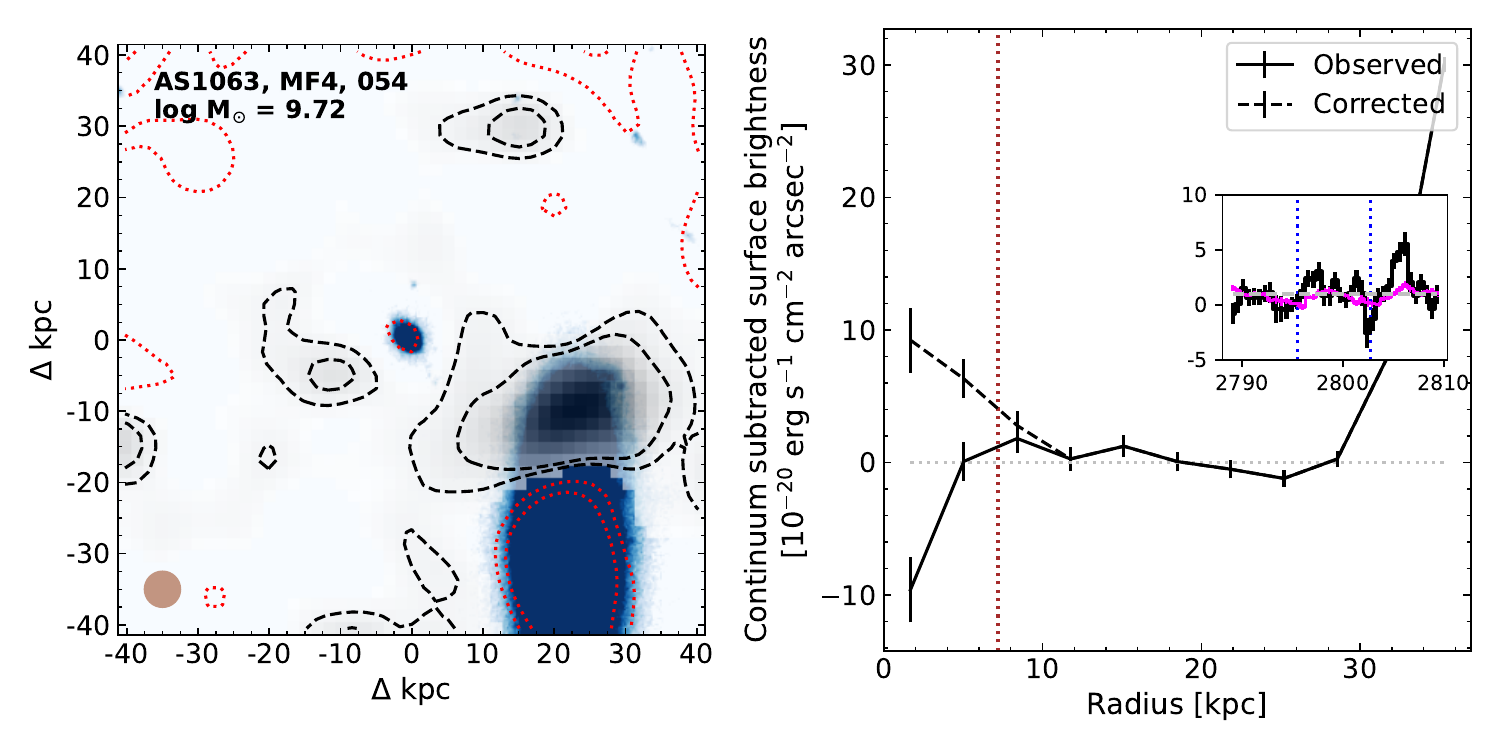}
    \end{minipage}%
    \begin{minipage}{0.49\textwidth}
        \centering
        \includegraphics[width=0.99\columnwidth]{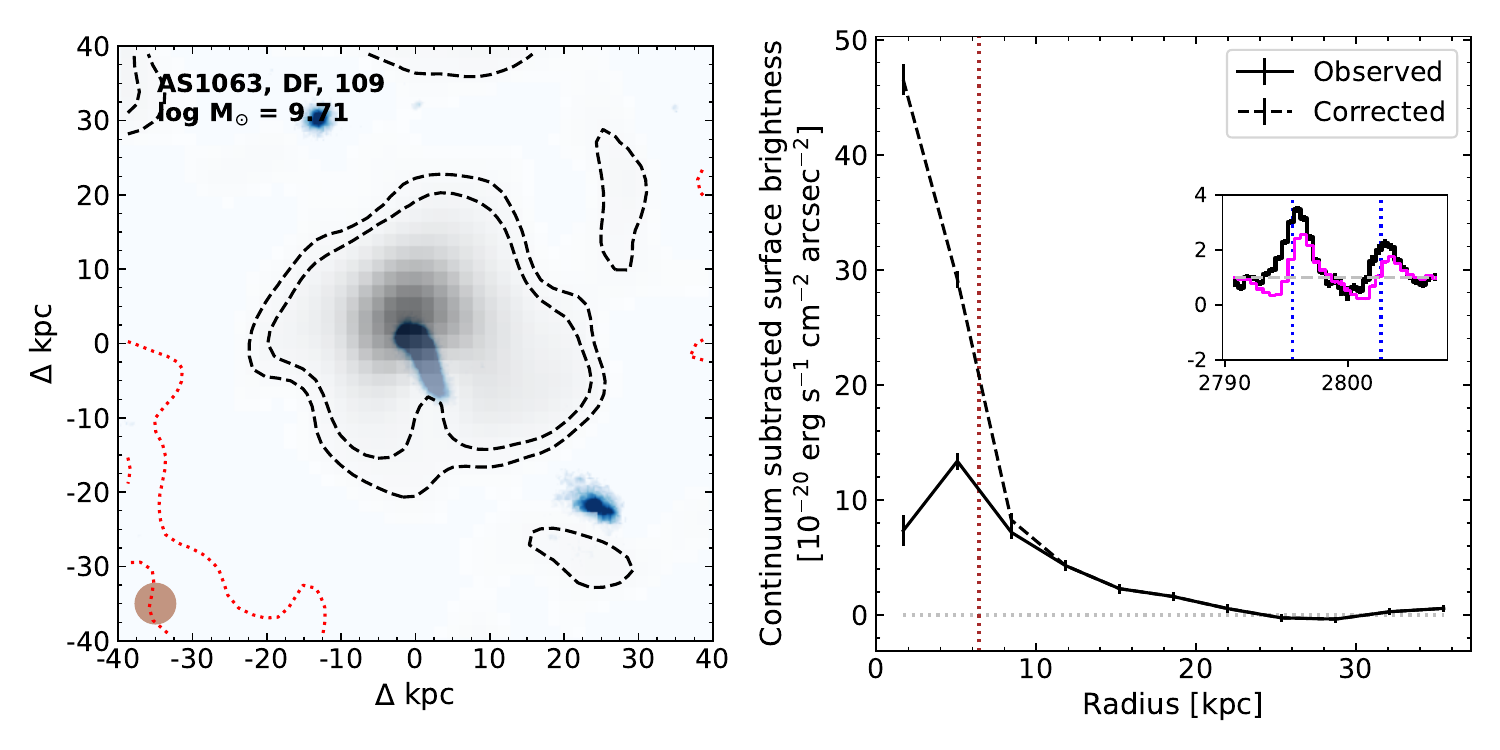}
    \end{minipage}
    
    \begin{minipage}{.49\textwidth}
        \centering
        \includegraphics[width=0.99\columnwidth]{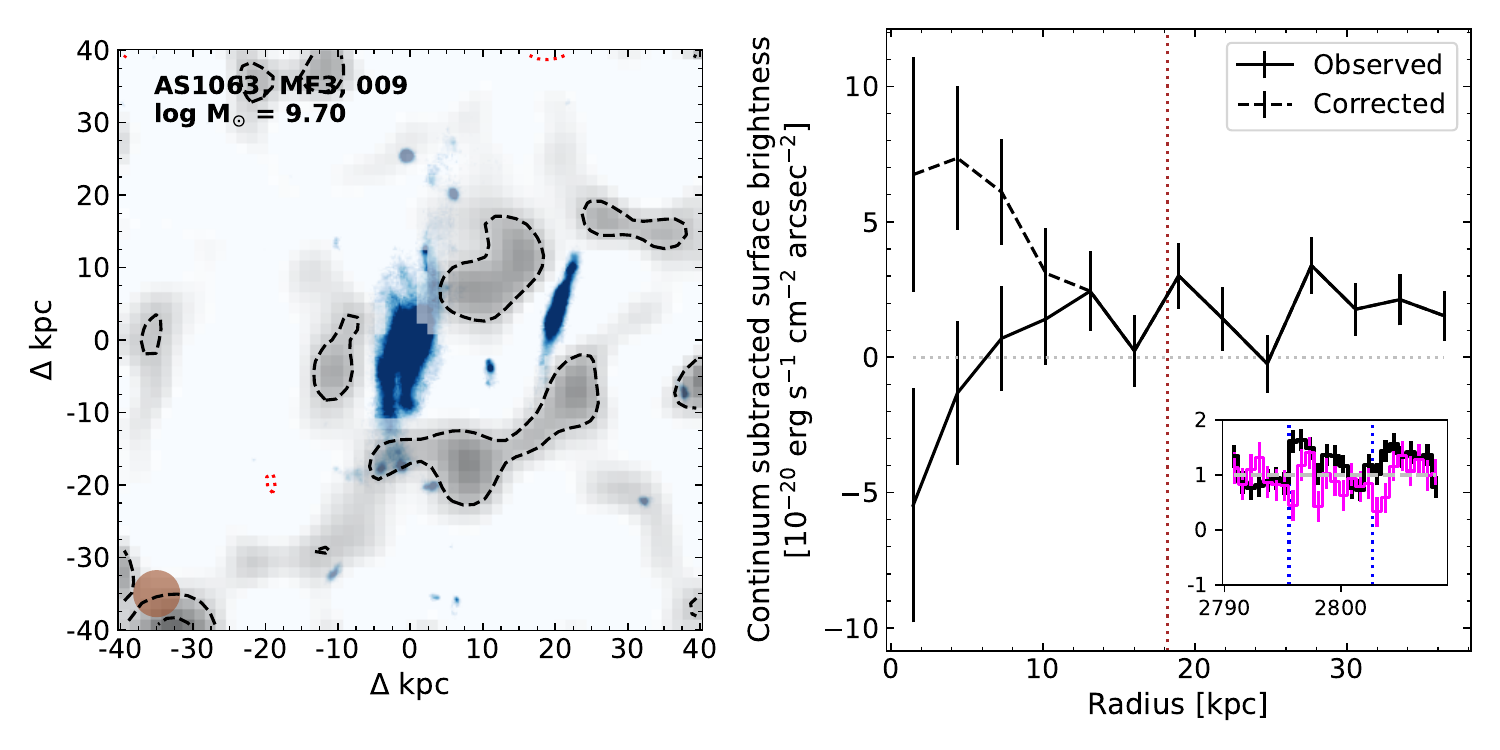}
    \end{minipage}%
    \begin{minipage}{0.49\textwidth}
        \centering
        \includegraphics[width=0.99\columnwidth]{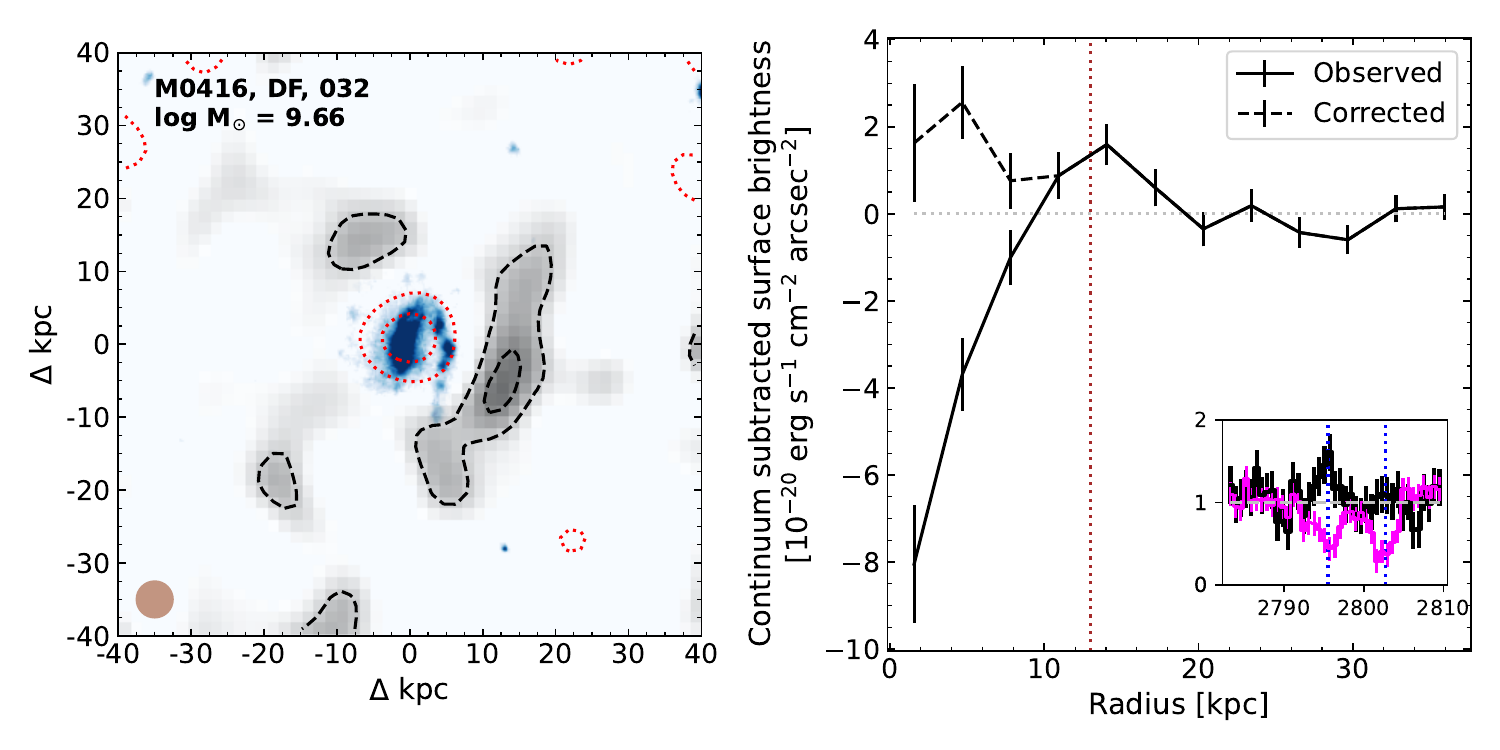}
    \end{minipage}
    
    \begin{minipage}{.49\textwidth}
        \centering
        \includegraphics[width=0.99\columnwidth]{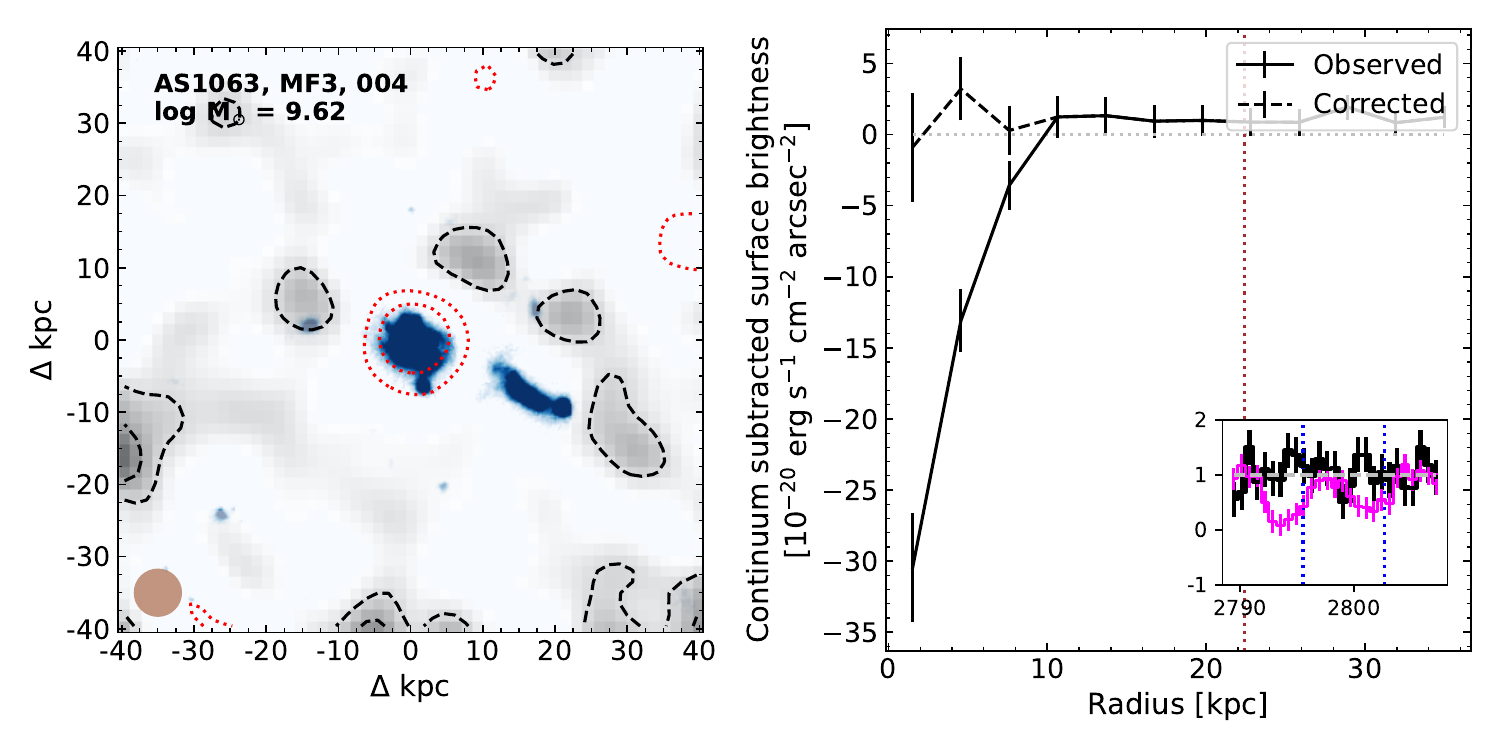}
    \end{minipage}%
    \begin{minipage}{0.49\textwidth}
        \centering
        \includegraphics[width=0.99\columnwidth]{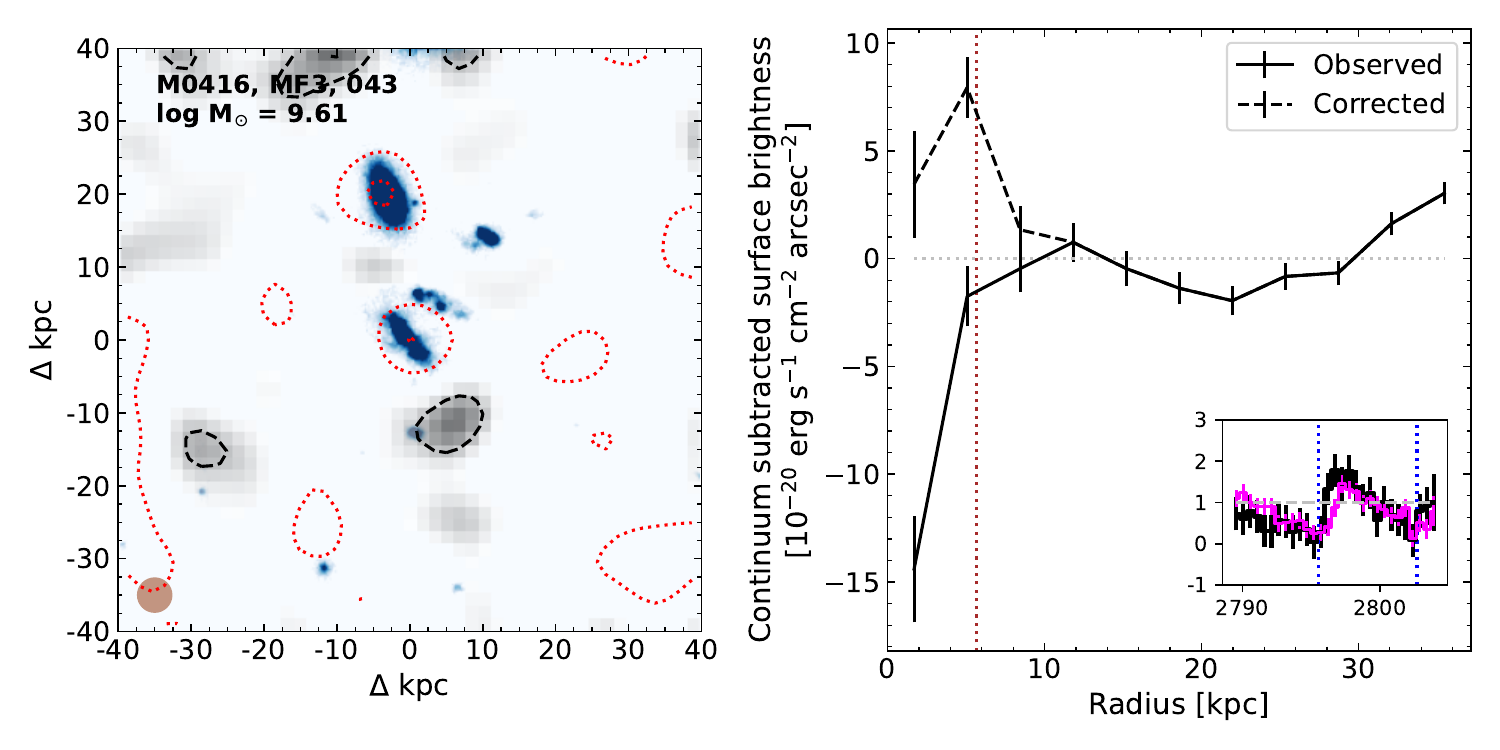}
    \end{minipage}

    \begin{minipage}{.49\textwidth}
        \centering
        \includegraphics[width=0.99\columnwidth]{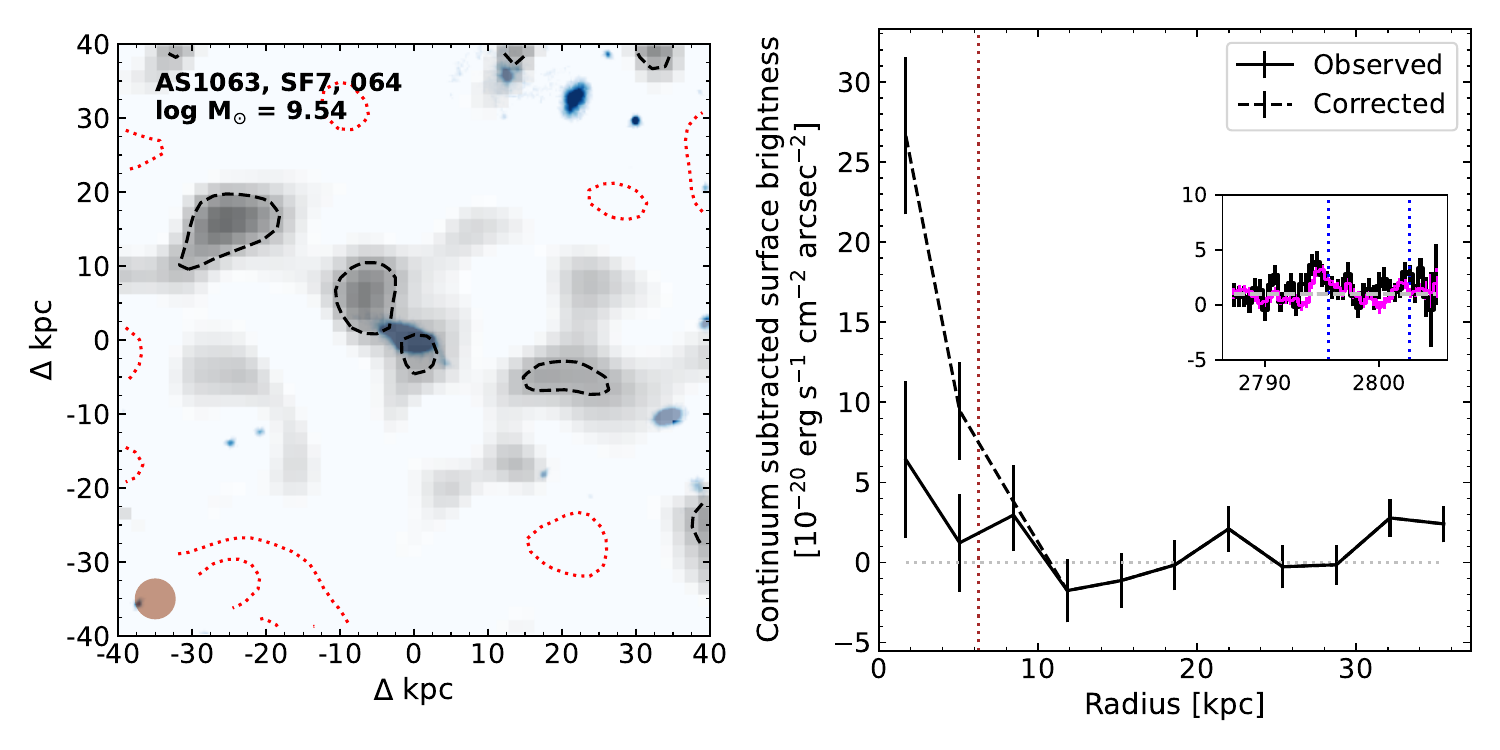}
    \end{minipage}%
    \begin{minipage}{0.49\textwidth}
        \centering
        \includegraphics[width=0.99\columnwidth]{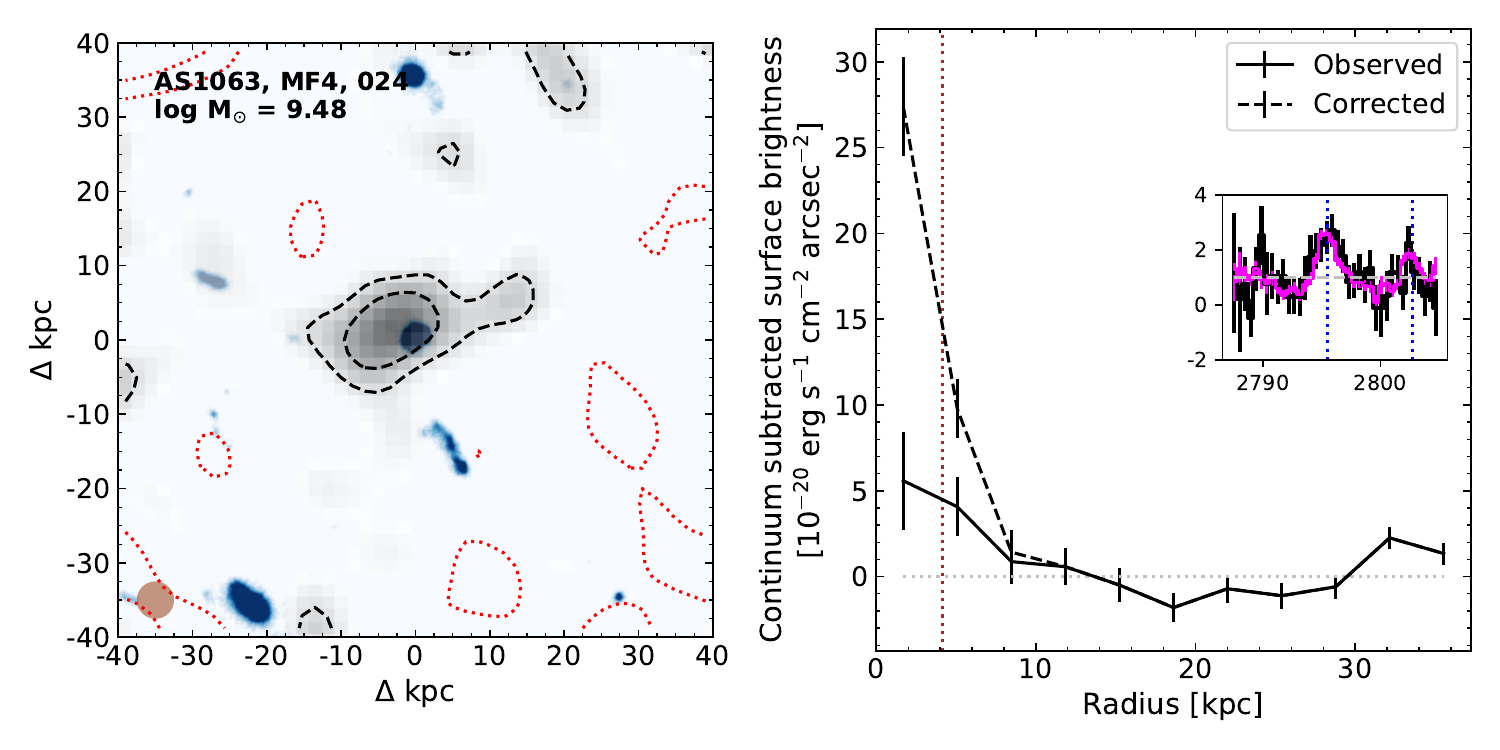}
    \end{minipage}

    \begin{minipage}{0.49\textwidth}
        \centering
        \includegraphics[width=0.99\columnwidth]{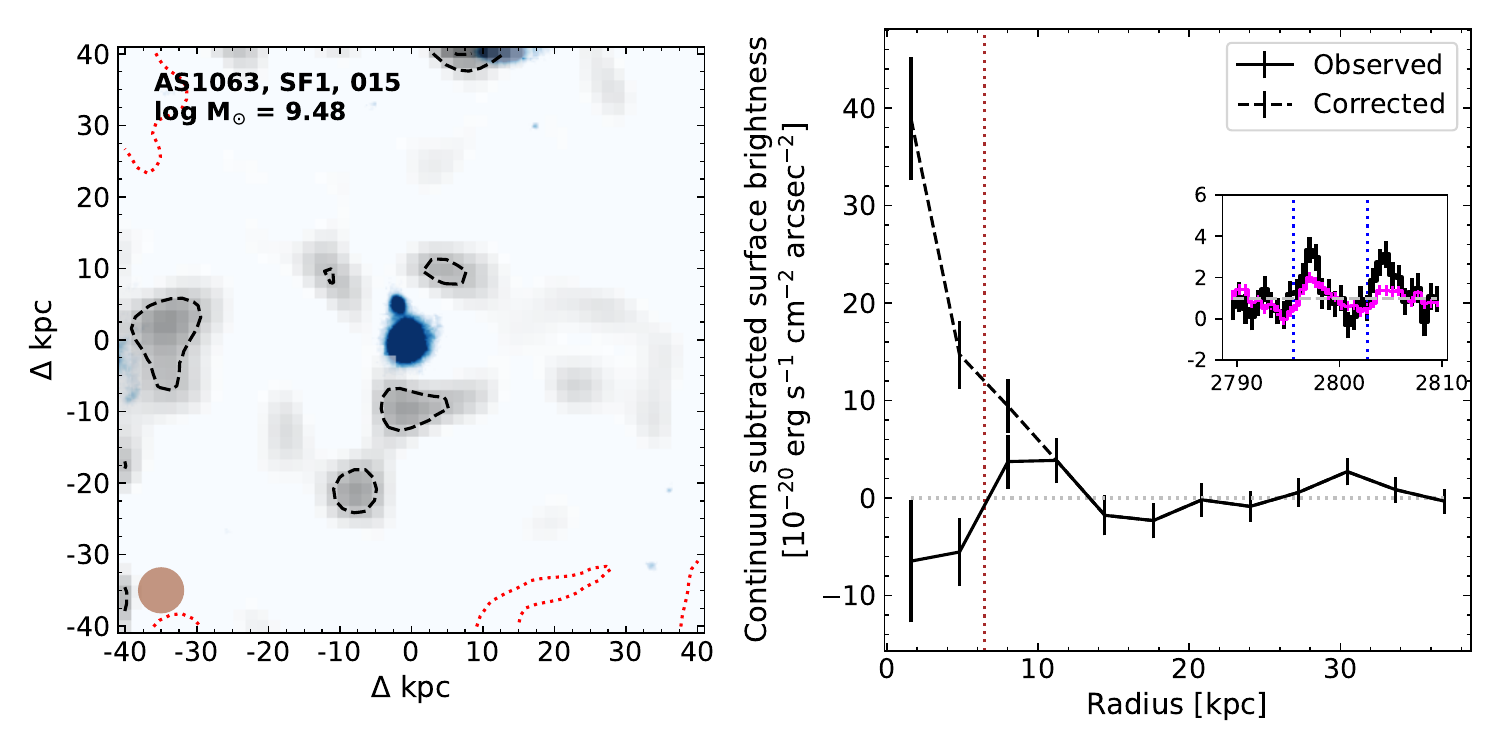}
    \end{minipage}
    \begin{minipage}{0.49\textwidth}
        \centering
        \includegraphics[width=0.99\columnwidth]{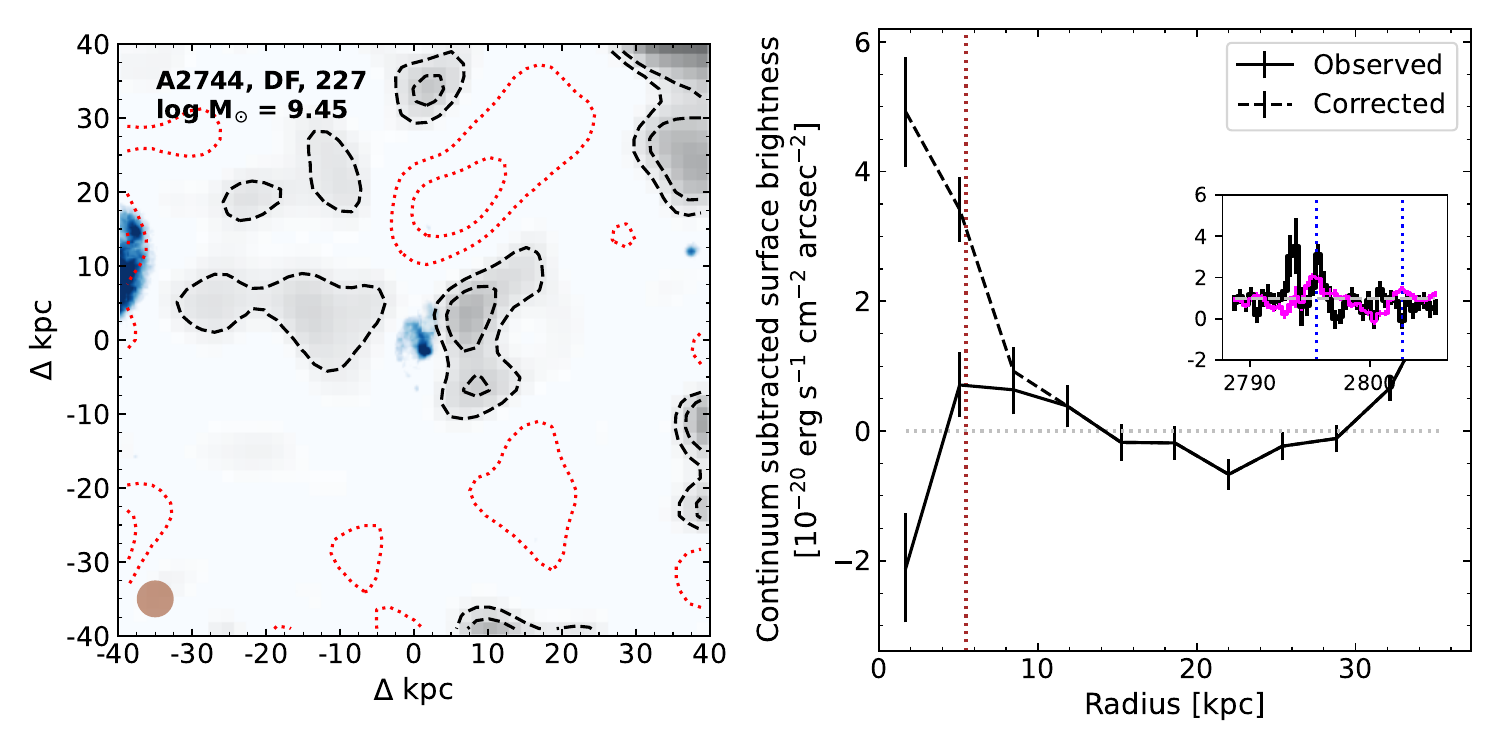}
    \end{minipage}%
    
    \caption{(continued)}
\end{figure*}

\addtocounter{figure}{-1}
\begin{figure*}[!htb]
    \centering

    \begin{minipage}{0.49\textwidth}
        \centering
        \includegraphics[width=0.99\columnwidth]{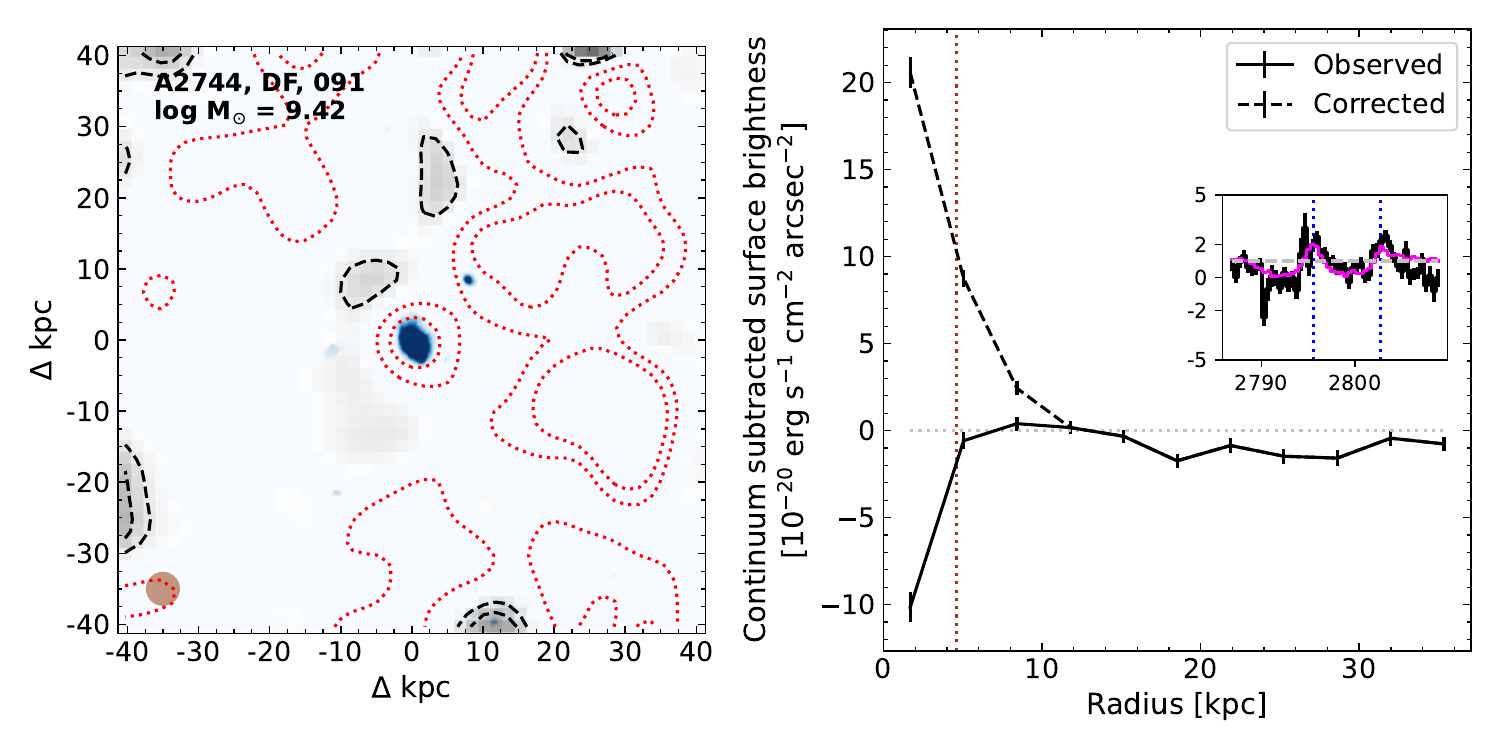}
    \end{minipage}
    \begin{minipage}{.49\textwidth}
        \centering
        \includegraphics[width=0.99\columnwidth]{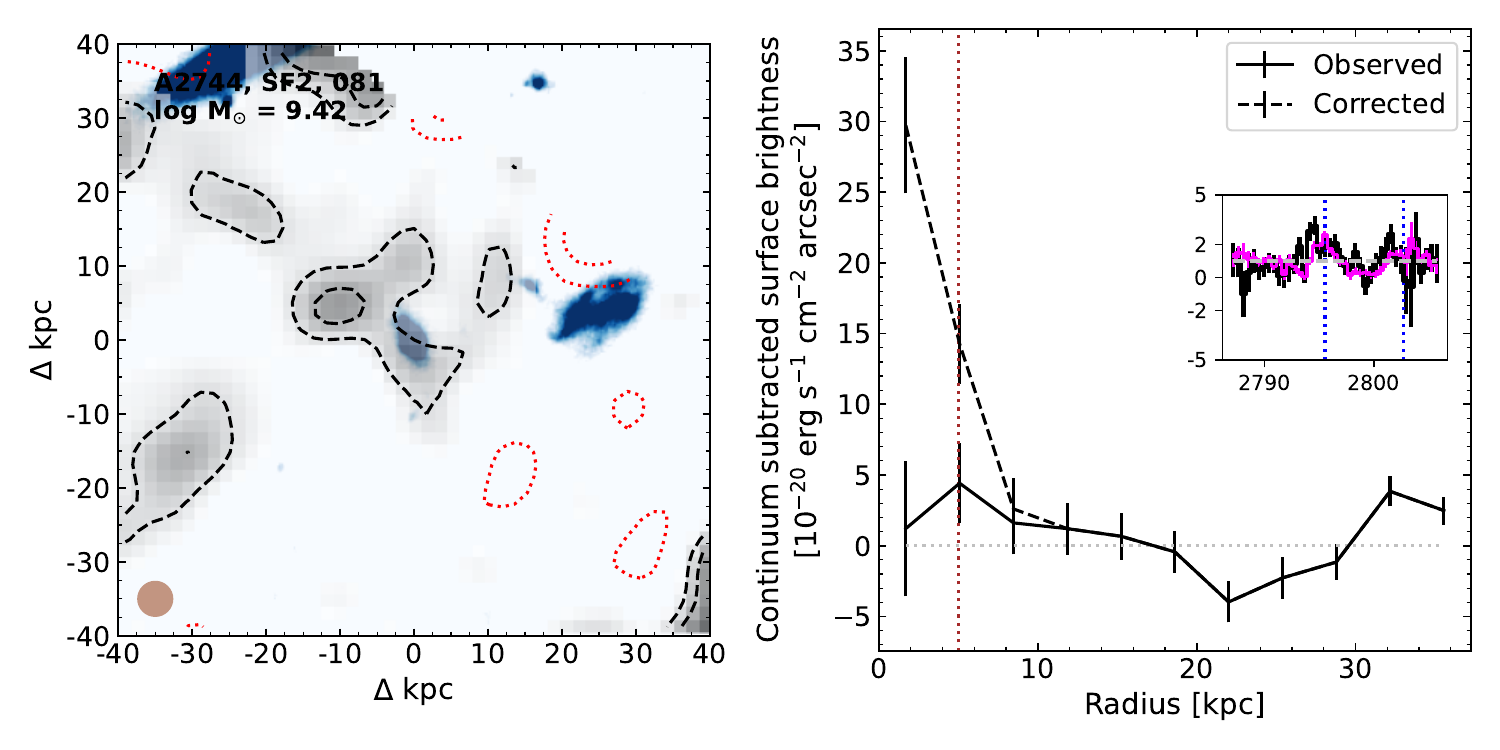}
    \end{minipage}
    
    \begin{minipage}{0.49\textwidth}
        \centering
        \includegraphics[width=0.99\columnwidth]{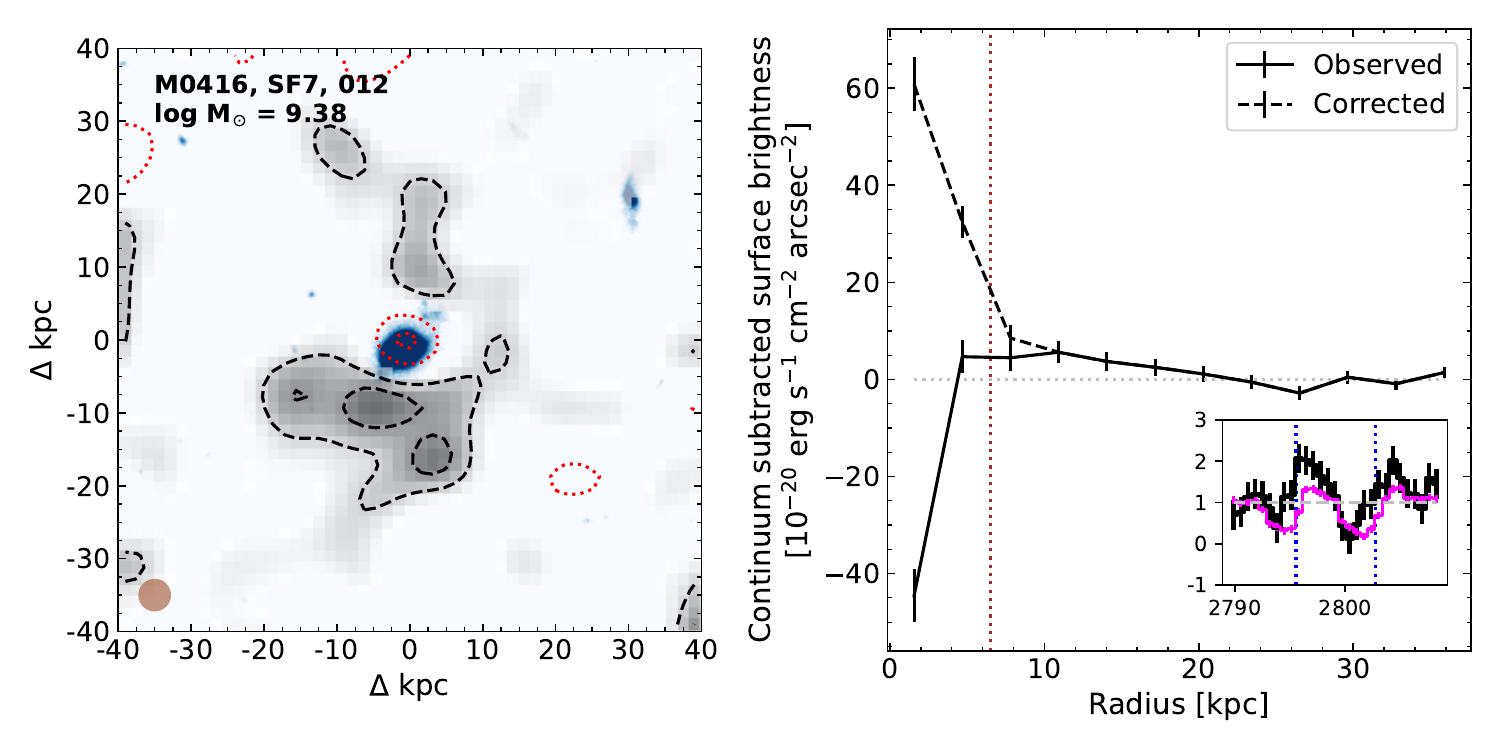}
    \end{minipage} 
    \begin{minipage}{.49\textwidth}
        \centering
        \includegraphics[width=0.99\columnwidth]{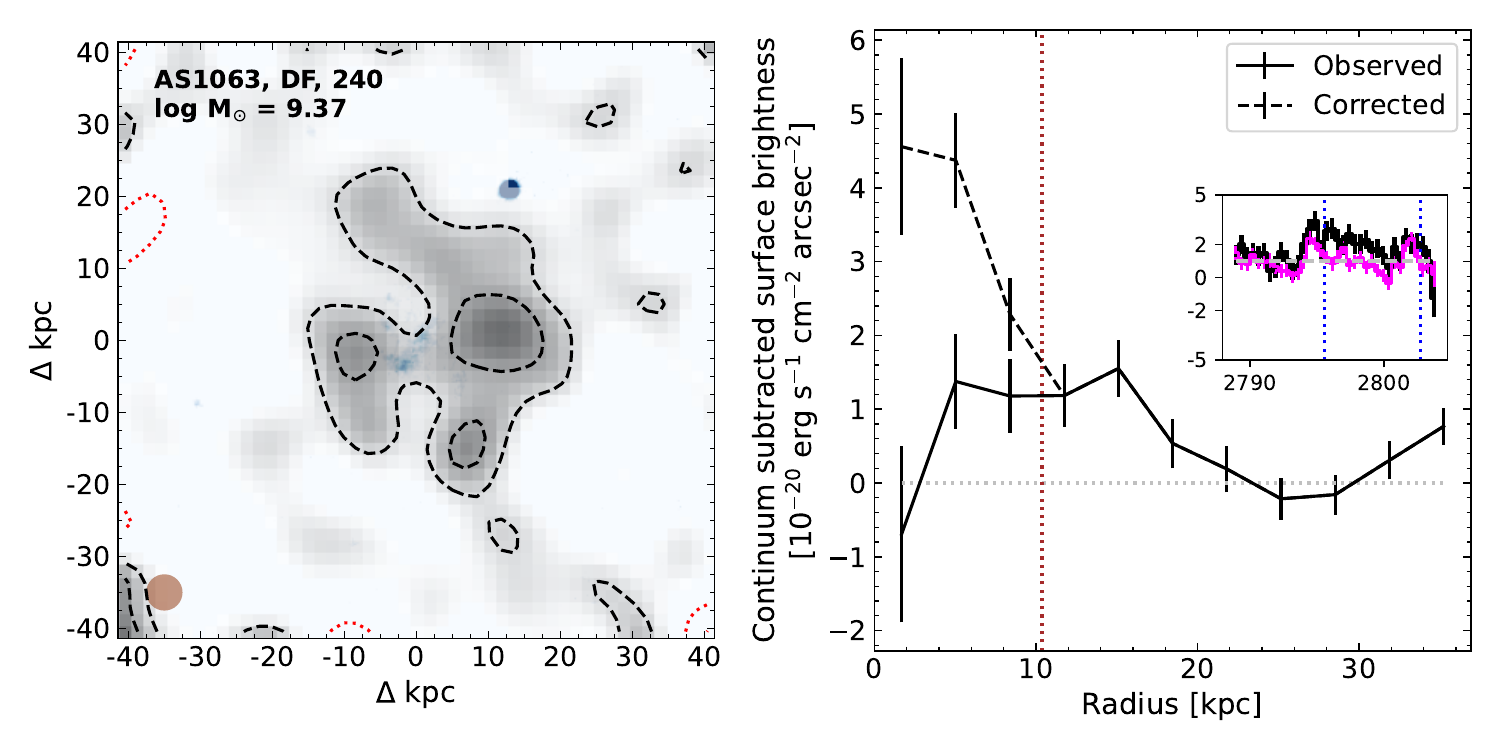}
    \end{minipage}%

    \begin{minipage}{0.49\textwidth}
        \centering
        \includegraphics[width=0.99\columnwidth]{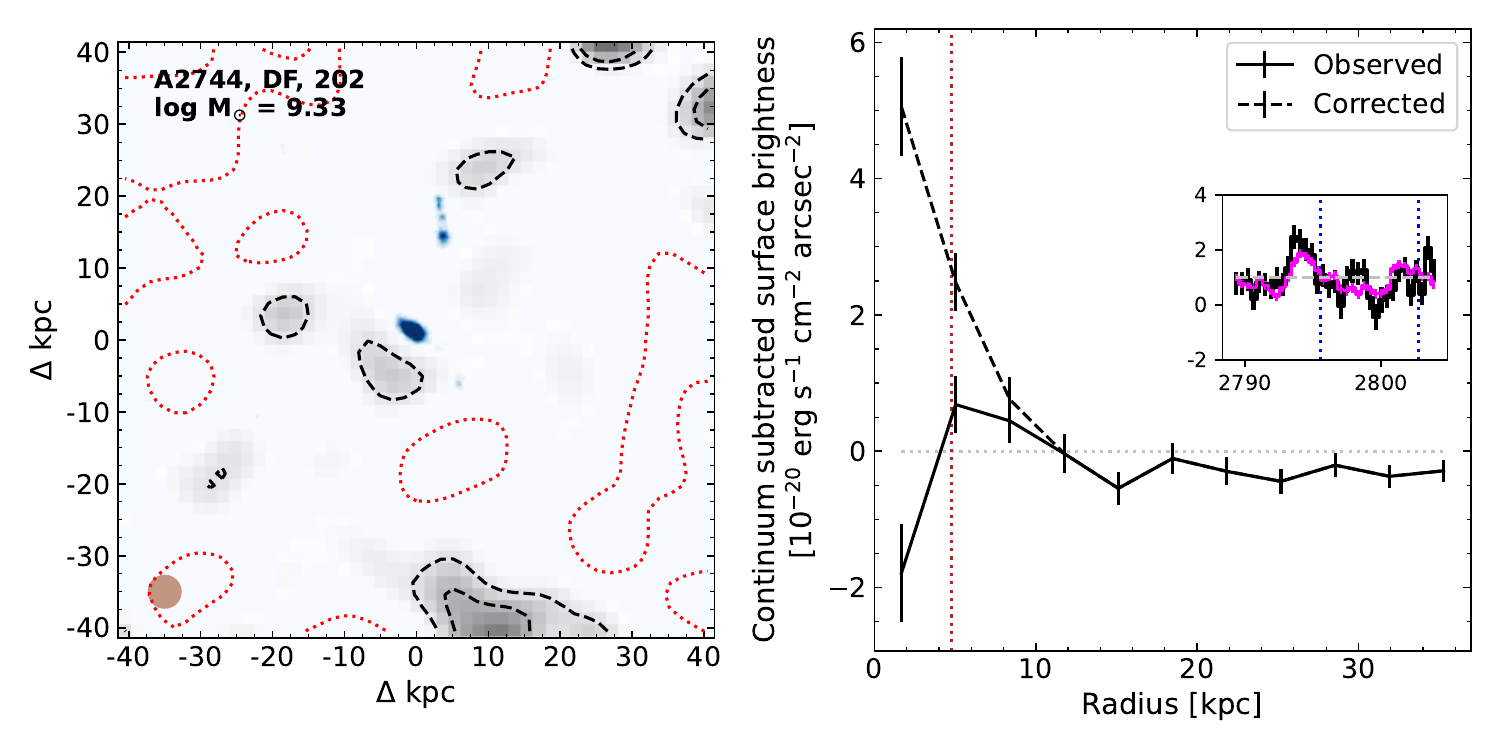}
    \end{minipage}
    \begin{minipage}{.49\textwidth}
        \centering
        \includegraphics[width=0.99\columnwidth]{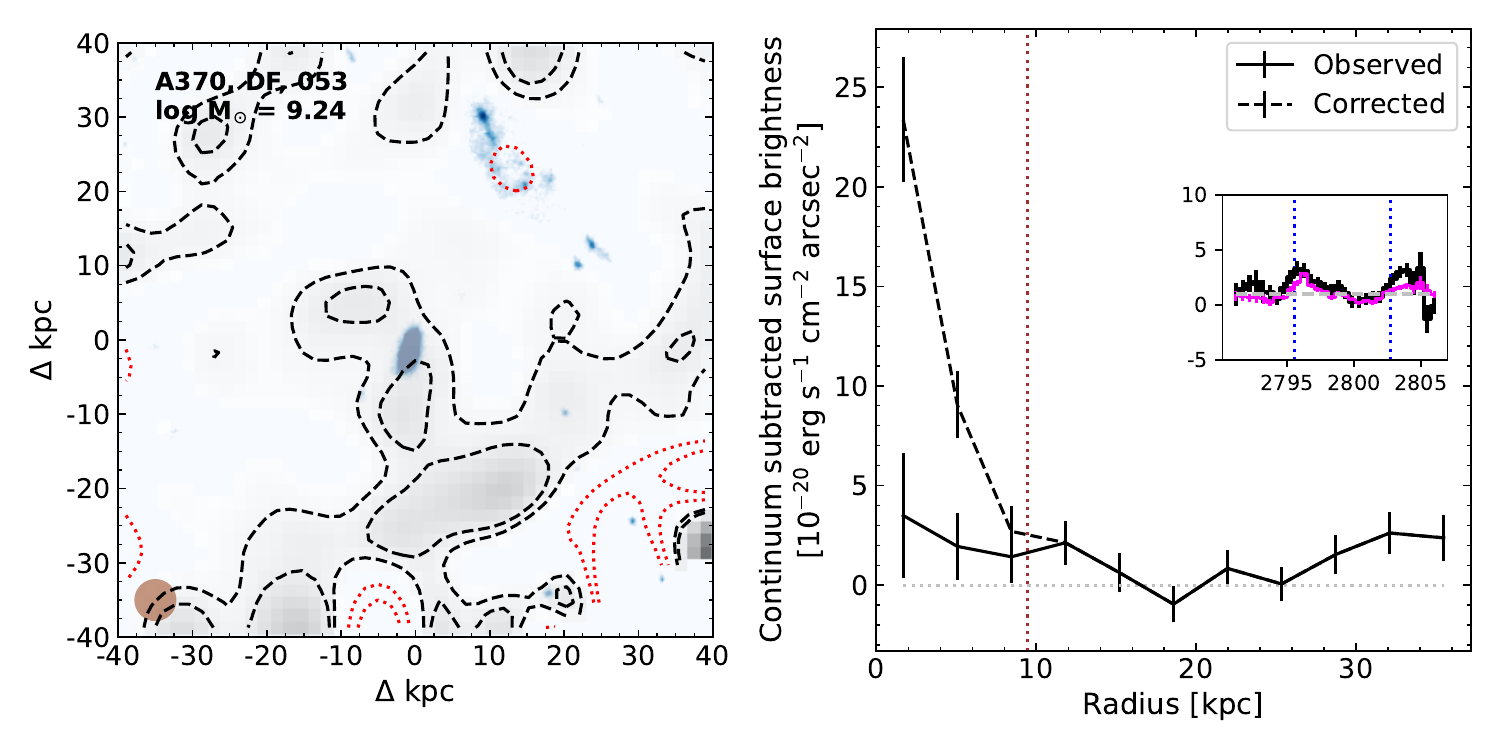}
    \end{minipage}%

    \begin{minipage}{0.49\textwidth}
        \centering
        \includegraphics[width=0.99\columnwidth]{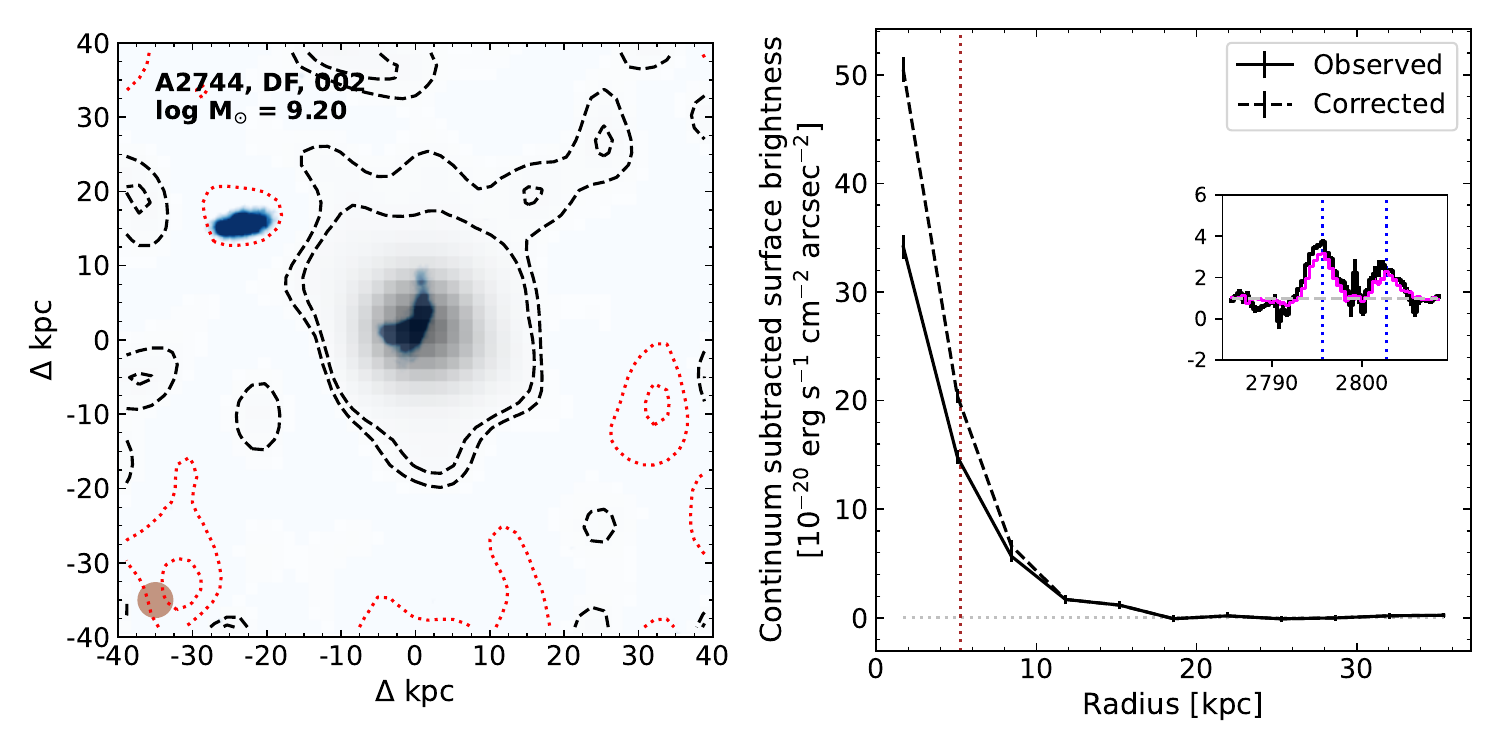}
    \end{minipage}
    \begin{minipage}{.49\textwidth}
        \centering
        \includegraphics[width=0.99\columnwidth]{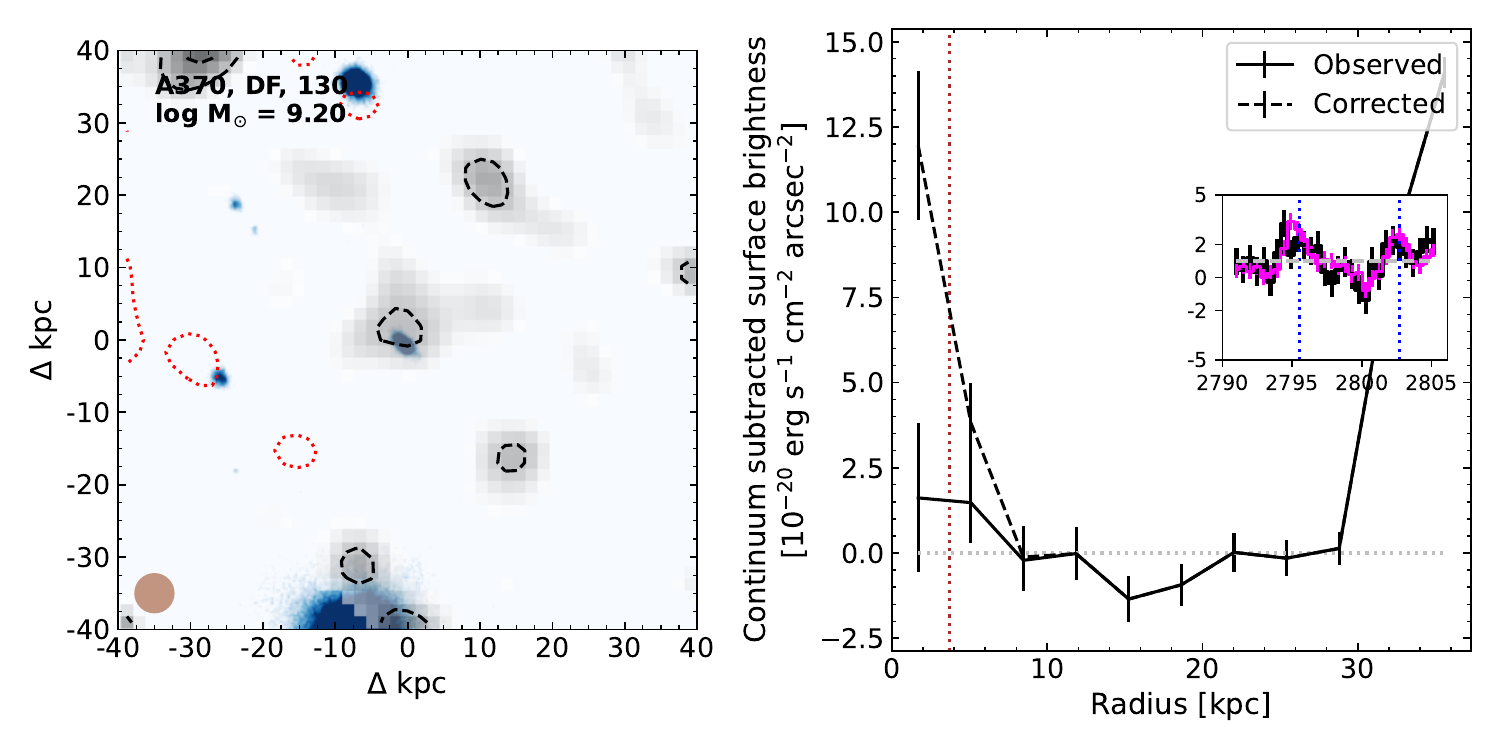}
    \end{minipage}%

    \begin{minipage}{0.49\textwidth}
        \centering
        \includegraphics[width=0.99\columnwidth]{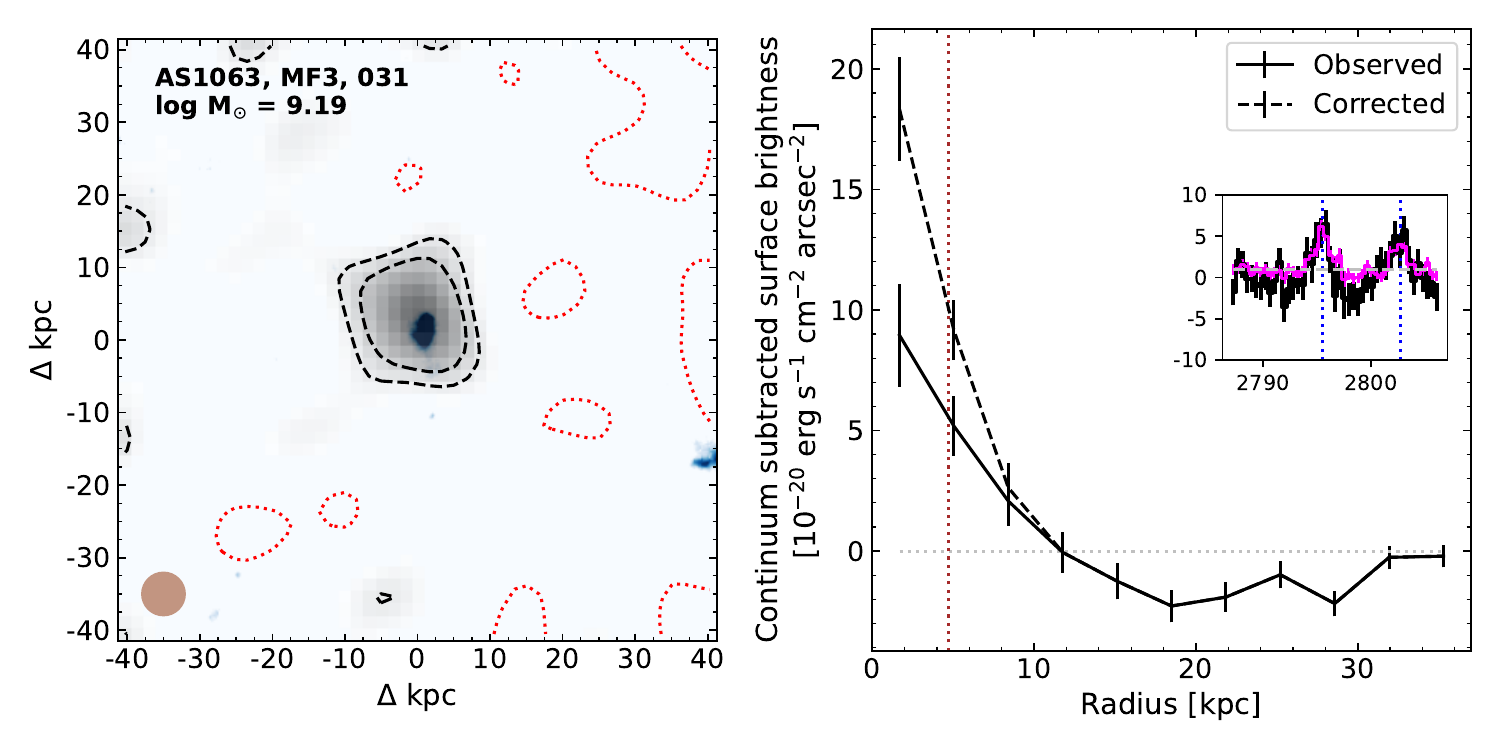}
    \end{minipage}
    \begin{minipage}{.49\textwidth}
        \centering
        \includegraphics[width=0.99\columnwidth]{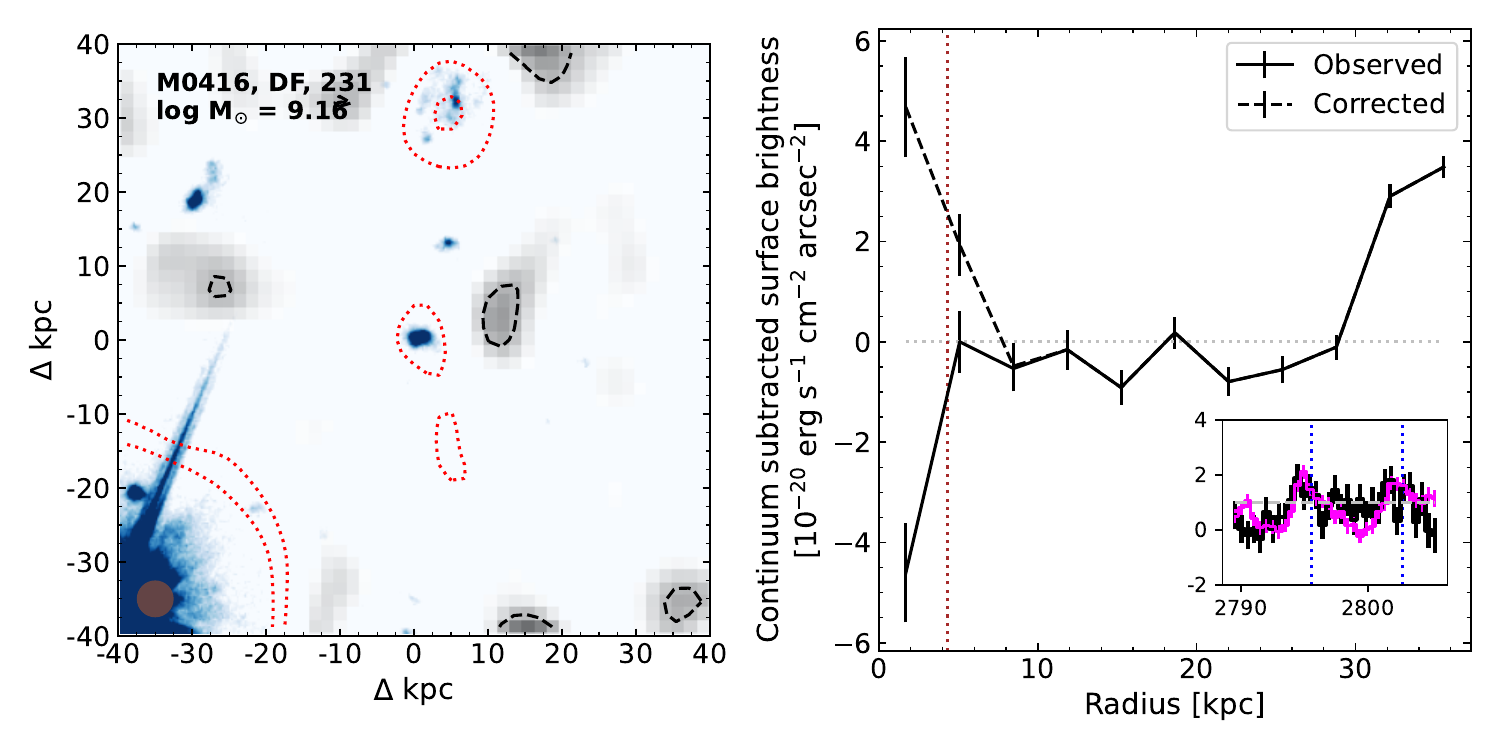}
    \end{minipage}%

    \caption{(continued)}
\end{figure*}

\addtocounter{figure}{-1}
\begin{figure*}[!htb]
    \centering    

    \begin{minipage}{0.49\textwidth}
        \centering
        \includegraphics[width=0.99\columnwidth]{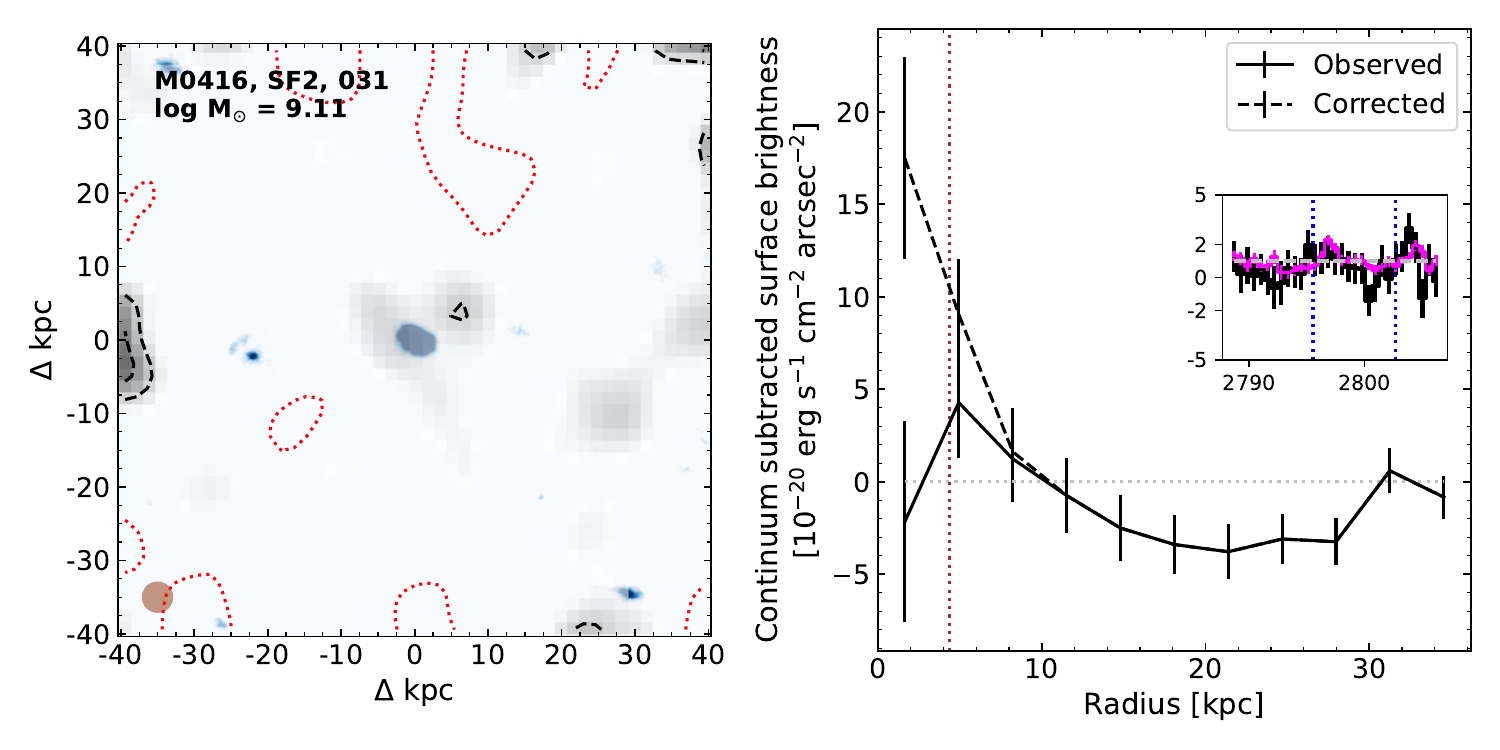}
    \end{minipage}    
    \begin{minipage}{.49\textwidth}
        \centering
        \includegraphics[width=0.99\columnwidth]{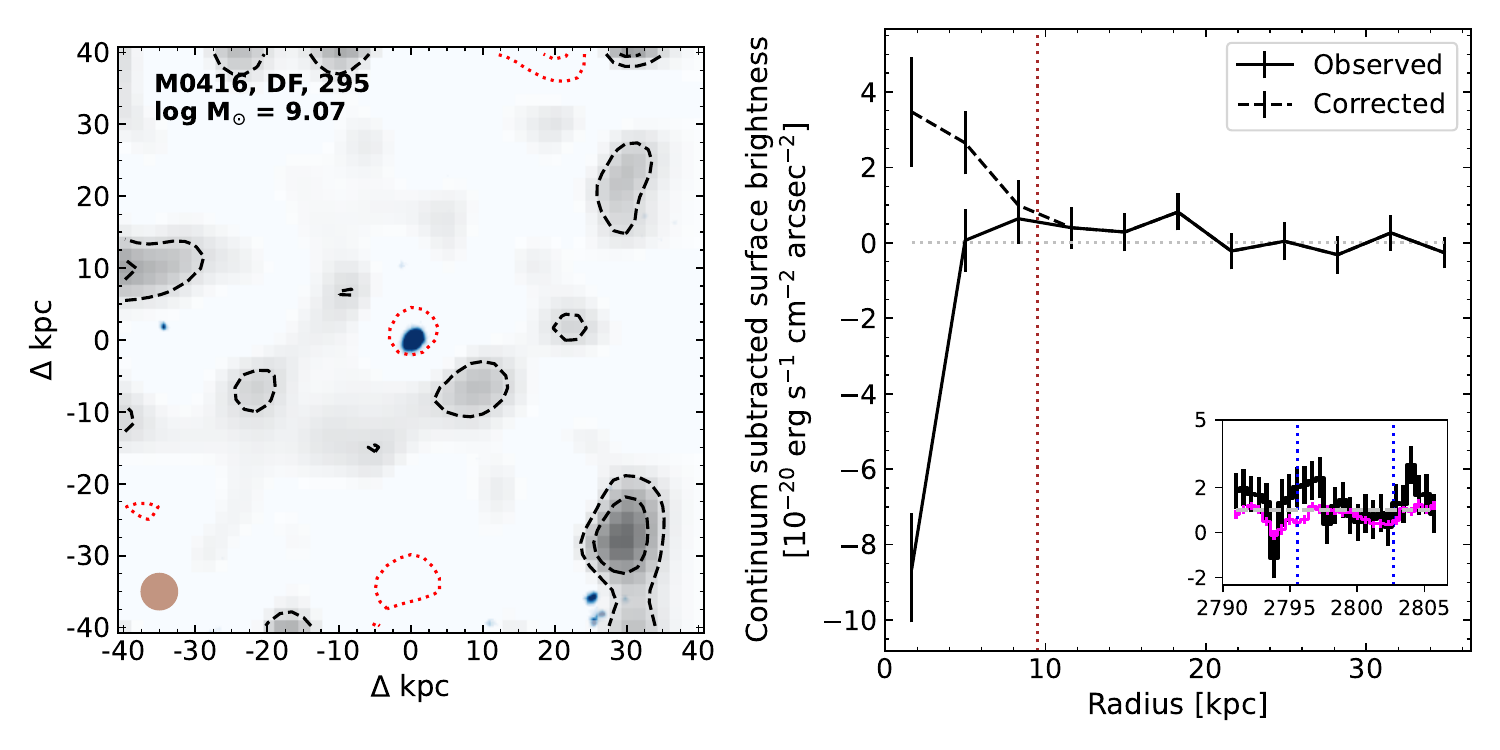}
    \end{minipage}%

    \begin{minipage}{0.49\textwidth}
        \centering
        \includegraphics[width=0.99\columnwidth]{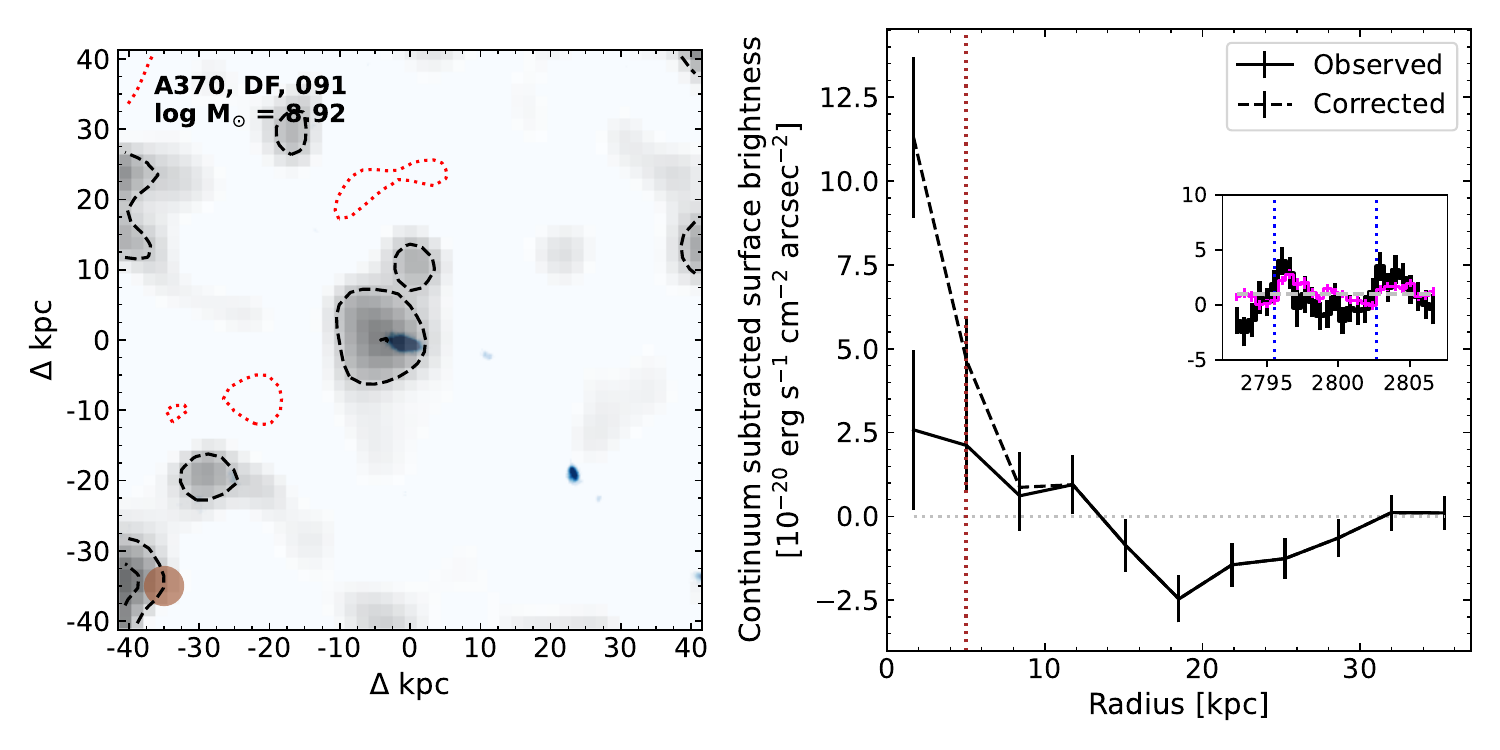}
    \end{minipage}
    \begin{minipage}{.49\textwidth}
        \centering
        \includegraphics[width=0.99\columnwidth]{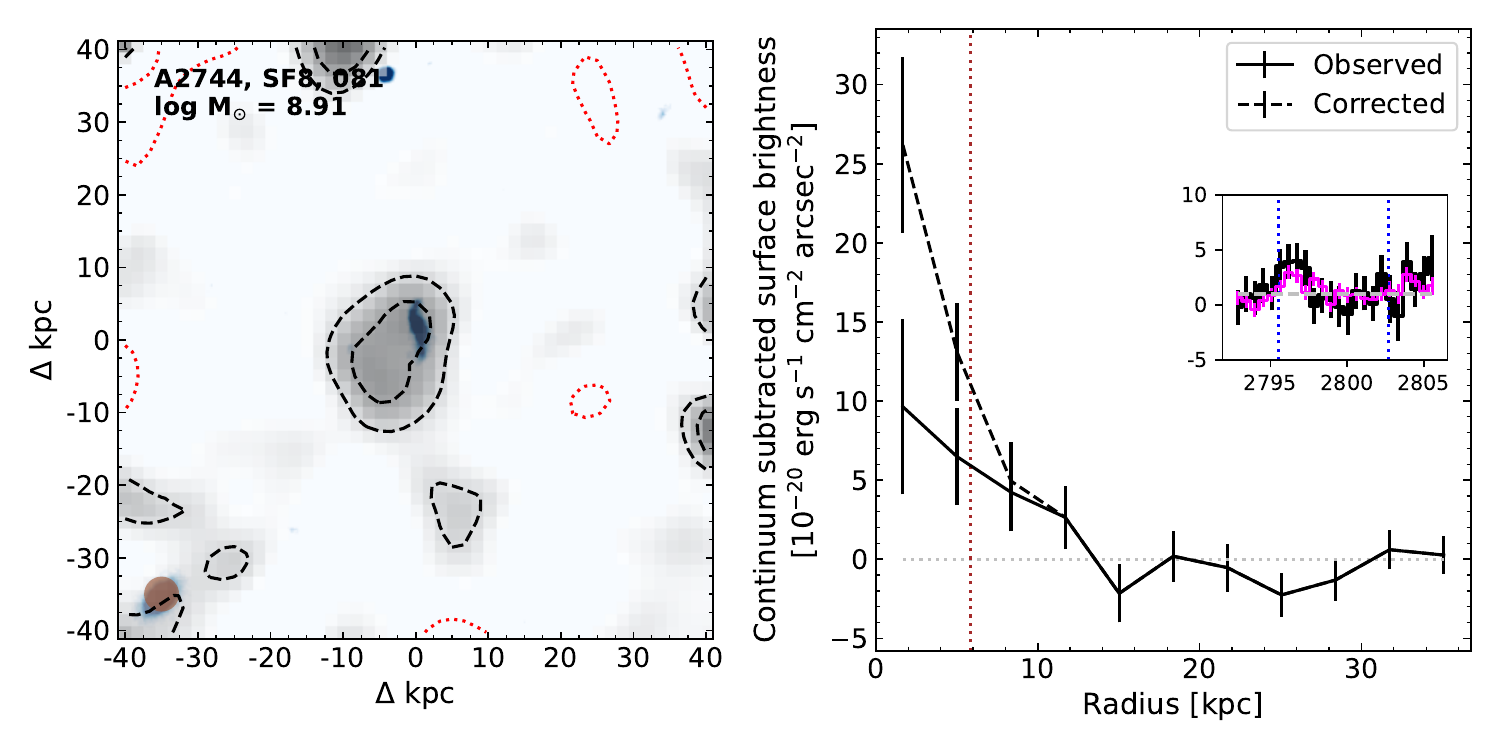}
    \end{minipage}%

    \begin{minipage}{0.49\textwidth}
        \centering
        \includegraphics[width=0.99\columnwidth]{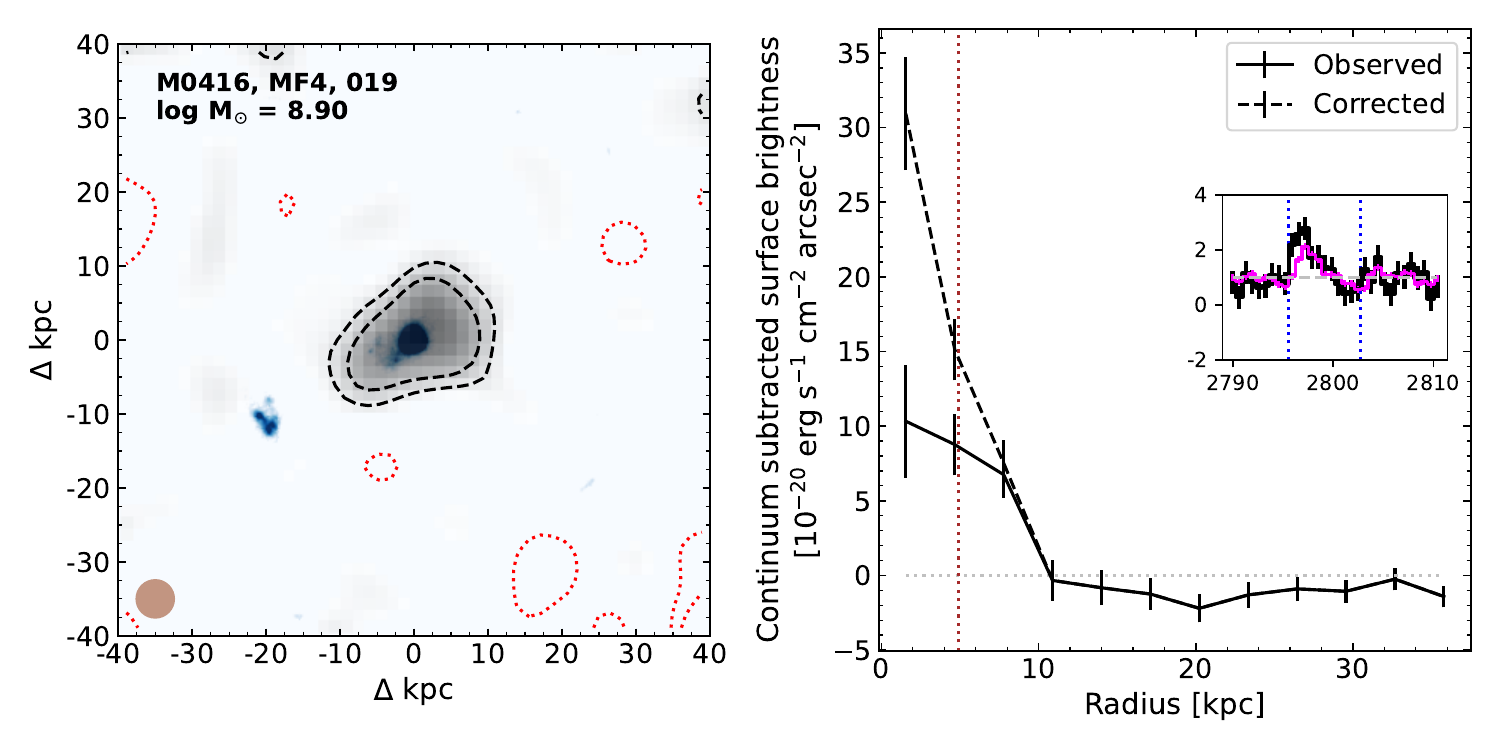}
    \end{minipage}
    \begin{minipage}{.49\textwidth}
        \centering
        \includegraphics[width=0.99\columnwidth]{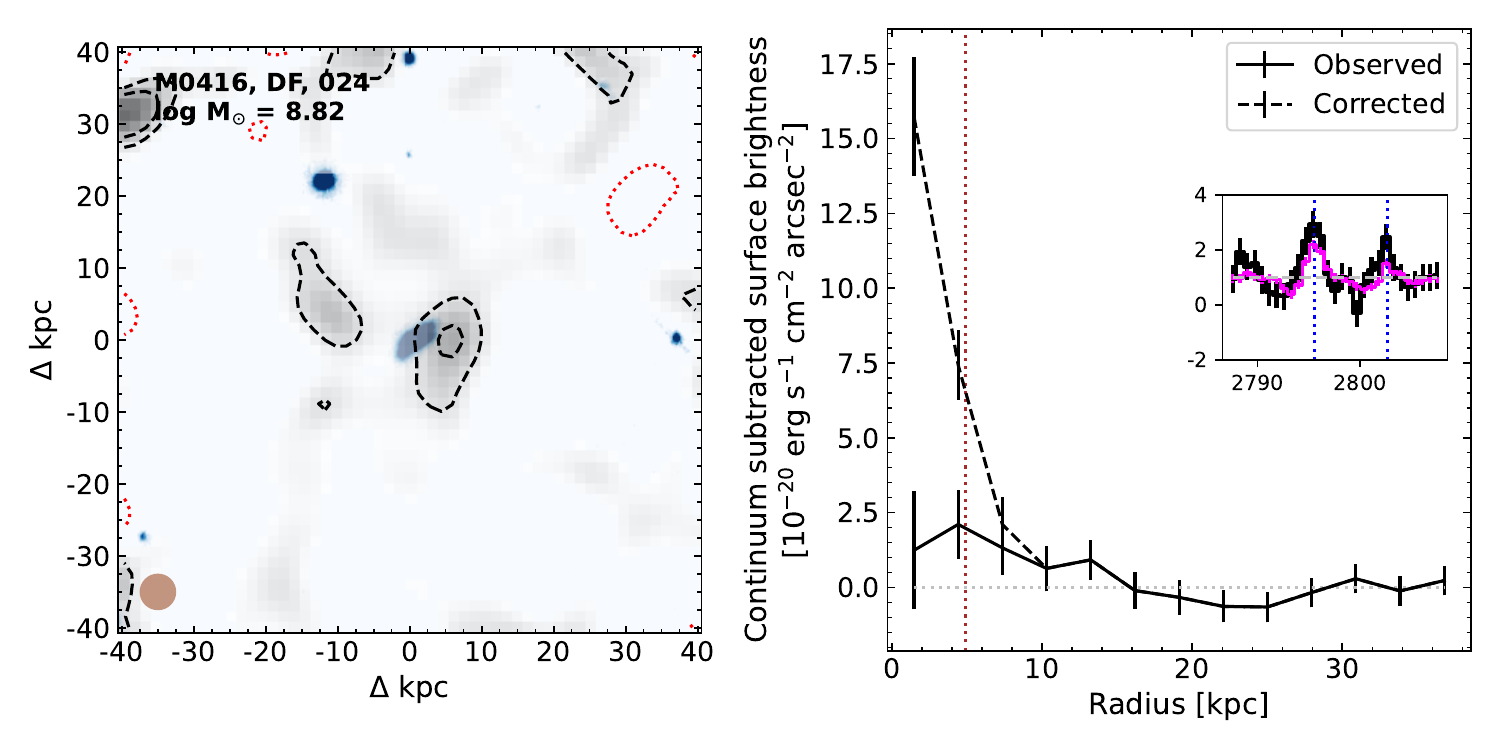}
    \end{minipage}%

    \begin{minipage}{0.49\textwidth}
        \centering
        \includegraphics[width=0.99\columnwidth]{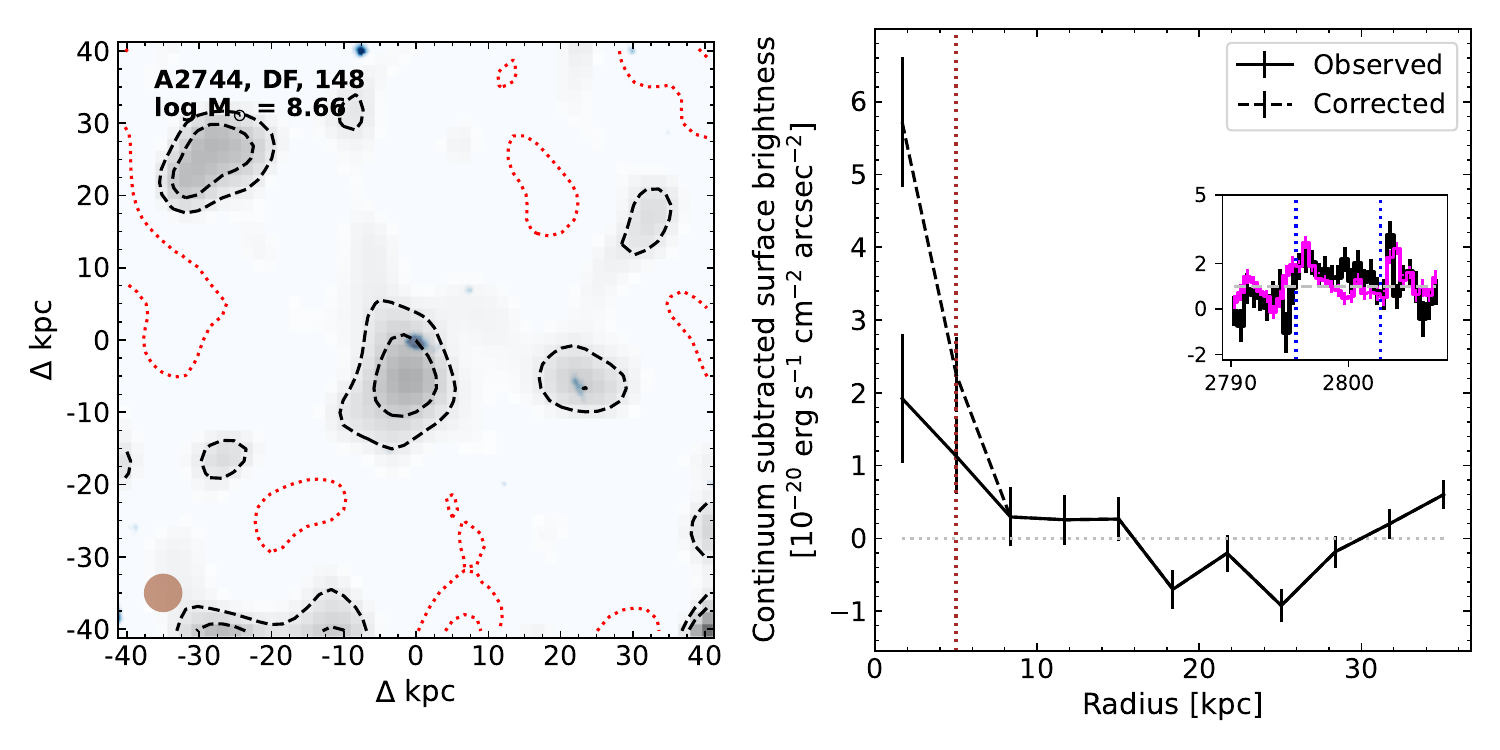}
    \end{minipage}
    \begin{minipage}{.49\textwidth}
        \centering
        \includegraphics[width=0.99\columnwidth]{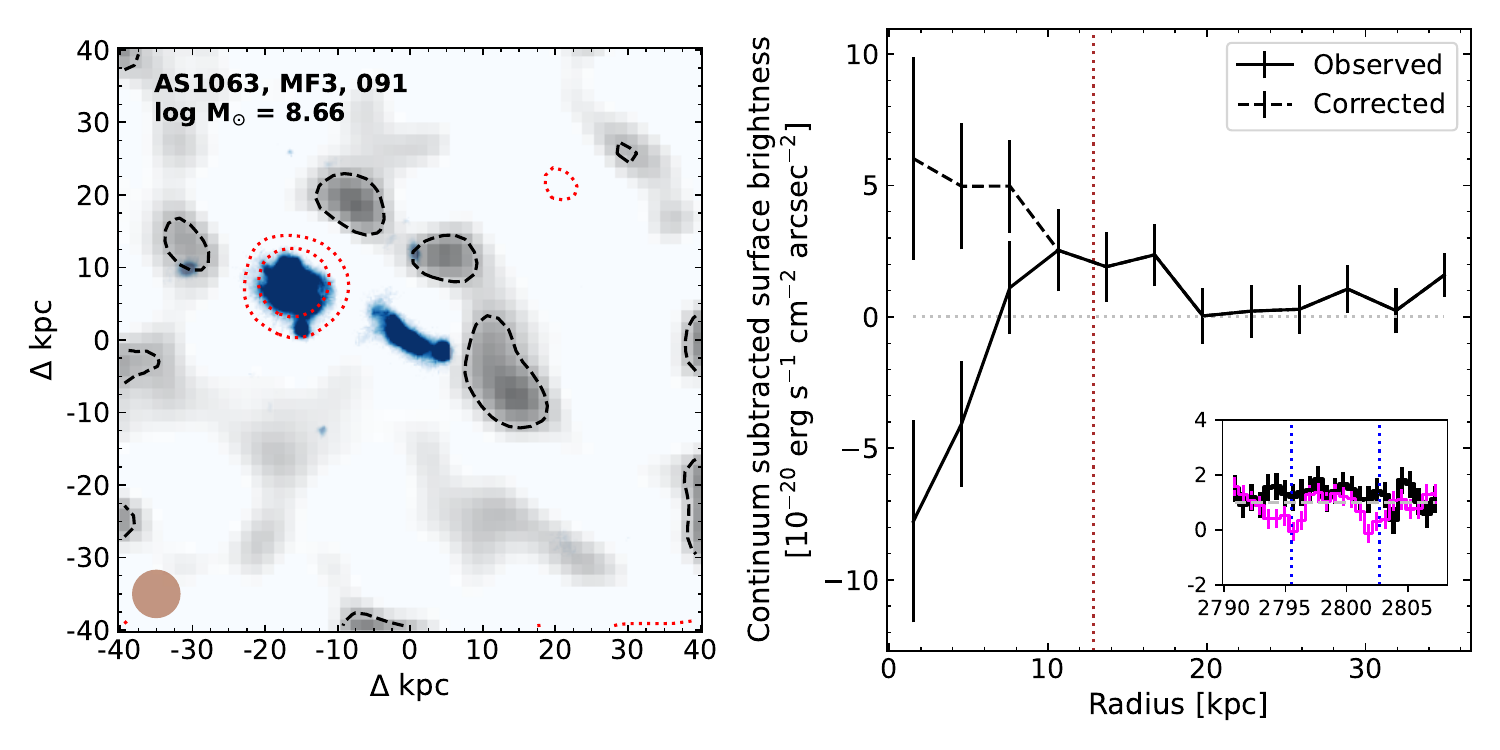}
    \end{minipage}%

    \begin{minipage}{0.49\textwidth}
        \centering
        \includegraphics[width=0.99\columnwidth]{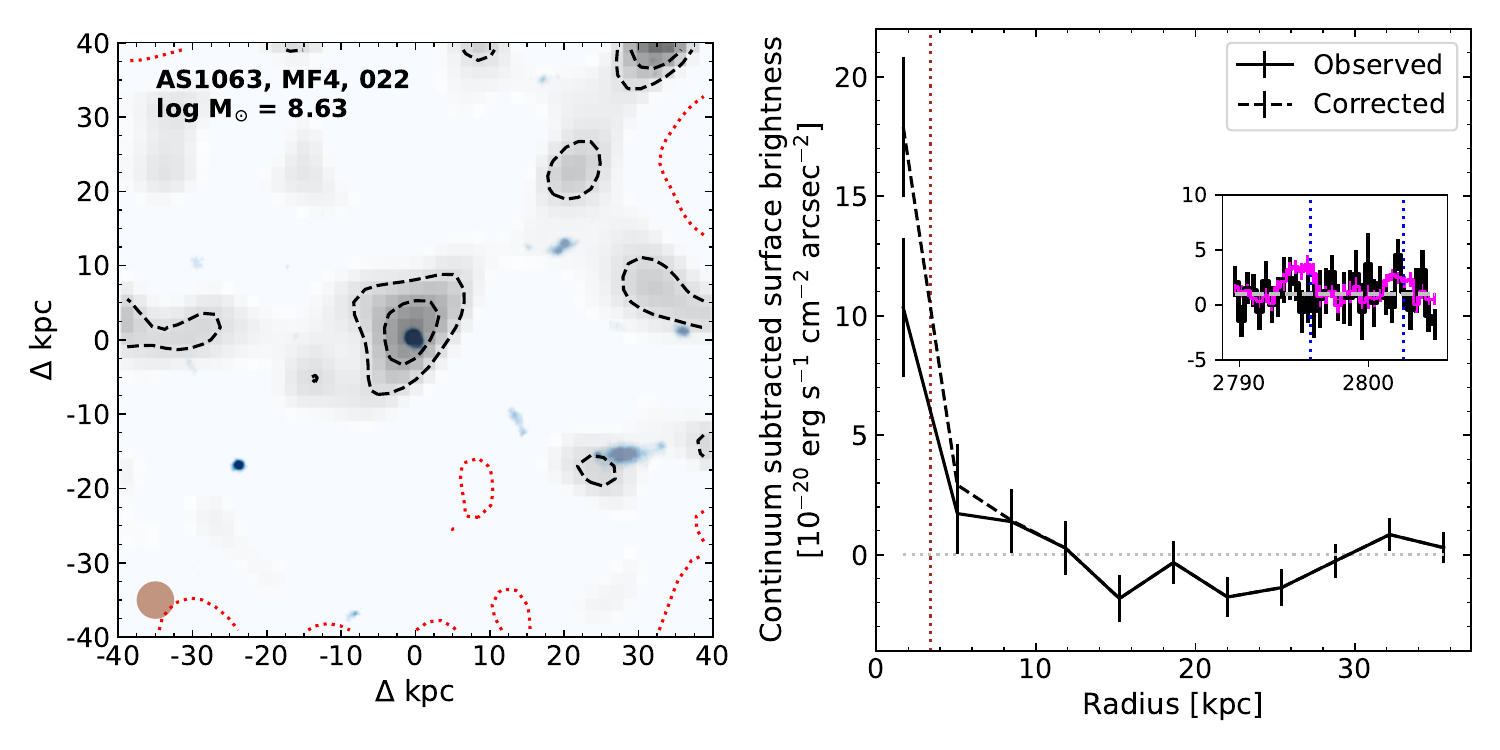}
    \end{minipage}
    \begin{minipage}{.49\textwidth}
        \centering
        \includegraphics[width=0.99\columnwidth]{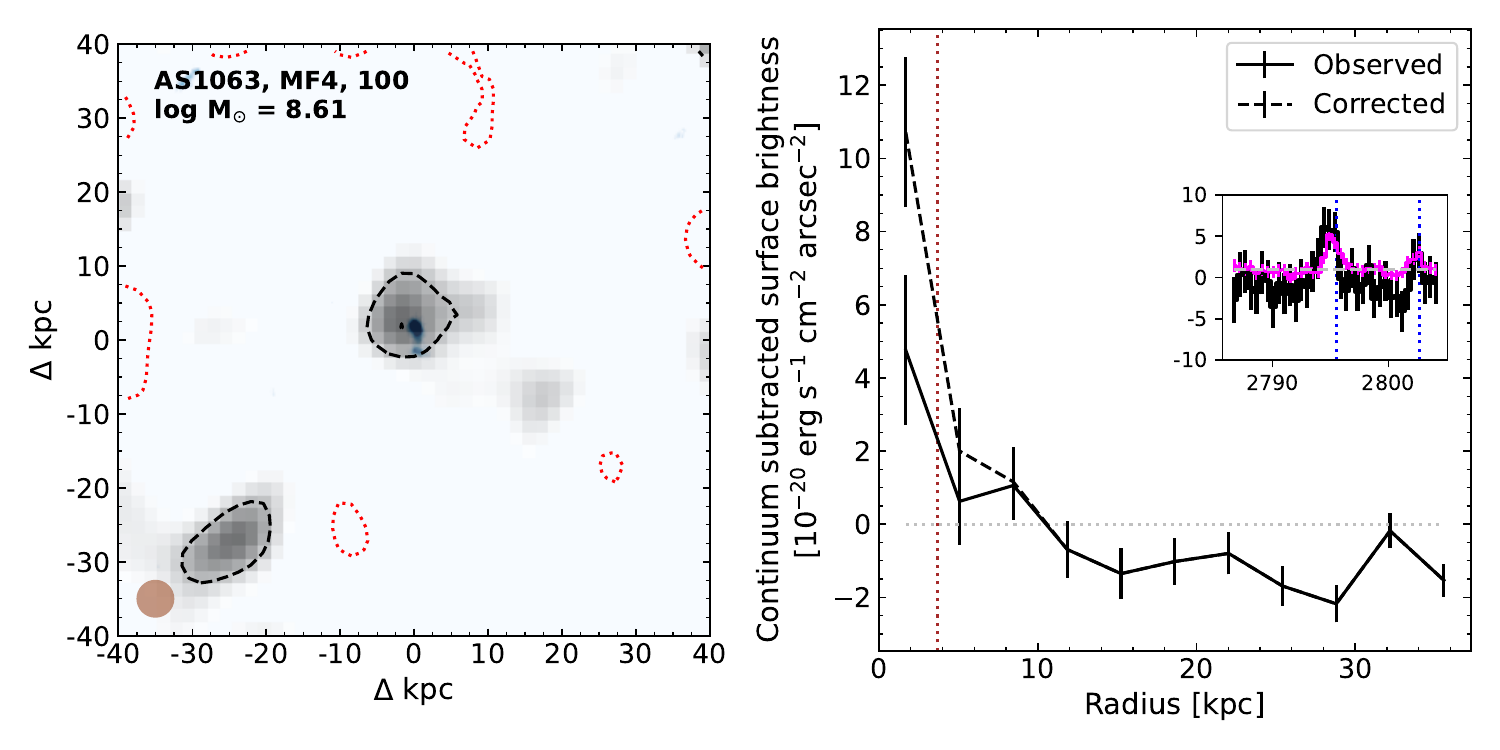}
    \end{minipage}%

    \caption{(continued)}
\end{figure*}

\addtocounter{figure}{-1}
\begin{figure*}[!htb]
    \centering    

    \begin{minipage}{0.49\textwidth}
        \centering
        \includegraphics[width=0.99\columnwidth]{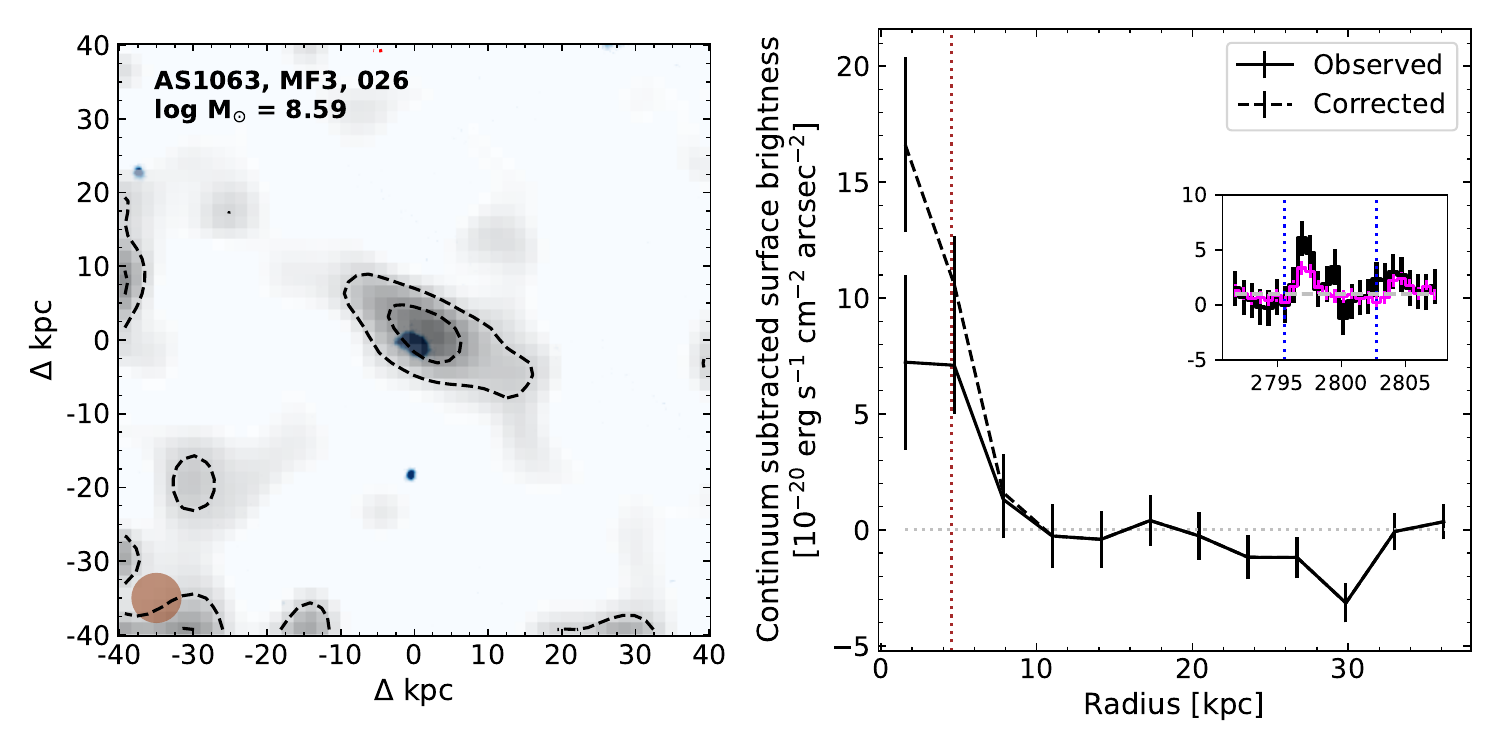}
    \end{minipage}
    \begin{minipage}{.49\textwidth}
        \centering
        \includegraphics[width=0.99\columnwidth]{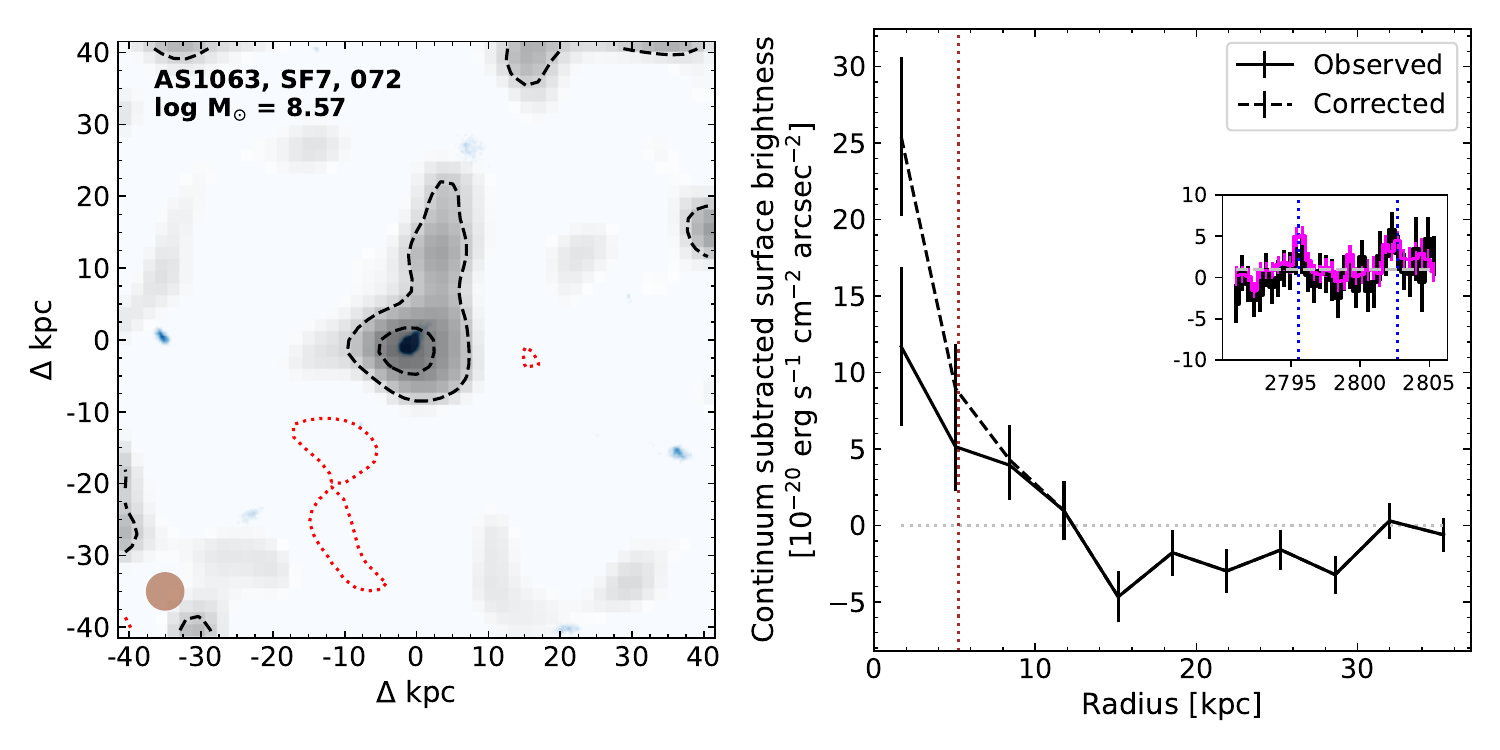}
    \end{minipage}%

    \begin{minipage}{0.49\textwidth}
        \centering
        \includegraphics[width=0.99\columnwidth]{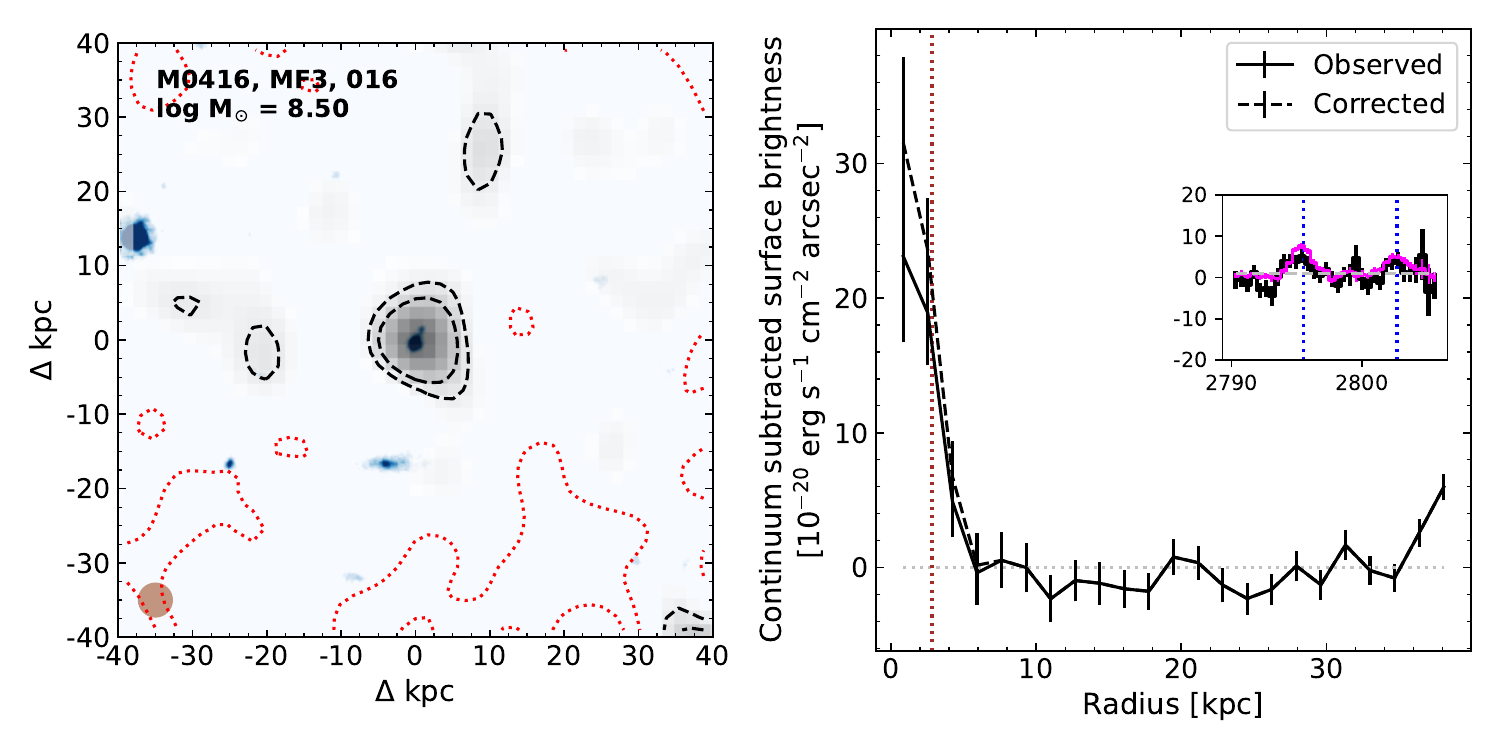}
    \end{minipage}
    \begin{minipage}{.49\textwidth}
        \centering
        \includegraphics[width=0.99\columnwidth]{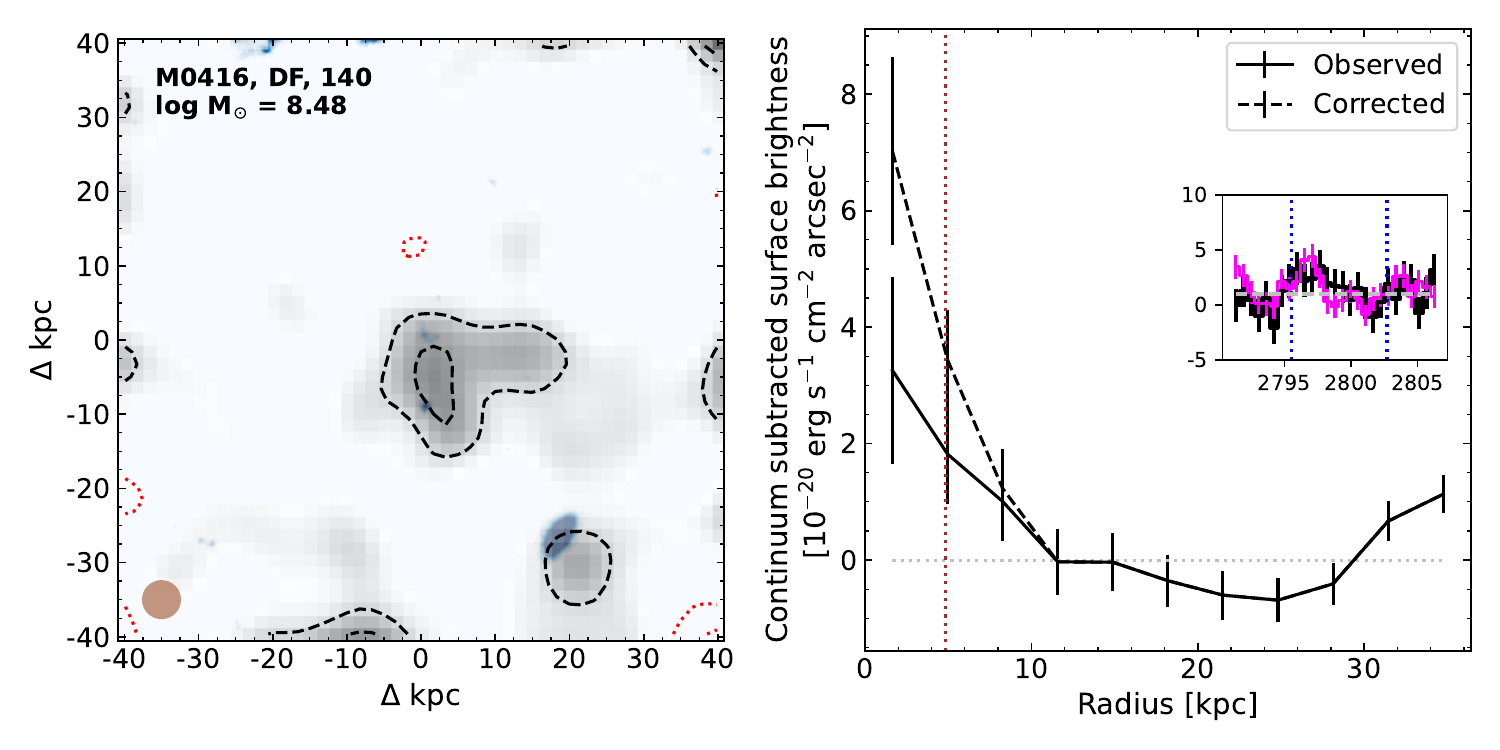}
    \end{minipage}%

    \begin{minipage}{0.49\textwidth}
        \centering
        \includegraphics[width=0.99\columnwidth]{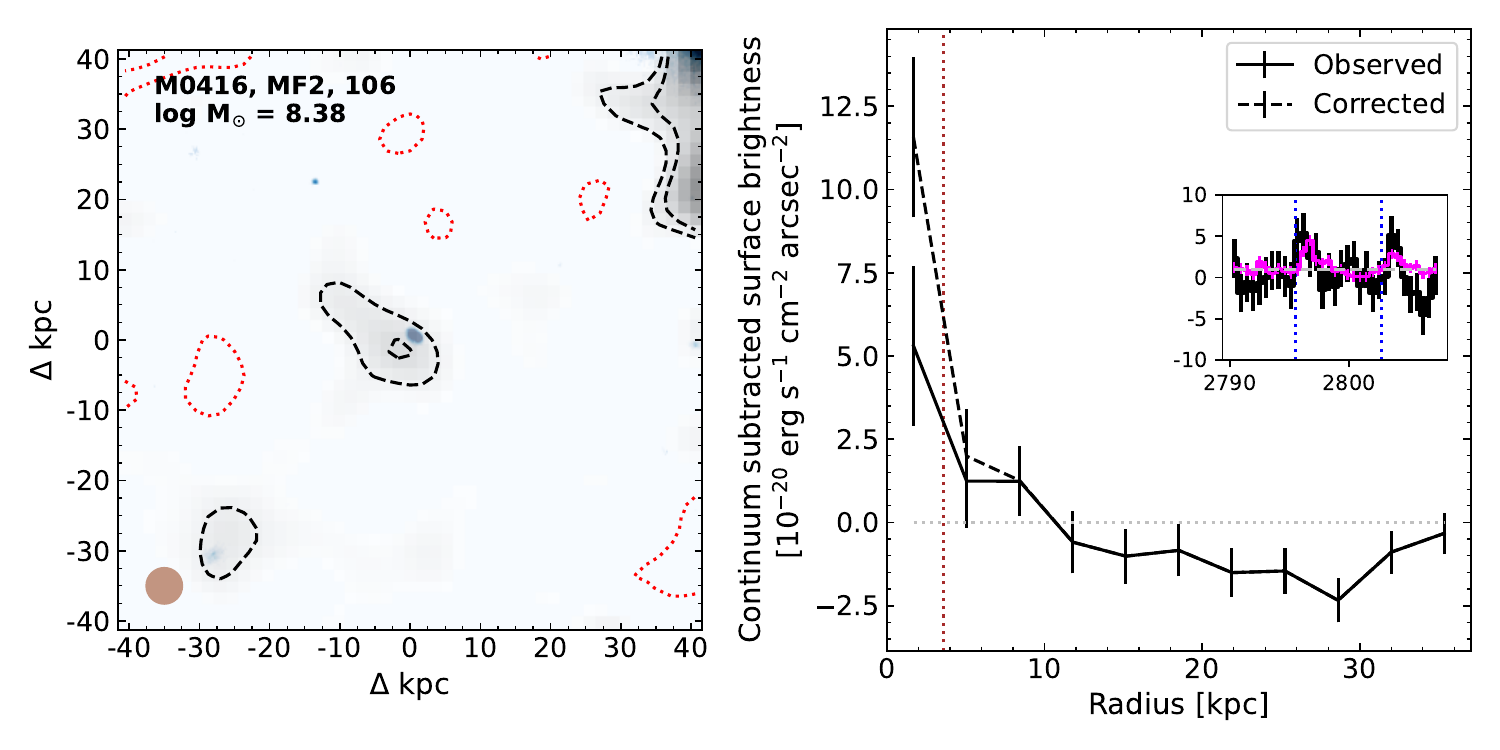}
    \end{minipage}
    \begin{minipage}{.49\textwidth}
        \centering
        \includegraphics[width=0.99\columnwidth]{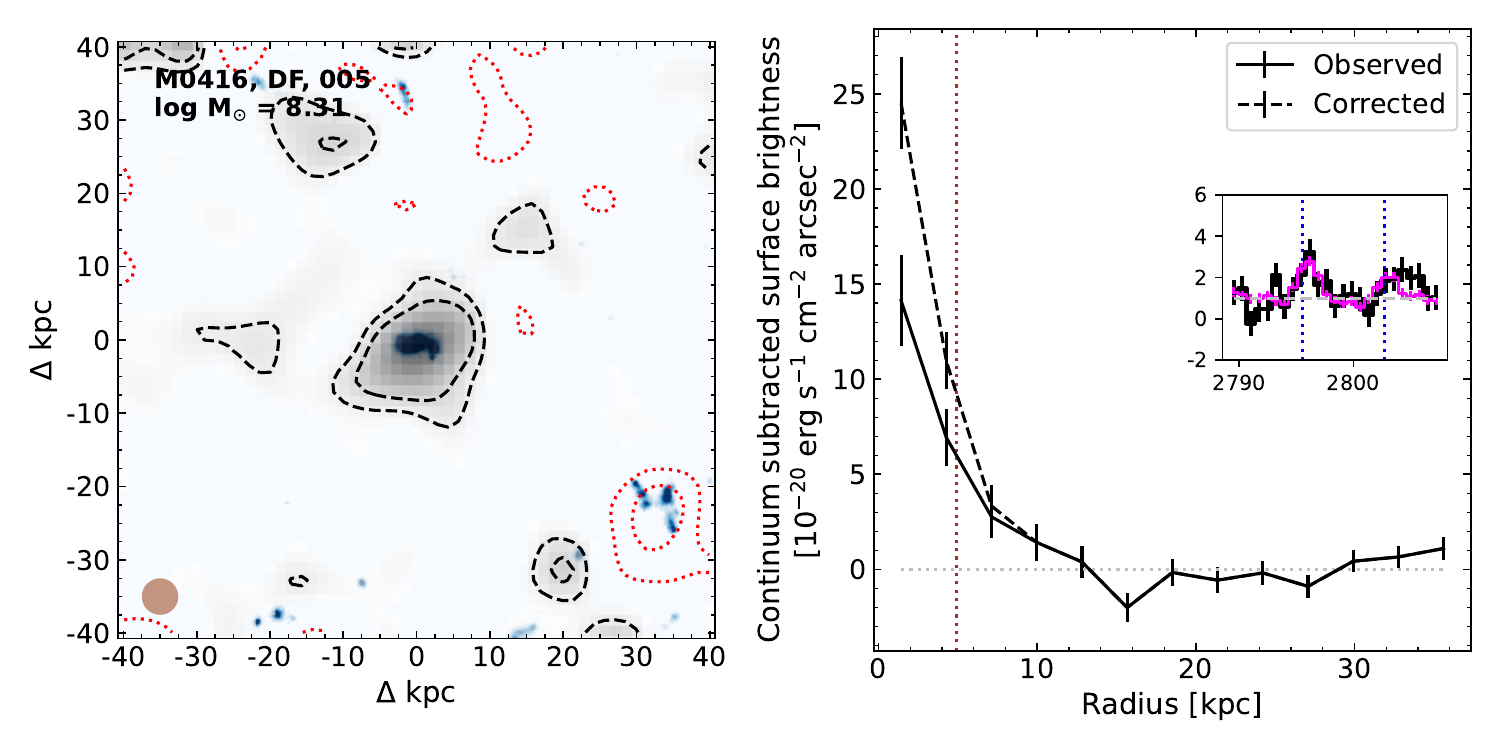}
    \end{minipage}%

    \begin{minipage}{0.49\textwidth}
        \centering
        \includegraphics[width=0.99\columnwidth]{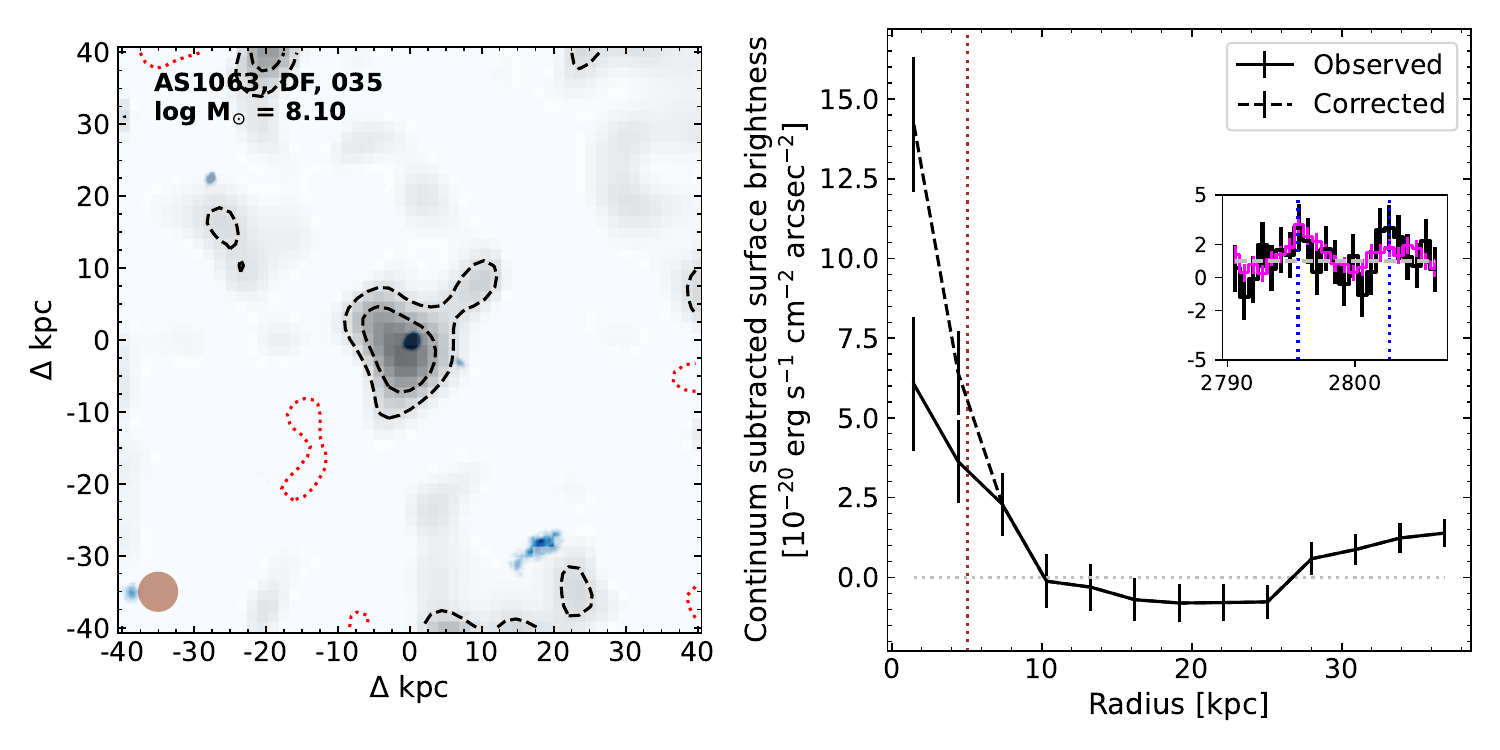}
    \end{minipage}
    \begin{minipage}{.49\textwidth}
        \centering
        \includegraphics[width=0.99\columnwidth]{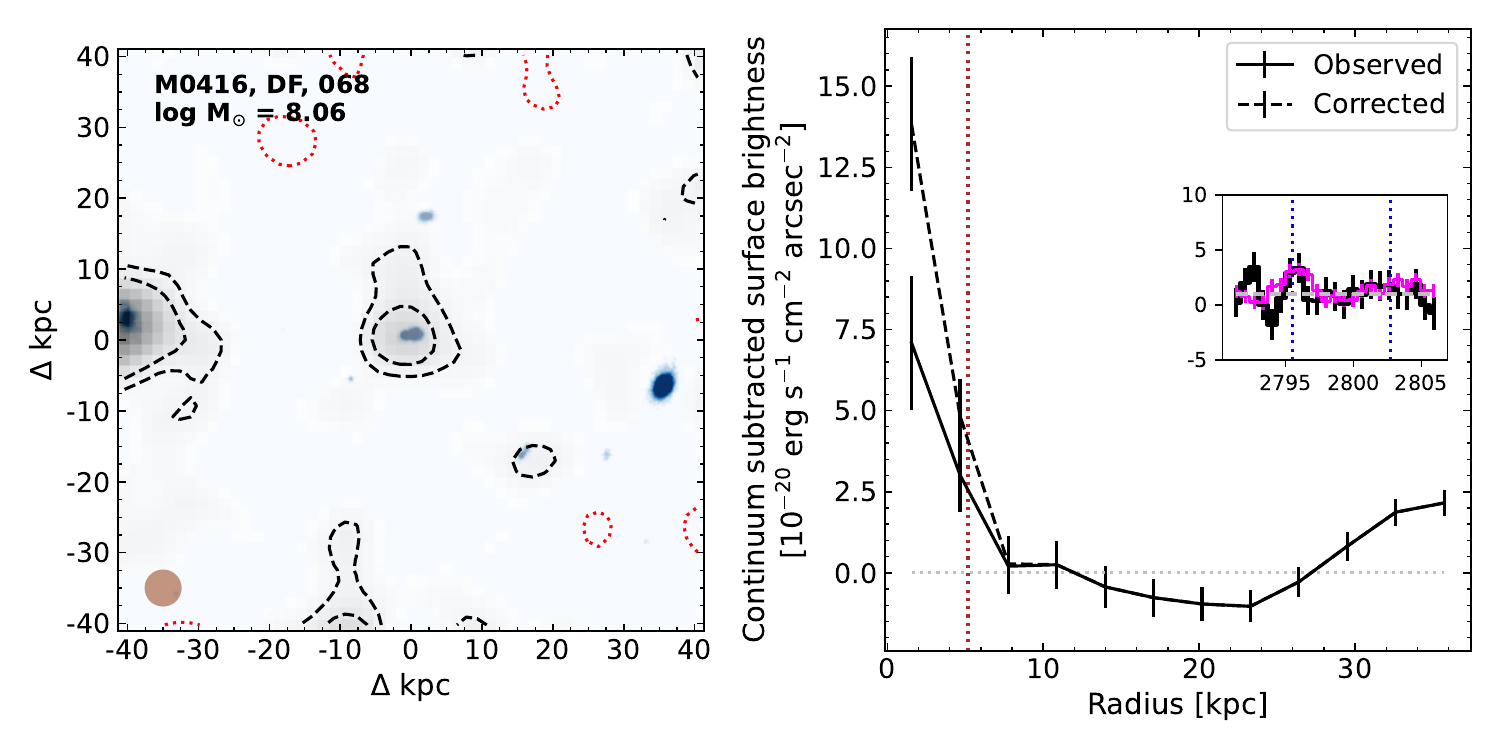}
    \end{minipage}%

    \begin{minipage}{0.49\textwidth}
        \centering
        \includegraphics[width=0.99\columnwidth]{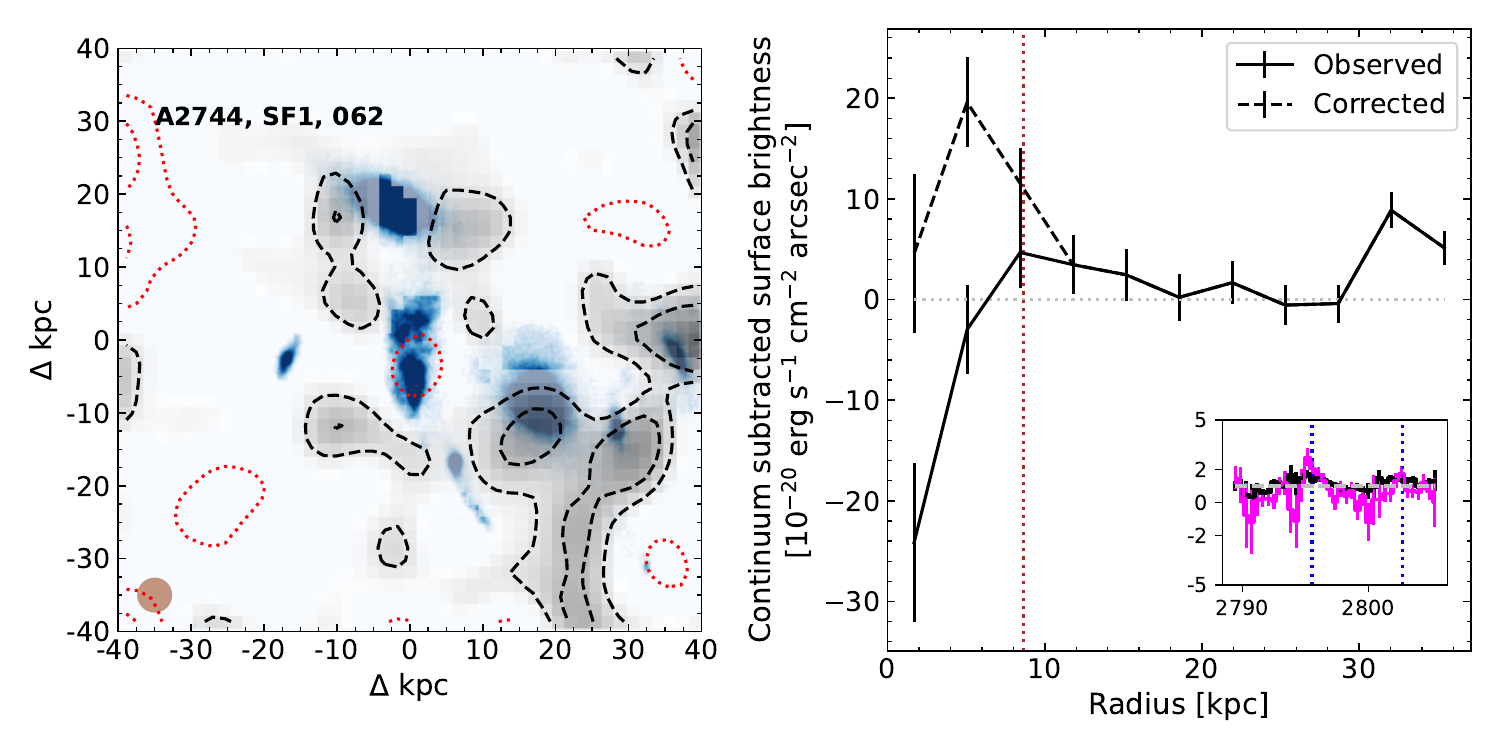}
    \end{minipage}
    \begin{minipage}{.49\textwidth}
        \centering
    \end{minipage}%
    
    \caption{(continued)}
\end{figure*}

\FloatBarrier

\subsubsection{Independent SFR measurement for our sample galaxies}
\label{sec:SFR_OII}

\begin{figure}
\includegraphics[width=\columnwidth]{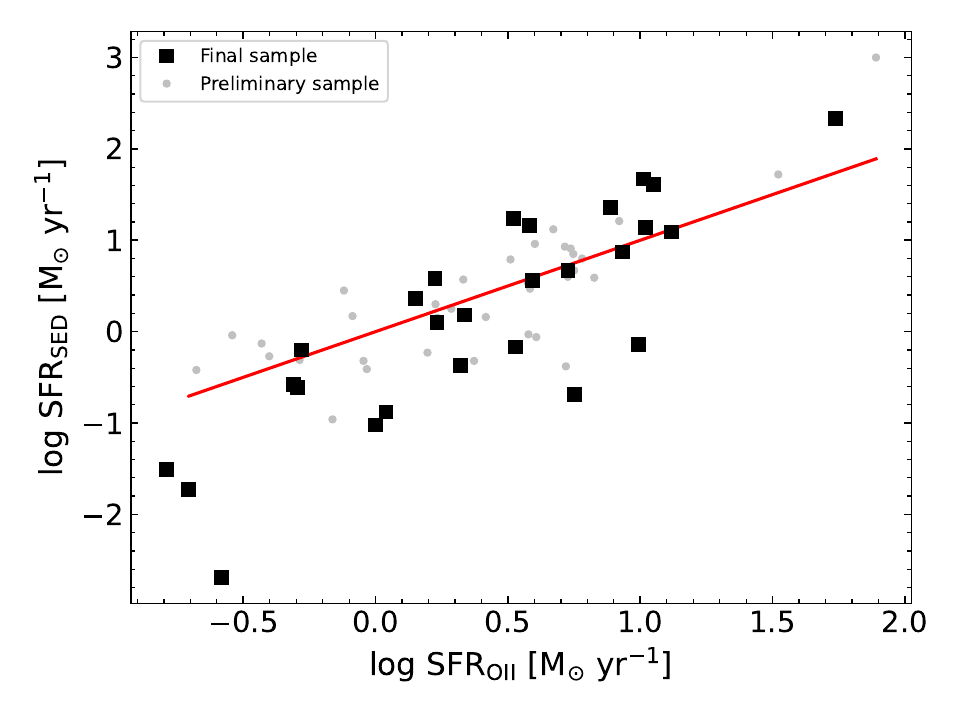}
\caption{Comparison of the SFR values derived for our sample galaxies obtained via SED fitting (see Sec.~\ref{sec:photometry} for details) and from the [\ion{O}{II}] luminosity of each galaxy (see Sec.~\ref{sec:SFR_OII}). The black squares show the SFR values for galaxies in our final sample, and the gray dots show the SFR values for galaxies in the preliminary sample, for the subset of galaxies at $z\lesssim1.48$, where the [\ion{O}{II}] doublet falls well within the MUSE wavelength range. The red line shows the identity for reference. Table~\ref{tab:sample_table} provides these measurements for each galaxy in our final sample.}
\label{fig:SFR_comparison}
\end{figure}

For galaxies at redshift lower than 1.48, the [\ion{O}{II}] $\lambda$$\lambda$3727, 3729 doublet falls comfortably within the MUSE wavelength range, and thus, for these galaxies, we can compute an independent estimation of their SFR from their [\ion{O}{II}] $\lambda$$\lambda$3727, 3729 emission, which traces star formation on timescales of a few megayears. 

To compute the SFR values from their [\ion{O}{II}] $\lambda$$\lambda$3727, 3729 luminosity, we used the following relation from \citet{Kewley2004}:
\begin{equation}
    \mathrm{SFR}_\mathrm{OII} \, (M_{\odot} \mathrm{yr}^{-1}) = 6.58 \times 10^{-42} \, \mathrm{L}_\mathrm{OII} \,(\mathrm{erg}\,\mathrm{s}^{-1})
\end{equation}

The spectra of the sample galaxies have been corrected by dust extinction before computing their SFR$_\mathrm{OII}$, using the extinction derived from the SED fitting \citep[generally, the emission lines will be subject to higher extinction values than the stellar continuum, so this mainly provides a lower limit on the emission line extinction, see, e.g.,][]{Hemmati2015, Emsellem2022}. Also, because of the lack of additional emission lines to perform further diagnosis of the ionization mechanism \citep[e.g., Baldwin–Phillips–Terlevich diagram][]{BPT}, we assume that all the [\ion{O}{II}] emission of these galaxies traces SFR, which might not be entirely true, as some [\ion{O}{II}] emission can be produced by other mechanisms such as shock ionization \citep[see, e.g.,][]{Reynaldi2013}. Figure~\ref{fig:SFR_comparison} shows the comparison of the SFR estimated from the SED fitting with the values determined from the [\ion{O}{II}] luminosity for the galaxies where [\ion{O}{II}] is available. Despite not accounting for non star-formation ionization, and using the dust extinction derived from the SED fitting, we find that there is a relatively good agreement between both independent estimations, with some higher dispersion for the galaxies on the low-end of the SFR distribution, which is not unexpected because, in addition to the intrinsic random and systematic uncertainties in both measurements, they also trace SFR on different timescales. Table~\ref{tab:sample_table} provides the SFR values measured from the [\ion{O}{II}] line for those galaxies where [\ion{O}{II}] is available, given their redshift and the MUSE wavelength range (There are no galaxies in our sample at $z<1.48$ where the [\ion{O}{II}] line is undetected).

\begin{table*}
\fontsize{8pt}{8pt}\selectfont
\caption{Summary of the main properties of the galaxies from our final sample (see Sec.~\ref{sec:sample_selection} for details on our sample selection).}
\centering
\begin{tabular}{llccccccccr}
\hline
ID & \multicolumn{1}{c}{Name} &RA& DEC & $z$ & log(M$_{*}$) & log(SFR)$_\mathrm{SED}$& log(Age)  & log(SFR)$_{\mathrm{OII}}$ & HLR$_{\mathrm{halo}}$ & HLR$_{\mathrm{cont}}$  \\ 
 & & [$^{\circ}$] & [$^{\circ}$] & & [$M_{\odot}$] &[$M_{\odot}$ yr$^{-1}$]&[yr] & [$M_{\odot}$ yr$^{-1}$] & [kpc] & [kpc]  \\ 
(1) & \multicolumn{1}{c}{(2)} & (3) & (4) & (5) & (6) & (7) &(8) & (9) & (10) & (11) \\ 
\hline
1 & M0416-MF2-167& 64.13975& -24.09870 & 1.36317 & 10.44 & 1.14& 9.0 & 1.04 & 7.1$\pm$1.3 & 4.1 \\ 
2 & A2744-DF-092& 3.46440& -30.36788 & 1.88845 & 10.05 & 2.65& 7.3 & - & 5.7$\pm$0.8 & 3.6 \\ 
3 & AS1063-MF3-055& 342.32953& -44.55687 & 1.55457 & 9.75 & 1.56& 8.3 & - & 6.6$\pm$1.2 & 3.7 \\ 
4 & A2744-DF-013& 3.47013& -30.36504 & 0.75432 & 9.75 & 0.87& 9.0 & 0.96 & 15.9$\pm$0.2 & 3.6 \\ 
5 & AS1063-MF2-011& 342.31253& -44.52657 & 1.53938 & 9.75 & 0.14& 7.8 & - & 5.9$\pm$0.6 & 2.4 \\ 
6 & A2744-MF4-064& 3.46049& -30.38623 & 1.17375 & 9.74 & 1.61& 8.1 & 1.07 & 6.0$\pm$0.5 & 3.7 \\ 
7 & AS1063-DF-029& 342.32190& -44.53876 & 1.71289 & 9.73 & -0.57& 7.9 & - & 11.1$\pm$1.7 & 4.0 \\ 
8 & M0416-DF-147& 64.12998& -24.12085 & 1.33893 & 9.73 & 1.24& 8.5 & 0.54 & 14.6$\pm$2.1 & 6.8 \\ 
9 & AS1063-MF4-054& 342.30835& -44.54379 & 1.33864 & 9.72 & 2.33& 7.2 & 1.76 & 7.2$\pm$1.6 & 2.7 \\ 
10 & AS1063-DF-109& 342.33005& -44.54077 & 1.45095 & 9.71 & 1.09& 8.1 & 1.12 & 6.4$\pm$0.2 & 3.7 \\ 
11 & AS1063-MF3-009& 342.33896& -44.54645 & 0.73470 & 9.70 & 1.16& 8.0 & 0.61 & 18.2$\pm$2.3 & 7.2 \\ 
12 & M0416-DF-032& 64.12423& -24.12187 & 0.90712 & 9.66 & 0.67& 8.0 & 0.75 & 13.0$\pm$1.5 & 4.9 \\ 
13 & AS1063-MF3-004& 342.32480& -44.55129 & 0.83087 & 9.62 & -0.17& 8.0 & 0.55 & 22.4$\pm$3.5 & 3.4 \\ 
14 & M0416-MF3-043& 64.15156& -24.11929 & 1.43364 & 9.61 & -0.69& 7.9 & 0.75 & 5.7$\pm$0.8 & 4.2 \\ 
15 & AS1063-SF7-064& 342.36584& -44.54976 & 1.74978 & 9.54 & -0.16& 7.9 & - & 6.2$\pm$3.9 & 2.7 \\ 
16 & AS1063-MF4-024& 342.30585& -44.55262 & 1.69904 & 9.48 & 0.83& 8.3 & - & 4.1$\pm$0.7 & 1.9 \\ 
17 & AS1063-SF1-015& 342.32565& -44.51646 & 1.00275 & 9.48 & -0.14& 7.8 & 1.02 & 6.5$\pm$1.2 & 2.1 \\ 
18 & A2744-DF-227& 3.47288& -30.37379 & 1.69302 & 9.45 & 0.26& 8.5 & - & 5.5$\pm$0.6 & 2.7 \\ 
19 & A2744-DF-091& 3.45532& -30.37054 & 1.88615 & 9.42 & 2.15& 7.2 & - & 4.6$\pm$0.2 & 1.8 \\ 
20 & A2744-SF2-081& 3.45478& -30.36284 & 1.68376 & 9.42 & 1.14& 7.5 & - & 5.0$\pm$0.9 & 2.3 \\ 
21 & M0416-SF7-012& 64.15696& -24.12107 & 0.90457 & 9.38 & 1.67& 7.8 & 1.04 & 6.5$\pm$1.4 & 1.7 \\ 
22 & AS1063-DF-240& 342.31955& -44.54046 & 1.95230 & 9.37 & -1.89& 8.1 & - & 10.4$\pm$1.3 & 4.7 \\ 
23 & A2744-DF-202& 3.47359& -30.37656 & 1.92721 & 9.33 & 0.67& 8.3 & - & 4.8$\pm$0.6 & 2.5 \\ 
24 & A370-DF-053& 40.06268& -1.62771 & 1.44661 & 9.24 & -0.37& 7.8 & 0.32 & 9.4$\pm$2.5 & 2.5 \\ 
25 & A2744-DF-002& 3.46596& -30.37785 & 1.82717 & 9.20 & 1.95& 7.2 & - & 5.3$\pm$0.2 & 3.0 \\ 
26 & A370-DF-130& 40.07022& -1.62454 & 1.56360 & 9.20 & 0.62& 8.5 & - & 3.7$\pm$2.3 & 2.4 \\ 
27 & AS1063-MF3-031& 342.32333& -44.55754 & 1.91240 & 9.19 & 0.37& 8.1 & - & 4.7$\pm$0.4 & 2.5 \\ 
28 & M0416-DF-231& 64.12701& -24.11898 & 1.54673 & 9.16 & 0.26& 7.9 & - & 4.3$\pm$1.7 & 3.2 \\ 
29 & M0416-SF2-031& 64.13725& -24.09095 & 1.14035 & 9.11 & 0.56& 8.3 & 0.62 & 4.4$\pm$0.8 & 1.8 \\ 
30 & M0416-DF-295& 64.12247& -24.12288 & 1.20755 & 9.07 & 0.58& 8.4 & 0.25 & 9.5$\pm$3.2 & 1.7 \\ 
31 & A370-DF-091& 40.06786& -1.62676 & 1.36091 & 8.92 & 0.36& 8.4 & 0.17 & 5.0$\pm$1.1 & 2.0 \\ 
32 & A2744-SF8-081& 3.49565& -30.39407 & 1.26738 & 8.91 & 0.10& 8.3 & 0.25 & 5.8$\pm$1.2 & 2.9 \\ 
33 & M0416-MF4-019& 64.13194& -24.12798 & 0.89519 & 8.90 & 0.18& 8.0 & 0.36 & 4.9$\pm$0.5 & 2.1 \\ 
34 & M0416-DF-024& 64.13852& -24.10866 & 0.75516 & 8.82 & -0.88& 7.9 & 0.06 & 4.9$\pm$0.9 & 2.3 \\ 
35 & A2744-DF-148& 3.46561& -30.38096 & 1.26500 & 8.66 & -0.20& 8.8 & -0.26 & 5.0$\pm$1.3 & 2.5 \\ 
36 & AS1063-MF3-091& 342.32413& -44.55154 & 0.83008 & 8.66 & 1.36& 7.2 & 0.91 & 12.9$\pm$2.7 & 5.1 \\ 
37 & AS1063-MF4-022& 342.31525& -44.54934 & 1.67279 & 8.63 & 0.10& 8.0 & - & 3.4$\pm$2.1 & 1.7 \\ 
38 & AS1063-MF4-100& 342.31503& -44.55623 & 1.57880 & 8.61 & -0.12& 8.0 & - & 3.7$\pm$1.5 & 2.6 \\ 
39 & AS1063-MF3-026& 342.33447& -44.55628 & 0.92758 & 8.59 & -1.02& 7.8 & 0.02 & 4.5$\pm$1.0 & 1.9 \\ 
40 & AS1063-SF7-072& 342.35016& -44.54823 & 1.88961 & 8.57 & 1.41& 7.2 & - & 5.2$\pm$1.0 & 1.0 \\ 
41 & M0416-MF3-016& 64.14242& -24.13184 & 1.69953 & 8.50 & -0.03& 8.0 & - & 2.8$\pm$2.9 & 2.5 \\ 
42 & M0416-DF-140& 64.13330& -24.11985 & 1.18174 & 8.48 & -2.69& 8.0 & -0.56 & 4.8$\pm$1.1 & 3.5 \\ 
43 & M0416-MF2-106& 64.12798& -24.09401 & 1.37333 & 8.38 & -0.61& 8.0 & -0.27 & 3.6$\pm$1.3 & 1.7 \\ 
44 & M0416-DF-005& 64.12878& -24.10865 & 0.69428 & 8.31 & -0.58& 7.9 & -0.29 & 4.9$\pm$1.0 & 2.5 \\ 
45 & AS1063-DF-035& 342.31863& -44.53286 & 0.75768 & 8.10 & -1.51& 7.8 & -0.77 & 5.1$\pm$1.1 & 1.3 \\ 
46 & M0416-DF-068& 64.13641& -24.11698 & 0.88842 & 8.06 & -1.73& 8.0 & -0.68 & 5.2$\pm$2.8 & 1.5 \\ 
47 & A2744-SF1-062& 3.47818& -30.35135 & 1.83840 & - & -& - & - & 8.6$\pm$2.2 & - \\ 
\hline
\end{tabular}
\tablefoot{The name of each galaxy shown in column (2) indicates the field and depth level of the data. Galaxies are sorted by stellar mass in descending order. Columns (3) and (4) show the coordinates of the galaxy in degrees. Column (5) shows the redshift of each galaxy, with an expected uncertainty of $\sim50$ km s$^{-1}$. The stellar masses, SFRs, and ages in columns (6), (7), and (8) have been derived via SED fitting, as detailed in Sec.~\ref{sec:photometry}. The uncertainties in stellar mass, SFR, and ages are on the order of $0.2$, $0.3$, and $0.2$ dex, respectively. The SFRs derived from the [\ion{O}{II}] luminosity for those galaxies where the [\ion{O}{II}] line is inside the MUSE wavelength are shown in column (9), and are calculated following \citet{Kewley2004}, with typical uncertainties of $0.1$ dex. Column (10) shows the half-light radius of the absorption-corrected \ion{Mg}{II} emission halos, calculated as detailed in Sec.~\ref{sec:reconstruction}. Column (11) shows the half-light radius of the continuum emission, calculated by fitting a S\'ersic model to the HST F160W image using \textsc{GALFIT}. The uncertainties in these measurements are lower than $0.1$ kpc. For A2744-SF1-062, we were not able to unambiguously determine its HST counterpart, and thus, the quantities in columns  (6), (7), (8), and (11) are not provided.}
\label{tab:sample_table}

\end{table*}

\subsubsection{Comparison of final sample with parent sample galaxies}
\label{sec:paren_sample_comparison}

In this subsection, we compare different properties of our final sample of galaxies derived via SED fitting with those from the original parent sample, aiming to study whether there are some peculiarities in our sample galaxies relative to the parent sample. In other words, if there is something special about our sample galaxies that could correlate with the existence of galactic-scale outflows.

\begin{figure}
\centering
\includegraphics[width=\columnwidth]{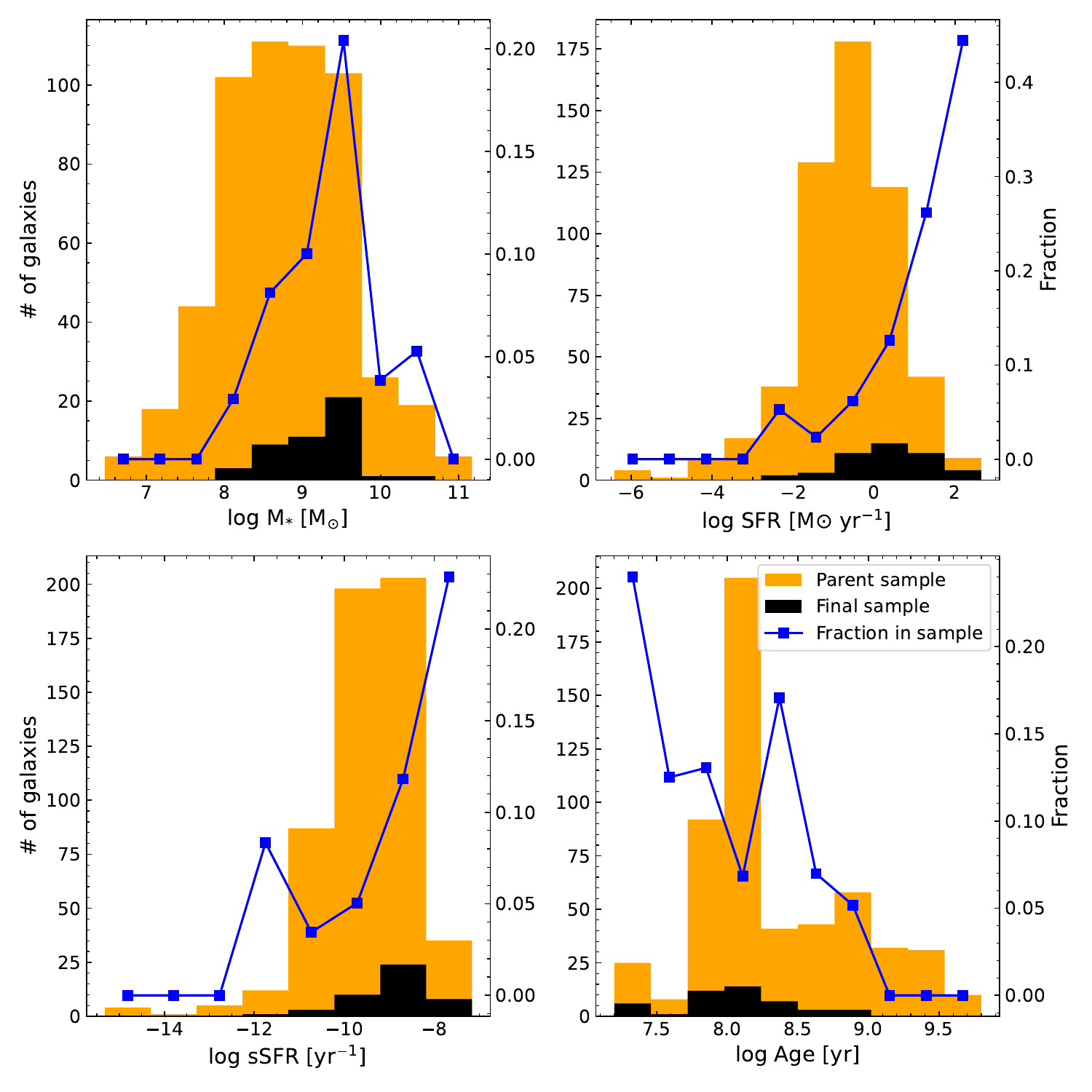}
\caption{Comparison of the stellar mass, SFR, sSFR, and age distributions for our final sample of galaxies (black) and our parent sample (orange). The blue line shows the fraction of galaxies from the parent sample in our final sample, in each bin of the relevant quantity for each panel. The right-side $y$-axis indicates the fraction shown by the blue line.}
\label{fig:parent_sample_comparison}
\end{figure}

Figure~\ref{fig:parent_sample_comparison} shows the distribution of stellar mass, SFR, sSFR, and stellar age for our final sample of galaxies, as well as for the original parent sample.

In terms of stellar mass, there is no clear preference for our sample galaxies. The fraction of galaxies from the parent sample inside our sample peaks at log M$_{*}\sim9.5$, with a sharp decrease toward higher and lower stellar masses. The sharp declines toward the extremes of the mass distribution could be, at least partially, driven by our preselection of galaxies that exhibit a P-Cygni profile in their \ion{Mg}{II} line, which, as discussed earlier in this section, are preferentially associated with galaxies of intermediate stellar masses. 

On the other hand, there seems to be some preference in the SFR, sSFR, and age values occupied by our sample galaxies, where the fraction of galaxies included in our sample tends to increase toward higher SFR and sSFR bins (that also correspond to younger age bins). We have performed a Kolmogorov-Smirnov (K-S) test to compare the distributions of our parent and final samples, for the four quantities shown in Fig.~\ref{fig:parent_sample_comparison} in a more quantitative manner. We found that for all of them, the p-value is lower than $0.02$, meaning that the distributions are significantly different.  

This is consistent with the expectations, since galactic-scale outflows on the stellar mass range relevant for our sample are expected to be primarily powered by stellar feedback \citep{Dylan2019b}, and the strength of stellar feedback is directly connected to the galaxy SFR, sSFR, and thus, age. 

However, even in these high SFR/sSFR and young bins, the fraction of galaxies that present an outflow traced by their \ion{Mg}{II} emission is not extremely high ($\sim20-40\%$), meaning that having a certain SFR, or a certain number of young stars, does not strictly correlate with the presence of cool-gas outflows (although as mentioned in Sec.~\ref{sec:sample_selection}, there could also be outflowing galaxies that do not present a P-Cygni profile, for instance, an edge-on biconical outflow that does not produce central absorption). There must also be other physical conditions required, such as ISM density or temperature. Alternatively, it might also be a matter of timescales, that is, the time when a cool-gas outflow is detectable could be a relatively small window after the last star-forming episode. In a future paper, we will further investigate in more detail different aspects of our sample galaxies, aiming at shedding light on what is fundamentally determining the presence of an outflow in a given galaxy. 

Note that our final sample contains galaxies from the three different depth levels of the MUSCATEL survey. Thus, the heterogeneous exposure times could also potentially factor into the fraction of galaxies from the parent sample included in our final sample. Indeed, the total fraction of galaxies in our final sample per depth level is the highest for the deep field ($\sim14\%$), and the lowest for the shallow field ($\sim4\%$), with the medium field in between ($\sim9\%$). However, the trends with the host galaxy properties explored in this section remain the same for the different depth levels, that is, higher fractions for young, star-forming galaxies (albeit with higher scatter, due to the lower statistics when considering individual depth levels only). On the other side, the fraction of galaxies in the youngest and highest star-forming bins from the parent sample in our sample is significantly higher in the deep and medium fields ($\sim80\%$ in the highest SFR bin, and $\sim30-40\%$ in the second highest bin, for both deep and medium field, individually), compared to a $\sim16\%$ and $\sim11\%$ fraction in the highest and second highest SFR bin for the shallow field. Thus, although the dependence on host galaxy properties is robust, the absolute fraction varies significantly for different depth levels (see Appendix~\ref{sec:frac_depth_levels}). This is not surprising, and it is the core reason of why previous studies of galactic-scale outflows traced by extended \ion{Mg}{II} emission have focused mainly on single-objects analyses, for which deep (> 10 hrs exposure time) integral field data is available \cite[see, e.g.,][]{Burchett2021, Zabl2021, Leclercq2022, Pessa2024} or stacking data \citep[see, e.g.,][]{Guo2023b, Dutta2023}.

\section{Reconstruction of the Mg II halos}

\label{sec:reconstruction}
As discussed in Sec.~\ref{sec:sample_selection}, by construction, we built our sample including those galaxies that show both emission and absorption of \ion{Mg}{II} in their spectra. This implies that the observed surface brightness profiles of the \ion{Mg}{II} emission halos are different from the intrinsic profiles that would be observed if there were no self-absorption of the central source by the halo, meaning that any measurement of size or morphology strongly depends on the geometry and orientation of the halo with respect to the line of sight, since face-on outflows will be more affected by central absorption than more edge-on outflows \citep[see, e.g.,][]{Guo2023b}.

However, as described in Sec.~\ref{sec:sample_selection}, we can use the best-fitting models for each galaxy to correct for the \ion{Mg}{II} absorption of the central galaxy spectrum, and then infer the absorption-corrected intrinsic surface brightness profile of the \ion{Mg}{II} emission halos. Figure~\ref{fig:sample_1} shows both the observed and absorption-corrected surface brightness profiles of the \ion{Mg}{II} pseudo-narrowband image. Since, by construction, all the galaxies present some level of central self-absorption, the corrected profile is always brighter in the center than the observed one. The corrected profile is always positive in the central region, with the exception of A2744-DF-013, where the model underestimates the strength of the absorption, and thus, after correcting for absorption, it remains negative (this particular galaxy is discussed in more detail in Appendix.~\ref{sec:particular_cases}).

Once we have corrected for self-absorption, we proceed to characterize the physical extent of the \ion{Mg}{II} halos using their half-light radius (HLR). To minimize the possible contamination by nearby sources, we compute the HLR as the radius at which half of the total \ion{Mg}{II} emission within a radius of 30 kpc from the central source is contained. The HLR of each galaxy in our sample is shown as a vertical brown dotted line in the panels that show the surface brightness profiles in Fig.~\ref{fig:sample_1}, and it is provided in Table~\ref{tab:sample_table}. The errors of the HLRs have been computed by performing 100 Monte Carlo iterations, perturbing the radial profile according to the errors of each radial bin, in each iteration. We have used the standard deviation of these 100 iterations as the error of the HLR measurement.

We stress that these HLR values are measured from the MUSE data, and thus, they include the contribution from the MUSE PSF. Nevertheless, the \ion{Mg}{II} emission of our sample galaxies is generally significantly more extended than the MUSE PSF (see Fig.~\ref{fig:sample_1}), and thus, our measurements are not strongly affected by the PSF contribution. Another caveat of the HLR is that since it is an azimuthally averaged quantity, it is not an optimal metric for non-isotropic configurations. For instance, galaxies such as M0416-MF2-106 and AS1063-MF4-100 exhibit some of the lowest HLRs among our sample galaxies (both of about $3.6$ kpc). However, in the corresponding panels of Fig.~\ref{fig:sample_1}, it is clear that they present an elongated shape whose extent is not properly captured by an azimuthally averaged metric, and are somewhat off-center with respect to their stellar counterpart. Due to the difficulties of parameterizing the shape of the observed \ion{Mg}{II} halos, we stick to the HLR metric to quantify sizes, as it does provide a useful reference value, but we caution the reader about this caveat.

\begin{figure}
\centering
\includegraphics[width=\columnwidth]{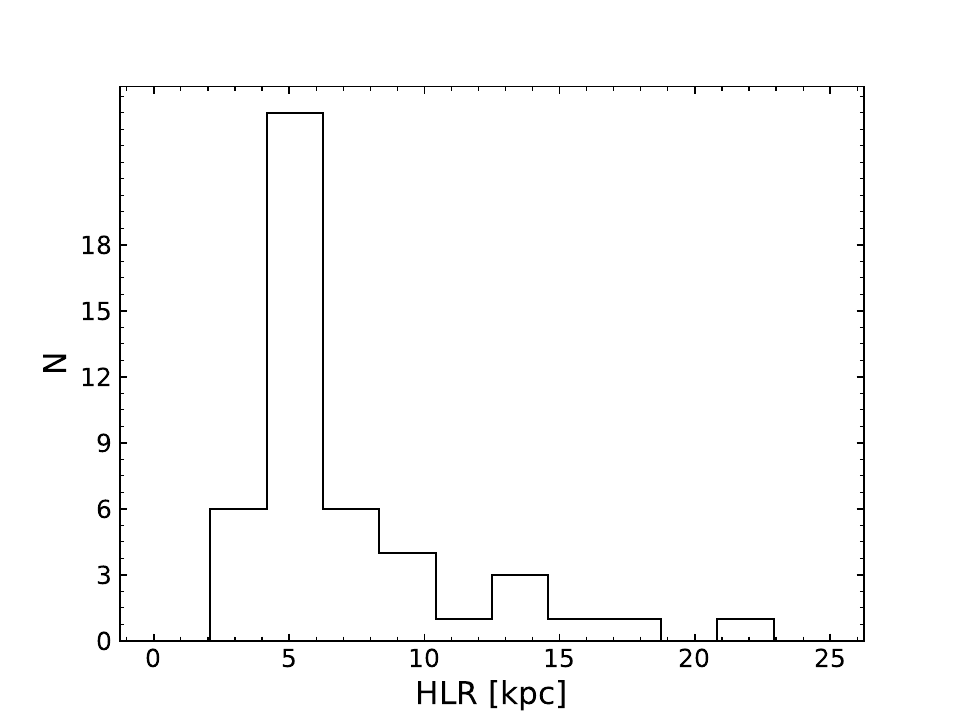}
\caption{Distribution of the sizes (in terms of HLR) of the \ion{Mg}{II} halos. The halos have been corrected by self-absorption before measuring their HLR using the absorption component of the best-fitting model. The radial profiles of the absorption-corrected \ion{Mg}{II} emission halos are shown with a dashed line in Fig.~\ref{fig:sample_1}. The HLR of each galaxy is provided in Table~\ref{tab:sample_table}.}
\label{fig:halo_size_dist}
\end{figure}

Figure~\ref{fig:halo_size_dist} shows the distribution of the HLRs of the \ion{Mg}{II} halos in our sample galaxies. Most galaxies have an HLR of approximately 5 kpc. However, the distribution shows an extended tail toward larger HLRs, reaching up to sizes of $\sim20$ kpc. This behavior is similar to that reported by \citet{Nelson2021} for simulated \ion{Mg}{II} halos in the TNG50 cosmological magnetohydrodynamical simulation \citep{Pillepich2019}, part of the IllustrisTNG project \citep{Dylan2019}.

\FloatBarrier
\section{Fitting results}
\label{sec:fitting_results}

\subsection{Best fitting models}
\label{sec:best_fitting_models}

In Secs.~\ref{sec:sample_selection} and~\ref{sec:reconstruction}, we have used the model only to correct for the self absorption in the observed spectra, and infer the intrinsic surface brightness profile and size of the \ion{Mg}{II} halos. For this correction, it is only relevant how well the spectral profile of the absorption line is modeled. However, we can also model the full \ion{Mg}{II} absorption plus emission across a larger aperture to infer physical properties of galactic winds. 

In this section, we provide an overview of the performance of our outflow model at reproducing the MUSE observations of the \ion{Mg}{II} halos, as well as the overall distribution of the best-fitting parameters. The model has been introduced in \citet{Pessa2024}, and we refer the reader to that paper for a complete description of our modeling scheme, its limitations, and main assumptions. In Appendix~\ref{sec:model}, we include a summarized description of the model used, as well as our fitting approach.

In a few words, the outflow is modeled as an ensemble of spherical shells, where the velocity of each shell increases with radius, such that continuum photons produced by the central source will interact with the outflowing material only at
the specific radius where the absorbing \ion{Mg}{II} ions are at resonance (due to their Doppler shift), an assumption commonly known as the ``Sobolev'' approximation. The gas density (and thus, optical depth) also varies radially, following the velocity field (assuming mass conservation).

The main free parameters that describe the wind properties are the launching velocity of the wind ($v_0$), the launching radius of the wind ($R_0$), its central optical depth ($\tau_{0}$), radial acceleration rate ($\gamma$), and terminal velocity ($v_\mathrm{max}$). The biconical geometry is described through three additional parameters: the opening angle (O.A.) of the bicone, a rotation angle (R.A.) that measures the rotation of the outflow perpendicular to the plane of the sky, toward the observer, and a position angle (P.A.) that corresponds to the angle of the outflow with respect to the horizontal axis, in the plane of the sky. Lastly, an additional contribution of nebular emission to the ion{Mg}{II} is described by the $f_{C}$ parameter.

\subsubsection{Data and best-fitting models comparison}

\begin{figure}
\includegraphics[width=\columnwidth]{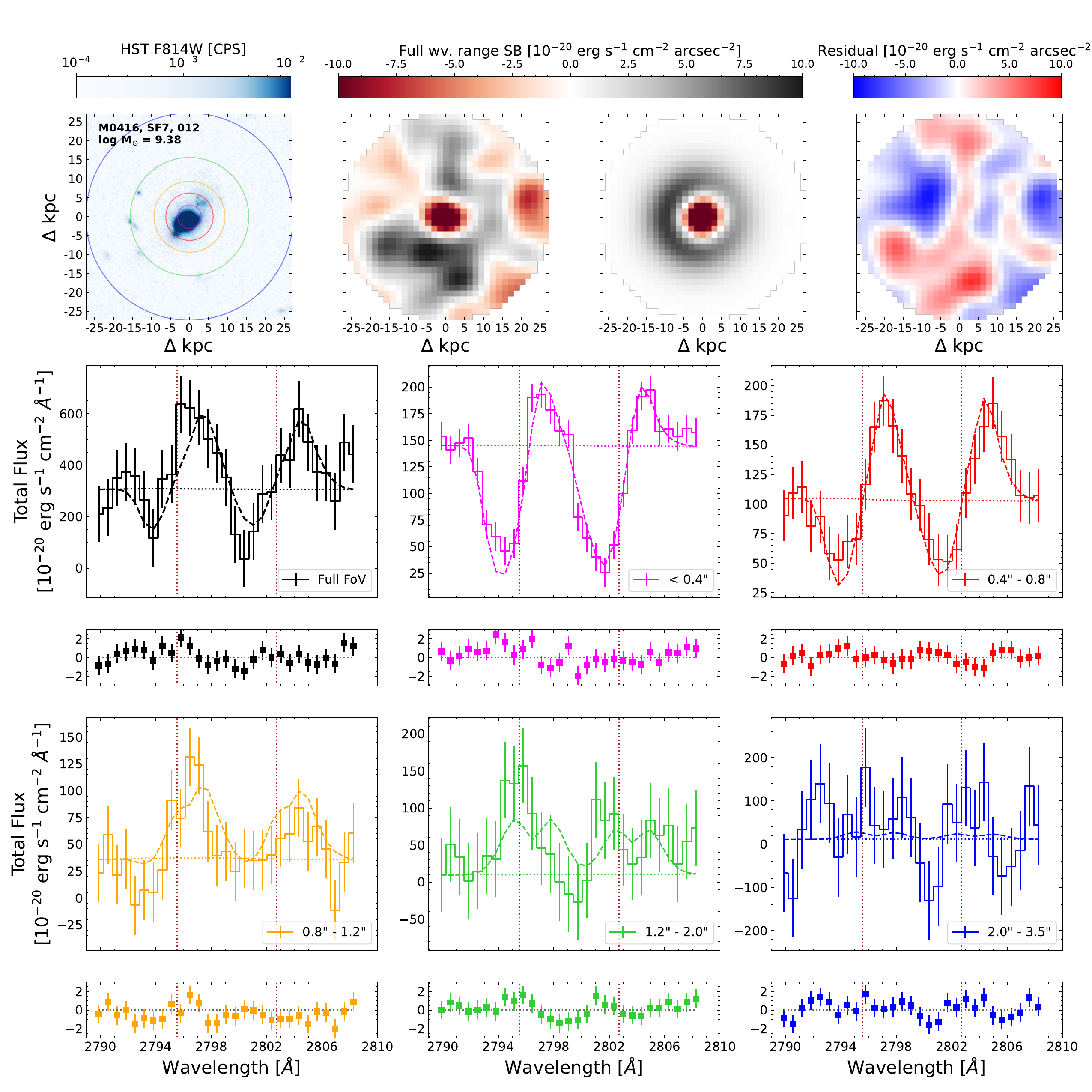}
\caption{Summary of the comparison of the observed and best-fitting model \ion{Mg}{II} spectral profile and continuum-subtracted pseudo-narrowband images, for an example galaxy in our final sample (M0416-SF7-012). The top-left panel shows the cutout of the HST F814W image around the galaxy, of 7 arcsec$^{2}$ size. The top-middle panels show the observed (middle-left) and best-fitting model (middle-right) continuum-subtracted \ion{Mg}{II} pseudo-narrowband images. The pseudo-narrowband images are created by collapsing the cubes across the wavelength axis, for wavelengths that enclose the full P-Cygni profile of the \ion{Mg}{II} doublet. The size of the pseudo-narrowband images is the same as that of the HST F814W cutout. The two panels show net \ion{Mg}{II} emission in black, and net \ion{Mg}{II} absorption in red. The residuals between both images are shown in the top-right panel. The rest of the panels show the integrated \ion{Mg}{II} spectra on different annular (and circular) apertures for the data (solid) and best-fitting model (dashed) cubes. The residuals are shown in the small panels below, in units of standard deviations. The vertical brown dotted lines indicate the rest-frame wavelength of the \ion{Mg}{II} doublet. The blue circle in the top-left panel shows the size of the full modeled region, and the black spectrum corresponds to the spectrum integrated over it. The rest of the panels that show colored spectra (and residuals) correspond to the comparison of the spectra extracted from the data and model cubes, in different annular and circular (in the case of the magenta spectra) apertures. In the legend of each panel, the inner and outer radii of the annular apertures are indicated. The color of the spectra in each panel matches the color of the circle in the top-left panel, which corresponds to the outer radii of its annular (or circular) aperture. The nearly horizontal dotted lines in each panel indicate the approximate continuum level.}
\label{fig:M0416_SF07_012_Master_figure}
\end{figure}

Figure~\ref{fig:M0416_SF07_012_Master_figure} shows a detailed comparison between the MUSE data and the best-fitting models, in terms of the continuum-subtracted \ion{Mg}{II} pseudo-narrowband images and the spectra extracted from annular apertures, for a galaxy drawn from one of the MUSCATEL fields. The continuum-subtracted pseudo-narrowband image exhibits a highly irregular morphology that the model cannot faithfully reproduce. However, when comparing the spectra extracted from azimuthally integrated apertures, it is clear that there is broad agreement between the data and the model, and that the line profiles of the \ion{Mg}{II} absorption and emission are well captured by the outflow model. The inner apertures show a clear P-Cygni profile with a prominent absorption, and some emission that becomes progressively more dominant toward the outer apertures.

While we show here only an illustrative representative case from our sample, we include the modeling results for all our sample galaxies in  Appendix~\ref{sec:model_results_all_galaxies}. Additionally, in Sec.~\ref{sec:deviation}, we discuss a subset of particular objects whose observed properties could indicate that their \ion{Mg}{II} emission is not entirely described by our simple outflow model. 

\FloatBarrier

\begin{table*}
\fontsize{8pt}{8pt}\selectfont
\caption{Summary of the best-fitting model parameters obtained for each one of our sample galaxies.}
\centering
\begin{tabular}{llccccccccccr}
\hline
ID & \multicolumn{1}{c}{Name} & log $\tau_{0}$ & $\gamma$ & $v_0$ & $v_\mathrm{max}$& $\Delta v$ & log $f_{C}$ & O.A. & R.A. & P.A. & $R_0$  & log $f$ \\ 
& & & & [km s$^{-1}$]&[km s$^{-1}$]&[km s$^{-1}$]& & [$^{\circ}$]&[$^{\circ}$]&[$^{\circ}$]& [kpc]& \\ 
(1) & \multicolumn{1}{c}{(2)} & (3) & (4) & (5) & (6) & (7) & (8) & (9) & (10) & (11) & (12) & (13) \\ 
\hline
1 & M0416-MF2-167  & 2.6$\pm$0.9 & 0.9$\pm$0.5 & 74$\pm$41 & 426$\pm$120 & 367$\pm$27 & -2.1$\pm$0.7 & 61$\pm$16 & 57$\pm$21 & 105$\pm$43 & 4.7$\pm$0.8 & -3.0$\pm$1.2 \\ 
2 & A2744-DF-092  & 2.9$\pm$0.1 & 0.6$\pm$0.1 & 80$\pm$8 & 542$\pm$13 & 50$\pm$6 & -2.1$\pm$0.5 & 25$\pm$3 & 71$\pm$3 & 2$\pm$3 & 3.0$\pm$0.2 & -2.9$\pm$1.1 \\ 
3 & AS1063-MF3-055  & 2.6$\pm$0.4 & 1.8$\pm$0.4 & 47$\pm$19 & 578$\pm$57 & 171$\pm$12 & -1.9$\pm$0.6 & 44$\pm$8 & 72$\pm$8 & 69$\pm$31 & 3.8$\pm$1.2 & -3.3$\pm$1.0 \\ 
4 & A2744-DF-013  & 3.0$\pm$0.1 & 0.9$\pm$0.0 & 23$\pm$2 & 565$\pm$57 & 231$\pm$1 & -2.7$\pm$0.4 & 60$\pm$2 & 80$\pm$2 & 52$\pm$12 & 3.9$\pm$0.0 & -1.0$\pm$0.0 \\ 
5 & AS1063-MF2-011  & 2.2$\pm$0.2 & 0.5$\pm$0.1 & 61$\pm$10 & 546$\pm$90 & 263$\pm$3 & -0.3$\pm$0.1 & 82$\pm$4 & 35$\pm$20 & 93$\pm$61 & 1.2$\pm$0.3 & -3.5$\pm$0.9 \\ 
6 & A2744-MF4-064  & 2.9$\pm$0.6 & 1.0$\pm$0.3 & 33$\pm$11 & 446$\pm$41 & 136$\pm$8 & -0.9$\pm$0.7 & 16$\pm$11 & 82$\pm$7 & 36$\pm$40 & 4.5$\pm$0.9 & -1.8$\pm$1.3 \\ 
7 & AS1063-DF-029  & 3.2$\pm$0.3 & 2.4$\pm$0.1 & 23$\pm$8 & 685$\pm$13 & 68$\pm$12 & -2.3$\pm$0.5 & 72$\pm$5 & 61$\pm$8 & 14$\pm$11 & 5.1$\pm$0.6 & -3.4$\pm$1.0 \\ 
8 & M0416-DF-147  & 2.4$\pm$1.0 & 1.3$\pm$0.4 & 73$\pm$25 & 555$\pm$113 & 111$\pm$14 & -0.4$\pm$0.7 & 73$\pm$9 & 6$\pm$3 & 32$\pm$45 & 6.2$\pm$0.8 & -3.1$\pm$1.0 \\ 
9 & AS1063-MF4-054  & 2.6$\pm$0.9 & 0.9$\pm$0.3 & 97$\pm$26 & 487$\pm$47 & 236$\pm$11 & -1.8$\pm$0.7 & 84$\pm$7 & 41$\pm$24 & 67$\pm$49 & 3.6$\pm$0.6 & -3.2$\pm$1.1 \\ 
10 & AS1063-DF-109  & 3.0$\pm$0.2 & 1.3$\pm$0.1 & 29$\pm$5 & 368$\pm$9 & 34$\pm$1 & 0.2$\pm$0.1 & 84$\pm$1 & 85$\pm$1 & 107$\pm$4 & 3.0$\pm$0.2 & -1.0$\pm$0.0 \\ 
11 & AS1063-MF3-009  & 3.3$\pm$1.1 & 0.5$\pm$0.3 & 158$\pm$41 & 490$\pm$121 & 158$\pm$16 & -2.0$\pm$0.6 & 28$\pm$7 & 38$\pm$6 & 83$\pm$9 & 8.9$\pm$1.1 & -3.0$\pm$1.0 \\ 
12 & M0416-DF-032  & 3.2$\pm$0.3 & 1.7$\pm$0.3 & 13$\pm$4 & 585$\pm$78 & 114$\pm$7 & -1.0$\pm$0.7 & 63$\pm$12 & 45$\pm$19 & 137$\pm$46 & 6.2$\pm$1.2 & -3.4$\pm$0.9 \\ 
13 & AS1063-MF3-004  & 3.0$\pm$0.4 & 1.2$\pm$0.2 & 27$\pm$11 & 430$\pm$79 & -0$\pm$15 & -2.2$\pm$0.6 & 59$\pm$15 & 61$\pm$18 & 65$\pm$49 & 6.3$\pm$1.3 & -3.2$\pm$1.1 \\ 
14 & M0416-MF3-043  & 2.0$\pm$0.5 & 1.2$\pm$0.4 & 53$\pm$25 & 582$\pm$70 & 160$\pm$12 & -1.4$\pm$0.8 & 58$\pm$11 & 57$\pm$14 & 123$\pm$32 & 4.1$\pm$0.7 & -3.2$\pm$1.0 \\ 
15 & AS1063-SF7-064  & 2.6$\pm$0.7 & 0.9$\pm$0.3 & 50$\pm$25 & 554$\pm$110 & -100$\pm$9 & 0.0$\pm$0.4 & 63$\pm$9 & 37$\pm$11 & 29$\pm$15 & 2.7$\pm$0.8 & -3.0$\pm$1.1 \\ 
16 & AS1063-MF4-024  & 2.9$\pm$0.5 & 1.5$\pm$0.3 & 32$\pm$14 & 519$\pm$78 & -52$\pm$6 & 0.1$\pm$0.1 & 61$\pm$8 & 51$\pm$11 & 13$\pm$13 & 2.1$\pm$0.7 & -3.2$\pm$1.0 \\ 
17 & AS1063-SF1-015  & 2.6$\pm$0.5 & 0.6$\pm$0.3 & 63$\pm$27 & 574$\pm$101 & 148$\pm$10 & -0.1$\pm$0.2 & 79$\pm$9 & 41$\pm$20 & 108$\pm$43 & 3.2$\pm$0.8 & -3.0$\pm$1.1 \\ 
18 & A2744-DF-227  & 2.8$\pm$0.3 & 1.2$\pm$0.3 & 32$\pm$8 & 401$\pm$41 & -67$\pm$6 & -0.6$\pm$0.5 & 47$\pm$8 & 60$\pm$8 & 103$\pm$19 & 2.2$\pm$0.8 & -3.2$\pm$1.1 \\ 
19 & A2744-DF-091  & 3.6$\pm$0.4 & 1.9$\pm$0.1 & 44$\pm$6 & 620$\pm$15 & -28$\pm$5 & -0.1$\pm$0.1 & 31$\pm$4 & 70$\pm$4 & 178$\pm$3 & 3.4$\pm$0.4 & -2.9$\pm$1.1 \\ 
20 & A2744-SF2-081  & 2.4$\pm$0.3 & 0.7$\pm$0.3 & 70$\pm$20 & 589$\pm$63 & -56$\pm$9 & -0.1$\pm$0.5 & 55$\pm$12 & 73$\pm$12 & 69$\pm$41 & 2.6$\pm$0.7 & -3.3$\pm$1.1 \\ 
21 & M0416-SF7-012  & 2.4$\pm$0.6 & 0.8$\pm$0.2 & 66$\pm$22 & 425$\pm$15 & 103$\pm$6 & -1.6$\pm$0.7 & 71$\pm$7 & 85$\pm$7 & 34$\pm$37 & 2.0$\pm$0.5 & -3.5$\pm$1.0 \\ 
22 & AS1063-DF-240  & 2.4$\pm$0.4 & 1.1$\pm$0.3 & 43$\pm$18 & 479$\pm$99 & -96$\pm$6 & -0.5$\pm$0.6 & 85$\pm$5 & 52$\pm$30 & 74$\pm$56 & 3.4$\pm$0.9 & -3.4$\pm$1.0 \\ 
23 & A2744-DF-202  & 2.9$\pm$0.4 & 1.5$\pm$0.3 & 19$\pm$9 & 555$\pm$81 & -202$\pm$8 & -0.2$\pm$0.1 & 31$\pm$12 & 69$\pm$13 & 174$\pm$39 & 3.2$\pm$0.8 & -3.2$\pm$1.1 \\ 
24 & A370-DF-053  & 1.8$\pm$0.5 & 0.7$\pm$0.3 & 96$\pm$40 & 547$\pm$88 & 56$\pm$12 & 0.1$\pm$0.2 & 75$\pm$9 & 56$\pm$20 & 94$\pm$46 & 4.1$\pm$1.1 & -3.3$\pm$1.0 \\ 
25 & A2744-DF-002  & 1.1$\pm$0.1 & 1.8$\pm$0.1 & 146$\pm$4 & 718$\pm$23 & -36$\pm$5 & 0.3$\pm$0.1 & 77$\pm$5 & 76$\pm$27 & 22$\pm$11 & 3.5$\pm$0.1 & -1.0$\pm$0.0 \\ 
26 & A370-DF-130  & 4.0$\pm$0.6 & 0.4$\pm$0.2 & 51$\pm$16 & 407$\pm$114 & -39$\pm$8 & 0.2$\pm$0.1 & 61$\pm$11 & 68$\pm$16 & 134$\pm$46 & 1.9$\pm$1.0 & -3.2$\pm$1.0 \\ 
27 & AS1063-MF3-031  & 3.8$\pm$0.8 & 0.6$\pm$0.2 & 62$\pm$21 & 501$\pm$83 & -24$\pm$7 & 0.1$\pm$0.2 & 84$\pm$7 & 31$\pm$20 & 84$\pm$45 & 0.6$\pm$0.2 & -3.5$\pm$1.1 \\ 
28 & M0416-DF-231  & 3.9$\pm$0.6 & 0.3$\pm$0.2 & 63$\pm$18 & 436$\pm$48 & -83$\pm$11 & -0.3$\pm$0.4 & 44$\pm$11 & 76$\pm$12 & 32$\pm$29 & 2.9$\pm$0.9 & -3.2$\pm$1.1 \\ 
29 & M0416-SF2-031  & 1.8$\pm$0.9 & 1.7$\pm$0.6 & 58$\pm$36 & 544$\pm$114 & 116$\pm$29 & -0.2$\pm$0.6 & 65$\pm$19 & 53$\pm$23 & 96$\pm$50 & 2.9$\pm$0.8 & -3.0$\pm$1.1 \\ 
30 & M0416-DF-295  & 2.3$\pm$0.6 & 0.7$\pm$0.3 & 71$\pm$26 & 399$\pm$53 & 99$\pm$16 & -1.9$\pm$0.7 & 70$\pm$13 & 50$\pm$21 & 34$\pm$24 & 3.5$\pm$0.8 & -3.1$\pm$1.1 \\ 
31 & A370-DF-091  & 3.3$\pm$1.1 & 0.6$\pm$0.3 & 78$\pm$34 & 303$\pm$84 & 114$\pm$12 & 0.2$\pm$0.2 & 74$\pm$16 & 46$\pm$22 & 106$\pm$53 & 2.5$\pm$0.9 & -2.9$\pm$1.1 \\ 
32 & A2744-SF8-081  & 3.3$\pm$1.1 & 0.7$\pm$0.5 & 126$\pm$42 & 396$\pm$125 & 88$\pm$17 & -0.0$\pm$0.6 & 68$\pm$16 & 31$\pm$18 & 71$\pm$39 & 1.9$\pm$0.8 & -3.1$\pm$1.1 \\ 
33 & M0416-MF4-019  & 1.5$\pm$0.9 & 1.1$\pm$0.4 & 128$\pm$19 & 410$\pm$112 & 155$\pm$8 & 0.1$\pm$0.1 & 55$\pm$12 & 16$\pm$10 & 36$\pm$22 & 3.8$\pm$0.5 & -3.3$\pm$1.0 \\ 
34 & M0416-DF-024  & 2.4$\pm$0.6 & 0.6$\pm$0.3 & 46$\pm$17 & 519$\pm$108 & -30$\pm$7 & 0.3$\pm$0.1 & 79$\pm$8 & 37$\pm$20 & 114$\pm$50 & 2.7$\pm$0.9 & -3.1$\pm$1.0 \\ 
35 & A2744-DF-148  & 2.3$\pm$0.8 & 1.5$\pm$0.4 & 32$\pm$41 & 519$\pm$102 & 71$\pm$10 & 0.4$\pm$0.1 & 62$\pm$19 & 44$\pm$19 & 94$\pm$43 & 4.1$\pm$1.0 & -3.2$\pm$1.1 \\ 
36 & AS1063-MF3-091  & 2.6$\pm$0.9 & 1.4$\pm$0.4 & 47$\pm$31 & 319$\pm$91 & 118$\pm$27 & -1.8$\pm$0.8 & 72$\pm$13 & 65$\pm$23 & 134$\pm$37 & 5.5$\pm$1.0 & -3.1$\pm$1.1 \\ 
37 & AS1063-MF4-022  & 2.3$\pm$1.2 & 1.0$\pm$0.5 & 73$\pm$41 & 357$\pm$138 & -108$\pm$16 & 0.3$\pm$0.2 & 74$\pm$16 & 38$\pm$24 & 101$\pm$55 & 1.8$\pm$0.8 & -3.2$\pm$1.1 \\ 
38 & AS1063-MF4-100  & 2.5$\pm$0.9 & 0.9$\pm$0.5 & 53$\pm$37 & 490$\pm$140 & -49$\pm$10 & 0.5$\pm$0.1 & 78$\pm$17 & 43$\pm$22 & 73$\pm$53 & 2.0$\pm$1.4 & -2.8$\pm$1.1 \\ 
39 & AS1063-MF3-026  & 3.0$\pm$0.8 & 0.4$\pm$0.3 & 64$\pm$28 & 372$\pm$133 & 171$\pm$11 & 0.1$\pm$0.8 & 82$\pm$9 & 40$\pm$22 & 109$\pm$46 & 0.7$\pm$0.8 & -3.2$\pm$1.1 \\ 
40 & AS1063-SF7-072  & 3.2$\pm$1.0 & 1.0$\pm$0.4 & 74$\pm$33 & 554$\pm$124 & -44$\pm$23 & -0.8$\pm$1.0 & 80$\pm$14 & 35$\pm$24 & 85$\pm$24 & 1.1$\pm$0.6 & -3.5$\pm$1.1 \\ 
41 & M0416-MF3-016  & 3.5$\pm$1.0 & 1.0$\pm$0.4 & 53$\pm$30 & 305$\pm$77 & -29$\pm$7 & 0.6$\pm$0.2 & 77$\pm$11 & 38$\pm$21 & 69$\pm$51 & 1.0$\pm$0.6 & -3.0$\pm$1.1 \\ 
42 & M0416-DF-140  & 3.4$\pm$1.0 & 1.2$\pm$0.4 & 100$\pm$43 & 482$\pm$75 & 116$\pm$21 & 0.3$\pm$0.6 & 71$\pm$12 & 46$\pm$20 & 120$\pm$41 & 3.0$\pm$0.9 & -3.0$\pm$1.1 \\ 
43 & M0416-MF2-106  & 3.0$\pm$1.0 & 0.5$\pm$0.4 & 69$\pm$35 & 334$\pm$104 & 105$\pm$10 & 0.5$\pm$0.8 & 80$\pm$11 & 37$\pm$21 & 123$\pm$52 & 0.9$\pm$0.7 & -3.3$\pm$1.1 \\ 
44 & M0416-DF-005  & 1.7$\pm$0.5 & 0.9$\pm$0.4 & 62$\pm$22 & 419$\pm$145 & 9$\pm$9 & 0.3$\pm$0.1 & 75$\pm$15 & 54$\pm$23 & 128$\pm$49 & 1.9$\pm$0.7 & -3.2$\pm$1.1 \\ 
45 & AS1063-DF-035  & 2.6$\pm$1.1 & 1.2$\pm$0.5 & 82$\pm$28 & 466$\pm$129 & 26$\pm$26 & 0.4$\pm$0.5 & 68$\pm$19 & 17$\pm$26 & 60$\pm$53 & 1.9$\pm$1.0 & -3.0$\pm$1.1 \\ 
46 & M0416-DF-068  & 2.4$\pm$1.4 & 0.8$\pm$0.4 & 111$\pm$49 & 403$\pm$131 & 4$\pm$21 & 0.4$\pm$0.3 & 68$\pm$22 & 32$\pm$25 & 117$\pm$56 & 1.5$\pm$0.7 & -2.8$\pm$1.1 \\ 
47 & A2744-SF1-062  & 3.6$\pm$0.8 & 2.0$\pm$0.3 & 35$\pm$29 & 503$\pm$44 & -69$\pm$13 & -0.1$\pm$0.2 & 54$\pm$12 & 56$\pm$12 & 85$\pm$27 & 7.6$\pm$2.2 & -3.4$\pm$1.0 \\  
\hline
\end{tabular}
\tablefoot{Galaxies are sorted by stellar mass in descending order. The first two columns are the same as in Table~\ref{tab:sample_table}. Columns (3) to (13) indicate the best-fitting value and associated error for each one of the parameters of the model described in Appendix~\ref{sec:model}. The best-fitting value and associated error correspond to the median and standard deviation of the posterior distribution computed by our MCMC fitting (see Appendix~\ref{sec:fitting_desc} for details).}
\label{tab:best_fit_par}
\end{table*}

\subsection{Distribution of best-fitting model parameters}
\label{sec:best_fitting_model_parameters}

In Sec.~\ref{sec:best_fitting_models}, we show some examples of the best-fitting model obtained for a subset of our sample galaxies. By fitting our outflow model to our entire sample, we can obtain, for the first time, a population-level distribution of galactic wind properties derived from the \ion{Mg}{II} emission halos, moving forward from reporting the discovery and modeling of individual objects. Figure~\ref{fig:best_fitting_parameter_distribution} shows the distribution of the best-fitting model parameters derived for our sample galaxies. The best-fitting parameters obtained for each individual galaxy, and their corresponding uncertainties, are provided in Table~\ref{tab:best_fit_par}.

The distribution of central optical depth $\tau_0$ is centered around high values of $2.5 \lesssim \log \tau_0 \lesssim 3.0$, meaning that our model predicts a completely optically thick medium in the inner shells of the CGM. We find a median and standard deviation of $\log \tau_0 = 2.6\pm0.6$. We can also use Eq.~\ref{eq:tau_0_n_0}, together with the best fitting values of $r_0$ and $v_0$ to infer the distribution of central \ion{Mg}{II} number density $n_{0,\ion{Mg}{II}}$, obtaining a median and standard deviation values of $\log n_{0,\ion{Mg}{II}} =  -6.3\pm0.7$ cm$^{-3}$. This number is lower than the value derived for the galaxy UDF884 in \citet{Pessa2024} of $-5.2\pm0.2$ cm$^{-3}$, although consistent within less than two standard deviations. However, translating this value into a gas density is not straightforward since it would require knowledge about the underlying gas metallicity, ionization correction, and level of dust depletion of the \ion{Mg}{II} ions.

The distribution of the index of the velocity power law $\gamma$ exhibits median and standard deviation values of $1.0\pm0.5$, being generally lower than two. The fact that most galaxies show $\gamma\sim1$ suggests a linear relation between radius and velocity, and as a consequence, a density that falls as $n\propto r^{-3}$. However, a wind velocity that increases with radius does not unambiguously imply an accelerating wind. If there is a distribution of velocities for the gas leaving the ISM, gas with higher velocities will tend to dominate at larger distances, even if the wind is not accelerating or decelerating at all radii \citep[see also hydrodynamical simulations of SNe powered outflows from ][]{DallaVecchia2008}. We caution the reader about this possible degeneracy in order to avoid an overinterpretation of our modeling results.

A density that decreases as $n\propto r^{-3}$ represents a steeper decrease with radius than that estimated from absorption line studies, which find a slope closer to  $n\propto r^{-2}$ ({Bouch{\'e} et al. in preparation). While this result, taken at face value, might seem like a discrepancy, it can be explained by the fact that here we are only accounting for the currently outflowing gas that dominates the inner CGM. Additional gas in the CGM that is not outflowing, or that is connected to past outflow events, would not contribute to the inferred density profile. Thus, it is not unexpected that the density profile inferred from absorption line studies that account for all the absorbing gas present at some impact parameter from the central galaxy (regardless of their origin) is flatter than the one derived here. This additional contribution of non-outflowing gas would produce a profile that decreases more slowly than the one accounting for outflowing gas only.

The launching velocity of the winds is more widely distributed, with a median and standard deviation value of $62\pm32$ km s$^{-1}$. This velocity increases radially up to maximum velocities of $v_\mathrm{max} = 490\pm95$ km s$^{-1}$. These results are consistent with predictions from three-dimensional hydrodynamic simulations of supernovae-driven galactic winds \cite[see, e.g.,][]{Fielding2017}.

A maximum velocity of $\sim490$ km s$^{-1}$ is significantly higher than the outflow velocities derived from absorption line studies \citep[see, e.g.,][]{Schroetter:2016bl, Schroetter:2019es, Schroetter2024, Bouche2025}. \citet{Bouche2025} report outflow velocities that are generally closer to $\sim200$ km s$^{-1}$, even in sightlines at impact parameters comparable with the size of the modeled regions in this work. However, despite this apparent disagreement, one must consider the nature of each measurement. While here, we infer the maximum radial velocity of the gas; absorption-line studies measure the line-of-sight projected velocity, averaged along a sight line at a specific impact parameter. This measurement has to be lower than our inferred maximum velocity, on one hand, because it only accounts for the component of the velocity parallel to the line of sight, and on the other, because it is an average (weighted by column density) along the line of sight, instead of strictly a maximum value.

The velocity offset $\Delta v$ is distributed around relatively small offsets, with a median and standard deviation value of $56\pm112$ km s$^{-1}$. This offset is shown, for instance, in the innermost apertures of Fig.~\ref{fig:AS1063_MF02_011_Master_figure}, where the rest-frame velocity of the \ion{Mg}{II} doublet coincides with the absorption component of the P-Cygni profile, rather than with the peak of the emission. This offset could be explained either by the simplicity of the outflow model kinematics not being able to capture the actual kinematics of the galactic winds, as well as by an inaccurate redshift measurement that has been computed via a (supervised) template matching, rather than by modeling the continuum and/or emission lines of the spectra. In that line, we estimate that the accuracy of the redshift determination should be generally within $\sim50$ km s$^{-1}$.

The additional nebular emission contribution, parametrized by the parameter $\log$ $f_{C}$, presents a mildly bimodal distribution, where some galaxies essentially do not present any significant level of nebular emission (low $f_{C}$). However, in most of them, some nebular emission is required to reproduce the observations.

Regarding the geometry of the outflows, we find mostly outflows with wide opening angles, with a median opening angle of $68^\circ \pm 17^\circ$. These wide opening angles are, at least in part, a selection effect. Since by construction, our sample is composed of galaxies that exhibit both \ion{Mg}{II} emission and absorption, and a wide opening angle makes it more likely to produce this spectral feature.

For the rotation angle in the direction of the sightline R.A., we find predominantly outflows pointing toward the observer, with a median and standard deviation of $51^\circ \pm 18^\circ$ (R.A. = $90^{\circ}$ represents a face-on outflow). We only consider those outflows with an O.A. $< 75^{\circ}$, since in the isotropic cases, where O.A. $\sim 90^{\circ}$, the definition of R.A. is meaningless. Similarly to the O.A., this is also, at least partially, a selection effect, since outflows pointing toward the observer are more likely to exhibit a P-Cygni profile and, thus, be a part of our sample. Outflows that lie nearly perpendicular to the line of sight would only be able to produce an absorption if they also have a very broad opening angle. Otherwise, they would be seen only in emissions.

For the distribution of the position angle in the plane of the sky, P.A., we consider only those cases that are not isotropic (O.A.  $< 75^{\circ}$) and that are not essentially face-on outflows (R.A.  $< 75^{\circ}$), because in both of these cases, the P.A. is essentially a physically meaningless quantity. For the remaining galaxies, although our statistics are low, we find a distribution without a strongly preferred orientation, consistent with what one would expect from the intrinsically random orientation of the outflow in the plane of the sky.

For the launching radii of the wind, as discussed in Appendix~\ref{sec:fitting_desc}, we use as prior the effective radius of each galaxy, measured with \textsc{GALFIT}, so it is not an entirely independent measured quantity. Nevertheless, the $r_0$ values measured are consistent with the expected sizes of star-forming galaxies at $1.0 \lesssim z \lesssim 2.0$ for the stellar mass range of our sample \citep[see, e.g.,][]{VanDerWel2014, Mowla2019, Nedkova2021}.

Finally, the parameter $f$, which scales the additional variance contribution, is generally small, with values mainly in the range $\log f < -2.5$, indicating that a significant additional variance is not required to model our data.

We note that while the MUSE datacubes are extremely information-rich, some covariance between model parameters in the posterior distribution is present. These parameter covariances are discussed in greater detail in \citet{Pessa2024}. For instance, we find a negative covariance between $\gamma$ and $v_0$, indicating that winds launched with higher initial velocities require a smaller radial increase in velocity to reproduce the data. Similarly, $\Delta v$ shows covariance with both $\gamma$ and $v_0$, since the velocity offset is closely linked to the adopted velocity law. Along the same lines, because the optical depth gradient is directly related to the velocity gradient through the assumption of mass conservation, $\tau_0$ exhibits covariance with the parameters that govern the velocity (and hence optical-depth) gradient, namely $\gamma$ and $v_0$.

Nevertheless, while moderate parameter covariances are present, the model parameters do not exhibit strong degeneracies, in the sense that the model remains identifiable and distinct parameter combinations do not lead to statistically indistinguishable solutions. A multi-modality in the posterior distribution of the model parameters could also point to a degeneracy between the parameters, however, we do not observe this behavior in the obtained posterior distributions. Accordingly, the posterior distributions generally display well-defined modes, indicating that the parameters are well constrained, with the widths of the posterior distributions quantifying the associated uncertainties.

\begin{figure*}
\centering
\includegraphics[width=0.8\textwidth]{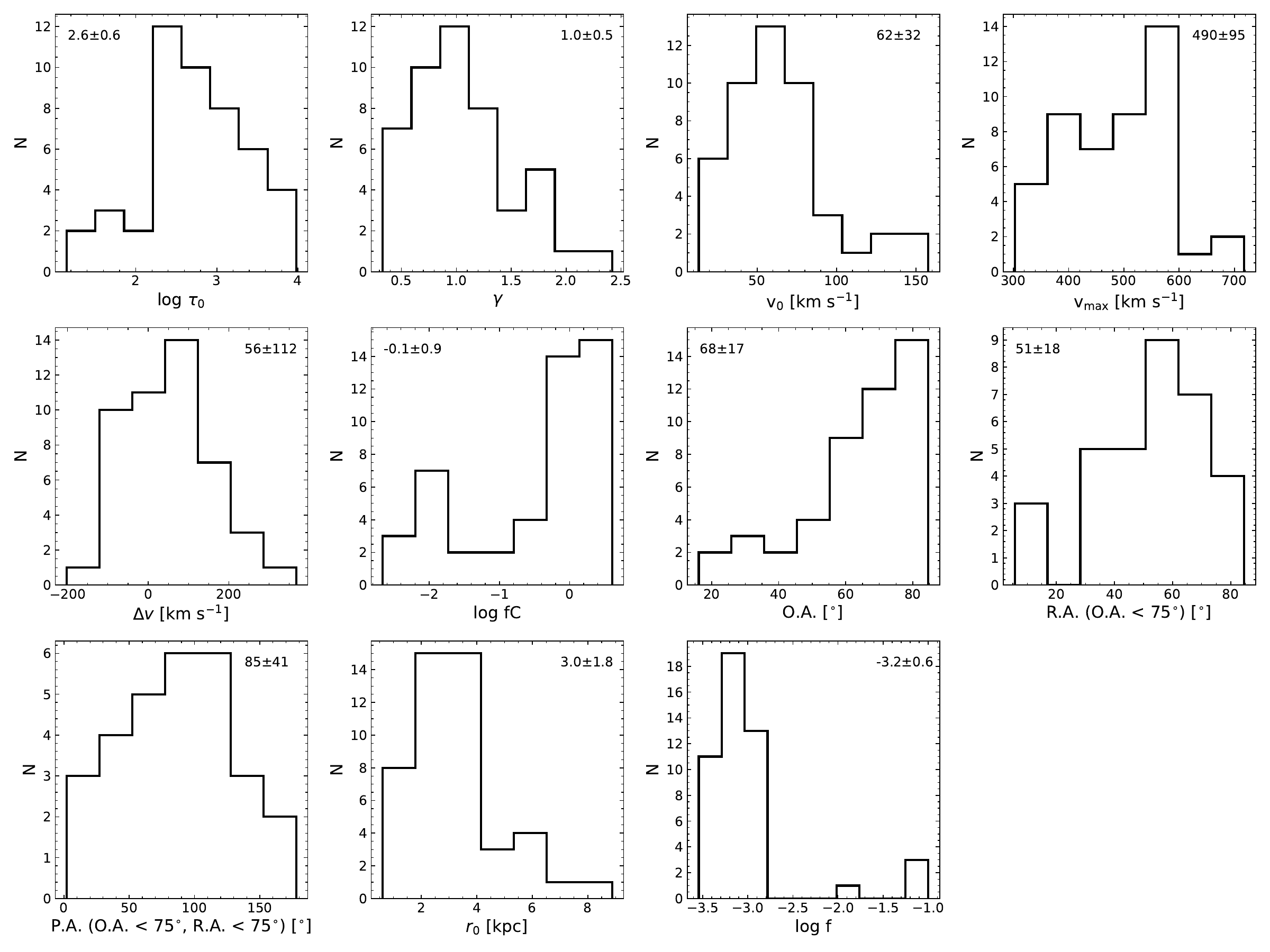}
\caption{Distribution of the best-fitting model parameters for our sample galaxies. The parameters of the model are described in detail in Appendix~\ref{sec:model}. For R.A., we only include those galaxies with non-isotropic outflows (O.A. $< 75^{\circ}$), and for P.A., we consider only galaxies with non-isotropic outflows (O.A. $< 75^{\circ}$) that are also not seen nearly face on (R.A. $< 75^{\circ}$). The median and standard deviation of each distribution are indicated in each corresponding panel. The best-fitting parameters and their uncertainties obtained for each galaxy are provided in Table~\ref{tab:best_fit_par}.}
\label{fig:best_fitting_parameter_distribution}
\end{figure*}

\FloatBarrier
\section{Discussion}
\label{sec:discussion}

\subsection{Prevalence of cool-gas outflows traced by their extended \ion{Mg}{II} emission}

In Sec.~\ref{sec:paren_sample_comparison} we show that, relative to our parent sample, galaxies in our final sample have higher SFR, higher sSFR, and younger age bins (see Fig.~\ref{fig:parent_sample_comparison}). While this seems like an intuitive result, from the perspective of stellar-feedback driven outflows, there is no clear consensus in the literature on how the presence of galactic outflows depends on the galaxy properties.

Our results contrast with those of \citet{Rubin2014}, who find a much higher fraction of galaxies with cool-gas outflows ($66\pm5\%$), traced by the presence of \ion{Mg}{II} absorption in integrated spectra of 105 star-forming galaxies. In their study, the outflow detection rate primarily depends on galaxy inclination (with higher fractions in face-on galaxies) and shows little dependence on intrinsic galaxy properties. This difference could be due to the fact that we focus on outflows traced by extended emission rather than absorption. Outflows detected in \ion{Mg}{II} absorption may trace different (and more common) gas structures than those that produce extended emission halos. The latter may require the outflowing material to escape to circumgalactic scales, rather than simply being launched, potentially linking extended emission to higher SFR/sSFR/younger stellar populations. In support of this, \citet{Rubin2014} also report that the inferred strength and velocity of cool-gas outflows are correlated with host galaxy properties, which aligns with our findings.

Additionally, the stellar mass and SFR range of the \citet{Rubin2014} sample overlaps mainly with the high end of our distribution ($\log M_{*} > 9.8$ $M_{\odot}$ and $\log \mathrm{SFR} \gtrsim 0$ $M_{\odot},\mathrm{yr}^{-1}$). When we include galaxies in our sample that show \ion{Mg}{II} only in absorption and restrict our analysis to the same stellar mass and star formation ranges as \citet{Rubin2014}, we find that any dependence on intrinsic galaxy properties largely disappears.

In a more recent work, \citet{Das2025} search for galaxies associated with intervening \ion{Mg}{II} absorbers over a redshift range of $0.4 < z < 1.0$. By identifying $\sim$270 \ion{Mg}{II}–galaxy systems, they find that only a small fraction ($\sim10\%$) are associated with suppressed star formation (log(sSFR) $< -10.6,\mathrm{yr}^{-1}$), implying a strong dependence of \ion{Mg}{II} absorption on intrinsic galaxy properties, consistent with our results. This is further supported by several studies that find stronger cool-gas absorbers associated with star-forming galaxies than with passive ones \citep[e.g.,][]{Bordoloi2014, Harvey2025}.

In this line, \citet{Schroetter2024} use data from the MusE GAs FLOw and Wind survey \citep[MEGAFLOW,][]{Schroetter:2016bl, Bouche2025} to explore correlations between the inferred galatic outflows properties probed by background quasars and host galaxy properties, and find that the velocity of the outflows is correlated with the host galaxy SFR and SFR surface density. Furthermore, they report a distinction between strong and weak outflows, based on the wind momentum compared to the momentum injection rate from SFR \citep[following the formalism from ][]{Heckman:2000du}, where strong outflows show a tighter correlation with galaxy properties.

Lastly, as detailed in Sec.~\ref{sec:paren_sample_comparison}, the fraction of galaxies with galactic-scale outflows in our sample not only depends on galaxy properties, but also on the depth of the data, varying significantly between the deep and shallow levels of MUSCATEL, especially for the highest SFR galaxies. In Appendix~\ref{sec:frac_depth_levels} we show the prevalence of galactic-scale outflows traced by \ion{Mg}{II} for the different depth levels of the MUSCATEL survey (i.e., SF, MF, and DF). Figure~\ref{fig:SFR_vs_depth} shows that the prevalence of outflows for high-SFR galaxies is significantly higher for galaxies drawn from the DF and MF, compared to the SF. This implies that galactic outflows are probably a much more common feature among high-SFR galaxies than what Fig.~\ref{fig:parent_sample_comparison} could initially suggest.

Overall, our findings are broadly consistent with previous work. However, it is important to keep in mind that galaxy populations exhibiting outflows traced in CGM emission and absorption may not be directly comparable, particularly when different stellar mass or star formation rate ranges are involved.

\subsection{Correlation between host galaxy properties and outflow parameters}

In this section, we show how the parameters of our best-fitting outflow model for each galaxy correlate with the host galaxy properties, specifically, their stellar mass, SFR, specific SFR, and SFR surface density.

Figure~\ref{fig:parameters_vs_mass_corr} shows the derived outflow parameters as a function of the stellar mass of each galaxy. Overall, there are no clear correlations with stellar mass in our sample, meaning that the mechanisms that set the outflow properties are likely too complex to be described by a single quantity.

However, there are some tentative trends between stellar mass and certain model parameters. For instance, more massive galaxies seem to more frequently (although not exclusively) show higher acceleration rates ($\gamma$), and their outflows reach, in general, higher maximum velocities. The Pearson correlation coefficient $\rho$ for this trend between v$_\mathrm{max}$ and stellar mass is 0.43, generally interpreted as a weak to moderate correlation. A correlation between outflow velocity and stellar mass is also consistent with the results from \citet{Rubin2014}, where the authors find a similar dependency, inferring the outflow velocity from the \ion{Mg}{II} absorption line profile in the integrated spectra of galaxies.

The nebular emission contribution appears to correlate with halo size ($\rho \approx -0.66$), rather than with galaxy stellar mass, as more compact halos generally exhibit a higher nebular emission contribution (although galaxies with stronger nebular emission contributions seem to be predominantly low-mass galaxies). Nebular emission is, by construction, concentrated in the central region, following the continuum emission. Thus, it is a self-consistent result that galaxies with relatively stronger nebular emission (higher $f_{C}$) also tend to be more compact.

Regarding the geometry of the outflows, we find that while less massive galaxies exhibit primarily isotropic outflows, more massive galaxies show a wider variety, with both low and high opening angles. In principle, this could be explained as the outflowing material escaping more easily through the ISM of less massive galaxies, compared to the most massive ones. Interestingly, A similar trend was reported by \citet{Nelson2021} for simulated \ion{Mg}{II} halos, where the authors find axis ratios around one (i.e., nearly isotropic emission) for low-mass galaxies, and increasing values, with higher scatter for the more massive ones.

Finally, the central optical depth $\tau_0$ does not show a priori any sign of correlation with the host galaxy properties investigated here. However, we note that for several of our sample galaxies, $\tau_0$ is poorly constrained, and this is reflected in the error bars shown in the corresponding panel of Fig.~\ref{fig:parameters_vs_mass_corr}. To explore whether these poorly constrained measurements could be hindering the identification of a trend, we have split our sample according to the uncertainty in their derived $\tau_0$. Those galaxies with a $\tau_0$ uncertainty lower than the median $\tau_0$ uncertainty of the full sample ($\sim0.6$ dex) represent the subset of galaxies (half of the sample) with a better constrained $\tau_0$, and are marked with blue squares in the corresponding panels of Fig.~\ref{fig:parameters_vs_mass_corr}. If we restrict our analysis to these galaxies, we then identify a tentative trend between stellar mass and $\tau_0$, where more massive galaxies appear to exhibit higher central optical depths ($\rho = 0.39$, interpreted as a moderate correlation).

A higher optical depth in more massive galaxies could be, at face value, interpreted as more massive galaxies having a more massive and denser CGM. However, there are additional relevant factors that would play a significant role in the relation between $\tau_0$ and the actual density of the outflow. The metallicity, dust depletion, and ionization correction will have a direct impact on the \ion{Mg}{II} optical depth, at a given gas density \citep[see, e.g.,][]{Martin:2013ho}. Furthermore, these mechanisms are also likely correlated with stellar mass, for instance, through the mass vs. gas-phase metallicity correlation \citep{Tremonti2004}. Thus, a higher optical depth in a more massive galaxy could also be qualitatively explained by a higher metallicity, leading to a higher relative number of \ion{Mg}{II} ions at a given gas density.

Nevertheless, keeping only those galaxies with a relatively more robust constraint on $\tau_0$ comes with the clear caveat of dropping mostly galaxies on the low-mass end of the distribution. It could be the case that either the higher uncertainty of $\tau_0$ in low-masss galaxies hinders the identification of a trend between stellar mass and central optical depth in the full sample, or that due to some astrophysical mechanism related, for instance, to the interplay between the strength of galactic winds and the gravitational potential of galaxies, low-mass galaxies do not follow the same trend as higher-mass galaxies. Unfortunately, with our current data and analyses, we can not exclude any of those scenarios.

None of the other model parameters show a trend with galaxy stellar mass, except for $r_0$, but this parameter is, by construction, tied to the continuum size of each galaxy. Therefore, a correlation between $r_0$ and stellar mass is not unexpected.

Figure~\ref{fig:parameters_vs_ssfr_corr} shows the correlations of the outflow parameters with the sSFR of our sample galaxies. The trends with sSFR do not show any clear correlation. For SFR (not shown here), we observe trends similar to those seen with stellar mass, albeit with higher scatter. 

The star formation rate surface density ($\Sigma_{\rm SFR}$) is also a relevant quantity in the context of SF-driven outflows. \citet{Heckman2015} find that the velocity of outflows measured from absorption lines in a sample of starburst galaxies correlates with $\Sigma_{\rm SFR}$. In the same line, \citet{Verhamme2017} find that high $\Sigma_{\rm SFR}$ values promote low-density channels that facilitate the escape of Lyman continuum photons. Thus, we also explore possible correlations between the outflow properties and $\Sigma_{\rm SFR}$, using SFR/$\pi$HLR$_{\rm cont}^2$ as a tracer of $\Sigma_{\rm SFR}$. Figure~\ref{fig:parameters_vs_SigmaSFR_corr} shows these trends. However, there are no clear correlations. Perhaps similar trends to those observed with stellar mass, although with a higher scatter (similar to that observed with SFR). An intriguing aspect is that we find a broad range of $\Sigma_{\rm SFR}$ for our sample galaxies, between $\log \Sigma_{\rm SFR}\sim -4$ and $\log \Sigma_{\rm SFR}\sim 1$. Some previous works have suggested minimum $\Sigma_{\rm SFR}$ values required to launch strong galactic-scale outflows on the order of $\log \Sigma_{\rm SFR}\sim -1$ \citep{Heckman2015} and $\log \Sigma_{\rm SFR}\sim -2$ \citep{ReichardtChu2025}. On the other hand, while most ($\sim83\%$) of our sample galaxies exhibit $\log \Sigma_{\rm SFR}\gtrsim -2$, this is not strictly always the case. Naturally, this comparison comes with the caveat that the $\Sigma_{\rm SFR}$ of each galaxy is calculated globally, which does not rule out that locally, around the outflowing star-forming regions, $\Sigma_{\rm SFR}$ could reach significantly higher values. Local physical conditions (e.g., ISM density,  $\Sigma_{\rm SFR}$) are likely a key driver of star-forming driven outflows, besides global galactic properties.

In principle, it would be expected that the outflow properties correlate more closely with SFR, sSFR, and $\Sigma_{\rm SFR}$) than with stellar mass, since these quantities are more directly related to the amount of energy injected into the ISM by stellar feedback, powering the outflows. Furthermore, in Sec.~\ref{sec:paren_sample_comparison}, we show that our sample galaxies tend to be preferentially at higher SFR and sSFR values, compared to our parent sample. Nevertheless, several possibilities exist that could explain this higher scatter. First of all, the large systematic uncertainties in the SFR determination that arise, for instance, from the assumption on the shape of the SFH of galaxies \citep[see, e.g.,][]{Wang2024}. And while our SFR estimates are generally consistent with our independent measurement from the [\ion{O}{II}] emission line, the latter is also subject to relevant systematic uncertainties, as it is a single-line-based measurement, such as non-star-formation ionization or dust reddening. 

Finally, given the kinematics of the gas predicted by our modeling scheme, the timescales involved in the traveling of the outflowing gas from the launching radius into the observed \ion{Mg}{II} halos are of the order of hundreds of megayears \citep[][]{Pessa2024}. This means that the present-day SFR of a galaxy might not be directly correlated with the outflow properties that we see today. Instead, they would be more closely connected to past episodes of star formation. However, past bursts of SFR would not be properly quantified either by the SFR measured from the SED fitting (which assumes an underlying smooth SFH) or by the [\ion{O}{II}] based SFR measurement (which probes SFR in much shorter timescales of $<10$ Myr). In this line, in a recent work, \citet{Harvey2025} stack the \ion{Mg}{II} absorption lines in the spectra of background quasars for a sample of qusiescent, star-forming, and post-starburst galaxies, and find that post-starburst galaxies exhibit significantly stronger \ion{Mg}{II} absorption within 1 Mpc than star-forming or quiescent galaxies of the same stellar mass. In a future paper, we will aim to link the outflow properties to the recent SFH of galaxies, exploring if it is possible to connect features of the \ion{Mg}{II} halos with a specific star formation event.

\begin{figure*}
\centering
\includegraphics[width=0.8\textwidth]{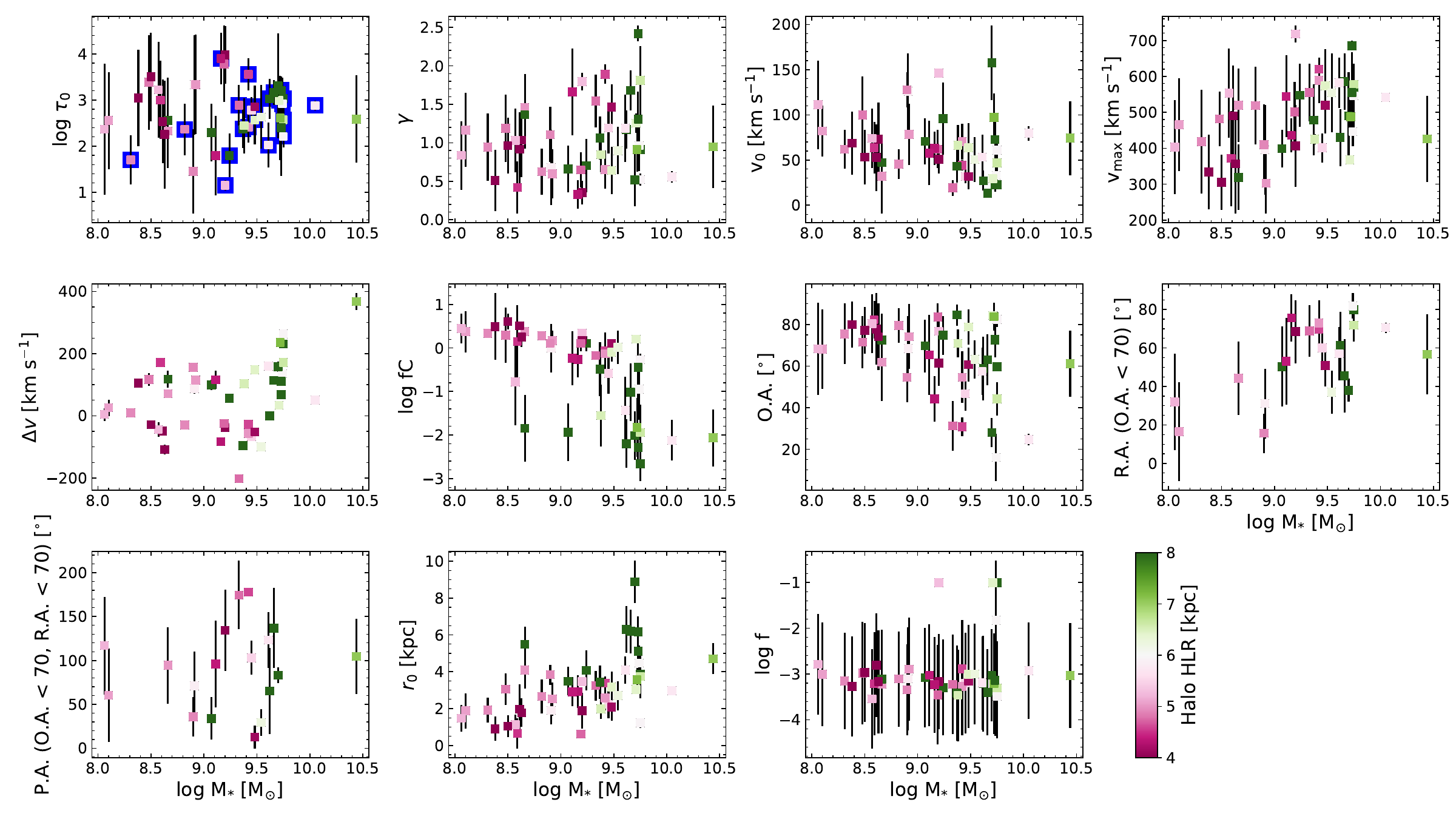}
\caption{Correlation of the best-fitting model parameters obtained with our outflow modeling scheme (see Appendix~\ref{sec:model}) with the stellar mass of each galaxy, derived via SED fitting (see Sec.~\ref{sec:photometry}). The plots are color-coded by the size of the \ion{Mg}{II} halo of each galaxy, reported in Sec.~\ref{sec:reconstruction}, with green colors indicating halos in the extended regime of the distribution, and pink colors indicating more compact halos. The blue squares in the top left panel indicate galaxies with a more robustly constrained $\tau_0$. For R.A., we only include those galaxies with non-isotropic outflows (O.A. $< 75^{\circ}$), and for P.A., we consider only galaxies with non-isotropic outflows (O.A. $< 75^{\circ}$) that are also not seen nearly face on (R.A. $< 75^{\circ}$).}
\label{fig:parameters_vs_mass_corr}
\end{figure*}

\begin{figure*}
\centering
\includegraphics[width=0.8\textwidth]{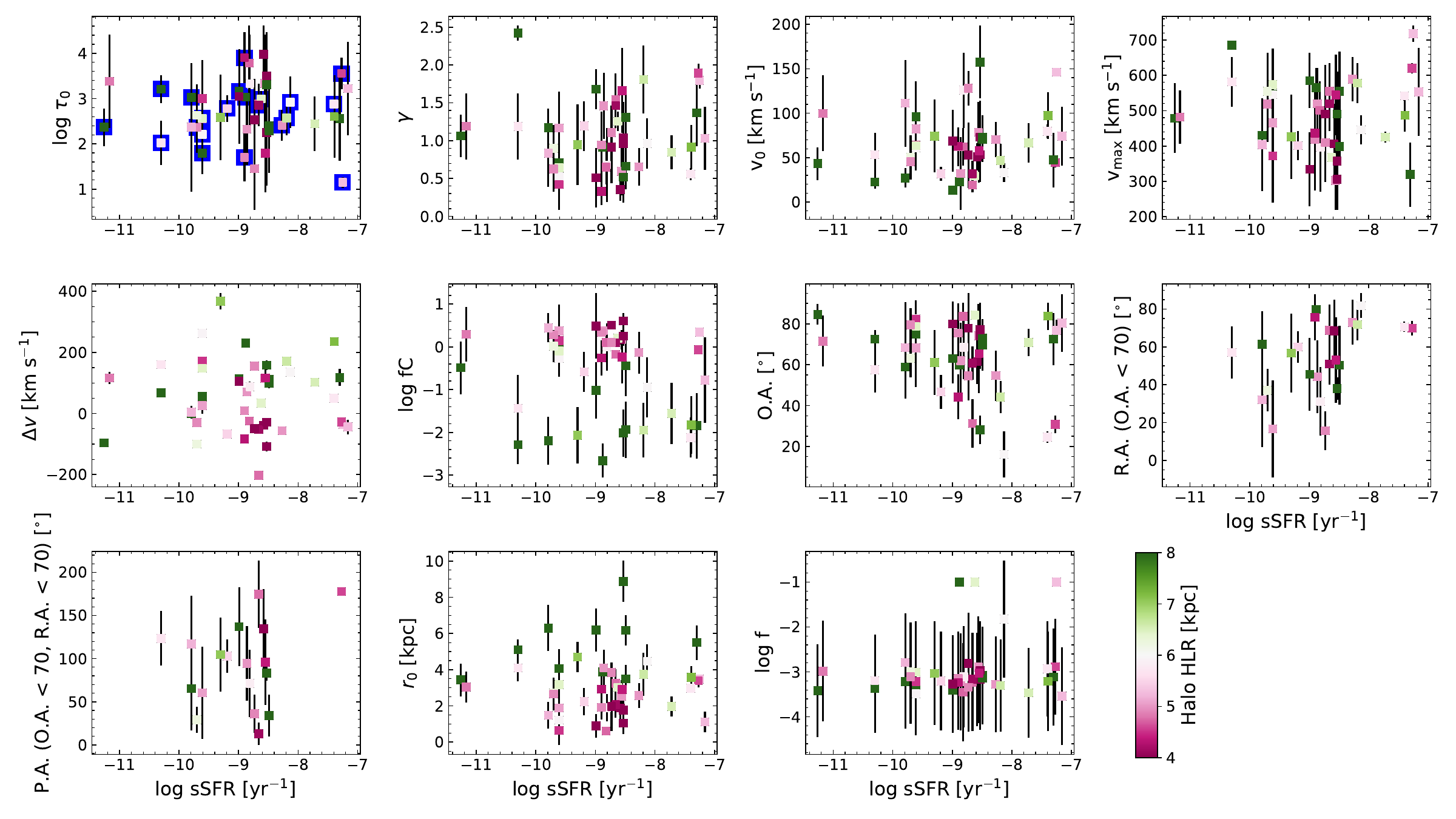}
\caption{Same as Fig.~\ref{fig:parameters_vs_mass_corr}, for the sSFR derived via SED fitting for our sample galaxies.}
\label{fig:parameters_vs_ssfr_corr}
\end{figure*}

\begin{figure*}
\centering
\includegraphics[width=0.8\textwidth]{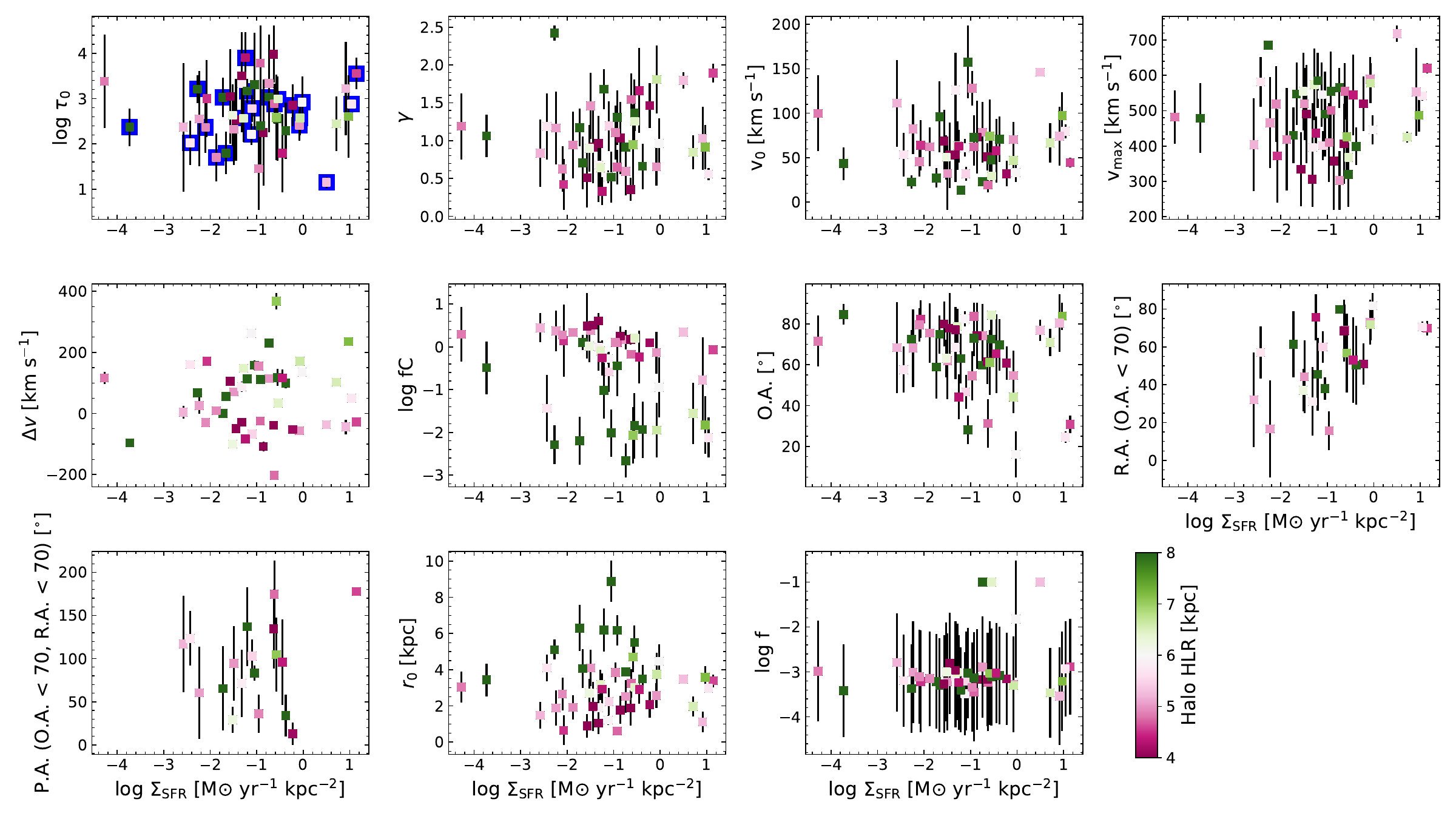}
\caption{Same as Fig.~\ref{fig:parameters_vs_mass_corr}, for the $\Sigma_{\rm SFR}$ estimated for our sample galaxies.}
\label{fig:parameters_vs_SigmaSFR_corr}
\end{figure*}

\subsection{Drivers of the size of the \ion{Mg}{II} halos}
\label{sec:sizes}

In Sec.~\ref{sec:reconstruction}, we have described how the size of the \ion{Mg}{II} halos is measured and shown the distribution of their HLRs. In this section, we explore for correlations between these sizes and properties of the host galaxies.

The left panel of Fig.~\ref{fig:halo_size_and_ratio_versus_mass} presents the HLRs of the \ion{Mg}{II} emission halos as a function of each galaxy's stellar mass. While there isn't any clear correlation between the HLRs and stellar masses (also not with SFRs, not shown here), it could be the case that the size of the more compact \ion{Mg}{II} halos (HLR $< 8$ kpc) correlates with the stellar mass of the host galaxy (see inset in left panel of Fig.~\ref{fig:halo_size_and_ratio_versus_mass}). Indeed, the Pearson correlation coefficient between HLR and stellar mass, for galaxies with HLRs smaller than 8 kpc is $\rho \approx 0.62$,  interpreted as a significant correlation.

This could potentially point out that while the \ion{Mg}{II} emission closer to the galaxy is likely driven by mechanisms that correlate with stellar mass (e.g., outflows), the \ion{Mg}{II} emission of more extended halos could also be powered by additional mechanisms, such as inflows, or the presence of satellites, that do not correlate directly with stellar mass. This would be consistent with our findings from \citet{Pessa2024}, where we find larger differences between the observed \ion{Mg}{II} emission and the prediction from our outflow-only model toward larger galactocentric distances for a galaxy in the MHUDF mosaic, potentially indicating the presence of additional mechanisms shaping the CGM of galaxies at larger spatial scales.

Furthermore, some of the galaxies with the most extended halos are actually not isolated systems. The objects AS1063-MF3-004 and AS1063-MF3-091 exhibit large (HLR $> 10$ kpc) \ion{Mg}{II} halos, and they are a likely gravitationally interacting system (see discussion in Sec.~\ref{sec:interacting_system}). The object AS1063-MF3-009 also presents an extended halo, and a relatively nearby galaxy ($\sim20$ kpc in projected distance and a velocity offset of $\sim1000$ km s$^{-1}$). In this case, the potential companion does not show a P-Cygni profile, and thus, it is not part of our sample. The object A2744-SF1-062, for which we could not unambiguously determine a HST counterpart is also a possible ongoing merge, and exhibits a relatively large HLR of 8.6 kpc, although we can not confirm this because the system is unresolved in the MUSE data, but it does present a disturbed morphology in the HST images.

This is probably not the only driver of the most extended halos, as the rest of the halos with HLR > $8$ kpc do not show clear evidence of companion galaxies, given our detection limits, although it does suggest that gravitational interaction between galaxies can be a relevant factor setting the observed properties of the \ion{Mg}{II} halos.

However, the fact that we see mostly compact halos (HLR$\lesssim5$ kpc) for galaxies on the low-mass end of the stellar mass distribution could also be (at least partially) caused by a selection effect. As pointed out in Sec.~\ref{sec:sample_selection}, due to unavoidable selection effects, we could be missing the fainter and more extended halos, since they are intrinsically harder to detect. 

We also explore trends between galaxy properties and the relative size of the halo with respect to the stellar counterpart. The right panel of Figure~\ref{fig:halo_size_and_ratio_versus_mass} shows the ratio between the halo size and the stellar size (i.e., continuum size) as a function of the stellar mass of our sample galaxies. 

We find a wide range of values for this ratio, with halos being up to $\sim7$ times more extended than their stellar counterpart. However, we do not see any trend between the size ratio and stellar mass (nor for SFR, not shown here), implying that differences in size with respect to the stellar counterpart are not driven by mechanisms that correlate with stellar mass.

\citet{Leclercq:2017cp} perform a similar measurement, for Ly$\alpha$ halos, also probing neutral gas halos, and find that the size of gaseous halos scales roughly linearly with the size of the stellar counterpart, being on average a factor 10 larger. In contrast, Fig.~\ref{fig:halo_size_and_ratio_versus_mass} shows that \ion{Mg}{II} halos are on average a factor of about $\sim2.5$ larger than their stellar size (albeit with large scatter, and with several halos significantly more extended than this). It is not unexpected that halos traced by their \ion{Mg}{II} emission are significantly smaller than those observed via their Ly$\alpha$ emission. Essentially, the lower optical depth of \ion{Mg}{II} means that for a given density gradient, the CGM will become optically thin for \ion{Mg}{II} earlier (at a lower galactocentric distance) than for Ly$\alpha$ photons, leading to more compact halos.

Lastly, for some galaxies the HLR of their \ion{Mg}{II} emission is consistent with that of their stellar counterpart (i.e., a ratio of $\sim1$ in the right panel of Fig.~\ref{fig:halo_size_and_ratio_versus_mass}). However, as detailed in Sec.~\ref{sec:reconstruction}, this does not necessarily mean that the \ion{Mg}{II} emission distribution is consistent with that from the stellar counterpart. This is because the computation of the HLR involves azimuthally averaging the \ion{Mg}{II} emission, whereas some halos show elongated shapes, and their outer areas would not significantly contribute to the azimuthally averaged profile. Furthermore, since by construction, we selected galaxies that exhibit central absorption of \ion{Mg}{II}, this implies that the spatial distribution of the observed \ion{Mg}{II} emission does not follow that of the continuum emission for any of our sample galaxies. For instance, for one of the most compact \ion{Mg}{II} emitters in our sample (M0416-MF2-106, HLR $\sim3.6$ kpc), the corresponding panels of Fig.~\ref{fig:master_11} show that while the central apertures show both, \ion{Mg}{II} absorption and emission, beyond $0\farcs8$ (yellow spectrum), there is essentially no continuum emission, but still clear \ion{Mg}{II} emission. Thus, by construction, the spatial distribution of the \ion{Mg}{II} emission will not be consistent with a scaled version of the stellar counterpart, for any of our sample galaxies, including the more compact ones.

\begin{figure}
\centering
\includegraphics[width=\columnwidth]{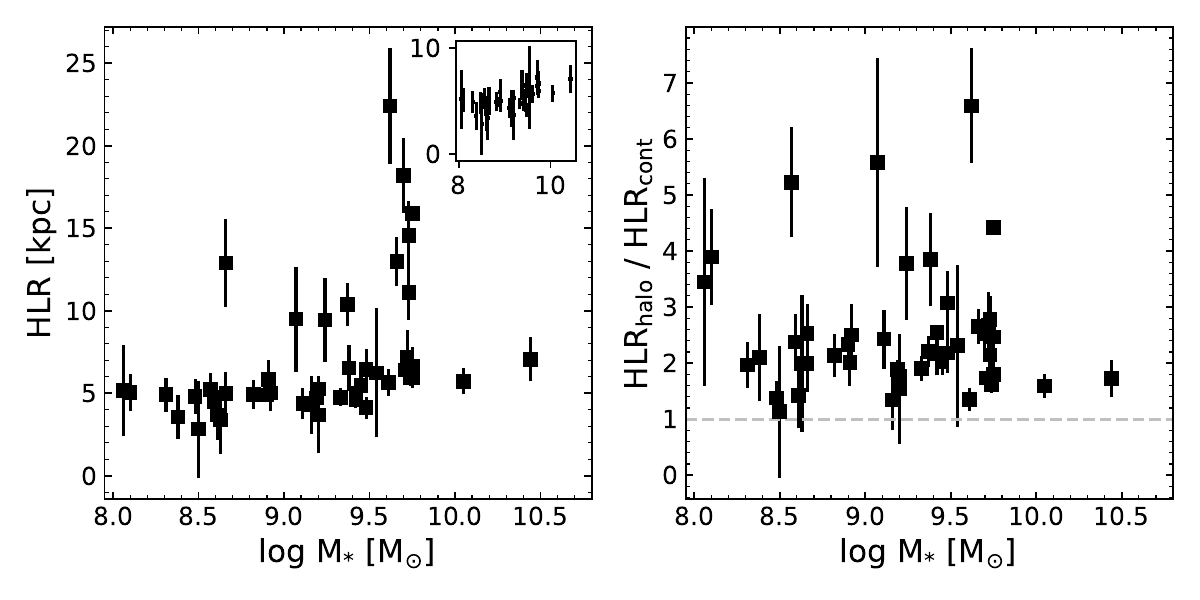}
\caption{Size of the \ion{Mg}{II} emission halos, after correcting by self-absorption as described in Sec.~\ref{sec:reconstruction}. The left panel shows the sizes of the \ion{Mg}{II} emission halos as a function of the stellar mass of each galaxy, derived via SED fitting. The inset in the left panel shows a zoom-in to the HLR vs. stellar mass relation for only those galaxies with HLR $<8$ kpc. The right panel shows the ratio between the size of the \ion{Mg}{II} emission halos and the size of the stellar component, as a function of the stellar mass of our sample galaxies. The gray horizontal dashed line shows the ratio of one for reference.}
\label{fig:halo_size_and_ratio_versus_mass}
\end{figure}

\FloatBarrier
\subsection{Cases that potentially deviate from a simple outflow model}
\label{sec:deviation}
We find that the MUSE data from most of our sample galaxies can be well reproduced by a simple outflow model. However, the kinematics and morphology of the \ion{Mg}{II} halos will always be more complex than the representation that the model provides, firstly because, outflows are not the only mechanisms at play in the CGM of galaxies, and secondly, because outflows themselves are much more complex than our implementation, in terms of kinematics, small-scale structure, etc \citep[see extended discussion about this in][]{Pessa2024}.

In Appendix~\ref{sec:particular_cases}, we discuss in detail some particular cases that could potentially deviate from a simple outflow model. One of them is a seemingly gravitationally interacting system, where tidal forces could be actively shaping the properties of the \ion{Mg}{II} emission halo. Another is a galaxy with a highly complex \ion{Mg}{II} spectral profile, for which the outflow model performs poorly but may instead reflect a combination of inflow and outflow occurring simultaneously. Lastly, we discuss two galaxies whose \ion{Mg}{II} emission toward large impact parameters is notoriously blueshifted with respect to the emission closer to the galaxy, and with respect to the expectations from our simple outflow scenario, pointing to the presence of more complex physics shaping the emission from their \ion{Mg}{II} halo.

\subsection{Outflow vs. galaxy orientation}

A relevant aspect of the model is the geometry and orientation of the outflow, parameterized by three angles (O.A., R.A., and P.A., see Appendix~\ref{sec:model}). In this section, we compare the outflow orientation inferred from the \ion{Mg}{II} emission with that derived for the stellar counterpart of the same galaxies.

We use \textsc{GALFIT} to fit a 2D S\'ersic model to the HST F814W image of each galaxy, which is dominated by stellar continuum emission, and measure their spatial orientation. Specifically, \textsc{GALFIT} provides the semi-axis ratio ($q$) and position angle of the S\'ersic model. 

From $q$, we calculate the inclination with respect to the line of sight as:
\begin{equation}
    I = 90 - arccosine(q)
\end{equation}

With $I$ being the inclination of the galaxy in degrees, such that for $q = 1$ (face-on disk), $I = 90^{\circ}$, and for $q \sim 0$ (edge-on disk), $I \approx 0^{\circ}$, consistent with the definition of R.A. in our model.

We compare the inclination inferred for each galaxy with its corresponding R.A.. For perfectly polar outflows, the inclination should be roughly consistent with the R.A.. We have excluded nearly isotropic outflows (O.A. > $75^{\circ}$), as well as galaxies where $q$ was too poorly constrained due to their irregular geometry.

Figure~\ref{fig:geometry_comparison} shows R.A. as a function of the inclination inferred for the stellar counterpart. We do not find a systematic agreement between them, and we discuss several possible reasons for this.

First of all, the formal uncertainties of the galaxy inclinations provided by \textsc{GALFIT} are small ($<1^{\circ}$) and are not shown in the figure. However, \textsc{GALFIT} orientations are subject to errors that are not captured by the formal uncertainties, for instance, due to irregular morphologies that are not consistent with a simple  S\'ersic model (even though the more extreme cases were excluded as mentioned above). For compact galaxies, estimating the inclination from the observed axis ratio is also uncertain. Lastly, since \textsc{GALFIT} essentially performs a light-weighted fit of the surface brightness, particularly bright star-forming regions can strongly bias the fit (e.g., a very bright star-forming center can lead to a rounder model, or a bright star-forming region in the disk of a face-on galaxy can produce an elongated geometry). Furthermore, it is not only that \textsc{GALFIT} can lead to biased results for the inclination. An actual inclination (and disk-plane or polar direction) could not even be well defined at all for some galaxies, as their definition relies on the assumption of an underlying disky geometry. And since we selected our sample based on the presence of large galactic-scale outflows, it might be the case that this feature also correlates with more disturbed morphological properties.

With this caveat in mind, our results do not support a clear alignment between the \ion{Mg}{II} outflow and the stellar disk. For nearly half of the galaxies, the direction of the outflow is not consistent with the inclination of the galaxy. This is also the case if we consider only galaxies with continuum HLRs larger than 2 kpc (indicated with orange squares in the figure).

While polar outflows are consistent with results from absorption line studies \citep[e.g.,][]{Bouche:2013il} and simulations \citep[e.g., ][]{Peroux2020}, the small number of studies of \ion{Mg}{II} halos in emission makes it difficult to unambiguously assess the orientation of outflows with respect to their host galaxies.

For instance, \citet{Leclercq2022} find that the \ion{Mg}{II} nebula in a group of five galaxies is well aligned with the semi-major axis of the most massive group member. On the other hand, \citet{Zabl2021} report a \ion{Mg}{II} halo that is aligned with the semi-minor axis of its host galaxy, as would be expected from a polar outflow. 

From stacking studies, \citet{Guo2023b} find that edge-on galaxies present an excess of \ion{Mg}{II} emission along their semi-minor axes, compared to their semi-major axes. However, comparing results from stacking with individual objects is not straightforward. We have selected the sample used in this work based on the presence of a P-Cygni profile in the \ion{Mg}{II} line. In the context of polar outflows, this inevitably biases the sample toward more face-on orientations, excluding edge-on cases. Thus, our current sample, by construction, will lack the edge-on cases that produce the signature reported in \citet{Guo2023b}.

In that sense, the misalignment between galaxies and their outflows is also supported by the fact that we do not see a bias toward face-on orientations in the inclinations derived with \textsc{GALFIT}. Indeed, the highest inclination that we find among the sample (including those with nearly isotropic outflows, which are not shown in Fig.~\ref{fig:geometry_comparison}) is $\sim64^{\circ}$. On the other hand, this expected bias toward face-on geometries is clearly seen in the distribution of the outflow R.A.. This difference between the outflows and galaxies' inclination distributions already suggests that these two quantities are not strictly connected. Furthermore, in another stacking study, \citet{Dutta2023} does not find a clear correlation between the surface brightness of \ion{Mg}{II} halos and azimuthal angle, for a stack of $\sim40$ nearly edge-on galaxies.

A lack of correspondence between R.A. and galaxy inclination would suggest that the geometry of outflows is more complex than simply being ejected along the polar direction. In the context of star-forming driven outflows, this is not entirely unexpected. Star-forming regions are distributed across the galaxy disk, and they form stars in a stochastic manner. The ejection of ISM material by stellar feedback would thus, naturally, depend on the location of the star-forming region inside the disk, as well as the ISM conditions surrounding it (in addition to the energy input by the stellar feedback itself). And while generally, star-formation surface density (and thus, feedback from star formation) is higher at the central regions of late-type galaxies \citep[see, e.g.,][]{Ellison2021, Pessa2021}, there is evidence that indicates that the radial mass assembly of low-mass galaxies (M$_{*} \lesssim 9.3$ $M_{\odot}$) is more diverse, compared to more massive galaxies \citep{Ibarra2016, Smith2022, Pessa2023}. Moreover, in \citet{Guo2023b}, the authors find that for galaxies less massive than $9.5$ $M_{\odot}$, the differences between the edge-on and face-on stack samples essentially vanish.

Altogether, we find that the orientation of outflows with respect to their galaxy is not aligned with their semi-minor axis, suggesting that the interaction between the star-formation feedback and the surrounding ISM is complex and leads to a wider variety of geometries, rather than outflows being always ejected in the polar direction. However, a definitive assessment in this regard would require a more careful determination of the galaxy orientation, taking into account the stellar kinematics (which is beyond the scope of this paper), as well as accounting for differences between high- and low-mass galaxies.

\begin{figure}
\centering
\includegraphics[width=0.95\columnwidth]{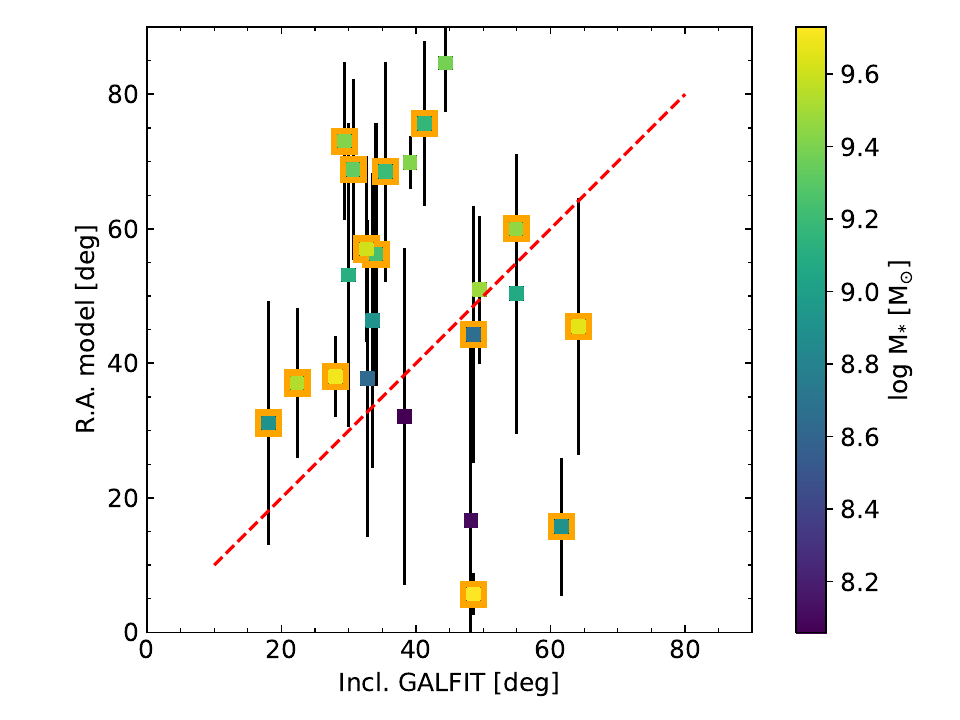}
\caption{Comparison of the orientation derived for the outflow (R.A.) with that inferred for the stellar disk of each galaxy from the HST F814W image using \textsc{GALFIT}. The red dashed line shows the identity, that is, when the inclination of the galaxy and the inclination of the outflow are the same. For a face-on polar outflow, both inclinations would be $\sim90^{\circ}$, and for an edge-on polar outflow, both would be $\sim0^{\circ}$. We have excluded from the figure galaxies that exhibit a nearly isotropic outflow (O.A. $> 75^{\circ}$), as well as galaxies for which the inclination measured with \textsc{GALFIT} was too uncertain due to their complex surface brightness profile. The orange squares indicate galaxies with a measured continuum HLRs  $> 2$ kpc. The formal uncertainties of the inclination provided by \textsc{GALFIT} are $< 1^{\circ}$.} 
\label{fig:geometry_comparison}
\end{figure}

\section{Summary and conclusions}
\label{sec:summary}

We have presented, for the first time, a population-level analysis of cool-gas galactic-scale outflows traced by their \ion{Mg}{II} emission. To this end, we have used deep MUSE data from the MUSCATEL survey and built a sample of 47 galaxies where we detect significant \ion{Mg}{II} extended emission and identify a P-Cygni spectral profile in the \ion{Mg}{II} doublet, a characteristic signature of resonant scattering in an optically
thick outflow.

Using this sample, we were able to study in a systematic manner the properties of their extended \ion{Mg}{II} emission, such as morphology and spatial extent, and we have used an outflow modeling scheme to infer the properties of the galactic winds powering the extended emission. Using available HST data, we have also been able to derive properties from the host galaxies, and explore for correlations between galactic winds properties derived from the extended \ion{Mg}{II} emission and host-galaxy properties. Our main findings are the following:

\begin{itemize}
    \item Galaxies with extended \ion{Mg}{II} emission tracing galactic-scale outflows are not necessarily significantly above the SFMS of galaxies, or exclusively associated with starbursts. A fraction of them lies significantly above the relation, and could be catalogued as starbursting galaxies; however, most of them lie closer to (or even below) the SFMS.

    \item Relative to our original parent sample, galaxies in our final sample are preferentially located in the high-end of the SFR and sSFR distributions, and show commonly younger ages. However, even in the young and high SFR/sSFR regimes, the fraction of galaxies that present an outflow traced by their Mg II emission is low ($\sim 20-30\%$). Altogether, this implies that having a certain SFR, or a certain number of young stars, does not strictly correlate with the presence of cool-gas outflows. Thus, there must also be other physical conditions required to produce a detectable outflow signal. 

    \item We have corrected the \ion{Mg}{II} emission for the self-absorption of the central galaxy, and we have been able to derive the intrinsic size of the \ion{Mg}{II} emission halos, in terms of their HLR. We find that most of our sample galaxies exhibit halos with HLRs of $\sim5$ kpc, but the distribution also presents an extended tail of larger HLRs.

    \item We do not identify any clear trend between the size of the \ion{Mg}{II} emission halos and host-galaxy properties. However, we find a tentative correlation between the size of the halos on the compact side of the distribution (HLR $\lesssim 8$ kpc) and the stellar mass of their host galaxy, potentially hinting at a difference in the mechanisms powering compact and extended halos.

    \item Our outflow model is able to reproduce many of the features seen in the data, in terms of spectral profile and general morphology. At the same time, residuals of the model also contain valuable information to put constraints on the physical mechanism shaping the CGM, such as additional secondary outflow components, or even hint at the presence of inflows.

    \item We have used our outflow modeling to derive the distribution of the galactic-wind properties and we found that the inner regions of the wind are completely optically thick ($\log \tau_0 \sim 2.6$), the data are consistent with winds that accelerate nearly linearly with radius, from launching velocities of $\sim60$ km s$^{-1}$ up to maximum velocities of $\sim 490$ km s$^{-1}$.

    \item We have investigated correlations between the outflow properties and host galaxy properties, and found tentative trends, such that more massive galaxies show preferentially higher maximum outflow velocities, compared to less massive ones, and perhaps also higher central optical depths than less massive ones. Regarding the geometry of the outflows, we find that low-mass galaxies show preferentially wider opening angles, while more massive galaxies show a much wider range of opening angles, which could be indicative of outflows escaping more easily through the ISM of less massive galaxies, compared to that of more massive ones.

    \item Lastly, by comparing the inclination of our sample galaxies with those derived for the outflows, we find that the direction of the outflows is not strictly perpendicular to the disk plane, which could be due to the non-homogeneous distribution of SF regions across the galactic disk and the interaction between stellar feedback and the ISM. However, a more rigorous determination of the galaxy orientations with respect to the line of sight is needed to make a more robust claim in this regard.
\end{itemize}

These findings represent a significant step forward in understanding the occurrence and nature of cool-gas galactic outflows traced by \ion{Mg}{II} emission. By combining deep integral-field spectroscopy with detailed outflow modeling and host-galaxy characterization, we have highlighted both the diversity and the complexity of mechanisms driving these winds. The presence of outflows in galaxies that are not strongly offset from the SFMS, the moderate detection fraction even among high-sSFR systems, and the tentative correlations with stellar mass all point to a multifaceted interplay between internal galaxy properties and the conditions necessary to launch and sustain cool-gas outflows. 

Future studies expanding the sample size, incorporating multi-wavelength constraints, more sophisticated modeling (e.g., radiative transfer simulations), and taking into account the recent star-formation history of galaxies will be crucial to fully disentangle the physical processes at play and to place these outflows in the broader context of galaxy evolution and baryon cycling in the CGM.

\begin{acknowledgements}

We thank the referee for their insightful and constructive comments, which have improved the paper. This project has received funding from the European Research Council (ERC) under the European Union's Horizon 2020 research and innovation programme (grant agreement 101020943, SPECMAP-CGM). 


LW acknowledges funding by the Deutsche Forschungsgemeinschaft, Grant Wi 1369/32-1.

JP and LW acknowledge funding by the Deutsche Forschungsgemeinschaft, Grant Wi 1369/31-1.

We thank contributors to {\sc SciPy} \citep{Virtanen2020}, {\sc Matplotlib} \citep{Hunter2007}, {\sc NumPy} \citep{Harris2020}, {\sc Astropy} \citep{Astropy, Astropy2018}, and the PYTHON programming language; the free and open-source community; and the NASA Astrophysics Data System for software and services. We also thank contributors to linetools \citep{Prochaska2016} and PyMUSE \citep{PyMUSE}.

\end{acknowledgements}  

\bibliographystyle{aa}
\bibliography{MgII_halo}

@ARTICLE{Scuderi1992,
       author = {{Scuderi}, S. and {Bonanno}, G. and {di Benedetto}, R. and {Spadaro}, D. and {Panagia}, N.},
        title = "{H alpha Observations of Early-Type Stars}",
      journal = {\apj},
         year = 1992,
        month = jun,
       volume = {392},
        pages = {201},
          doi = {10.1086/171418},
       adsurl = {https://ui.adsabs.harvard.edu/abs/1992ApJ...392..201S}
}

@ARTICLE{Hemmati2015,
       author = {{Hemmati}, Shoubaneh and {Mobasher}, Bahram and {Darvish}, Behnam and {Nayyeri}, Hooshang and {Sobral}, David and {Miller}, Sarah},
        title = "{Nebular and Stellar Dust Extinction Across the Disk of Emission-line Galaxies on Kiloparsec Scales}",
      journal = {\apj},
         year = 2015,
        month = nov,
       volume = {814},
       number = {1},
          eid = {46},
        pages = {46},
          doi = {10.1088/0004-637X/814/1/46},
archivePrefix = {arXiv},
       eprint = {1510.02506},
 primaryClass = {astro-ph.GA},
       adsurl = {https://ui.adsabs.harvard.edu/abs/2015ApJ...814...46H}
}

@ARTICLE{Pessa2024,
       author = {{Pessa}, Ismael and {Wisotzki}, Lutz and {Urrutia}, Tanya and {Pharo}, John and {Augustin}, Ramona and {Bouch{\'e}}, Nicolas F. and {Feltre}, Anna and {Guo}, Yucheng and {Kozlova}, Daria and {Krajnovic}, Davor and {Kusakabe}, Haruka and {Leclercq}, Floriane and {Salas}, H{\'e}ctor and {Schaye}, Joop and {Verhamme}, Anne},
        title = "{A galactic outflow traced by its extended Mg II emission out to a {\ensuremath{\sim}}30 kpc radius in the Hubble Ultra Deep Field with MUSE}",
      journal = {\aap},
         year = 2024,
        month = nov,
       volume = {691},
          eid = {A5},
        pages = {A5},
          doi = {10.1051/0004-6361/202450547},
archivePrefix = {arXiv},
       eprint = {2408.16067},
 primaryClass = {astro-ph.GA},
       adsurl = {https://ui.adsabs.harvard.edu/abs/2024A&A...691A...5P}
}

@ARTICLE{Wang2024,
       author = {{Wang}, Bingjie and {Leja}, Joel and {Atek}, Hakim and {Labb{\'e}}, Ivo and {Li}, Yijia and {Bezanson}, Rachel and {Brammer}, Gabriel and {Cutler}, Sam E. and {Dayal}, Pratika and {Furtak}, Lukas J. and {Greene}, Jenny E. and {Kokorev}, Vasily and {Pan}, Richard and {Price}, Sedona H. and {Suess}, Katherine A. and {Weaver}, John R. and {Whitaker}, Katherine E. and {Williams}, Christina C.},
        title = "{Quantifying the Effects of Known Unknowns on Inferred High-redshift Galaxy Properties: Burstiness, IMF, and Nebular Physics}",
      journal = {\apj},
         year = 2024,
        month = mar,
       volume = {963},
       number = {1},
          eid = {74},
        pages = {74},
          doi = {10.3847/1538-4357/ad187c},
archivePrefix = {arXiv},
       eprint = {2310.06781},
 primaryClass = {astro-ph.GA},
       adsurl = {https://ui.adsabs.harvard.edu/abs/2024ApJ...963...74W}
}

@ARTICLE{Galfit_b,
       author = {{Peng}, Chien Y. and {Ho}, Luis C. and {Impey}, Chris D. and {Rix}, Hans-Walter},
        title = "{Detailed Decomposition of Galaxy Images. II. Beyond Axisymmetric Models}",
      journal = {\aj},
         year = 2010,
        month = jun,
       volume = {139},
       number = {6},
        pages = {2097-2129},
          doi = {10.1088/0004-6256/139/6/2097},
archivePrefix = {arXiv},
       eprint = {0912.0731},
 primaryClass = {astro-ph.CO},
       adsurl = {https://ui.adsabs.harvard.edu/abs/2010AJ....139.2097P}
}

@ARTICLE{Galfit_a,
       author = {{Peng}, Chien Y. and {Ho}, Luis C. and {Impey}, Chris D. and {Rix}, Hans-Walter},
        title = "{Detailed Structural Decomposition of Galaxy Images}",
      journal = {\aj},
         year = 2002,
        month = jul,
       volume = {124},
       number = {1},
        pages = {266-293},
          doi = {10.1086/340952},
archivePrefix = {arXiv},
       eprint = {astro-ph/0204182},
 primaryClass = {astro-ph},
       adsurl = {https://ui.adsabs.harvard.edu/abs/2002AJ....124..266P}
}

@ARTICLE{Reynaldi2013,
       author = {{Reynaldi}, V. and {Feinstein}, C.},
        title = "{Shock ionization in the extended emission-line region of 3C 305: the last piece of the (optical) puzzle}",
      journal = {\mnras},
         year = 2013,
        month = oct,
       volume = {435},
       number = {2},
        pages = {1350-1357},
          doi = {10.1093/mnras/stt1377},
archivePrefix = {arXiv},
       eprint = {1307.7489},
 primaryClass = {astro-ph.CO},
       adsurl = {https://ui.adsabs.harvard.edu/abs/2013MNRAS.435.1350R}
}

@ARTICLE{BPT,
       author = {{Baldwin}, J.~A. and {Phillips}, M.~M. and {Terlevich}, R.},
        title = "{Classification parameters for the emission-line spectra of extragalactic objects.}",
      journal = {\pasp},
         year = 1981,
        month = feb,
       volume = {93},
        pages = {5-19},
          doi = {10.1086/130766},
       adsurl = {https://ui.adsabs.harvard.edu/abs/1981PASP...93....5B}
}

@ARTICLE{Emsellem2022,
       author = {{Emsellem}, Eric and {Schinnerer}, Eva and {Santoro}, Francesco and {Belfiore}, Francesco and {Pessa}, Ismael and {McElroy}, Rebecca and {Blanc}, Guillermo A. and {Congiu}, Enrico and {Groves}, Brent and {Ho}, I. -Ting and {Kreckel}, Kathryn and {Razza}, Alessandro and {Sanchez-Blazquez}, Patricia and {Egorov}, Oleg and {Faesi}, Chris and {Klessen}, Ralf S. and {Leroy}, Adam K. and {Meidt}, Sharon and {Querejeta}, Miguel and {Rosolowsky}, Erik and {Scheuermann}, Fabian and {Anand}, Gagandeep S. and {Barnes}, Ashley T. and {Be{\v{s}}li{\'c}}, Ivana and {Bigiel}, Frank and {Boquien}, M{\'e}d{\'e}ric and {Cao}, Yixian and {Chevance}, M{\'e}lanie and {Dale}, Daniel A. and {Eibensteiner}, Cosima and {Glover}, Simon C.~O. and {Grasha}, Kathryn and {Henshaw}, Jonathan D. and {Hughes}, Annie and {Koch}, Eric W. and {Kruijssen}, J.~M. Diederik and {Lee}, Janice and {Liu}, Daizhong and {Pan}, Hsi-An and {Pety}, J{\'e}r{\^o}me and {Saito}, Toshiki and {Sandstrom}, Karin M. and {Schruba}, Andreas and {Sun}, Jiayi and {Thilker}, David A. and {Usero}, Antonio and {Watkins}, Elizabeth J. and {Williams}, Thomas G.},
        title = "{The PHANGS-MUSE survey. Probing the chemo-dynamical evolution of disc galaxies}",
      journal = {\aap},
         year = 2022,
        month = mar,
       volume = {659},
          eid = {A191},
        pages = {A191},
          doi = {10.1051/0004-6361/202141727},
archivePrefix = {arXiv},
       eprint = {2110.03708},
 primaryClass = {astro-ph.GA},
       adsurl = {https://ui.adsabs.harvard.edu/abs/2022A&A...659A.191E}
}

@ARTICLE{Kewley2004,
       author = {{Kewley}, Lisa J. and {Geller}, Margaret J. and {Jansen}, Rolf A.},
        title = "{[O II] as a Star Formation Rate Indicator}",
      journal = {\aj},
         year = 2004,
        month = apr,
       volume = {127},
       number = {4},
        pages = {2002-2030},
          doi = {10.1086/382723},
archivePrefix = {arXiv},
       eprint = {astro-ph/0401172},
 primaryClass = {astro-ph},
       adsurl = {https://ui.adsabs.harvard.edu/abs/2004AJ....127.2002K}
}

@BOOK{Ambartsumian1958,
       author = {{Ambartsumian}, Viktor Amazaspovich},
        title = "{Theoretical astrophysics.}",
         year = 1958,
       adsurl = {https://ui.adsabs.harvard.edu/abs/1958thas.book.....A}
}

@BOOK{Sobolev1960,
       author = {{Sobolev}, V.~V.},
        title = "{Moving Envelopes of Stars}",
         year = 1960,
          doi = {10.4159/harvard.9780674864658},
       adsurl = {https://ui.adsabs.harvard.edu/abs/1960mes..book.....S}
}

@ARTICLE{Grinin2001,
       author = {{Grinin}, V.~P.},
        title = "{Sobolev's Approximation}",
      journal = {Astrophysics},
         year = 2001,
        month = jul,
       volume = {44},
       number = {3},
        pages = {402-410},
          doi = {10.1023/A:1012884231706},
       adsurl = {https://ui.adsabs.harvard.edu/abs/2001Ap.....44..402G}
}

@ARTICLE{Castor1970,
       author = {{Castor}, J.~I.},
        title = "{Spectral line formation in Wolf-Rayet envelopes.}",
      journal = {\mnras},
         year = 1970,
        month = jan,
       volume = {149},
        pages = {111},
          doi = {10.1093/mnras/149.2.111},
       adsurl = {https://ui.adsabs.harvard.edu/abs/1970MNRAS.149..111C}
}

@ARTICLE{Xu2022,
       author = {{Xu}, Xinfeng and {Heckman}, Timothy and {Henry}, Alaina and {Berg}, Danielle A. and {Chisholm}, John and {James}, Bethan L. and {Martin}, Crystal L. and {Stark}, Daniel P. and {Aloisi}, Alessandra and {Amor{\'\i}n}, Ricardo O. and {Arellano-C{\'o}rdova}, Karla Z. and {Bordoloi}, Rongmon and {Charlot}, St{\'e}phane and {Chen}, Zuyi and {Hayes}, Matthew and {Mingozzi}, Matilde and {Sugahara}, Yuma and {Kewley}, Lisa J. and {Ouchi}, Masami and {Scarlata}, Claudia and {Steidel}, Charles C.},
        title = "{CLASSY III. The Properties of Starburst-driven Warm Ionized Outflows}",
      journal = {\apj},
         year = 2022,
        month = jul,
       volume = {933},
       number = {2},
          eid = {222},
        pages = {222},
          doi = {10.3847/1538-4357/ac6d56},
archivePrefix = {arXiv},
       eprint = {2204.09181},
 primaryClass = {astro-ph.GA},
       adsurl = {https://ui.adsabs.harvard.edu/abs/2022ApJ...933..222X}
}

@ARTICLE{Erb2023,
       author = {{Erb}, Dawn K. and {Li}, Zhihui and {Steidel}, Charles C. and {Chen}, Yuguang and {Gronke}, Max and {Strom}, Allison L. and {Trainor}, Ryan F. and {Rudie}, Gwen C.},
        title = "{The Circumgalactic Medium of Extreme Emission Line Galaxies at z 2: Resolved Spectroscopy and Radiative Transfer Modeling of Spatially Extended Ly{\ensuremath{\alpha}} Emission in the KBSS-KCWI Survey}",
      journal = {\apj},
         year = 2023,
        month = aug,
       volume = {953},
       number = {1},
          eid = {118},
        pages = {118},
          doi = {10.3847/1538-4357/acd849},
archivePrefix = {arXiv},
       eprint = {2210.02465},
 primaryClass = {astro-ph.GA},
       adsurl = {https://ui.adsabs.harvard.edu/abs/2023ApJ...953..118E}
}

@ARTICLE{Burchett2021,
       author = {{Burchett}, Joseph N. and {Rubin}, Kate H.~R. and {Prochaska}, J. Xavier and {Coil}, Alison L. and {Vaught}, Ryan Rickards and {Hennawi}, Joseph F.},
        title = "{Circumgalactic Mg II Emission from an Isotropic Starburst Galaxy Outflow Mapped by KCWI}",
      journal = {\apj},
         year = 2021,
        month = mar,
       volume = {909},
       number = {2},
          eid = {151},
        pages = {151},
          doi = {10.3847/1538-4357/abd4e0},
archivePrefix = {arXiv},
       eprint = {2005.03017},
 primaryClass = {astro-ph.GA},
       adsurl = {https://ui.adsabs.harvard.edu/abs/2021ApJ...909..151B}
}

@ARTICLE{Tumlinson2017,
       author = {{Tumlinson}, Jason and {Peeples}, Molly S. and {Werk}, Jessica K.},
        title = "{The Circumgalactic Medium}",
      journal = {\araa},
         year = 2017,
        month = aug,
       volume = {55},
       number = {1},
        pages = {389-432},
          doi = {10.1146/annurev-astro-091916-055240},
archivePrefix = {arXiv},
       eprint = {1709.09180},
 primaryClass = {astro-ph.GA},
       adsurl = {https://ui.adsabs.harvard.edu/abs/2017ARA&A..55..389T}
}

@ARTICLE{Fielding2017,
       author = {{Fielding}, Drummond and {Quataert}, Eliot and {Martizzi}, Davide and {Faucher-Gigu{\`e}re}, Claude-Andr{\'e}},
        title = "{How supernovae launch galactic winds?}",
      journal = {\mnras},
         year = 2017,
        month = sep,
       volume = {470},
       number = {1},
        pages = {L39-L43},
          doi = {10.1093/mnrasl/slx072},
archivePrefix = {arXiv},
       eprint = {1704.01579},
 primaryClass = {astro-ph.GA},
       adsurl = {https://ui.adsabs.harvard.edu/abs/2017MNRAS.470L..39F}
}

@ARTICLE{Mowla2019,
       author = {{Mowla}, Lamiya A. and {van Dokkum}, Pieter and {Brammer}, Gabriel B. and {Momcheva}, Ivelina and {van der Wel}, Arjen and {Whitaker}, Katherine and {Nelson}, Erica and {Bezanson}, Rachel and {Muzzin}, Adam and {Franx}, Marijn and {MacKenty}, John and {Leja}, Joel and {Kriek}, Mariska and {Marchesini}, Danilo},
        title = "{COSMOS-DASH: The Evolution of the Galaxy Size-Mass Relation since z {\ensuremath{\sim}} 3 from New Wide-field WFC3 Imaging Combined with CANDELS/3D-HST}",
      journal = {\apj},
         year = 2019,
        month = jul,
       volume = {880},
       number = {1},
          eid = {57},
        pages = {57},
          doi = {10.3847/1538-4357/ab290a},
archivePrefix = {arXiv},
       eprint = {1808.04379},
 primaryClass = {astro-ph.GA},
       adsurl = {https://ui.adsabs.harvard.edu/abs/2019ApJ...880...57M}
}

@ARTICLE{VanDerWel2014,
       author = {{van der Wel}, A. and {Franx}, M. and {van Dokkum}, P.~G. and {Skelton}, R.~E. and {Momcheva}, I.~G. and {Whitaker}, K.~E. and {Brammer}, G.~B. and {Bell}, E.~F. and {Rix}, H. -W. and {Wuyts}, S. and {Ferguson}, H.~C. and {Holden}, B.~P. and {Barro}, G. and {Koekemoer}, A.~M. and {Chang}, Yu-Yen and {McGrath}, E.~J. and {H{\"a}ussler}, B. and {Dekel}, A. and {Behroozi}, P. and {Fumagalli}, M. and {Leja}, J. and {Lundgren}, B.~F. and {Maseda}, M.~V. and {Nelson}, E.~J. and {Wake}, D.~A. and {Patel}, S.~G. and {Labb{\'e}}, I. and {Faber}, S.~M. and {Grogin}, N.~A. and {Kocevski}, D.~D.},
        title = "{3D-HST+CANDELS: The Evolution of the Galaxy Size-Mass Distribution since z = 3}",
      journal = {\apj},
         year = 2014,
        month = jun,
       volume = {788},
       number = {1},
          eid = {28},
        pages = {28},
          doi = {10.1088/0004-637X/788/1/28},
archivePrefix = {arXiv},
       eprint = {1404.2844},
 primaryClass = {astro-ph.GA},
       adsurl = {https://ui.adsabs.harvard.edu/abs/2014ApJ...788...28V}
}

@ARTICLE{Nedkova2021,
       author = {{Nedkova}, Kalina V. and {H{\"a}u{\ss}ler}, Boris and {Marchesini}, Danilo and {Dimauro}, Paola and {Brammer}, Gabriel and {Eigenthaler}, Paul and {Feinstein}, Adina D. and {Ferguson}, Henry C. and {Huertas-Company}, Marc and {Johnston}, Evelyn J. and {Kado-Fong}, Erin and {Kartaltepe}, Jeyhan S. and {Labb{\'e}}, Ivo and {Lange-Vagle}, Daniel and {Martis}, Nicholas S. and {McGrath}, Elizabeth J. and {Muzzin}, Adam and {Oesch}, Pascal and {Ordenes-Brice{\~n}o}, Yasna and {Puzia}, Thomas and {Shipley}, Heath V. and {Simmons}, Brooke D. and {Skelton}, Rosalind E. and {Stefanon}, Mauro and {van der Wel}, Arjen and {Whitaker}, Katherine E.},
        title = "{Extending the evolution of the stellar mass-size relation at z {\ensuremath{\leq}} 2 to low stellar mass galaxies from HFF and CANDELS}",
      journal = {\mnras},
         year = 2021,
        month = sep,
       volume = {506},
       number = {1},
        pages = {928-956},
          doi = {10.1093/mnras/stab1744},
archivePrefix = {arXiv},
       eprint = {2106.07663},
 primaryClass = {astro-ph.GA},
       adsurl = {https://ui.adsabs.harvard.edu/abs/2021MNRAS.506..928N}
}

@ARTICLE{Dylan2019,
       author = {{Nelson}, Dylan and {Springel}, Volker and {Pillepich}, Annalisa and {Rodriguez-Gomez}, Vicente and {Torrey}, Paul and {Genel}, Shy and {Vogelsberger}, Mark and {Pakmor}, Ruediger and {Marinacci}, Federico and {Weinberger}, Rainer and {Kelley}, Luke and {Lovell}, Mark and {Diemer}, Benedikt and {Hernquist}, Lars},
        title = "{The IllustrisTNG simulations: public data release}",
      journal = {Computational Astrophysics and Cosmology},
         year = 2019,
        month = may,
       volume = {6},
       number = {1},
          eid = {2},
        pages = {2},
          doi = {10.1186/s40668-019-0028-x},
archivePrefix = {arXiv},
       eprint = {1812.05609},
 primaryClass = {astro-ph.GA},
       adsurl = {https://ui.adsabs.harvard.edu/abs/2019ComAC...6....2N}
}

@ARTICLE{Leibundgut2017,
       author = {{Leibundgut}, B. and {Bacon}, R. and {Jaff{\'e}}, Y.~L. and {Johnston}, E. and {Kuntschner}, H. and {Selman}, F. and {Valenti}, E. and {Vernet}, J. and {Vogt}, F.},
        title = "{MUSE WFM AO Science Verification}",
      journal = {The Messenger},
         year = 2017,
        month = dec,
       volume = {170},
        pages = {20-25},
          doi = {10.18727/0722-6691/5050},
       adsurl = {https://ui.adsabs.harvard.edu/abs/2017Msngr.170...20L}
}

@ARTICLE{Dylan2019b,
       author = {{Nelson}, Dylan and {Pillepich}, Annalisa and {Springel}, Volker and {Pakmor}, R{\"u}diger and {Weinberger}, Rainer and {Genel}, Shy and {Torrey}, Paul and {Vogelsberger}, Mark and {Marinacci}, Federico and {Hernquist}, Lars},
        title = "{First results from the TNG50 simulation: galactic outflows driven by supernovae and black hole feedback}",
      journal = {\mnras},
         year = 2019,
        month = dec,
       volume = {490},
       number = {3},
        pages = {3234-3261},
          doi = {10.1093/mnras/stz2306},
archivePrefix = {arXiv},
       eprint = {1902.05554},
 primaryClass = {astro-ph.GA},
       adsurl = {https://ui.adsabs.harvard.edu/abs/2019MNRAS.490.3234N}
}

@ARTICLE{Chung2019,
       author = {{Chung}, Andrew S. and {Dijkstra}, Mark and {Ciardi}, Benedetta and {Kakiichi}, Koki and {Naab}, Thorsten},
        title = "{The circumgalactic medium in Lyman {\ensuremath{\alpha}}: a new constraint on galactic outflow models}",
      journal = {\mnras},
         year = 2019,
        month = apr,
       volume = {484},
       number = {2},
        pages = {2420-2432},
          doi = {10.1093/mnras/stz149},
archivePrefix = {arXiv},
       eprint = {1901.04015},
 primaryClass = {astro-ph.GA},
       adsurl = {https://ui.adsabs.harvard.edu/abs/2019MNRAS.484.2420C}
}

@article{Zabl:2020js,
author = {Zabl, Johannes and Bouch{\'e}, Nicolas F and Schroetter, Ilane and Wendt, Martin and Contini, Thierry and Schaye, Joop and Marino, Raffaella A and Muzahid, Sowgat and Pezzulli, Gabriele and Verhamme, Anne and Wisotzki, Lutz},
title = {{MusE GAs FLOw and Wind (MEGAFLOW) IV. A two sightline tomography of a galactic wind}},
journal = {MNRAS},
year = {2020},
volume = {492},
number = {3},
pages = {4576--4588},
month = jan
}

@ARTICLE{Guo2023b,
       author = {{Guo}, Yucheng and {Bacon}, Roland and {Bouch{\'e}}, Nicolas F. and {Wisotzki}, Lutz and {Schaye}, Joop and {Blaizot}, J{\'e}r{\'e}my and {Verhamme}, Anne and {Cantalupo}, Sebastiano and {Boogaard}, Leindert A. and {Brinchmann}, Jarle and {Cherrey}, Maxime and {Kusakabe}, Haruka and {Langan}, Ivanna and {Leclercq}, Floriane and {Matthee}, Jorryt and {Michel-Dansac}, L{\'e}o and {Schroetter}, Ilane and {Wendt}, Martin},
        title = "{Bipolar outflows out to 10 kpc for massive galaxies at redshift z {\ensuremath{\approx}} 1}",
      journal = {\nat},
         year = 2023,
        month = dec,
       volume = {624},
       number = {7990},
        pages = {53-56},
          doi = {10.1038/s41586-023-06718-w},
archivePrefix = {arXiv},
       eprint = {2312.05167},
 primaryClass = {astro-ph.GA},
       adsurl = {https://ui.adsabs.harvard.edu/abs/2023Natur.624...53G}
}

@ARTICLE{Kusakabe2024,
       author = {{Kusakabe}, Haruka and {Mauerhofer}, Valentin and {Verhamme}, Anne and {Garel}, Thibault and {Blaizot}, J{\'e}r{\'e}my and {Wisotzki}, Lutz and {Richard}, Johan and {Boogaard}, Leindert A. and {Leclercq}, Floriane and {Guo}, Yucheng and {Claeyssens}, Ad{\'e}la{\"\i}de and {Contini}, Thierry and {Herenz}, Edmund Christian and {Kerutt}, Josephine and {Maseda}, Michael V. and {Michel-Dansac}, Leo and {Nanayakkara}, Themiya and {Ouchi}, Masami and {Pessa}, Ismael and {Schaye}, Joop},
        title = "{The MUSE eXtremely Deep Field: Detections of circumgalactic Si II* emission at z {\ensuremath{\gtrsim}} 2}",
      journal = {\aap},
         year = 2024,
        month = nov,
       volume = {691},
          eid = {A255},
        pages = {A255},
          doi = {10.1051/0004-6361/202451009},
archivePrefix = {arXiv},
       eprint = {2406.04399},
 primaryClass = {astro-ph.GA},
       adsurl = {https://ui.adsabs.harvard.edu/abs/2024A&A...691A.255K}
}

@ARTICLE{Bacon2023,
       author = {{Bacon}, Roland and {Brinchmann}, Jarle and {Conseil}, Simon and {Maseda}, Michael and {Nanayakkara}, Themiya and {Wendt}, Martin and {Bacher}, Raphael and {Mary}, David and {Weilbacher}, Peter M. and {Krajnovi{\'c}}, Davor and {Boogaard}, Leindert and {Bouch{\'e}}, Nicolas F and {Contini}, Thierry and {Epinat}, Beno{\^\i}t and {Feltre}, Anna and {Guo}, Yucheng and {Herenz}, Christian and {Kollatschny}, Wolfram and {Kusakabe}, Haruka and {Leclercq}, Floriane and {Michel-Dansac}, L{\'e}o and {Pello}, Roser and {Richard}, Johan and {Roth}, Martin and {Salvignol}, Gregory and {Schaye}, Joop and {Steinmetz}, Matthias and {Tresse}, Laurence and {Urrutia}, Tanya and {Verhamme}, Anne and {Vitte}, Eloise and {Wisotzki}, Lutz and {Zoutendijk}, Sebastiaan L.},
        title = "{The MUSE Hubble Ultra Deep Field surveys: Data release II}",
      journal = {\aap},
         year = 2023,
        month = feb,
       volume = {670},
          eid = {A4},
        pages = {A4},
          doi = {10.1051/0004-6361/202244187},
archivePrefix = {arXiv},
       eprint = {2211.08493},
 primaryClass = {astro-ph.GA},
       adsurl = {https://ui.adsabs.harvard.edu/abs/2023A&A...670A...4B}
}

@ARTICLE{Leclercq2022,
       author = {{Leclercq}, Floriane and {Verhamme}, Anne and {Epinat}, Benoit and {Simmonds}, Charlotte and {Matthee}, Jorryt and {Bouch{\'e}}, Nicolas F. and {Garel}, Thibault and {Urrutia}, Tanya and {Wisotzki}, Lutz and {Zabl}, Johannes and {Bacon}, Roland and {Abril-Melgarejo}, Valentina and {Boogaard}, Leindert and {Brinchmann}, Jarle and {Cantalupo}, Sebastiano and {Contini}, Thierry and {Kerutt}, Josephine and {Kusakabe}, Haruka and {Maseda}, Michael and {Michel-Dansac}, L{\'e}o and {Muzahid}, Sowgat and {Nanayakkara}, Themiya and {Richard}, Johan and {Schaye}, Joop},
        title = "{The MUSE eXtremely deep field: first panoramic view of an Mg II emitting intragroup medium}",
      journal = {\aap},
         year = 2022,
        month = jul,
       volume = {663},
          eid = {A11},
        pages = {A11},
          doi = {10.1051/0004-6361/202142179},
archivePrefix = {arXiv},
       eprint = {2203.05614},
 primaryClass = {astro-ph.GA},
       adsurl = {https://ui.adsabs.harvard.edu/abs/2022A&A...663A..11L}
}

@ARTICLE{Zabl2021,
       author = {{Zabl}, Johannes and {Bouch{\'e}}, Nicolas F. and {Wisotzki}, Lutz and {Schaye}, Joop and {Leclercq}, Floriane and {Garel}, Thibault and {Wendt}, Martin and {Schroetter}, Ilane and {Muzahid}, Sowgat and {Cantalupo}, Sebastiano and {Contini}, Thierry and {Bacon}, Roland and {Brinchmann}, Jarle and {Richard}, Johan},
        title = "{MusE GAs FLOw and Wind (MEGAFLOW) VIII. Discovery of a MgII emission halo probed by a quasar sightline}",
      journal = {\mnras},
         year = 2021,
        month = nov,
       volume = {507},
       number = {3},
        pages = {4294-4315},
          doi = {10.1093/mnras/stab2165},
archivePrefix = {arXiv},
       eprint = {2105.14090},
 primaryClass = {astro-ph.GA},
       adsurl = {https://ui.adsabs.harvard.edu/abs/2021MNRAS.507.4294Z}
}

@article{Schroetter:2019es,
author = {Schroetter, Ilane and Bouch{\'e}, Nicolas F and Zabl, Johannes and Contini, Thierry and Wendt, Martin and Schaye, Joop and Mitchell, Peter and Muzahid, Sowgat and Marino, Raffaella A and Bacon, Roland and Lilly, Simon J and Richard, Johan and Wisotzki, Lutz},
title = {{MusE GAs FLOw and Wind (MEGAFLOW) - III. Galactic wind properties using background quasars}},
journal = {MNRAS},
year = {2019},
volume = {490},
number = {3},
pages = {4368--4381},
month = oct
}

@ARTICLE{Heckman2015,
       author = {{Heckman}, Timothy M. and {Alexandroff}, Rachel M. and {Borthakur}, Sanchayeeta and {Overzier}, Roderik and {Leitherer}, Claus},
        title = "{The Systematic Properties of the Warm Phase of Starburst-Driven Galactic Winds}",
      journal = {\apj},
         year = 2015,
        month = aug,
       volume = {809},
       number = {2},
          eid = {147},
        pages = {147},
          doi = {10.1088/0004-637X/809/2/147},
archivePrefix = {arXiv},
       eprint = {1507.05622},
 primaryClass = {astro-ph.GA},
       adsurl = {https://ui.adsabs.harvard.edu/abs/2015ApJ...809..147H}
}

@ARTICLE{Carr2022,
       author = {{Carr}, C. and {Scarlata}, C.},
        title = "{A Semianalytical Line Transfer Model. III. Galactic Inflows}",
      journal = {\apj},
         year = 2022,
        month = nov,
       volume = {939},
       number = {1},
          eid = {47},
        pages = {47},
          doi = {10.3847/1538-4357/ac93fa},
archivePrefix = {arXiv},
       eprint = {2209.14485},
 primaryClass = {astro-ph.GA},
       adsurl = {https://ui.adsabs.harvard.edu/abs/2022ApJ...939...47C}
}

@ARTICLE{Chang2025,
       author = {{Chang}, Seok-Jun and {Dutta}, Rajeshwari and {Gronke}, Max and {Fumagalli}, Michele and {Battaia}, Fabrizio Arrigoni and {Fossati}, Matteo},
        title = "{Modelling Mg II resonance doublet spectra from star-forming galaxies at z {\ensuremath{\sim}} 1}",
      journal = {\mnras},
         year = 2025,
        month = sep,
       volume = {542},
       number = {1},
        pages = {525-539},
          doi = {10.1093/mnras/staf1069},
archivePrefix = {arXiv},
       eprint = {2412.08837},
 primaryClass = {astro-ph.GA},
       adsurl = {https://ui.adsabs.harvard.edu/abs/2025MNRAS.542..525C}
}

@ARTICLE{Bezanson2024,
       author = {{Bezanson}, Rachel and {Labbe}, Ivo and {Whitaker}, Katherine E. and {Leja}, Joel and {Price}, Sedona H. and {Franx}, Marijn and {Brammer}, Gabriel and {Marchesini}, Danilo and {Zitrin}, Adi and {Wang}, Bingjie and {Weaver}, John R. and {Furtak}, Lukas J. and {Atek}, Hakim and {Coe}, Dan and {Cutler}, Sam E. and {Dayal}, Pratika and {van Dokkum}, Pieter and {Feldmann}, Robert and {F{\"o}rster Schreiber}, Natascha M. and {Fujimoto}, Seiji and {Geha}, Marla and {Glazebrook}, Karl and {de Graaff}, Anna and {Greene}, Jenny E. and {Juneau}, St{\'e}phanie and {Kassin}, Susan and {Kriek}, Mariska and {Khullar}, Gourav and {Maseda}, Michael and {Mowla}, Lamiya A. and {Muzzin}, Adam and {Nanayakkara}, Themiya and {Nelson}, Erica J. and {Oesch}, Pascal A. and {Pacifici}, Camilla and {Pan}, Richard and {Papovich}, Casey and {Setton}, David J. and {Shapley}, Alice E. and {Smit}, Renske and {Stefanon}, Mauro and {Taylor}, Edward N. and {Williams}, Christina C.},
        title = "{The JWST UNCOVER Treasury Survey: Ultradeep NIRSpec and NIRCam Observations before the Epoch of Reionization}",
      journal = {\apj},
         year = 2024,
        month = oct,
       volume = {974},
       number = {1},
          eid = {92},
        pages = {92},
          doi = {10.3847/1538-4357/ad66cf},
archivePrefix = {arXiv},
       eprint = {2212.04026},
 primaryClass = {astro-ph.GA},
       adsurl = {https://ui.adsabs.harvard.edu/abs/2024ApJ...974...92B}
}

@ARTICLE{Popesso2023,
       author = {{Popesso}, P. and {Concas}, A. and {Cresci}, G. and {Belli}, S. and {Rodighiero}, G. and {Inami}, H. and {Dickinson}, M. and {Ilbert}, O. and {Pannella}, M. and {Elbaz}, D.},
        title = "{The main sequence of star-forming galaxies across cosmic times}",
      journal = {\mnras},
         year = 2023,
        month = feb,
       volume = {519},
       number = {1},
        pages = {1526-1544},
          doi = {10.1093/mnras/stac3214},
archivePrefix = {arXiv},
       eprint = {2203.10487},
 primaryClass = {astro-ph.GA},
       adsurl = {https://ui.adsabs.harvard.edu/abs/2023MNRAS.519.1526P}
}

@ARTICLE{Noeske2007,
       author = {{Noeske}, K.~G. and {Weiner}, B.~J. and {Faber}, S.~M. and {Papovich}, C. and {Koo}, D.~C. and {Somerville}, R.~S. and {Bundy}, K. and {Conselice}, C.~J. and {Newman}, J.~A. and {Schiminovich}, D. and {Le Floc'h}, E. and {Coil}, A.~L. and {Rieke}, G.~H. and {Lotz}, J.~M. and {Primack}, J.~R. and {Barmby}, P. and {Cooper}, M.~C. and {Davis}, M. and {Ellis}, R.~S. and {Fazio}, G.~G. and {Guhathakurta}, P. and {Huang}, J. and {Kassin}, S.~A. and {Martin}, D.~C. and {Phillips}, A.~C. and {Rich}, R.~M. and {Small}, T.~A. and {Willmer}, C.~N.~A. and {Wilson}, G.},
        title = "{Star Formation in AEGIS Field Galaxies since z=1.1: The Dominance of Gradually Declining Star Formation, and the Main Sequence of Star-forming Galaxies}",
      journal = {\apjl},
         year = 2007,
        month = may,
       volume = {660},
       number = {1},
        pages = {L43-L46},
          doi = {10.1086/517926},
archivePrefix = {arXiv},
       eprint = {astro-ph/0701924},
 primaryClass = {astro-ph},
       adsurl = {https://ui.adsabs.harvard.edu/abs/2007ApJ...660L..43N}
}

@ARTICLE{Daddi2007,
       author = {{Daddi}, E. and {Dickinson}, M. and {Morrison}, G. and {Chary}, R. and {Cimatti}, A. and {Elbaz}, D. and {Frayer}, D. and {Renzini}, A. and {Pope}, A. and {Alexander}, D.~M. and {Bauer}, F.~E. and {Giavalisco}, M. and {Huynh}, M. and {Kurk}, J. and {Mignoli}, M.},
        title = "{Multiwavelength Study of Massive Galaxies at z\raisebox{-0.5ex}\textasciitilde2. I. Star Formation and Galaxy Growth}",
      journal = {\apj},
         year = 2007,
        month = nov,
       volume = {670},
       number = {1},
        pages = {156-172},
          doi = {10.1086/521818},
archivePrefix = {arXiv},
       eprint = {0705.2831},
 primaryClass = {astro-ph},
       adsurl = {https://ui.adsabs.harvard.edu/abs/2007ApJ...670..156D}
}

@ARTICLE{Brinchmann2004,
       author = {{Brinchmann}, J. and {Charlot}, S. and {White}, S.~D.~M. and {Tremonti}, C. and {Kauffmann}, G. and {Heckman}, T. and {Brinkmann}, J.},
        title = "{The physical properties of star-forming galaxies in the low-redshift Universe}",
      journal = {\mnras},
         year = 2004,
        month = jul,
       volume = {351},
       number = {4},
        pages = {1151-1179},
          doi = {10.1111/j.1365-2966.2004.07881.x},
archivePrefix = {arXiv},
       eprint = {astro-ph/0311060},
 primaryClass = {astro-ph},
       adsurl = {https://ui.adsabs.harvard.edu/abs/2004MNRAS.351.1151B}
}

@ARTICLE{Pharo2024,
       author = {{Pharo}, John and {Wisotzki}, Lutz and {Urrutia}, Tanya and {Bacon}, Roland and {Pessa}, Ismael and {Augustin}, Ramona and {Goovaerts}, Ilias and {Kozlova}, Daria and {Kusakabe}, Haruka and {Salas}, H{\'e}ctor and {Smirnov}, Daniil and {Thai}, Tran Thi and {Vitte}, Elo{\"\i}se},
        title = "{The intrinsic distribution of Lyman-{\ensuremath{\alpha}} halos}",
      journal = {\aap},
         year = 2024,
        month = oct,
       volume = {690},
          eid = {A343},
        pages = {A343},
          doi = {10.1051/0004-6361/202451318},
archivePrefix = {arXiv},
       eprint = {2409.04537},
 primaryClass = {astro-ph.GA},
       adsurl = {https://ui.adsabs.harvard.edu/abs/2024A&A...690A.343P}
}

@ARTICLE{Rihtar2025,
       author = {{Rihtar{\v{s}}i{\v{c}}}, G. and {Brada{\v{c}}}, M. and {Desprez}, G. and {Harshan}, A. and {Noirot}, G. and {Estrada-Carpenter}, V. and {Martis}, N.~S. and {Abraham}, R.~G. and {Asada}, Y. and {Brammer}, G. and {Iyer}, K.~G. and {Matharu}, J. and {Mowla}, L. and {Muzzin}, A. and {Sarrouh}, G.~T.~E. and {Sawicki}, M. and {Strait}, V. and {Willott}, C.~J. and {Gledhill}, R. and {Markov}, V. and {Tripodi}, R.},
        title = "{CANUCS: Constraining the MACS J0416.1-2403 strong lensing model with JWST NIRISS, NIRSpec, and NIRCam}",
      journal = {\aap},
         year = 2025,
        month = apr,
       volume = {696},
          eid = {A15},
        pages = {A15},
          doi = {10.1051/0004-6361/202451117},
       adsurl = {https://ui.adsabs.harvard.edu/abs/2025A&A...696A..15R}
}

@ARTICLE{Shipley2018,
       author = {{Shipley}, Heath V. and {Lange-Vagle}, Daniel and {Marchesini}, Danilo and {Brammer}, Gabriel B. and {Ferrarese}, Laura and {Stefanon}, Mauro and {Kado-Fong}, Erin and {Whitaker}, Katherine E. and {Oesch}, Pascal A. and {Feinstein}, Adina D. and {Labb{\'e}}, Ivo and {Lundgren}, Britt and {Martis}, Nicholas and {Muzzin}, Adam and {Nedkova}, Kalina and {Skelton}, Rosalind and {van der Wel}, Arjen},
        title = "{HFF-DeepSpace Photometric Catalogs of the 12 Hubble Frontier Fields, Clusters, and Parallels: Photometry, Photometric Redshifts, and Stellar Masses}",
      journal = {\apjs},
         year = 2018,
        month = mar,
       volume = {235},
       number = {1},
          eid = {14},
        pages = {14},
          doi = {10.3847/1538-4365/aaacce},
archivePrefix = {arXiv},
       eprint = {1801.09734},
 primaryClass = {astro-ph.GA},
       adsurl = {https://ui.adsabs.harvard.edu/abs/2018ApJS..235...14S}
}

@ARTICLE{Pagul2021,
       author = {{Pagul}, A. and {S{\'a}nchez}, F.~J. and {Davidzon}, I. and {Mobasher}, Bahram},
        title = "{Hubble Frontier Field Clusters and Their Parallel Fields: Photometric and Photometric Redshift Catalogs}",
      journal = {\apjs},
         year = 2021,
        month = oct,
       volume = {256},
       number = {2},
          eid = {27},
        pages = {27},
          doi = {10.3847/1538-4365/abea9d},
archivePrefix = {arXiv},
       eprint = {2103.01952},
 primaryClass = {astro-ph.GA},
       adsurl = {https://ui.adsabs.harvard.edu/abs/2021ApJS..256...27P}
}

@ARTICLE{Mitchell2021,
       author = {{Mitchell}, Peter D. and {Blaizot}, J{\'e}r{\'e}my and {Cadiou}, Corentin and {Dubois}, Yohan and {Garel}, Thibault and {Rosdahl}, Joakim},
        title = "{Tracing the simulated high-redshift circumgalactic medium with Lyman {\ensuremath{\alpha}} emission}",
      journal = {\mnras},
         year = 2021,
        month = mar,
       volume = {501},
       number = {4},
        pages = {5757-5775},
          doi = {10.1093/mnras/stab035},
archivePrefix = {arXiv},
       eprint = {2008.12790},
 primaryClass = {astro-ph.GA},
       adsurl = {https://ui.adsabs.harvard.edu/abs/2021MNRAS.501.5757M}
}

@ARTICLE{Chang2024,
       author = {{Chang}, Seok-Jun and {Gronke}, Max},
        title = "{Probing cold gas with Mg II and Ly {\ensuremath{\alpha}} radiative transfer}",
      journal = {\mnras},
         year = 2024,
        month = aug,
       volume = {532},
       number = {3},
        pages = {3526-3555},
          doi = {10.1093/mnras/stae1664},
archivePrefix = {arXiv},
       eprint = {2403.11524},
 primaryClass = {astro-ph.GA},
       adsurl = {https://ui.adsabs.harvard.edu/abs/2024MNRAS.532.3526C}
}

@ARTICLE{Barbani2025,
       author = {{Barbani}, Filippo and {Pascale}, Raffaele and {Marinacci}, Federico and {Torrey}, Paul and {Sales}, Laura V. and {Li}, Hui and {Vogelsberger}, Mark},
        title = "{Understanding the baryon cycle: Fueling star formation via inflows in Milky Way-like galaxies}",
      journal = {\aap},
         year = 2025,
        month = may,
       volume = {697},
          eid = {A121},
        pages = {A121},
          doi = {10.1051/0004-6361/202452608},
archivePrefix = {arXiv},
       eprint = {2504.01075},
 primaryClass = {astro-ph.GA},
       adsurl = {https://ui.adsabs.harvard.edu/abs/2025A&A...697A.121B}
}

@ARTICLE{Kacprzak2025,
       author = {{Kacprzak}, Glenn G. and {Oppenheimer}, Benjamin and {Nielsen}, Nikole and {Fern{\'a}ndez-Figueroa}, Antonia and {Murphy}, Michael T. and {Allen}, Rebecca and {Barone}, Tania and {Sameer}, Sameer and {Churchill}, Christopher W. and {Burchett}, Joseph and {Gupta}, Kaustubh R. and {Charlton}, Jane C. and {Platukis}, Caleb},
        title = "{COS-EDGES: Co-rotation and kinematic stratification of the multi-phase CGM around edge-on galaxies}",
      journal = {\pasa},
         year = 2025,
        month = sep,
       volume = {42},
          eid = {e128},
        pages = {e128},
          doi = {10.1017/pasa.2025.10091},
archivePrefix = {arXiv},
       eprint = {2507.11613},
 primaryClass = {astro-ph.GA},
       adsurl = {https://ui.adsabs.harvard.edu/abs/2025PASA...42..128K}
}

@ARTICLE{Astropy2018,
       author = {{Astropy Collaboration} and {Price-Whelan}, A.~M. and {Sip{\H{o}}cz}, B.~M. and {G{\"u}nther}, H.~M. and {Lim}, P.~L. and {Crawford}, S.~M. and {Conseil}, S. and {Shupe}, D.~L. and {Craig}, M.~W. and {Dencheva}, N. and {Ginsburg}, A. and {VanderPlas}, J.~T. and {Bradley}, L.~D. and {P{\'e}rez-Su{\'a}rez}, D. and {de Val-Borro}, M. and {Aldcroft}, T.~L. and {Cruz}, K.~L. and {Robitaille}, T.~P. and {Tollerud}, E.~J. and {Ardelean}, C. and {Babej}, T. and {Bach}, Y.~P. and {Bachetti}, M. and {Bakanov}, A.~V. and {Bamford}, S.~P. and {Barentsen}, G. and {Barmby}, P. and {Baumbach}, A. and {Berry}, K.~L. and {Biscani}, F. and {Boquien}, M. and {Bostroem}, K.~A. and {Bouma}, L.~G. and {Brammer}, G.~B. and {Bray}, E.~M. and {Breytenbach}, H. and {Buddelmeijer}, H. and {Burke}, D.~J. and {Calderone}, G. and {Cano Rodr{\'\i}guez}, J.~L. and {Cara}, M. and {Cardoso}, J.~V.~M. and {Cheedella}, S. and {Copin}, Y. and {Corrales}, L. and {Crichton}, D. and {D'Avella}, D. and {Deil}, C. and {Depagne}, {\'E}. and {Dietrich}, J.~P. and {Donath}, A. and {Droettboom}, M. and {Earl}, N. and {Erben}, T. and {Fabbro}, S. and {Ferreira}, L.~A. and {Finethy}, T. and {Fox}, R.~T. and {Garrison}, L.~H. and {Gibbons}, S.~L.~J. and {Goldstein}, D.~A. and {Gommers}, R. and {Greco}, J.~P. and {Greenfield}, P. and {Groener}, A.~M. and {Grollier}, F. and {Hagen}, A. and {Hirst}, P. and {Homeier}, D. and {Horton}, A.~J. and {Hosseinzadeh}, G. and {Hu}, L. and {Hunkeler}, J.~S. and {Ivezi{\'c}}, {\v{Z}}. and {Jain}, A. and {Jenness}, T. and {Kanarek}, G. and {Kendrew}, S. and {Kern}, N.~S. and {Kerzendorf}, W.~E. and {Khvalko}, A. and {King}, J. and {Kirkby}, D. and {Kulkarni}, A.~M. and {Kumar}, A. and {Lee}, A. and {Lenz}, D. and {Littlefair}, S.~P. and {Ma}, Z. and {Macleod}, D.~M. and {Mastropietro}, M. and {McCully}, C. and {Montagnac}, S. and {Morris}, B.~M. and {Mueller}, M. and {Mumford}, S.~J. and {Muna}, D. and {Murphy}, N.~A. and {Nelson}, S. and {Nguyen}, G.~H. and {Ninan}, J.~P. and {N{\"o}the}, M. and {Ogaz}, S. and {Oh}, S. and {Parejko}, J.~K. and {Parley}, N. and {Pascual}, S. and {Patil}, R. and {Patil}, A.~A. and {Plunkett}, A.~L. and {Prochaska}, J.~X. and {Rastogi}, T. and {Reddy Janga}, V. and {Sabater}, J. and {Sakurikar}, P. and {Seifert}, M. and {Sherbert}, L.~E. and {Sherwood-Taylor}, H. and {Shih}, A.~Y. and {Sick}, J. and {Silbiger}, M.~T. and {Singanamalla}, S. and {Singer}, L.~P. and {Sladen}, P.~H. and {Sooley}, K.~A. and {Sornarajah}, S. and {Streicher}, O. and {Teuben}, P. and {Thomas}, S.~W. and {Tremblay}, G.~R. and {Turner}, J.~E.~H. and {Terr{\'o}n}, V. and {van Kerkwijk}, M.~H. and {de la Vega}, A. and {Watkins}, L.~L. and {Weaver}, B.~A. and {Whitmore}, J.~B. and {Woillez}, J. and {Zabalza}, V. and {Astropy Contributors}},
        title = "{The Astropy Project: Building an Open-science Project and Status of the v2.0 Core Package}",
      journal = {\aj},
         year = 2018,
        month = sep,
       volume = {156},
       number = {3},
          eid = {123},
        pages = {123},
          doi = {10.3847/1538-3881/aabc4f},
archivePrefix = {arXiv},
       eprint = {1801.02634},
 primaryClass = {astro-ph.IM},
       adsurl = {https://ui.adsabs.harvard.edu/abs/2018AJ....156..123A}
}

@ARTICLE{Astropy,
       author = {{Astropy Collaboration} and {Robitaille}, Thomas P. and {Tollerud}, Erik J. and {Greenfield}, Perry and {Droettboom}, Michael and {Bray}, Erik and {Aldcroft}, Tom and {Davis}, Matt and {Ginsburg}, Adam and {Price-Whelan}, Adrian M. and {Kerzendorf}, Wolfgang E. and {Conley}, Alexander and {Crighton}, Neil and {Barbary}, Kyle and {Muna}, Demitri and {Ferguson}, Henry and {Grollier}, Fr{\'e}d{\'e}ric and {Parikh}, Madhura M. and {Nair}, Prasanth H. and {Unther}, Hans M. and {Deil}, Christoph and {Woillez}, Julien and {Conseil}, Simon and {Kramer}, Roban and {Turner}, James E.~H. and {Singer}, Leo and {Fox}, Ryan and {Weaver}, Benjamin A. and {Zabalza}, Victor and {Edwards}, Zachary I. and {Azalee Bostroem}, K. and {Burke}, D.~J. and {Casey}, Andrew R. and {Crawford}, Steven M. and {Dencheva}, Nadia and {Ely}, Justin and {Jenness}, Tim and {Labrie}, Kathleen and {Lim}, Pey Lian and {Pierfederici}, Francesco and {Pontzen}, Andrew and {Ptak}, Andy and {Refsdal}, Brian and {Servillat}, Mathieu and {Streicher}, Ole},
        title = "{Astropy: A community Python package for astronomy}",
      journal = {\aap},
         year = 2013,
        month = oct,
       volume = {558},
          eid = {A33},
        pages = {A33},
          doi = {10.1051/0004-6361/201322068},
archivePrefix = {arXiv},
       eprint = {1307.6212},
 primaryClass = {astro-ph.IM},
       adsurl = {https://ui.adsabs.harvard.edu/abs/2013A&A...558A..33A}
}

@MISC{Prochaska2016,
       author = {{Prochaska}, J. Xavier and {Tejos}, Nicolas and {Crighton}, Neil and {jnburchett} and {Tuo-Ji} and {tiffanyhsyu} and {ktirimba} and {jhennawi} and {O'Meara}, John and {Werk}, Jessica},
        title = "{Linetools/Linetools: Second Major Release}",
 howpublished = {Zenodo},
         year = 2016,
        month = nov,
          eid = {10.5281/zenodo.168270},
          doi = {10.5281/zenodo.168270},
      version = {v0.2},
    publisher = {Zenodo},
       adsurl = {https://ui.adsabs.harvard.edu/abs/2016zndo....168270P}
}

@INPROCEEDINGS{PyMUSE,
       author = {{Pessa}, Ismael and {Tejos}, Nicolas and {Moya1}, Cristobal},
        title = "{PyMUSE: a Python Package for VLT/MUSE Data}",
    booktitle = {Astronomical Data Analysis Software and Systems XXVII},
         year = 2020,
       editor = {{Ballester}, Pascal and {Ibsen}, Jorge and {Solar}, Mauricio and {Shortridge}, Keith},
       series = {Astronomical Society of the Pacific Conference Series},
       volume = {522},
        month = apr,
        pages = {61},
       adsurl = {https://ui.adsabs.harvard.edu/abs/2020ASPC..522...61P}
}

@Article{Harris2020,
 title         = {Array programming with {NumPy}},
 author        = {Charles R. Harris and K. Jarrod Millman and St{\'{e}}fan J.
                 van der Walt and Ralf Gommers and Pauli Virtanen and David
                 Cournapeau and Eric Wieser and Julian Taylor and Sebastian
                 Berg and Nathaniel J. Smith and Robert Kern and Matti Picus
                 and Stephan Hoyer and Marten H. van Kerkwijk and Matthew
                 Brett and Allan Haldane and Jaime Fern{\'{a}}ndez del
                 R{\'{i}}o and Mark Wiebe and Pearu Peterson and Pierre
                 G{\'{e}}rard-Marchant and Kevin Sheppard and Tyler Reddy and
                 Warren Weckesser and Hameer Abbasi and Christoph Gohlke and
                 Travis E. Oliphant},
 year          = {2020},
 month         = sep,
 journal       = {Nat},
 volume        = {585},
 number        = {7825},
 pages         = {357--362},
 doi           = {10.1038/s41586-020-2649-2},
 publisher     = {Springer Science and Business Media {LLC}},
 url           = {https://doi.org/10.1038/s41586-020-2649-2}
}

@ARTICLE{Virtanen2020,
  author  = {Virtanen, Pauli and Gommers, Ralf and Oliphant, Travis E. and
            Haberland, Matt and Reddy, Tyler and Cournapeau, David and
            Burovski, Evgeni and Peterson, Pearu and Weckesser, Warren and
            Bright, Jonathan and {van der Walt}, St{\'e}fan J. and
            Brett, Matthew and Wilson, Joshua and Millman, K. Jarrod and
            Mayorov, Nikolay and Nelson, Andrew R. J. and Jones, Eric and
            Kern, Robert and Larson, Eric and Carey, C J and
            Polat, {\.I}lhan and Feng, Yu and Moore, Eric W. and
            {VanderPlas}, Jake and Laxalde, Denis and Perktold, Josef and
            Cimrman, Robert and Henriksen, Ian and Quintero, E. A. and
            Harris, Charles R. and Archibald, Anne M. and
            Ribeiro, Ant{\^o}nio H. and Pedregosa, Fabian and
            {van Mulbregt}, Paul and {SciPy 1.0 Contributors}},
  title   = {{{SciPy} 1.0: Fundamental Algorithms for Scientific
            Computing in Python}},
  journal = {Nature Methods},
  year    = {2020},
  volume  = {17},
  pages   = {261--272},
  adsurl  = {https://rdcu.be/b08Wh},
  doi     = {10.1038/s41592-019-0686-2},
}

@Article{Hunter2007,
  Author    = {Hunter, J. D.},
  Title     = {Matplotlib: A 2D graphics environment},
  Journal   = {Computing in Science \& Engineering},
  Volume    = {9},
  Number    = {3},
  Pages     = {90--95},
  publisher = {IEEE COMPUTER SOC},
  doi       = {10.1109/MCSE.2007.55},
  year      = 2007
}

@article{Zabl:2019ija,
author = {Zabl, Johannes and Bouch{\'e}, Nicolas F and Schroetter, Ilane and Wendt, Martin and Finley, Hayley and Schaye, Joop and Conseil, Simon and Contini, Thierry and Marino, Raffaella A and Mitchell, Peter and Muzahid, Sowgat and Pezzulli, Gabriele and Wisotzki, Lutz},
title = {{MusE GAs FLOw and Wind (MEGAFLOW) II. A study of gas accretion around z $\approx$ 1 star-forming galaxies with background quasars}},
journal = {MNRAS},
year = {2019},
volume = {485},
number = {2},
pages = {1961--1980},
month = may
}

@ARTICLE{Tremonti2004,
       author = {{Tremonti}, Christy A. and {Heckman}, Timothy M. and {Kauffmann}, Guinevere and {Brinchmann}, Jarle and {Charlot}, St{\'e}phane and {White}, Simon D.~M. and {Seibert}, Mark and {Peng}, Eric W. and {Schlegel}, David J. and {Uomoto}, Alan and {Fukugita}, Masataka and {Brinkmann}, Jon},
        title = "{The Origin of the Mass-Metallicity Relation: Insights from 53,000 Star-forming Galaxies in the Sloan Digital Sky Survey}",
      journal = {\apj},
         year = 2004,
        month = oct,
       volume = {613},
       number = {2},
        pages = {898-913},
          doi = {10.1086/423264},
archivePrefix = {arXiv},
       eprint = {astro-ph/0405537},
 primaryClass = {astro-ph},
       adsurl = {https://ui.adsabs.harvard.edu/abs/2004ApJ...613..898T}
}

@ARTICLE{Kelleher2008,
       author = {{Kelleher}, D.~E. and {Podobedova}, L.~I.},
        title = "{Atomic Transition Probabilities of Sodium and Magnesium. A Critical Compilation}",
      journal = {Journal of Physical and Chemical Reference Data},
         year = 2008,
        month = mar,
       volume = {37},
       number = {1},
        pages = {267-706},
          doi = {10.1063/1.2735328},
       adsurl = {https://ui.adsabs.harvard.edu/abs/2008JPCRD..37..267K}
}

@ARTICLE{Guo2023,
       author = {{Guo}, Yucheng and {Bacon}, Roland and {Wisotzki}, Lutz and {Garel}, Thibault and {Blaizot}, J{\'e}r{\'e}my and {Schaye}, Joop and {Matthee}, Jorryt and {Leclercq}, Floriane and {Boogaard}, Leindert and {Richard}, Johan and {Verhamme}, Anne and {Brinchmann}, Jarle and {Michel-Dansac}, L{\'e}o and {Kusakabe}, Haruka},
        title = "{Spatially resolved spectroscopic analysis of Ly{\ensuremath{\alpha}} haloes: Radial evolution of the Ly{\ensuremath{\alpha}} line profile out to 60 kpc}",
      journal = {\aap},
         year = 2024,
        month = nov,
       volume = {691},
          eid = {A66},
        pages = {A66},
          doi = {10.1051/0004-6361/202347958},
archivePrefix = {arXiv},
       eprint = {2309.06311},
 primaryClass = {astro-ph.GA},
       adsurl = {https://ui.adsabs.harvard.edu/abs/2024A&A...691A..66G}
}

@ARTICLE{Bruzual2003,
       author = {{Bruzual}, G. and {Charlot}, S.},
        title = "{Stellar population synthesis at the resolution of 2003}",
      journal = {\mnras},
         year = 2003,
        month = oct,
       volume = {344},
       number = {4},
        pages = {1000-1028},
          doi = {10.1046/j.1365-8711.2003.06897.x},
archivePrefix = {arXiv},
       eprint = {astro-ph/0309134},
 primaryClass = {astro-ph},
       adsurl = {https://ui.adsabs.harvard.edu/abs/2003MNRAS.344.1000B}
}

@ARTICLE{Chabrier2003,
       author = {{Chabrier}, Gilles},
        title = "{Galactic Stellar and Substellar Initial Mass Function}",
      journal = {\pasp},
         year = 2003,
        month = jul,
       volume = {115},
       number = {809},
        pages = {763-795},
          doi = {10.1086/376392},
archivePrefix = {arXiv},
       eprint = {astro-ph/0304382},
 primaryClass = {astro-ph},
       adsurl = {https://ui.adsabs.harvard.edu/abs/2003PASP..115..763C}
}

@ARTICLE{Ellison2021,
       author = {{Ellison}, Sara L. and {Lin}, Lihwai and {Thorp}, Mallory D. and {Pan}, Hsi-An and {Scudder}, Jillian M. and {S{\'a}nchez}, Sebastian F. and {Bluck}, Asa F.~L. and {Maiolino}, Roberto},
        title = "{The ALMaQUEST Survey - V. The non-universality of kpc-scale star formation relations and the factors that drive them}",
      journal = {\mnras},
         year = 2021,
        month = mar,
       volume = {501},
       number = {4},
        pages = {4777-4797},
          doi = {10.1093/mnras/staa3822},
archivePrefix = {arXiv},
       eprint = {2012.04771},
 primaryClass = {astro-ph.GA},
       adsurl = {https://ui.adsabs.harvard.edu/abs/2021MNRAS.501.4777E}
}

@ARTICLE{Dutta2023,
       author = {{Dutta}, Rajeshwari and {Fossati}, Matteo and {Fumagalli}, Michele and {Revalski}, Mitchell and {Lofthouse}, Emma K. and {Nelson}, Dylan and {Papini}, Giulia and {Rafelski}, Marc and {Cantalupo}, Sebastiano and {Arrigoni Battaia}, Fabrizio and {Dayal}, Pratika and {Longobardi}, Alessia and {P{\'e}roux}, Celine and {Prichard}, Laura J. and {Prochaska}, J. Xavier},
        title = "{Metal line emission from galaxy haloes at z {\ensuremath{\approx}} 1}",
      journal = {\mnras},
         year = 2023,
        month = jun,
       volume = {522},
       number = {1},
        pages = {535-558},
          doi = {10.1093/mnras/stad1002},
archivePrefix = {arXiv},
       eprint = {2302.09087},
 primaryClass = {astro-ph.GA},
       adsurl = {https://ui.adsabs.harvard.edu/abs/2023MNRAS.522..535D}
}

@ARTICLE{Ibarra2016,
       author = {{Ibarra-Medel}, H{\'e}ctor J. and {S{\'a}nchez}, Sebasti{\'a}n F. and {Avila-Reese}, Vladimir and {Hern{\'a}ndez-Toledo}, H{\'e}ctor M. and {Gonz{\'a}lez}, J. Jes{\'u}s and {Drory}, Niv and {Bundy}, Kevin and {Bizyaev}, Dmitry and {Cano-D{\'\i}az}, Mariana and {Malanushenko}, Elena and {Pan}, Kaike and {Roman-Lopes}, Alexandre and {Thomas}, Daniel},
        title = "{SDSS IV MaNGA: the global and local stellar mass assemby histories of galaxies}",
      journal = {\mnras},
         year = 2016,
        month = dec,
       volume = {463},
       number = {3},
        pages = {2799-2818},
          doi = {10.1093/mnras/stw2126},
archivePrefix = {arXiv},
       eprint = {1609.01304},
 primaryClass = {astro-ph.GA},
       adsurl = {https://ui.adsabs.harvard.edu/abs/2016MNRAS.463.2799I}
}

@ARTICLE{Smith2022,
       author = {{Smith}, Madison V. and {van Zee}, L. and {Dale}, D.~A. and {Hunter}, L.~C. and {Staudaher}, S. and {Wrock}, T.},
        title = "{A multiwavelength study of star formation in nearby galaxies: evidence for inside-out growth of the stellar disc}",
      journal = {\mnras},
         year = 2022,
        month = sep,
       volume = {515},
       number = {3},
        pages = {3270-3298},
          doi = {10.1093/mnras/stac1974},
       adsurl = {https://ui.adsabs.harvard.edu/abs/2022MNRAS.515.3270S}
}

@ARTICLE{Pessa2023,
       author = {{Pessa}, I. and {Schinnerer}, E. and {Sanchez-Blazquez}, P. and {Belfiore}, F. and {Groves}, B. and {Emsellem}, E. and {Neumann}, J. and {Leroy}, A.~K. and {Bigiel}, F. and {Chevance}, M. and {Dale}, D.~A. and {Glover}, S.~C.~O. and {Grasha}, K. and {Klessen}, R.~S. and {Kreckel}, K. and {Kruijssen}, J.~M.~D. and {Pinna}, F. and {Querejeta}, M. and {Rosolowsky}, E. and {Williams}, T.~G.},
        title = "{Resolved stellar population properties of PHANGS-MUSE galaxies}",
      journal = {\aap},
         year = 2023,
        month = may,
       volume = {673},
          eid = {A147},
        pages = {A147},
          doi = {10.1051/0004-6361/202245673},
archivePrefix = {arXiv},
       eprint = {2303.13676},
 primaryClass = {astro-ph.GA},
       adsurl = {https://ui.adsabs.harvard.edu/abs/2023A&A...673A.147P}
}

@INPROCEEDINGS{Weilbacher2014,
       author = {{Weilbacher}, P.~M. and {Streicher}, O. and {Urrutia}, T. and {P{\'e}contal-Rousset}, A. and {Jarno}, A. and {Bacon}, R.},
        title = "{The MUSE Data Reduction Pipeline: Status after Preliminary Acceptance Europe}",
    booktitle = {Astronomical Data Analysis Software and Systems XXIII},
         year = 2014,
       editor = {{Manset}, N. and {Forshay}, P.},
       series = {Astronomical Society of the Pacific Conference Series},
       volume = {485},
        month = may,
        pages = {451},
          doi = {10.48550/arXiv.1507.00034},
archivePrefix = {arXiv},
       eprint = {1507.00034},
 primaryClass = {astro-ph.IM},
       adsurl = {https://ui.adsabs.harvard.edu/abs/2014ASPC..485..451W}
}

@ARTICLE{ReichardtChu2025,
       author = {{Reichardt Chu}, Bronwyn and {Fisher}, Deanne B. and {Chisholm}, John and {Berg}, Danielle and {Bolatto}, Alberto and {Cameron}, Alex J. and {Fielding}, Drummond B. and {Herrera-Camus}, Rodrigo and {Kacprzak}, Glenn G. and {Li}, Miao and {McLeod}, Anna F. and {McPherson}, Daniel K. and {Nielsen}, Nikole M. and {Rickards Vaught}, Ryan J. and {Ridolfo}, Sophia G. and {Sandstrom}, Karin},
        title = "{DUVET: sub-kiloparsec resolved star formation driven outflows in a sample of local starbursting disc galaxies}",
      journal = {\mnras},
         year = 2025,
        month = jan,
       volume = {536},
       number = {2},
        pages = {1799-1821},
          doi = {10.1093/mnras/stae2705},
archivePrefix = {arXiv},
       eprint = {2402.17830},
 primaryClass = {astro-ph.GA},
       adsurl = {https://ui.adsabs.harvard.edu/abs/2025MNRAS.536.1799R}
}

@ARTICLE{Figueroa2025,
       author = {{Fern{\'a}ndez-Figueroa}, Antonia and {Kacprzak}, Glenn G. and {Barone}, Tania M. and {Nielsen}, Nikole M. and {Rubin}, Kate H.~R. and {Pitts}, Andrew J. and {Mazzilli Ciraulo}, Barbara},
        title = "{Ultrastrong Mg II absorbers trace both inflowing and outflowing gas: insights from dual down-the-barrel and quasar sightlines}",
      journal = {\mnras},
         year = 2025,
        month = nov,
       volume = {544},
       number = {1},
        pages = {255-270},
          doi = {10.1093/mnras/staf1712},
archivePrefix = {arXiv},
       eprint = {2510.02708},
 primaryClass = {astro-ph.GA},
       adsurl = {https://ui.adsabs.harvard.edu/abs/2025MNRAS.544..255F}
}

@ARTICLE{Urrutia2019,
       author = {{Urrutia}, T. and {Wisotzki}, L. and {Kerutt}, J. and {Schmidt}, K.~B. and {Herenz}, E.~C. and {Klar}, J. and {Saust}, R. and {Werhahn}, M. and {Diener}, C. and {Caruana}, J. and {Krajnovi{\'c}}, D. and {Bacon}, R. and {Boogaard}, L. and {Brinchmann}, J. and {Enke}, H. and {Maseda}, M. and {Nanayakkara}, T. and {Richard}, J. and {Steinmetz}, M. and {Weilbacher}, P.~M.},
        title = "{The MUSE-Wide Survey: survey description and first data release}",
      journal = {\aap},
         year = 2019,
        month = apr,
       volume = {624},
          eid = {A141},
        pages = {A141},
          doi = {10.1051/0004-6361/201834656},
archivePrefix = {arXiv},
       eprint = {1811.06549},
 primaryClass = {astro-ph.GA},
       adsurl = {https://ui.adsabs.harvard.edu/abs/2019A&A...624A.141U}
}

@ARTICLE{Weilbacher2020,
       author = {{Weilbacher}, Peter M. and {Palsa}, Ralf and {Streicher}, Ole and {Bacon}, Roland and {Urrutia}, Tanya and {Wisotzki}, Lutz and {Conseil}, Simon and {Husemann}, Bernd and {Jarno}, Aur{\'e}lien and {Kelz}, Andreas and {P{\'e}contal-Rousset}, Arlette and {Richard}, Johan and {Roth}, Martin M. and {Selman}, Fernando and {Vernet}, Jo{\"e}l},
        title = "{The data processing pipeline for the MUSE instrument}",
      journal = {\aap},
         year = 2020,
        month = sep,
       volume = {641},
          eid = {A28},
        pages = {A28},
          doi = {10.1051/0004-6361/202037855},
archivePrefix = {arXiv},
       eprint = {2006.08638},
 primaryClass = {astro-ph.IM},
       adsurl = {https://ui.adsabs.harvard.edu/abs/2020A&A...641A..28W}
}

@ARTICLE{Pessa2021,
       author = {{Pessa}, I. and {Schinnerer}, E. and {Belfiore}, F. and {Emsellem}, E. and {Leroy}, A.~K. and {Schruba}, A. and {Kruijssen}, J.~M.~D. and {Pan}, H. -A. and {Blanc}, G.~A. and {Sanchez-Blazquez}, P. and {Bigiel}, F. and {Chevance}, M. and {Congiu}, E. and {Dale}, D. and {Faesi}, C.~M. and {Glover}, S.~C.~O. and {Grasha}, K. and {Groves}, B. and {Ho}, I. and {Jim{\'e}nez-Donaire}, M. and {Klessen}, R. and {Kreckel}, K. and {Koch}, E.~W. and {Liu}, D. and {Meidt}, S. and {Pety}, J. and {Querejeta}, M. and {Rosolowsky}, E. and {Saito}, T. and {Santoro}, F. and {Sun}, J. and {Usero}, A. and {Watkins}, E.~J. and {Williams}, T.~G.},
        title = "{Star formation scaling relations at {\ensuremath{\sim}}100 pc from PHANGS: Impact of completeness and spatial scale}",
      journal = {\aap},
         year = 2021,
        month = jun,
       volume = {650},
          eid = {A134},
        pages = {A134},
          doi = {10.1051/0004-6361/202140733},
archivePrefix = {arXiv},
       eprint = {2104.09536},
 primaryClass = {astro-ph.GA},
       adsurl = {https://ui.adsabs.harvard.edu/abs/2021A&A...650A.134P}
}

@ARTICLE{Peroux2020,
       author = {{P{\'e}roux}, C{\'e}line and {Nelson}, Dylan and {van de Voort}, Freeke and {Pillepich}, Annalisa and {Marinacci}, Federico and {Vogelsberger}, Mark and {Hernquist}, Lars},
        title = "{Predictions for the angular dependence of gas mass flow rate and metallicity in the circumgalactic medium}",
      journal = {\mnras},
         year = 2020,
        month = dec,
       volume = {499},
       number = {2},
        pages = {2462-2473},
          doi = {10.1093/mnras/staa2888},
archivePrefix = {arXiv},
       eprint = {2009.07809},
 primaryClass = {astro-ph.GA},
       adsurl = {https://ui.adsabs.harvard.edu/abs/2020MNRAS.499.2462P}
}

@article{Feltre:2018in,
author = {Feltre, Anna and Bacon, Roland and Tresse, Laurence and Finley, Hayley and Carton, David and Blaizot, J{\'e}r{\'e}my and Bouch{\'e}, Nicolas F and Garel, Thibault and Inami, Hanae and Boogaard, Leindert A and Brinchmann, Jarle and Charlot, St{\'e}phane and Chevallard, Jacopo and Contini, Thierry and Michel-Dansac, Leo and Mahler, Guillaume and Marino, Raffaella A and Maseda, Michael V and Richard, Johan and Schmidt, Kasper B and Verhamme, Anne},
title = {{The MUSE Hubble Ultra Deep Field Survey. XII. Mg II emission and absorption in star-forming galaxies}},
journal = {A{\&}A},
year = {2018},
volume = {617},
pages = {A62},
month = sep
}

@article{Erb:2018dw,
author = {Erb, Dawn K and Steidel, Charles C and Chen, Yuguang},
title = {{The Kinematics of Extended Ly $\alpha$ Emission in a Low-mass, Low-metallicity Galaxy at z= 2.3}},
journal = {ApJ},
year = {2018},
volume = {862},
number = {1},
pages = {L10},
month = jul
}

@ARTICLE{Rubin2014,
       author = {{Rubin}, Kate H.~R. and {Prochaska}, J. Xavier and {Koo}, David C. and {Phillips}, Andrew C. and {Martin}, Crystal L. and {Winstrom}, Lucas O.},
        title = "{Evidence for Ubiquitous Collimated Galactic-scale Outflows along the Star-forming Sequence at z \raisebox{-0.5ex}\textasciitilde 0.5}",
      journal = {\apj},
         year = 2014,
        month = oct,
       volume = {794},
       number = {2},
          eid = {156},
        pages = {156},
          doi = {10.1088/0004-637X/794/2/156},
archivePrefix = {arXiv},
       eprint = {1307.1476},
 primaryClass = {astro-ph.CO},
       adsurl = {https://ui.adsabs.harvard.edu/abs/2014ApJ...794..156R}
}

@ARTICLE{Schroetter2024,
       author = {{Schroetter}, Ilane and {Bouch{\'e}}, Nicolas F. and {Zabl}, Johannes and {Wendt}, Martin and {Cherrey}, Maxime and {Langan}, Ivanna and {Schaye}, Joop and {Contini}, Thierry},
        title = "{MusE GAs FLOw and Wind (MEGAFLOW). XI. Scaling relations between outflows and host galaxy properties}",
      journal = {\aap},
         year = 2024,
        month = jul,
       volume = {687},
          eid = {A39},
        pages = {A39},
          doi = {10.1051/0004-6361/202348725},
archivePrefix = {arXiv},
       eprint = {2404.03300},
 primaryClass = {astro-ph.GA},
       adsurl = {https://ui.adsabs.harvard.edu/abs/2024A&A...687A..39S}
}

@ARTICLE{Bouche2025,
       author = {{Bouch{\'e}}, Nicolas F. and {Wendt}, Martin and {Zabl}, Johannes and {Cherrey}, Maxime and {Schroetter}, Ilane and {Langan}, Ivanna and {Muzahid}, Sowgat and {Schaye}, Joop and {Epinat}, Beno{\^\i}t and {Wisotzki}, Lutz and {Contini}, Thierry and {Richard}, Johan and {Bacon}, Roland and {Weilbacher}, Peter M.},
        title = "{MusE GAs FLOw and Wind (MEGAFLOW): XII. Rationale and design of a Mg II survey of the cool circum-galactic medium with MUSE and UVES: The MEGAFLOW Survey}",
      journal = {\aap},
         year = 2025,
        month = feb,
       volume = {694},
          eid = {A67},
        pages = {A67},
          doi = {10.1051/0004-6361/202451093},
archivePrefix = {arXiv},
       eprint = {2411.07014},
 primaryClass = {astro-ph.GA},
       adsurl = {https://ui.adsabs.harvard.edu/abs/2025A&A...694A..67B}
}

@ARTICLE{Bordoloi2014,
       author = {{Bordoloi}, R. and {Lilly}, S.~J. and {Hardmeier}, E. and {Contini}, T. and {Kneib}, J. -P. and {Le Fevre}, O. and {Mainieri}, V. and {Renzini}, A. and {Scodeggio}, M. and {Zamorani}, G. and {Bardelli}, S. and {Bolzonella}, M. and {Bongiorno}, A. and {Caputi}, K. and {Carollo}, C.~M. and {Cucciati}, O. and {de la Torre}, S. and {de Ravel}, L. and {Garilli}, B. and {Iovino}, A. and {Kampczyk}, P. and {Kova{\v{c}}}, K. and {Knobel}, C. and {Lamareille}, F. and {Le Borgne}, J. -F. and {Le Brun}, V. and {Maier}, C. and {Mignoli}, M. and {Oesch}, P. and {Pello}, R. and {Peng}, Y. and {Perez Montero}, E. and {Presotto}, V. and {Silverman}, J. and {Tanaka}, M. and {Tasca}, L. and {Tresse}, L. and {Vergani}, D. and {Zucca}, E. and {Cappi}, A. and {Cimatti}, A. and {Coppa}, G. and {Franzetti}, P. and {Koekemoer}, A. and {Moresco}, M. and {Nair}, P. and {Pozzetti}, L.},
        title = "{The Dependence of Galactic Outflows on the Properties and Orientation of zCOSMOS Galaxies at z \raisebox{-0.5ex}\textasciitilde 1}",
      journal = {\apj},
         year = 2014,
        month = oct,
       volume = {794},
       number = {2},
          eid = {130},
        pages = {130},
          doi = {10.1088/0004-637X/794/2/130},
archivePrefix = {arXiv},
       eprint = {1307.6553},
 primaryClass = {astro-ph.CO},
       adsurl = {https://ui.adsabs.harvard.edu/abs/2014ApJ...794..130B}
}

@ARTICLE{Das2025,
       author = {{Das}, Sarbeswar and {Joshi}, Ravi and {Chaudhary}, Reena and {Fumagalli}, Michele and {Fossati}, Matteo and {P{\'e}roux}, Celine and {Ho}, Luis C.},
        title = "{Baryonic Ecosystem IN Galaxies (BEINGMgII): II. Unveiling the nature of galaxies harbouring cool gas reservoirs}",
      journal = {\aap},
         year = 2025,
        month = mar,
       volume = {695},
          eid = {A207},
        pages = {A207},
          doi = {10.1051/0004-6361/202452494},
archivePrefix = {arXiv},
       eprint = {2410.03824},
 primaryClass = {astro-ph.GA},
       adsurl = {https://ui.adsabs.harvard.edu/abs/2025A&A...695A.207D}
}

@ARTICLE{Harvey2025,
       author = {{Harvey}, Zoe and {Krishna}, Sahyadri and {Wild}, Vivienne and {Tojeiro}, Rita and {Hewett}, Paul},
        title = "{Cool Gas in the Circumgalactic Medium of Massive Post Starburst Galaxies}",
      journal = {The Open Journal of Astrophysics},
         year = 2025,
        month = nov,
       volume = {8},
        pages = {47836},
          doi = {10.33232/001c.147836},
archivePrefix = {arXiv},
       eprint = {2506.22287},
 primaryClass = {astro-ph.GA},
       adsurl = {https://ui.adsabs.harvard.edu/abs/2025OJAp....847836H}
}

@ARTICLE{Verhamme2017,
       author = {{Verhamme}, A. and {Orlitov{\'a}}, I. and {Schaerer}, D. and {Izotov}, Y. and {Worseck}, G. and {Thuan}, T.~X. and {Guseva}, N.},
        title = "{Lyman-{\ensuremath{\alpha}} spectral properties of five newly discovered Lyman continuum emitters}",
      journal = {\aap},
         year = 2017,
        month = jan,
       volume = {597},
          eid = {A13},
        pages = {A13},
          doi = {10.1051/0004-6361/201629264},
archivePrefix = {arXiv},
       eprint = {1609.03477},
 primaryClass = {astro-ph.GA},
       adsurl = {https://ui.adsabs.harvard.edu/abs/2017A&A...597A..13V}
}

@article{Henry:2018gd,
author = {Henry, Alaina and Berg, Danielle A and Scarlata, Claudia and Verhamme, Anne and Erb, Dawn},
title = {{A Close Relationship between Lyalpha and Mg II in Green Pea Galaxies}},
journal = {ApJ},
year = {2018},
volume = {855},
number = {2},
pages = {96},
month = mar
}

@ARTICLE{FernandezFigueroa2022,
       author = {{Fernandez-Figueroa}, A. and {Lopez}, S. and {Tejos}, N. and {Berg}, T.~A.~M. and {Ledoux}, C. and {Noterdaeme}, P. and {Afruni}, A. and {Barrientos}, L.~F. and {Gonzalez-Lopez}, J. and {Hamel}, M. and {Johnston}, E.~J. and {Katsianis}, A. and {Sharon}, K. and {Solimano}, M.},
        title = "{Orientation effects on cool gas absorption from gravitational-arc tomography of a z = 0.77 disc galaxy}",
      journal = {\mnras},
         year = 2022,
        month = dec,
       volume = {517},
       number = {2},
        pages = {2214-2220},
          doi = {10.1093/mnras/stac2851},
       adsurl = {https://ui.adsabs.harvard.edu/abs/2022MNRAS.517.2214F}
}

@article{Bacon:2017hn,
author = {Bacon, Roland and Conseil, Simon and Mary, David and Brinchmann, Jarle and Shepherd, Martin and Akhlaghi, Mohammad and Weilbacher, Peter M and Piqueras, Laure and Wisotzki, Lutz and Lagattuta, David and Epinat, Benoit and Guerou, Adrien and Inami, Hanae and Cantalupo, Sebastiano and Courbot, Jean Baptiste and Contini, Thierry and Richard, Johan and Maseda, Michael and Bouwens, Rychard and Bouch{\'e}, Nicolas F and Kollatschny, Wolfram and Schaye, Joop and Marino, Raffaella Anna and Pello, Roser and Herenz, Christian and Guiderdoni, Bruno and Carollo, Marcella},
title = {{The MUSE Hubble Ultra Deep Field Survey. I. Survey description, data reduction, and source detection}},
journal = {A{\&}A},
year = {2017},
volume = {608},
pages = {A1},
month = nov
}

@article{Leclercq:2017cp,
author = {Leclercq, Floriane and Bacon, Roland and Wisotzki, Lutz and Mitchell, Peter and Garel, Thibault and Verhamme, Anne and Blaizot, J{\'e}r{\'e}my and Hashimoto, Takuya and Herenz, Edmund Christian and Conseil, Simon and Cantalupo, Sebastiano and Inami, Hanae and Contini, Thierry and Richard, Johan and Maseda, Michael and Schaye, Joop and Marino, Raffaella Anna and Akhlaghi, Mohammad and Brinchmann, Jarle and Carollo, Marcella},
title = {{The MUSE Hubble Ultra Deep Field Survey. VIII. Extended Lyman-$\alpha$ haloes around high-z star-forming galaxies}},
journal = {A{\&}A},
year = {2017},
volume = {608},
pages = {A8},
month = nov
}

@article{Finley:2017eg,
author = {Finley, Hayley and Bouch{\'e}, Nicolas F and Contini, Thierry and Paalvast, Mieke and Boogaard, Leindert and Maseda, Michael and Bacon, Roland and Blaizot, J{\'e}r{\'e}my and Brinchmann, Jarle and Epinat, Benoit and Feltre, Anna and Marino, Raffaella Anna and Muzahid, Sowgat and Richard, Johan and Schaye, Joop and Verhamme, Anne and Weilbacher, Peter M and Wisotzki, Lutz},
title = {{The MUSE Hubble Ultra Deep Field Survey. VII. Fe II* emission in star-forming galaxies}},
journal = {A{\&}A},
year = {2017},
volume = {608},
pages = {A7},
month = nov
}

@article{Finley:2017gc,
author = {Finley, Hayley and Bouch{\'e}, Nicolas F and Contini, Thierry and Epinat, Benoit and Bacon, Roland and Brinchmann, Jarle and Cantalupo, Sebastiano and Erroz-Ferrer, Santiago and Marino, Raffaella Anna and Maseda, Michael and Richard, Johan and Schroetter, Ilane and Verhamme, Anne and Weilbacher, Peter M and Wendt, Martin and Wisotzki, Lutz},
title = {{Galactic winds with MUSE: A direct detection of Fe II* emission from a z = 1.29 galaxy}},
journal = {A{\&}A},
year = {2017},
volume = {605},
pages = {A118},
month = sep
}

@ARTICLE{Remolina2019,
       author = {{Remolina-Guti{\'e}rrez}, Maria Camila and {Forero-Romero}, Jaime E.},
        title = "{Lyman {\ensuremath{\alpha}} photons through rotating outflows}",
      journal = {\mnras},
         year = 2019,
        month = feb,
       volume = {482},
       number = {4},
        pages = {4553-4561},
          doi = {10.1093/mnras/sty3009},
archivePrefix = {arXiv},
       eprint = {1811.01266},
 primaryClass = {astro-ph.GA},
       adsurl = {https://ui.adsabs.harvard.edu/abs/2019MNRAS.482.4553R}
}

@article{Herenz:2017er,
author = {Herenz, Edmund Christian and Wisotzki, Lutz},
title = {{LSDCat: Detection and cataloguing of emission-line sources in integral-field spectroscopy datacubes}},
journal = {A{\&}A},
year = {2017},
volume = {602},
pages = {A111},
month = jun
}

@ARTICLE{DallaVecchia2008,
       author = {{Dalla Vecchia}, Claudio and {Schaye}, Joop},
        title = "{Simulating galactic outflows with kinetic supernova feedback}",
      journal = {\mnras},
         year = 2008,
        month = jul,
       volume = {387},
       number = {4},
        pages = {1431-1444},
          doi = {10.1111/j.1365-2966.2008.13322.x},
archivePrefix = {arXiv},
       eprint = {0801.2770},
 primaryClass = {astro-ph},
       adsurl = {https://ui.adsabs.harvard.edu/abs/2008MNRAS.387.1431D}
}

@ARTICLE{Kusakabe2022,
       author = {{Kusakabe}, Haruka and {Verhamme}, Anne and {Blaizot}, J{\'e}r{\'e}my and {Garel}, Thibault and {Wisotzki}, Lutz and {Leclercq}, Floriane and {Bacon}, Roland and {Schaye}, Joop and {Gallego}, Sofia G. and {Kerutt}, Josephine and {Matthee}, Jorryt and {Maseda}, Michael and {Nanayakkara}, Themiya and {Pell{\'o}}, Roser and {Richard}, Johan and {Tresse}, Laurence and {Urrutia}, Tanya and {Vitte}, Elo{\"\i}se},
        title = "{The MUSE eXtremely Deep Field: Individual detections of Ly{\ensuremath{\alpha}} haloes around rest-frame UV-selected galaxies at z ≃ 2.9-4.4}",
      journal = {\aap},
         year = 2022,
        month = apr,
       volume = {660},
          eid = {A44},
        pages = {A44},
          doi = {10.1051/0004-6361/202142302},
archivePrefix = {arXiv},
       eprint = {2201.07257},
 primaryClass = {astro-ph.GA},
       adsurl = {https://ui.adsabs.harvard.edu/abs/2022A&A...660A..44K}
}

@ARTICLE{Turner2017,
       author = {{Turner}, Monica L. and {Schaye}, Joop and {Crain}, Robert A. and {Rudie}, Gwen and {Steidel}, Charles C. and {Strom}, Allison and {Theuns}, Tom},
        title = "{A comparison of observed and simulated absorption from H I, C IV, and Si IV around z {\ensuremath{\approx}} 2 star-forming galaxies suggests redshift-space distortions are due to inflows}",
      journal = {\mnras},
         year = 2017,
        month = oct,
       volume = {471},
       number = {1},
        pages = {690-705},
          doi = {10.1093/mnras/stx1616},
archivePrefix = {arXiv},
       eprint = {1703.00086},
 primaryClass = {astro-ph.GA},
       adsurl = {https://ui.adsabs.harvard.edu/abs/2017MNRAS.471..690T}
}

@article{Schroetter:2016bl,
author = {Schroetter, I and Bouch{\'e}, N F and Wendt, M and Contini, T and Finley, H and Pell{\`o}, R and Bacon, R and Cantalupo, S and Marino, R A and Richard, J and Lilly, S J and Schaye, J and Soto, K and Steinmetz, M and Straka, L A and Wisotzki, L.},
title = {{Muse Gas Flow and Wind (MEGAFLOW). I. First MUSE Results on Background Quasars}},
journal = {ApJ},
year = {2016},
volume = {833},
number = {1},
pages = {39},
month = dec
}

@ARTICLE{Castor1979,
       author = {{Castor}, J.~I. and {Lamers}, H.~J.~G.~L.~M.},
        title = "{An atlas of theoretical P Cygni profiles.}",
      journal = {\apjs},
         year = 1979,
        month = apr,
       volume = {39},
        pages = {481-511},
          doi = {10.1086/190583},
       adsurl = {https://ui.adsabs.harvard.edu/abs/1979ApJS...39..481C}
}

@article{Wisotzki:2016hw,
author = {Wisotzki, L. and Bacon, R and Blaizot, J and Brinchmann, J and Herenz, E C and Schaye, J and Bouch{\'e}, N F and Cantalupo, S and Contini, T and Carollo, C M and Caruana, J and Courbot, J B and Emsellem, E and Kamann, S and Kerutt, J and Leclercq, F and Lilly, S J and Patricio, V and Sandin, C and Steinmetz, M and Straka, L A and Urrutia, T and Verhamme, A and Weilbacher, P M and Wendt, M},
title = {{Extended Lyman $\alpha$ haloes around individual high-redshift galaxies revealed by MUSE}},
journal = {A{\&}A},
year = {2016},
volume = {587},
pages = {A98},
month = feb
}

@article{Bouche:2013il,
author = {Bouch{\'e}, N F and Murphy, M T and Kacprzak, G G and Peroux, C. and Contini, T and Martin, C L and Dessauges-Zavadsky, M},
title = {{Signatures of Cool Gas Fueling a Star-Forming Galaxy at Redshift 2.3}},
journal = {Science},
year = {2013},
volume = {341},
number = {6141},
pages = {50--53},
month = jul
}

@article{Martin:2013ho,
author = {Martin, Crystal L and Shapley, Alice E and Coil, Alison L and Kornei, Katherine A and Murray, Norman and Pancoast, Anna},
title = {{SCATTERED EMISSION FROM z∼ 1 GALACTIC OUTFLOWS}},
journal = {ApJ},
year = {2013},
volume = {770},
number = {1},
pages = {41},
month = jun
}

@article{ForemanMackey:2013io,
author = {Foreman-Mackey, Daniel and Hogg, David W. and Lang, Dustin and Goodman, Jonathan},
title = {{emcee: The MCMC Hammer}},
journal = {PASP},
year = {2013},
volume = {125},
number = {925},
pages = {306--312},
month = mar
}

@article{Bouche:2012ho,
author = {Bouch{\'e}, N F and Hohensee, W and Vargas, R and Kacprzak, G G and Martin, C L and Cooke, J and Churchill, C W},
title = {{Physical properties of galactic winds using background quasars}},
journal = {Monthly Notice of the Royal Astronomical Society},
year = {2012},
volume = {426},
number = {2},
pages = {801--815},
month = oct
}

@ARTICLE{Fossati2019,
       author = {{Fossati}, M. and {Fumagalli}, M. and {Lofthouse}, E.~K. and {D'Odorico}, V. and {Lusso}, E. and {Cantalupo}, S. and {Cooke}, R.~J. and {Cristiani}, S. and {Haardt}, F. and {Morris}, S.~L. and {Peroux}, C. and {Prichard}, L.~J. and {Rafelski}, M. and {Smail}, I. and {Theuns}, T.},
        title = "{The MUSE Ultra Deep Field (MUDF). II. Survey design and the gaseous properties of galaxy groups at 0.5 < z < 1.5}",
      journal = {\mnras},
         year = 2019,
        month = nov,
       volume = {490},
       number = {1},
        pages = {1451-1469},
          doi = {10.1093/mnras/stz2693},
archivePrefix = {arXiv},
       eprint = {1909.04672},
 primaryClass = {astro-ph.GA},
       adsurl = {https://ui.adsabs.harvard.edu/abs/2019MNRAS.490.1451F}
}

@ARTICLE{Byrohl2020,
       author = {{Byrohl}, C. and {Gronke}, M.},
        title = "{Variations in shape among observed Lyman-{\ensuremath{\alpha}} spectra due to intergalactic absorption}",
      journal = {\aap},
         year = 2020,
        month = oct,
       volume = {642},
          eid = {L16},
        pages = {L16},
          doi = {10.1051/0004-6361/202038685},
archivePrefix = {arXiv},
       eprint = {2006.10041},
 primaryClass = {astro-ph.GA},
       adsurl = {https://ui.adsabs.harvard.edu/abs/2020A&A...642L..16B}
}

@ARTICLE{Song2020,
       author = {{Song}, Hyunmi and {Seon}, Kwang-Il and {Hwang}, Ho Seong},
        title = "{Ly{\ensuremath{\alpha}} Radiative Transfer: Modeling Spectrum and Surface Brightness Profiles of Ly{\ensuremath{\alpha}}-emitting Galaxies at Z = 3-6}",
      journal = {\apj},
         year = 2020,
        month = sep,
       volume = {901},
       number = {1},
          eid = {41},
        pages = {41},
          doi = {10.3847/1538-4357/abac02},
archivePrefix = {arXiv},
       eprint = {2007.08172},
 primaryClass = {astro-ph.GA},
       adsurl = {https://ui.adsabs.harvard.edu/abs/2020ApJ...901...41S}
}

@ARTICLE{Kusakabe2019,
       author = {{Kusakabe}, Haruka and {Shimasaku}, Kazuhiro and {Momose}, Rieko and {Ouchi}, Masami and {Nakajima}, Kimihiko and {Hashimoto}, Takuya and {Harikane}, Yuichi and {Silverman}, John D. and {Capak}, Peter L.},
        title = "{The dominant origin of diffuse Ly{\ensuremath{\alpha}} halos around Ly{\ensuremath{\alpha}} emitters explored by spectral energy distribution fitting and clustering analysis}",
      journal = {\pasj},
         year = 2019,
        month = jun,
       volume = {71},
       number = {3},
          eid = {55},
        pages = {55},
          doi = {10.1093/pasj/psz029},
archivePrefix = {arXiv},
       eprint = {1803.10265},
 primaryClass = {astro-ph.GA},
       adsurl = {https://ui.adsabs.harvard.edu/abs/2019PASJ...71...55K}
}

@ARTICLE{Werk2014,
       author = {{Werk}, Jessica K. and {Prochaska}, J. Xavier and {Tumlinson}, Jason and {Peeples}, Molly S. and {Tripp}, Todd M. and {Fox}, Andrew J. and {Lehner}, Nicolas and {Thom}, Christopher and {O'Meara}, John M. and {Ford}, Amanda Brady and {Bordoloi}, Rongmon and {Katz}, Neal and {Tejos}, Nicolas and {Oppenheimer}, Benjamin D. and {Dav{\'e}}, Romeel and {Weinberg}, David H.},
        title = "{The COS-Halos Survey: Physical Conditions and Baryonic Mass in the Low-redshift Circumgalactic Medium}",
      journal = {\apj},
         year = 2014,
        month = sep,
       volume = {792},
       number = {1},
          eid = {8},
        pages = {8},
          doi = {10.1088/0004-637X/792/1/8},
archivePrefix = {arXiv},
       eprint = {1403.0947},
 primaryClass = {astro-ph.CO},
       adsurl = {https://ui.adsabs.harvard.edu/abs/2014ApJ...792....8W}
}

@ARTICLE{Chisholm2020,
       author = {{Chisholm}, J. and {Prochaska}, J.~X. and {Schaerer}, D. and {Gazagnes}, S. and {Henry}, A.},
        title = "{Optically thin spatially resolved Mg II emission maps the escape of ionizing photons}",
      journal = {\mnras},
         year = 2020,
        month = oct,
       volume = {498},
       number = {2},
        pages = {2554-2574},
          doi = {10.1093/mnras/staa2470},
archivePrefix = {arXiv},
       eprint = {2008.06059},
 primaryClass = {astro-ph.GA},
       adsurl = {https://ui.adsabs.harvard.edu/abs/2020MNRAS.498.2554C}
}

@ARTICLE{Gelman1992,
       author = {{Gelman}, Andrew and {Rubin}, Donald B.},
        title = "{Inference from Iterative Simulation Using Multiple Sequences}",
      journal = {Statistical Science},
         year = 1992,
        month = jan,
       volume = {7},
        pages = {457-472},
          doi = {10.1214/ss/1177011136},
       adsurl = {https://ui.adsabs.harvard.edu/abs/1992StaSc...7..457G}
}

@article{Prochaska:2011eqa,
author = {Prochaska, J Xavier and Kasen, Daniel and Rubin, Kate},
title = {{Simple Models of Metal-line Absorption and Emission from Cool Gas Outflows}},
journal = {ApJ},
year = {2011},
volume = {734},
number = {1},
pages = {24},
month = jun
}

@article{Rubin:2011bm,
author = {Rubin, Kate H R and Prochaska, J Xavier and M{\'e}nard, Brice and Murray, Norman and Kasen, Daniel and Koo, David C and Phillips, Andrew C},
title = {{Low-ionization Line Emission from a Starburst Galaxy: A New Probe of a Galactic-scale Outflow}},
journal = {ApJ},
year = {2011},
volume = {728},
number = {1},
pages = {55},
month = feb
}

@inproceedings{Bacon:2010jn,
author = {Bacon, R and Accardo, M and Adjali, L and Anwand, H and Bauer, S and Biswas, I and Blaizot, J and Boudon, D and Brau-Nogue, S and Brinchmann, J and Caillier, P and Capoani, L and Carollo, C M and Contini, T and Couderc, P and Daguis{\'e}, E and Deiries, S and Delabre, B and Dreizler, S and Dubois, J and Dupieux, M and Dupuy, C and Emsellem, E and Fechner, T and Fleischmann, A and Fran{\c c}ois, M and Gallou, G and Gharsa, T and Glindemann, A and Gojak, D and Guiderdoni, B and Hansali, G and Hahn, T and Jarno, A and Kelz, A and Koehler, C and Kosmalski, J and Laurent, F and Le Floch, M and Lilly, S J and Lizon, J L and Loupias, M and Manescau, A and Monstein, C and Nicklas, H and Olaya, J C and Pares, L and Pasquini, L and P{\'e}contal-Rousset, A and Pell{\`o}, R and Petit, C and Popow, E and Reiss, R and Remillieux, A and Renault, E and Roth, M and Rupprecht, G and Serre, D and Schaye, J and Soucail, G and Steinmetz, M and Streicher, O and Stuik, R and Valentin, H and Vernet, J and Weilbacher, P and Wisotzki, L. and Yerle, N},
title = {{The MUSE second-generation VLT instrument}},
booktitle = {Astronomical Telescopes and Instrumentation},
year = {2010},
editor = {McLean, Ian S and Ramsay, Suzanne K and Takami, Hideki},
pages = {773508--773508--9},
publisher = {SPIE},
month = jul
}

@ARTICLE{Pillepich2019,
       author = {{Pillepich}, Annalisa and {Nelson}, Dylan and {Springel}, Volker and {Pakmor}, R{\"u}diger and {Torrey}, Paul and {Weinberger}, Rainer and {Vogelsberger}, Mark and {Marinacci}, Federico and {Genel}, Shy and {van der Wel}, Arjen and {Hernquist}, Lars},
        title = "{First results from the TNG50 simulation: the evolution of stellar and gaseous discs across cosmic time}",
      journal = {\mnras},
         year = 2019,
        month = dec,
       volume = {490},
       number = {3},
        pages = {3196-3233},
          doi = {10.1093/mnras/stz2338},
archivePrefix = {arXiv},
       eprint = {1902.05553},
 primaryClass = {astro-ph.GA},
       adsurl = {https://ui.adsabs.harvard.edu/abs/2019MNRAS.490.3196P}
}

@ARTICLE{Nelson2021,
       author = {{Nelson}, Dylan and {Byrohl}, Chris and {Peroux}, Celine and {Rubin}, Kate H.~R. and {Burchett}, Joseph N.},
        title = "{The cold circumgalactic medium in emission: Mg II haloes in TNG50}",
      journal = {\mnras},
         year = 2021,
        month = nov,
       volume = {507},
       number = {3},
        pages = {4445-4463},
          doi = {10.1093/mnras/stab2177},
archivePrefix = {arXiv},
       eprint = {2106.09023},
 primaryClass = {astro-ph.GA},
       adsurl = {https://ui.adsabs.harvard.edu/abs/2021MNRAS.507.4445N}
}

@article{Kriek:2009cs,
author = {Kriek, Mariska and van Dokkum, Pieter G and Labb{\'e}, Ivo and Franx, Marijn and Illingworth, Garth D and Marchesini, Danilo and Quadri, Ryan F},
title = {{An Ultra-Deep Near-Infrared Spectrum of a Compact Quiescent Galaxy at z = 2.2}},
journal = {ApJ},
year = {2009},
volume = {700},
number = {1},
pages = {221--231},
month = jul
}

@article{Heckman:2000du,
author = {Heckman, Timothy M. and Lehnert, Matthew D and Strickland, David K and Armus, Lee},
title = {{Absorption-Line Probes of Gas and Dust in Galactic Superwinds}},
journal = {ApJS},
year = {2000},
volume = {129},
number = {2},
pages = {493--516},
month = aug
}

@article{Calzetti:2000iy,
author = {Calzetti, Daniela and Armus, Lee and Bohlin, Ralph C and Kinney, Anne L and Koornneef, Jan and Storchi-Bergmann, Thaisa},
title = {{The Dust Content and Opacity of Actively Star-forming Galaxies*}},
journal = {ApJ},
year = {2000},
volume = {533},
number = {2},
pages = {682--695},
month = apr
}

\begin{appendix}

\onecolumn
\section{Model results for all our sample galaxies}
\label{sec:model_results_all_galaxies}

Here we show the figures that compare the data and best-fitting model cubes, analogous to the figures shown in Sec.~\ref{sec:best_fitting_models}, for all our sample galaxies. Figures~\ref{fig:master_1},~\ref{fig:master_2},~\ref{fig:master_3},~\ref{fig:master_4},~\ref{fig:master_5},~\ref{fig:master_6},~\ref{fig:master_7},~\ref{fig:master_8},~\ref{fig:master_9},~\ref{fig:master_10},~\ref{fig:master_11}, and~\ref{fig:master_12} summarize the modeling results for each galaxy in our sample. Galaxies are sorted by stellar mass, in descending order.

\begin{figure*}[!htb]
    \centering
    \begin{minipage}{.49\textwidth}
        \centering
        \includegraphics[width=0.99\columnwidth]{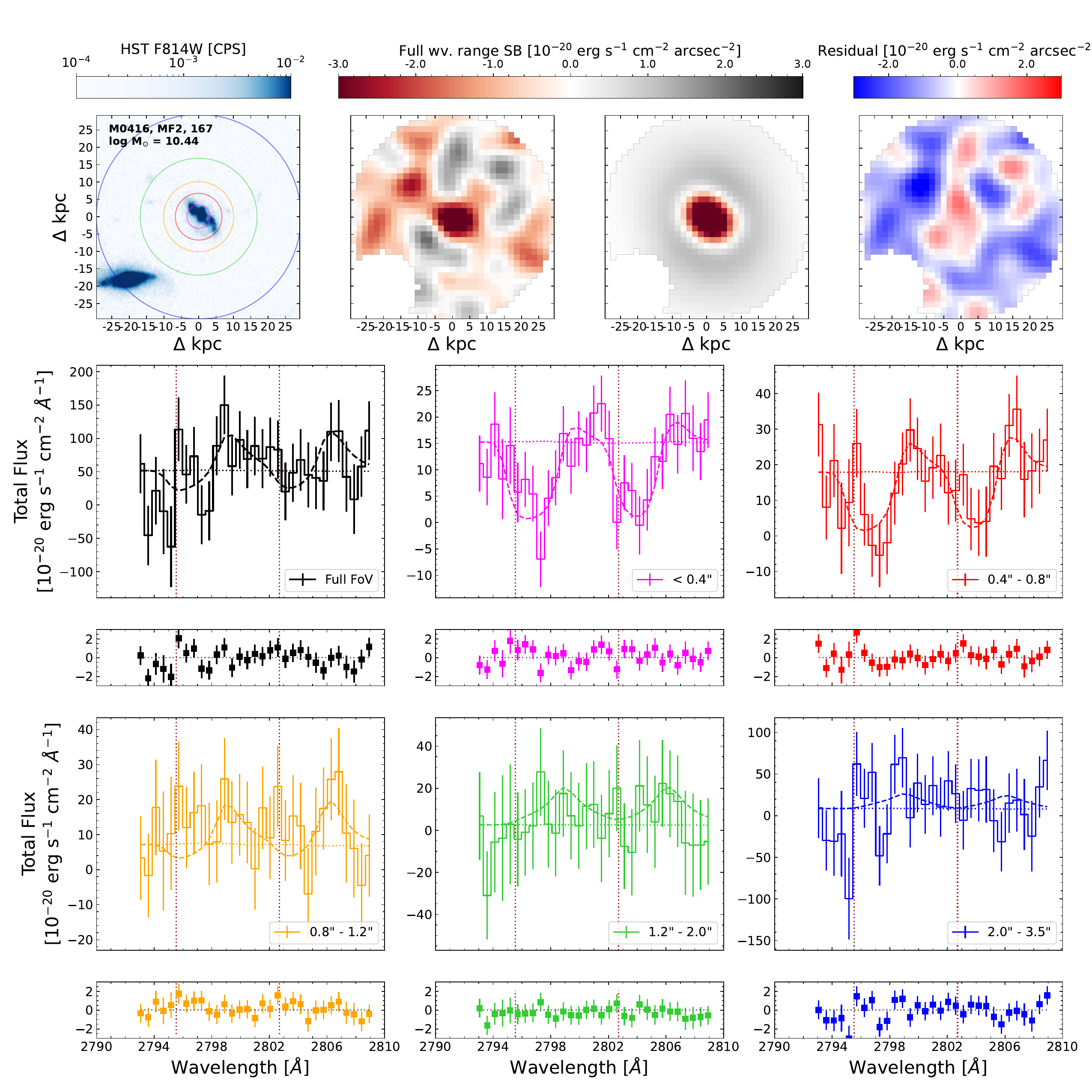}
    \end{minipage}%
    \begin{minipage}{0.49\textwidth}
        \centering
        \includegraphics[width=0.99\columnwidth]{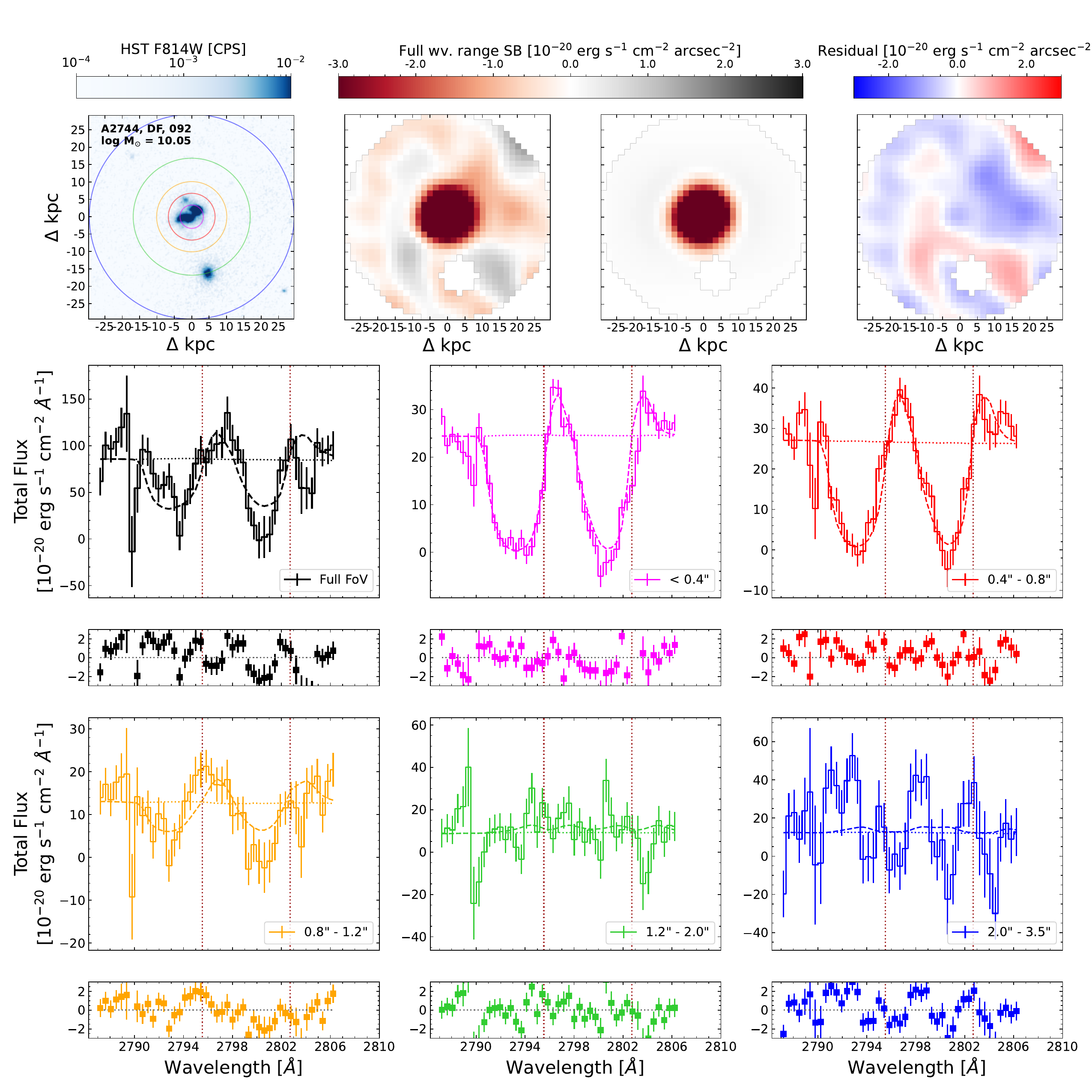}
    \end{minipage}

    \begin{minipage}{.49\textwidth}
        \centering
        \includegraphics[width=0.99\columnwidth]{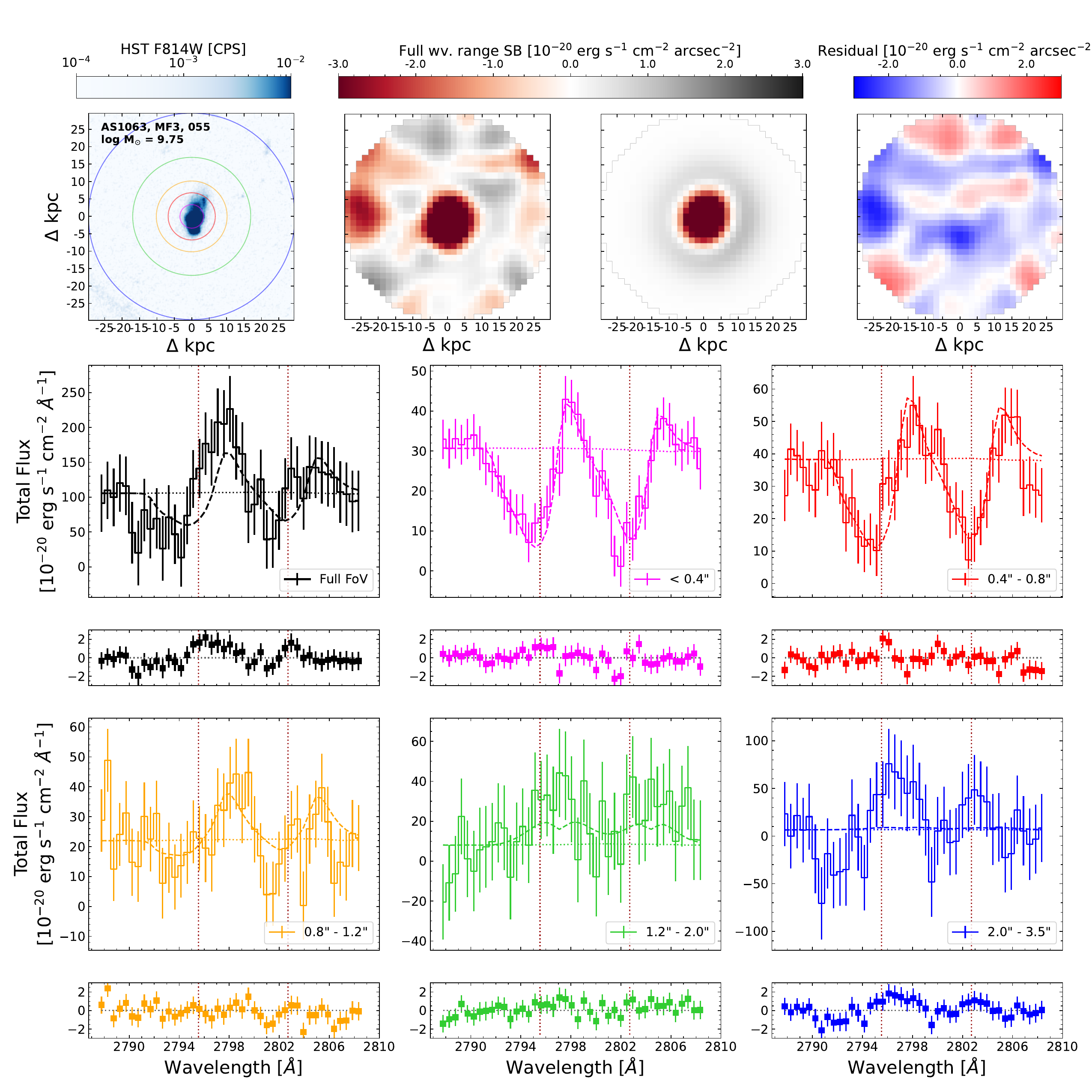}
    \end{minipage}%
    \begin{minipage}{0.49\textwidth}
        \centering
        \includegraphics[width=0.99\columnwidth]{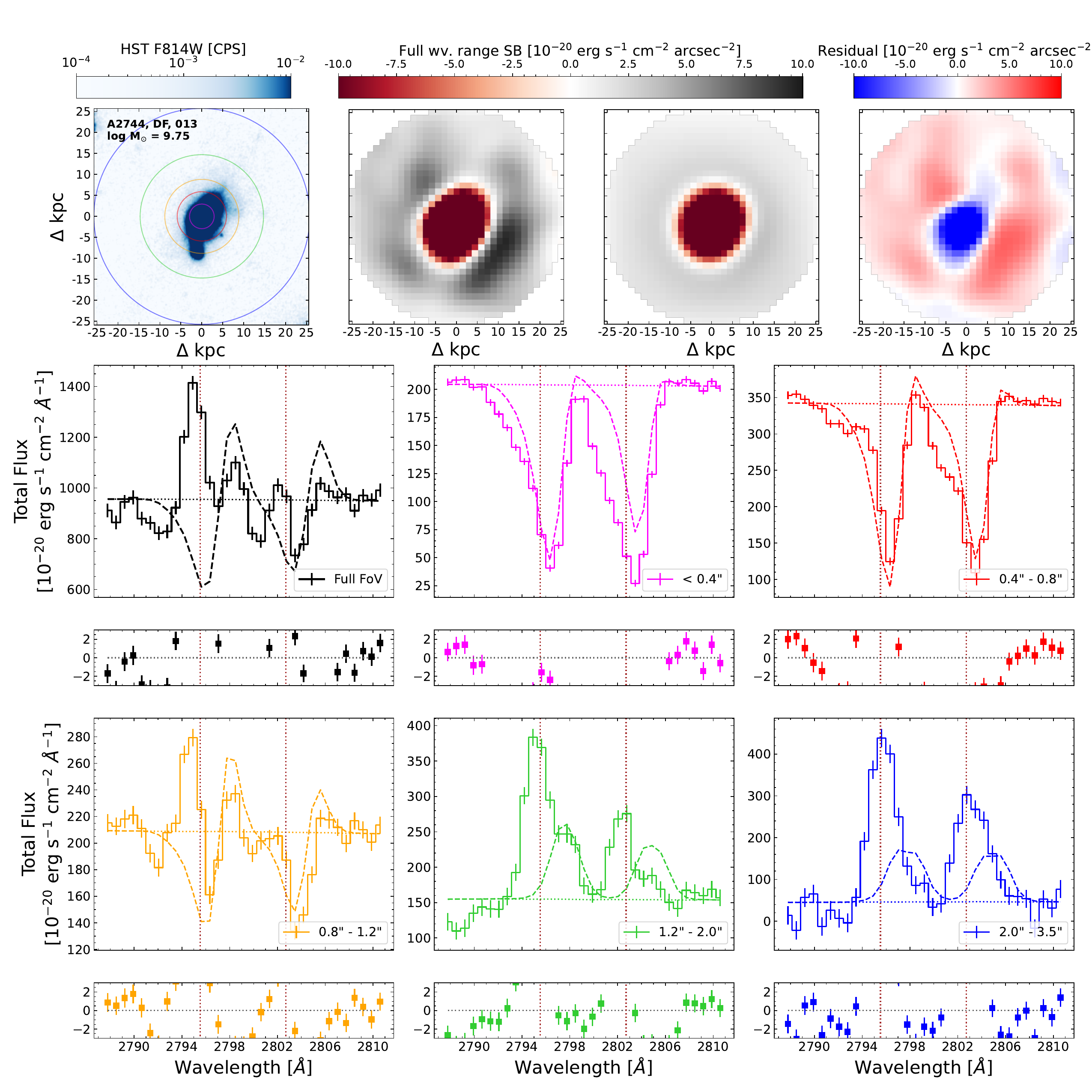}
    \end{minipage}
    \caption{Comparison between the observed and best-fitting model cubes, in terms of the continuum-subtracted pseudo-narrowband images and spectra extracted from annular apertures for four sample galaxies. Each set of panels corresponds to one galaxy, following the same layout as in Fig.~\ref{fig:M0416_SF07_012_Master_figure} (see caption of Fig.~\ref{fig:M0416_SF07_012_Master_figure} for a full description of each panel).}
    \label{fig:master_1}
\end{figure*}    
\begin{figure*}[!htb]
    \centering

    \begin{minipage}{.49\textwidth}
        \centering
        \includegraphics[width=0.99\columnwidth]{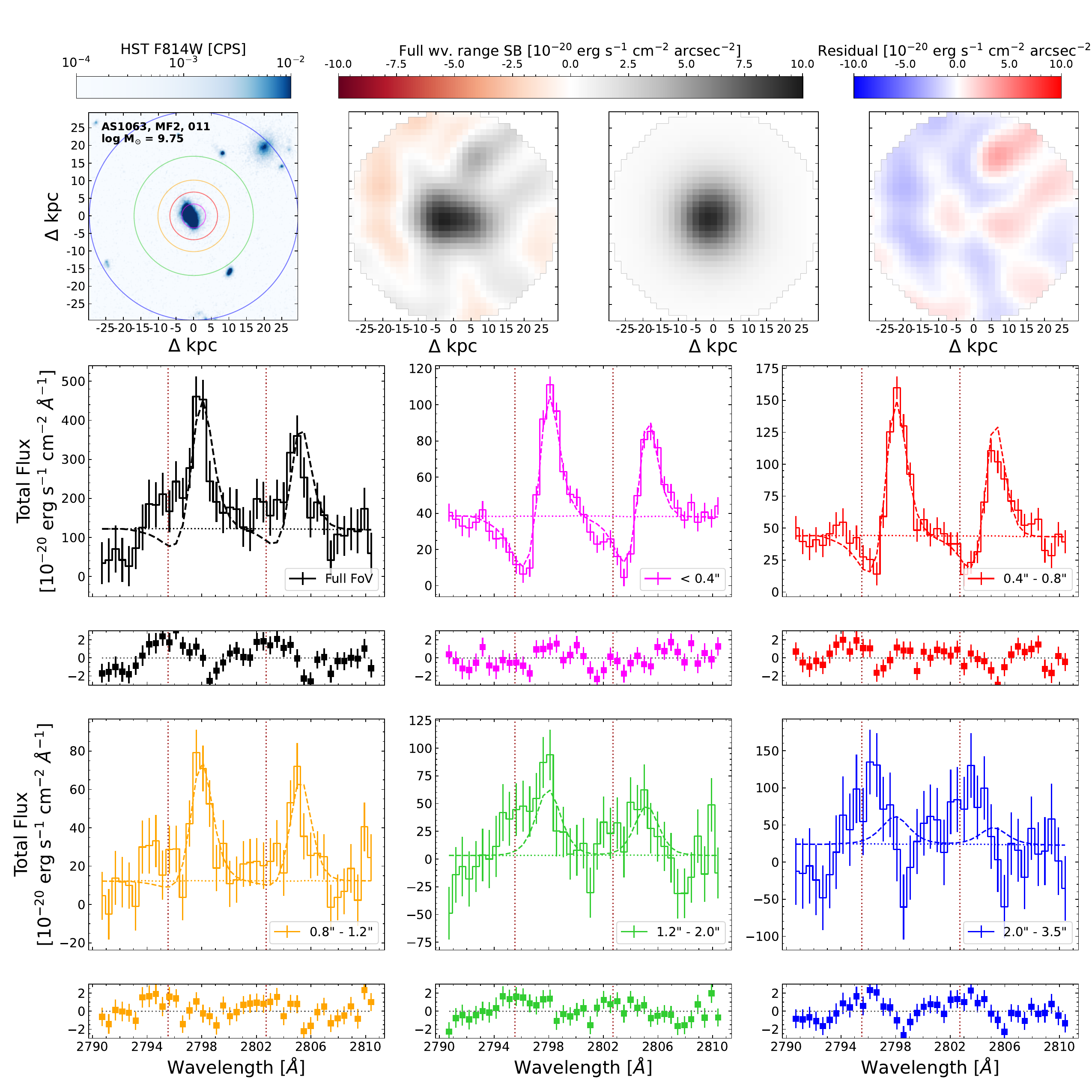}
    \end{minipage}%
    \begin{minipage}{0.49\textwidth}
        \centering
        \includegraphics[width=0.99\columnwidth]{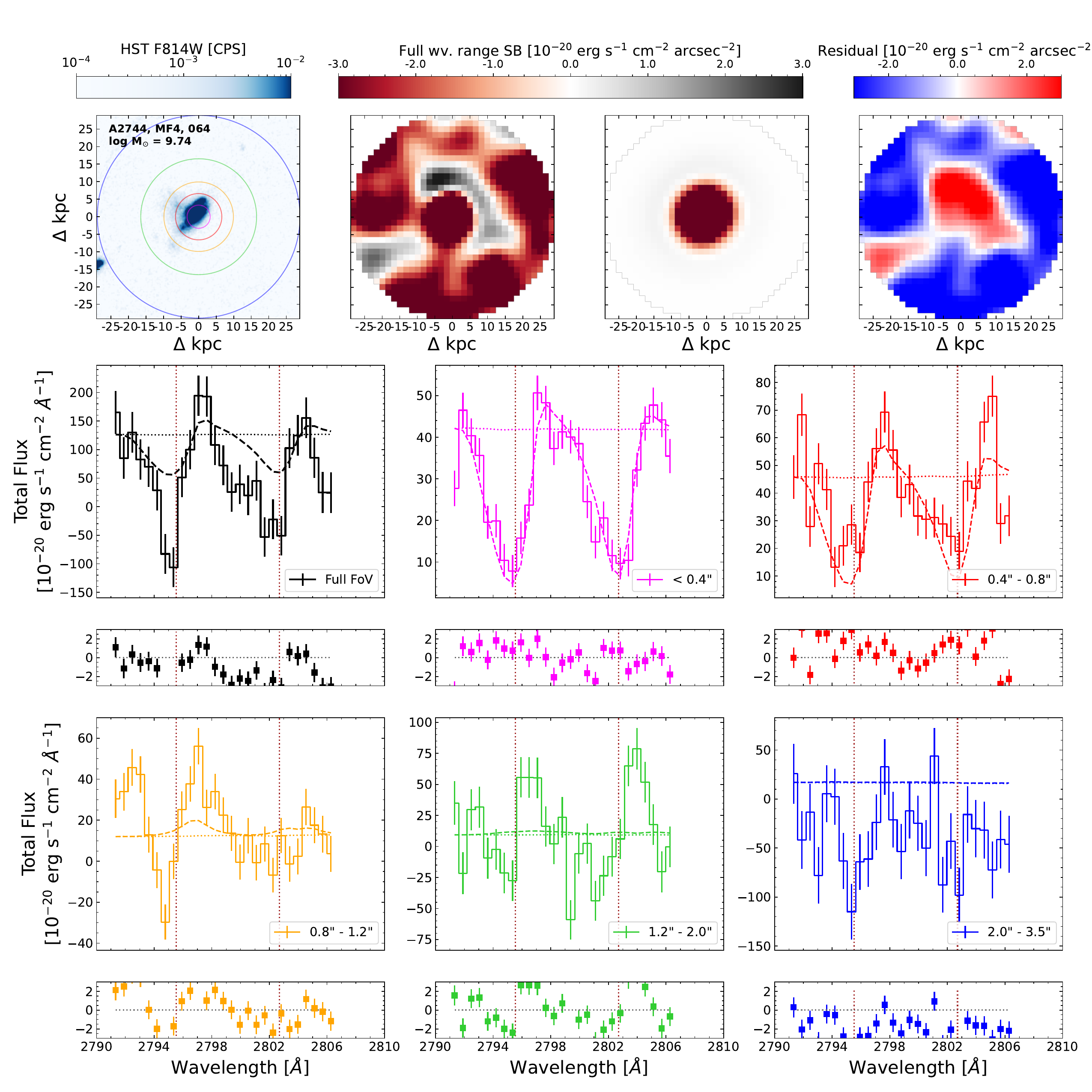}
    \end{minipage}

    \begin{minipage}{.49\textwidth}
        \centering
        \includegraphics[width=0.99\columnwidth]{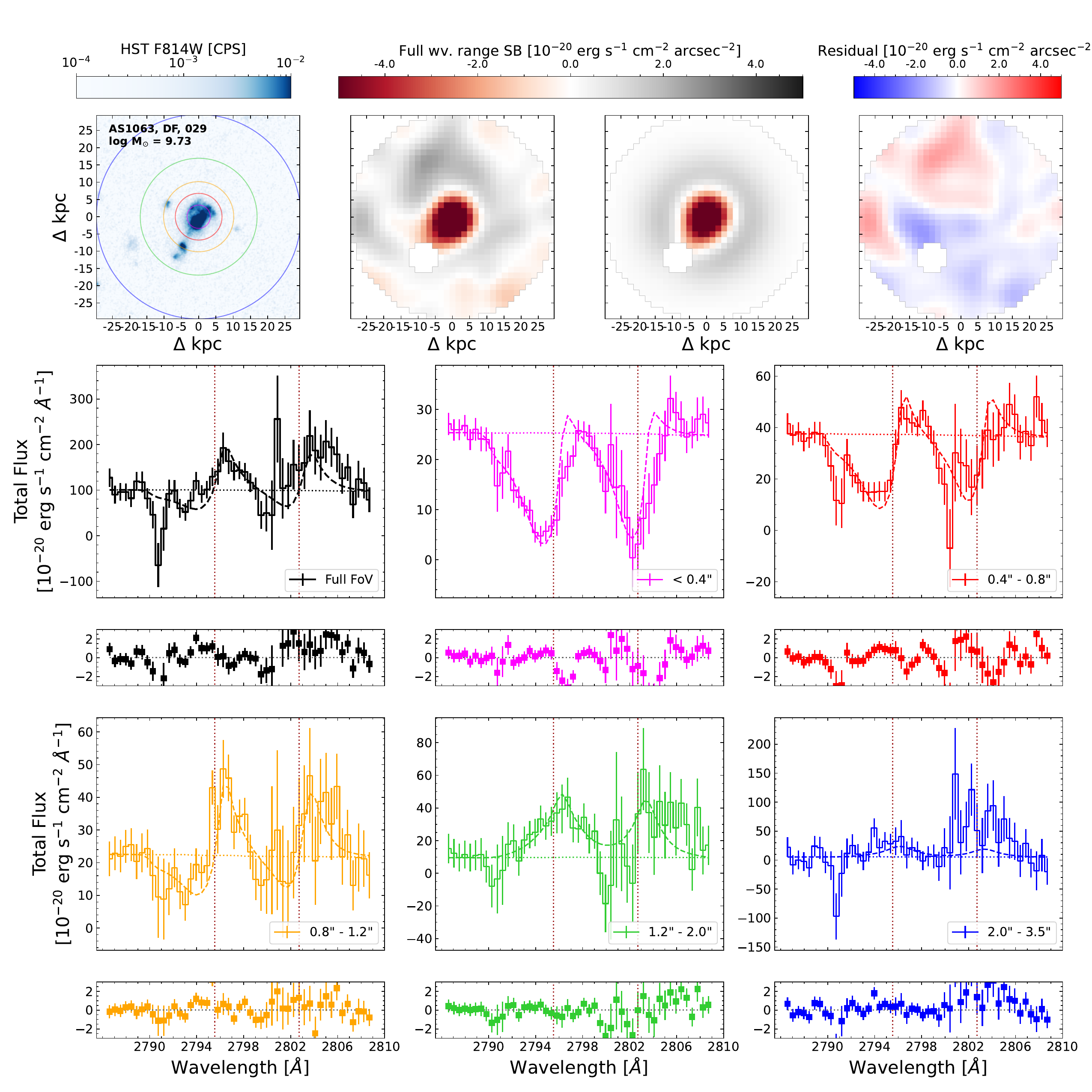}
    \end{minipage}%
    \begin{minipage}{0.49\textwidth}
        \centering
        \includegraphics[width=0.99\columnwidth]{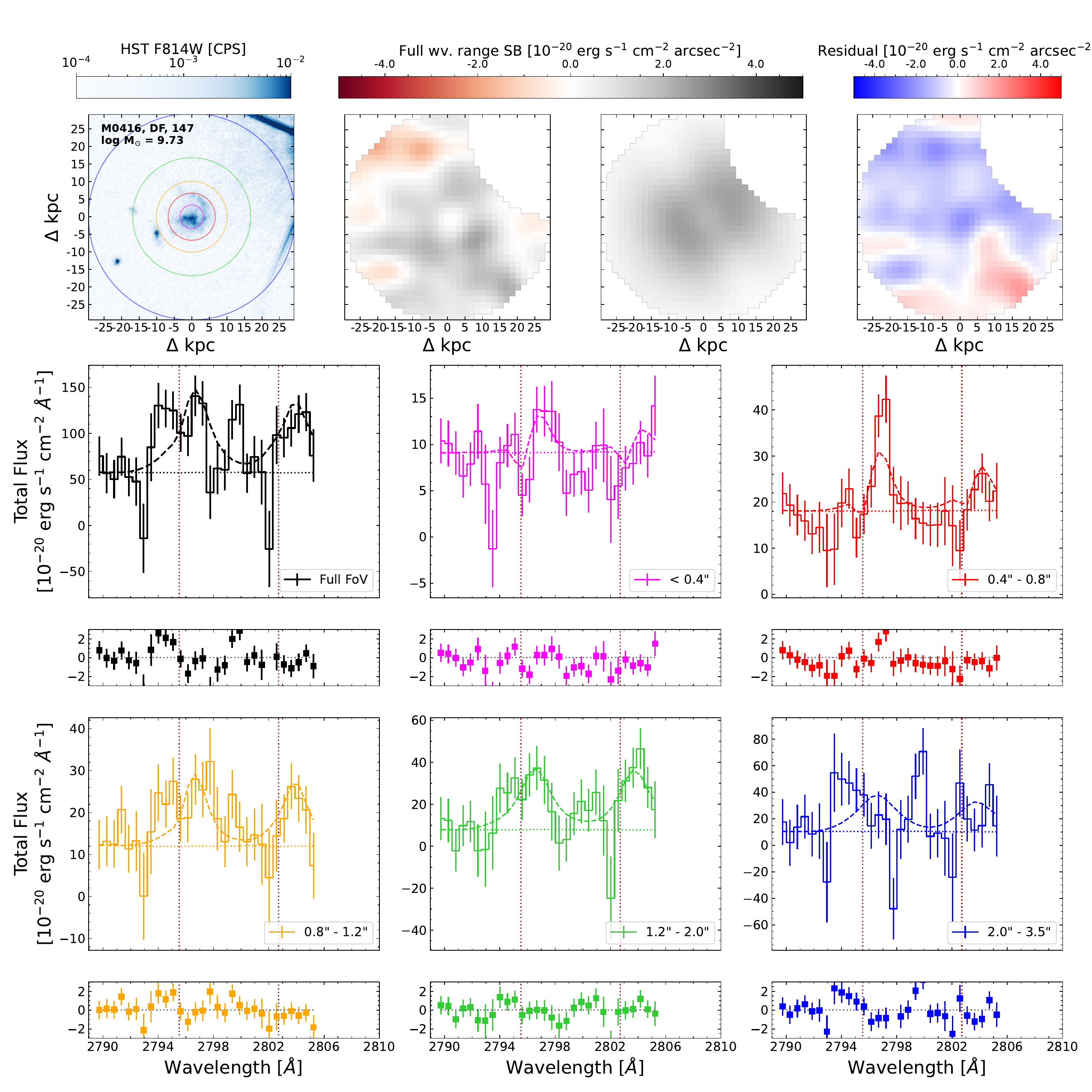}
    \end{minipage}

    \caption{Same as Fig.~\ref{fig:master_1}, for other four galaxies in our sample.}
    \label{fig:master_2}
\end{figure*}    
\begin{figure*}[!htb]
    \centering
    
        \begin{minipage}{.49\textwidth}
        \centering
        \includegraphics[width=0.99\columnwidth]{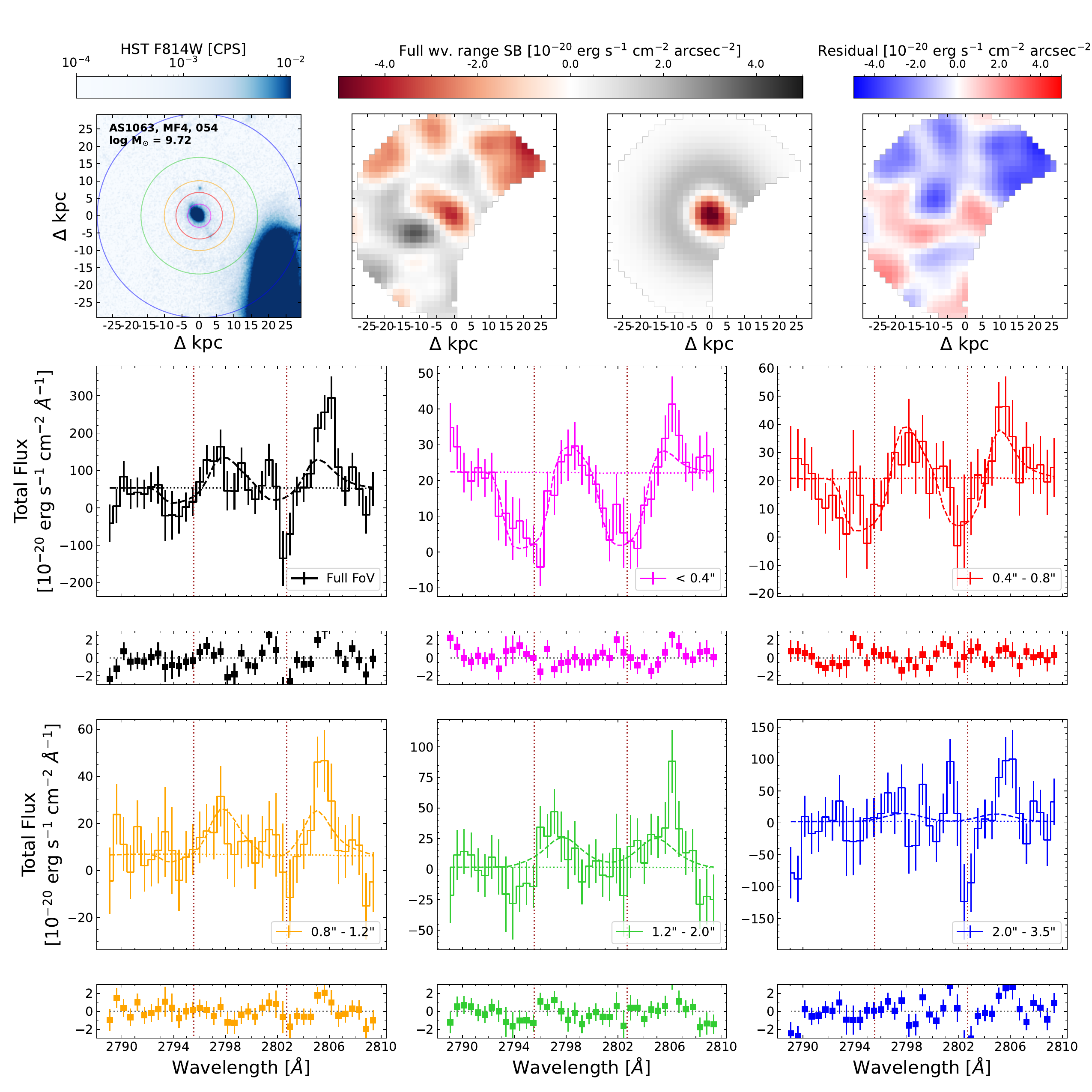}
    \end{minipage}%
    \begin{minipage}{0.49\textwidth}
        \centering
        \includegraphics[width=0.99\columnwidth]{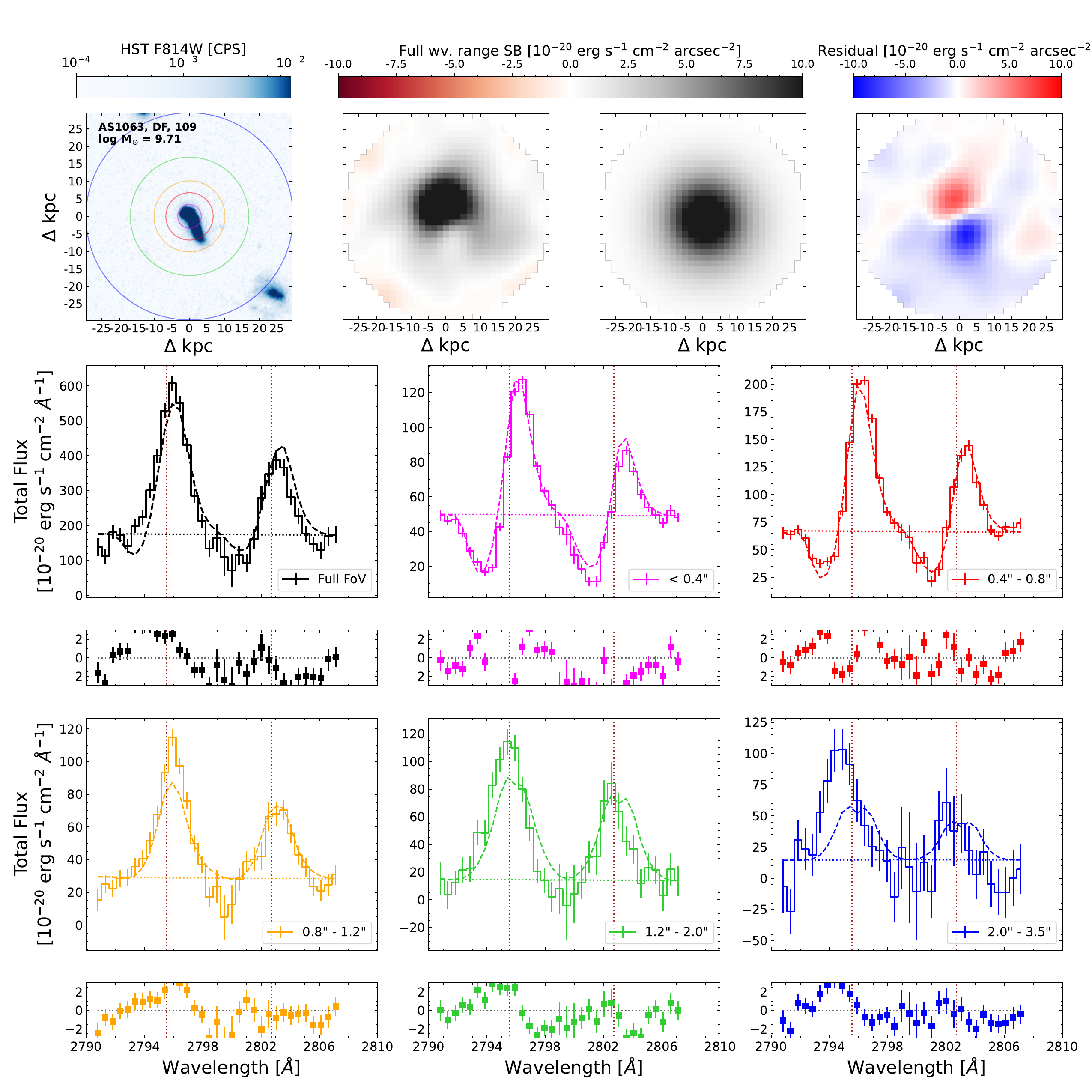}
    \end{minipage}

    \begin{minipage}{.49\textwidth}
        \centering
        \includegraphics[width=0.99\columnwidth]{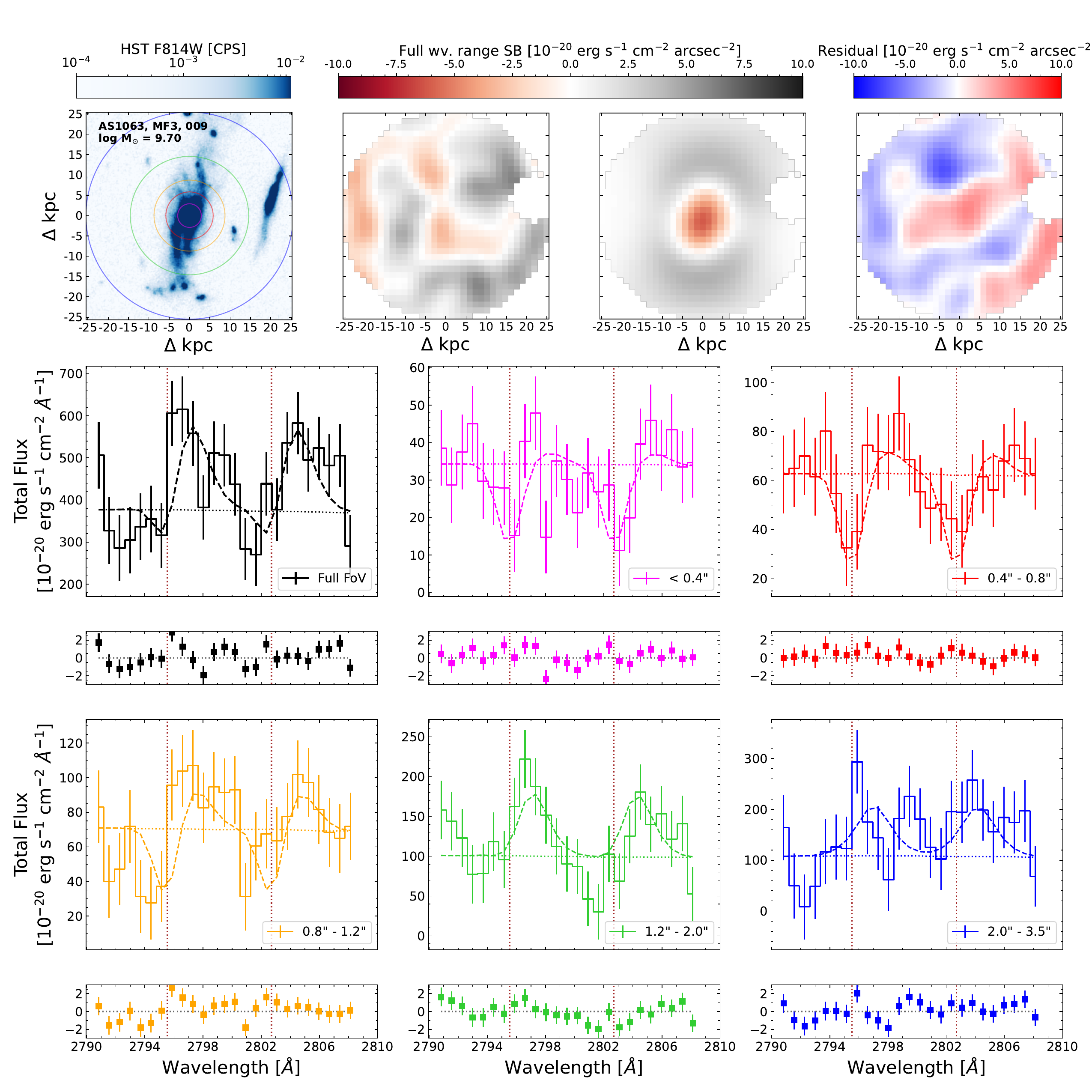}
    \end{minipage}%
    \begin{minipage}{0.49\textwidth}
        \centering
        \includegraphics[width=0.99\columnwidth]{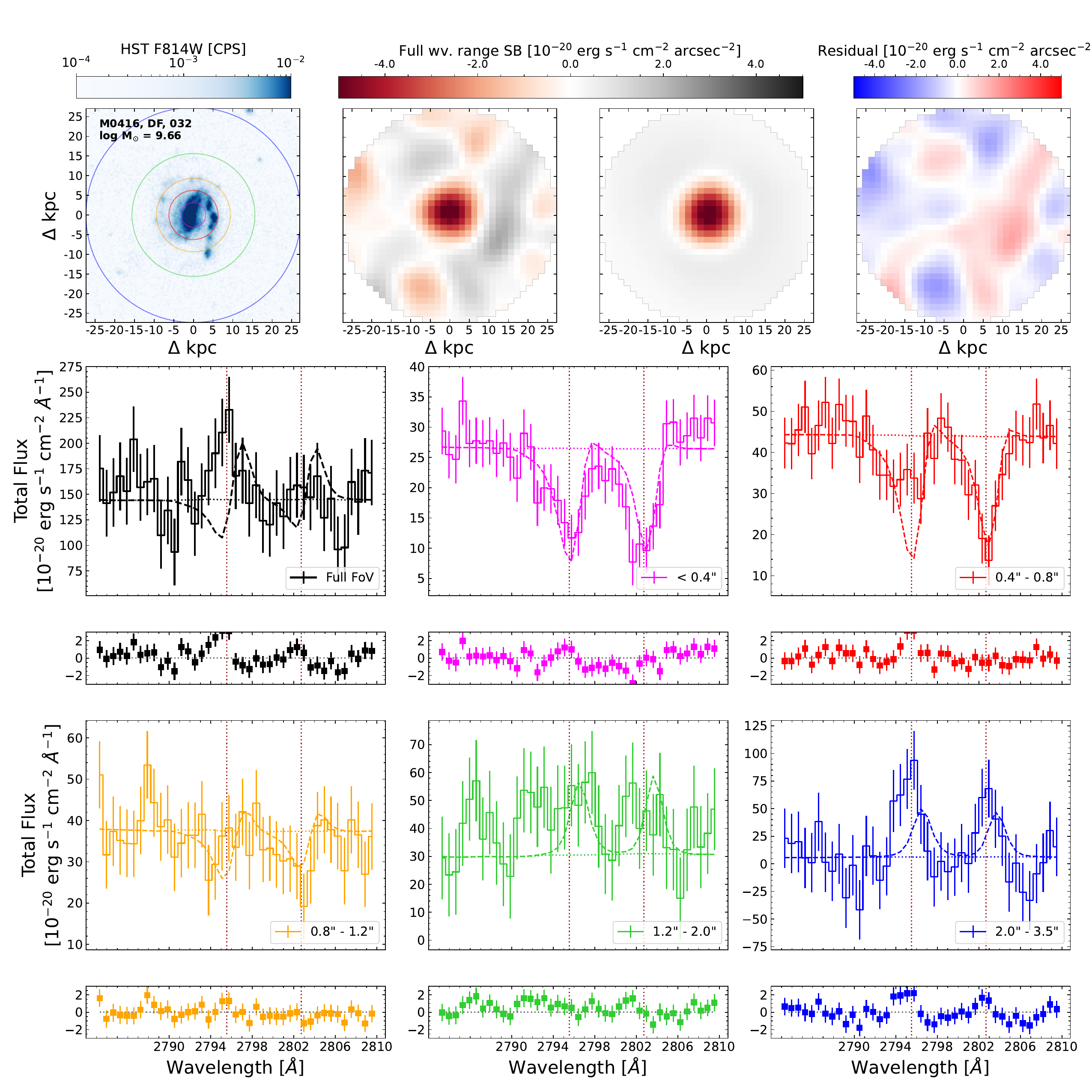}
    \end{minipage}

    \caption{Same as Fig.~\ref{fig:master_1}, for other four galaxies in our sample.}
    \label{fig:master_3}
\end{figure*}    
\begin{figure*}[!htb]
    \centering

    \begin{minipage}{.49\textwidth}
        \centering
        \includegraphics[width=0.99\columnwidth]{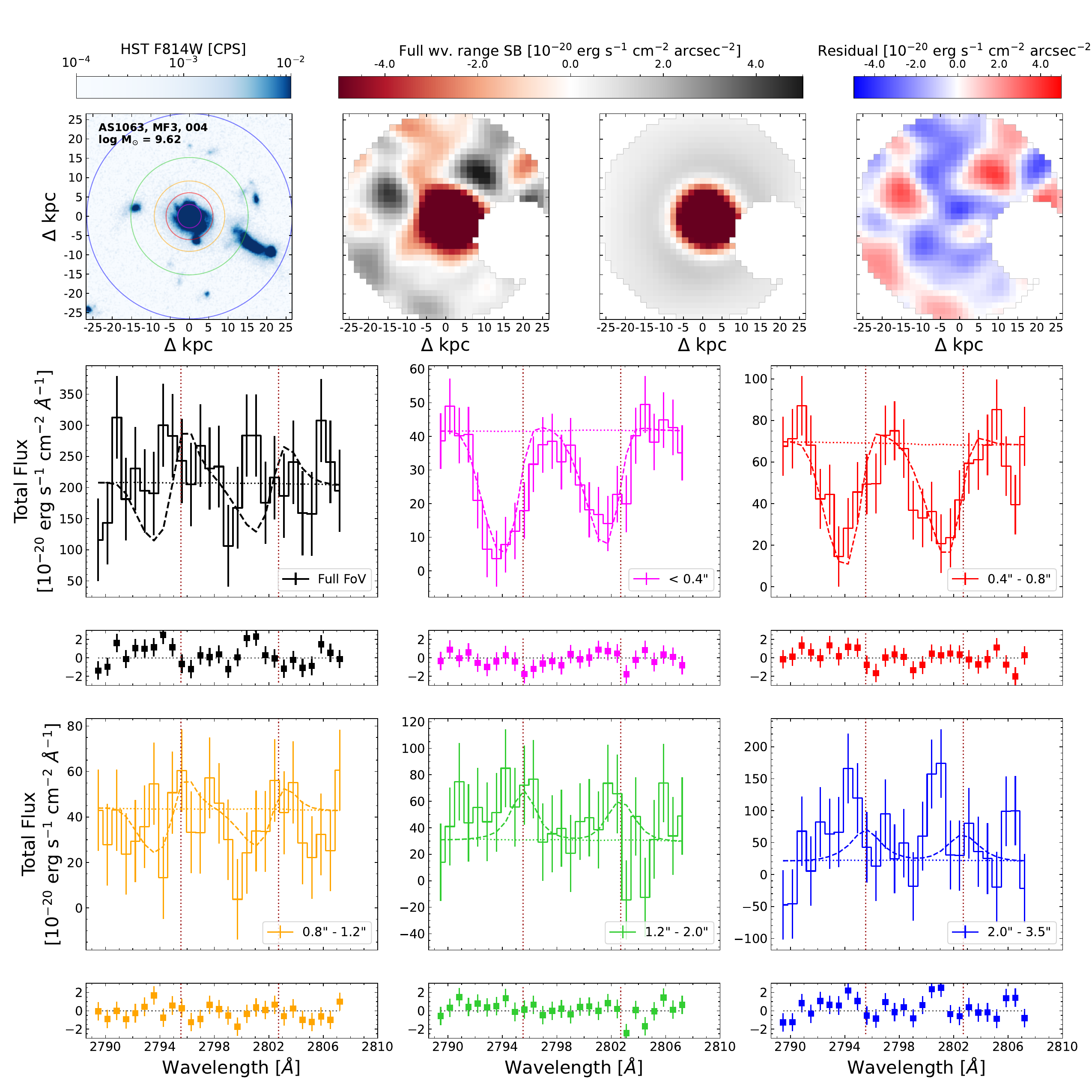}
    \end{minipage}%
    \begin{minipage}{0.49\textwidth}
        \centering
        \includegraphics[width=0.99\columnwidth]{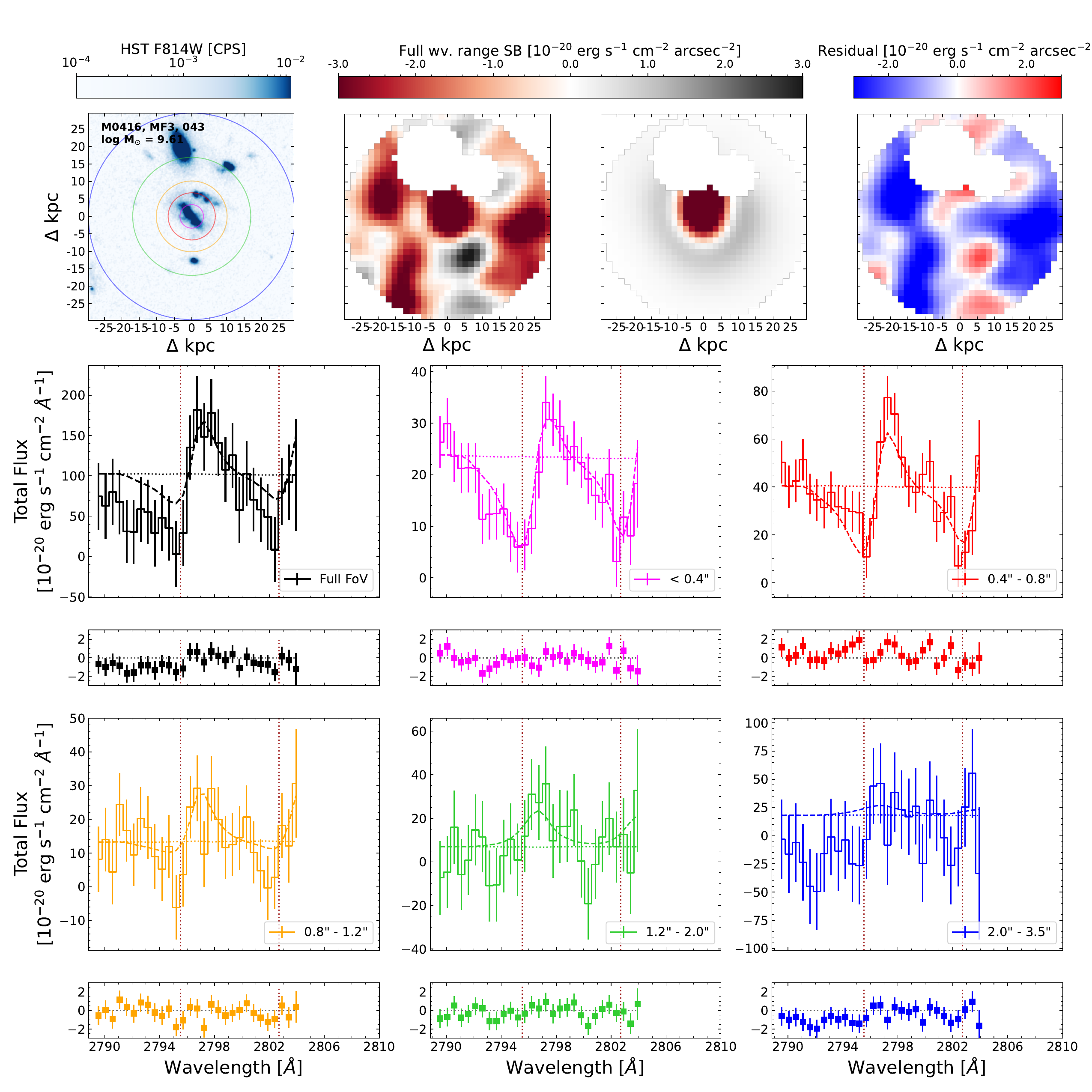}
    \end{minipage}

    \begin{minipage}{.49\textwidth}
        \centering
        \includegraphics[width=0.99\columnwidth]{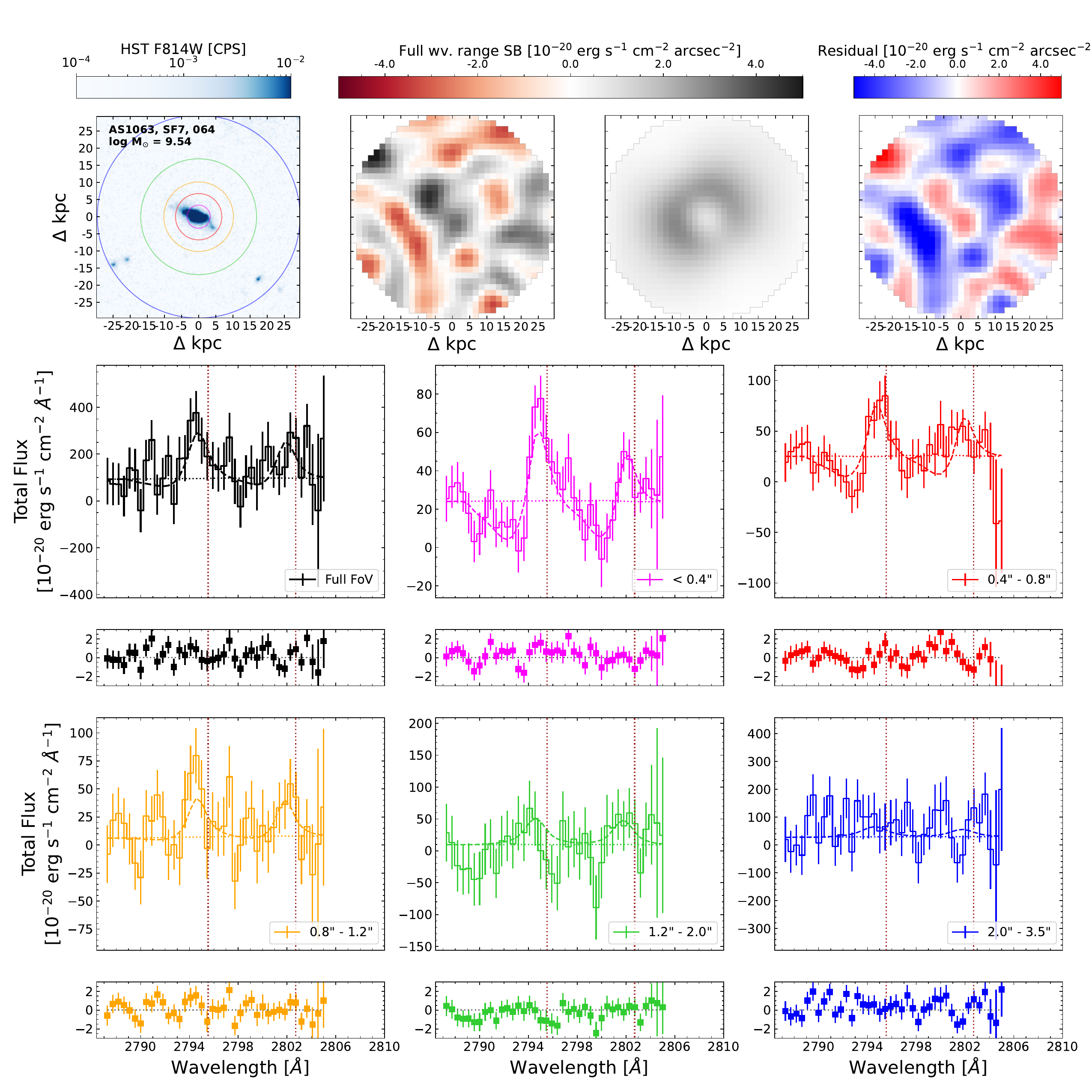}
    \end{minipage}%
    \begin{minipage}{0.49\textwidth}
        \centering
        \includegraphics[width=0.99\columnwidth]{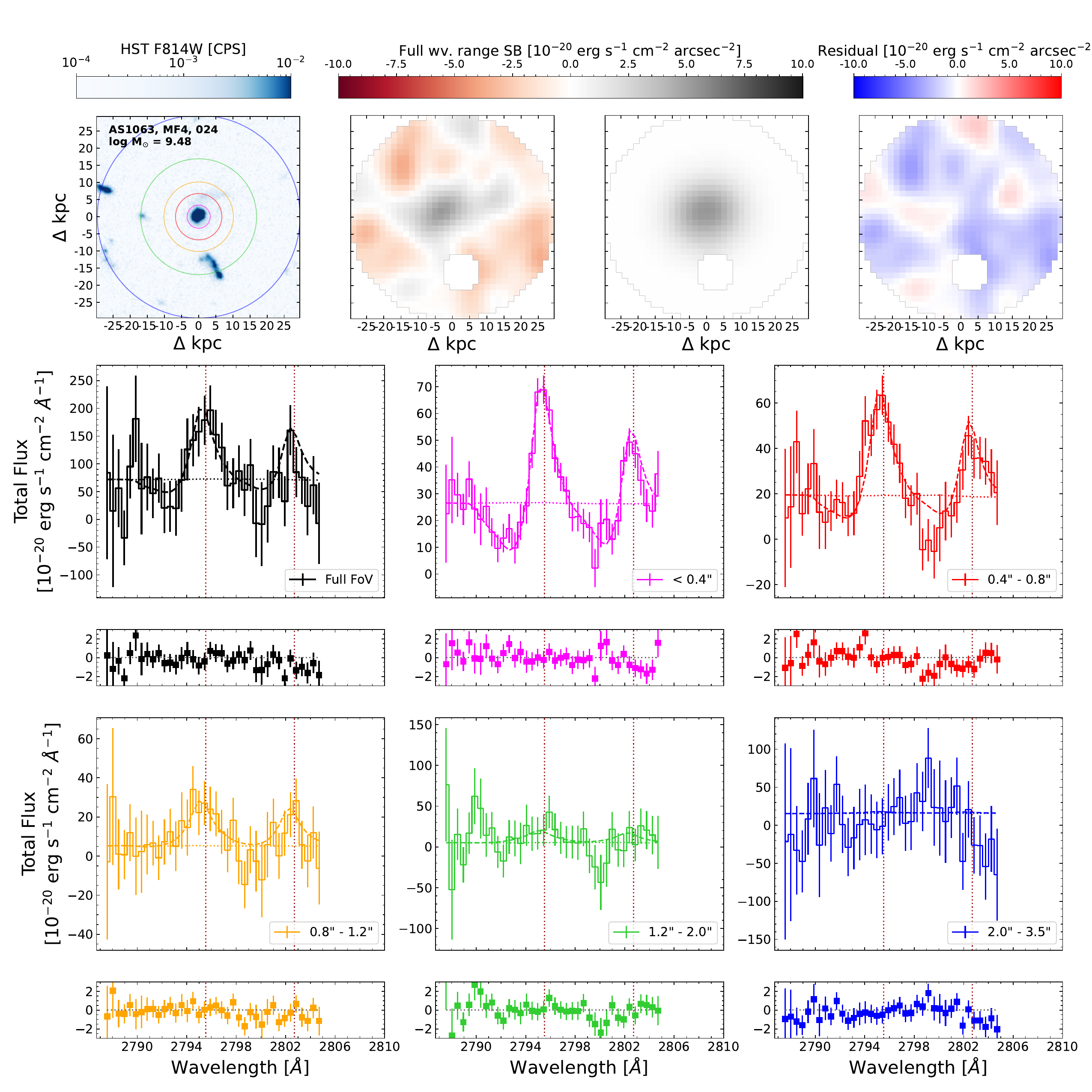}
    \end{minipage}

    \caption{Same as Fig.~\ref{fig:master_1}, for other four galaxies in our sample.}
    \label{fig:master_4}
\end{figure*}    
\begin{figure*}[!htb]
    \centering

    \begin{minipage}{0.49\textwidth}
        \centering
        \includegraphics[width=0.99\columnwidth]{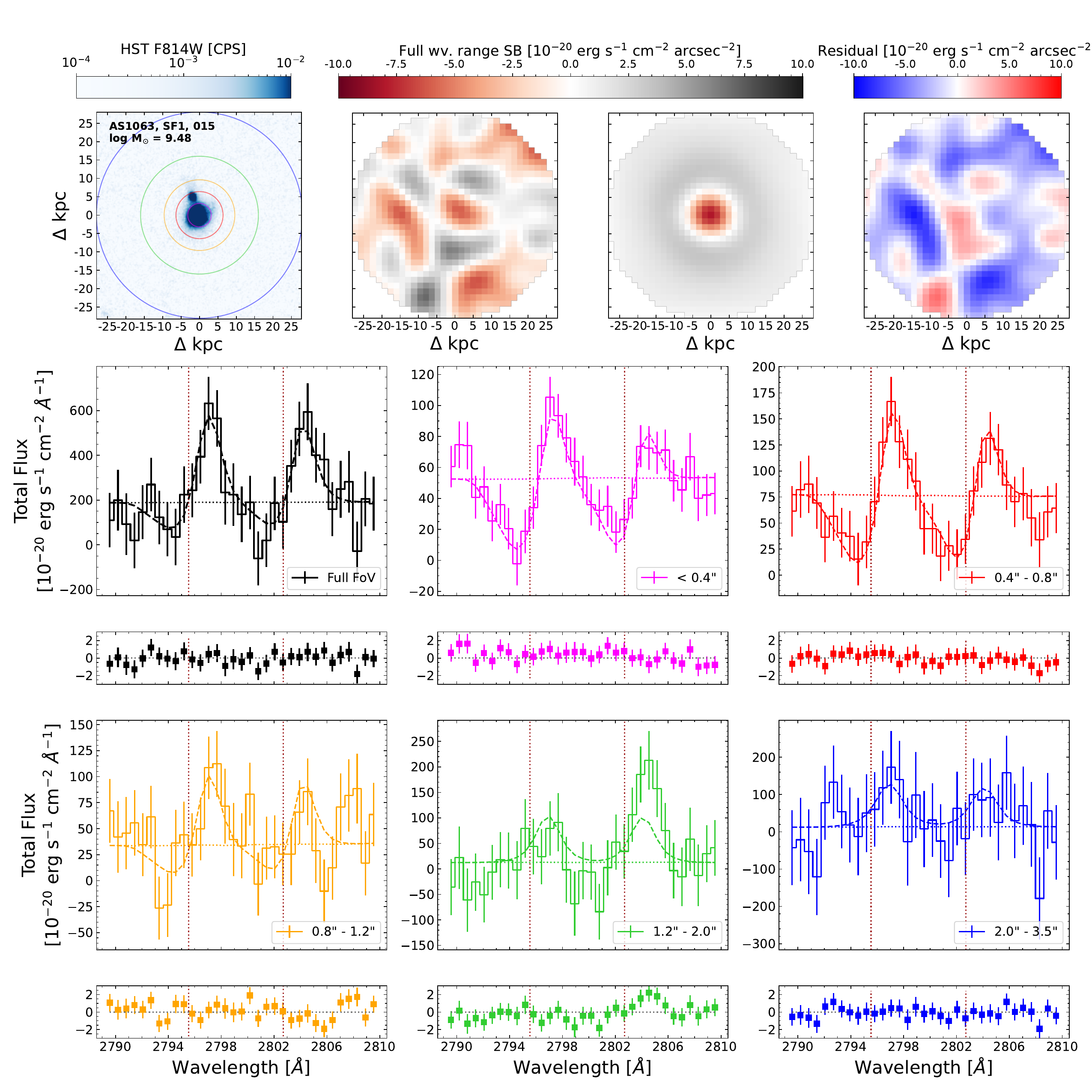}
    \end{minipage}
    \begin{minipage}{.49\textwidth}
        \centering
        \includegraphics[width=0.99\columnwidth]{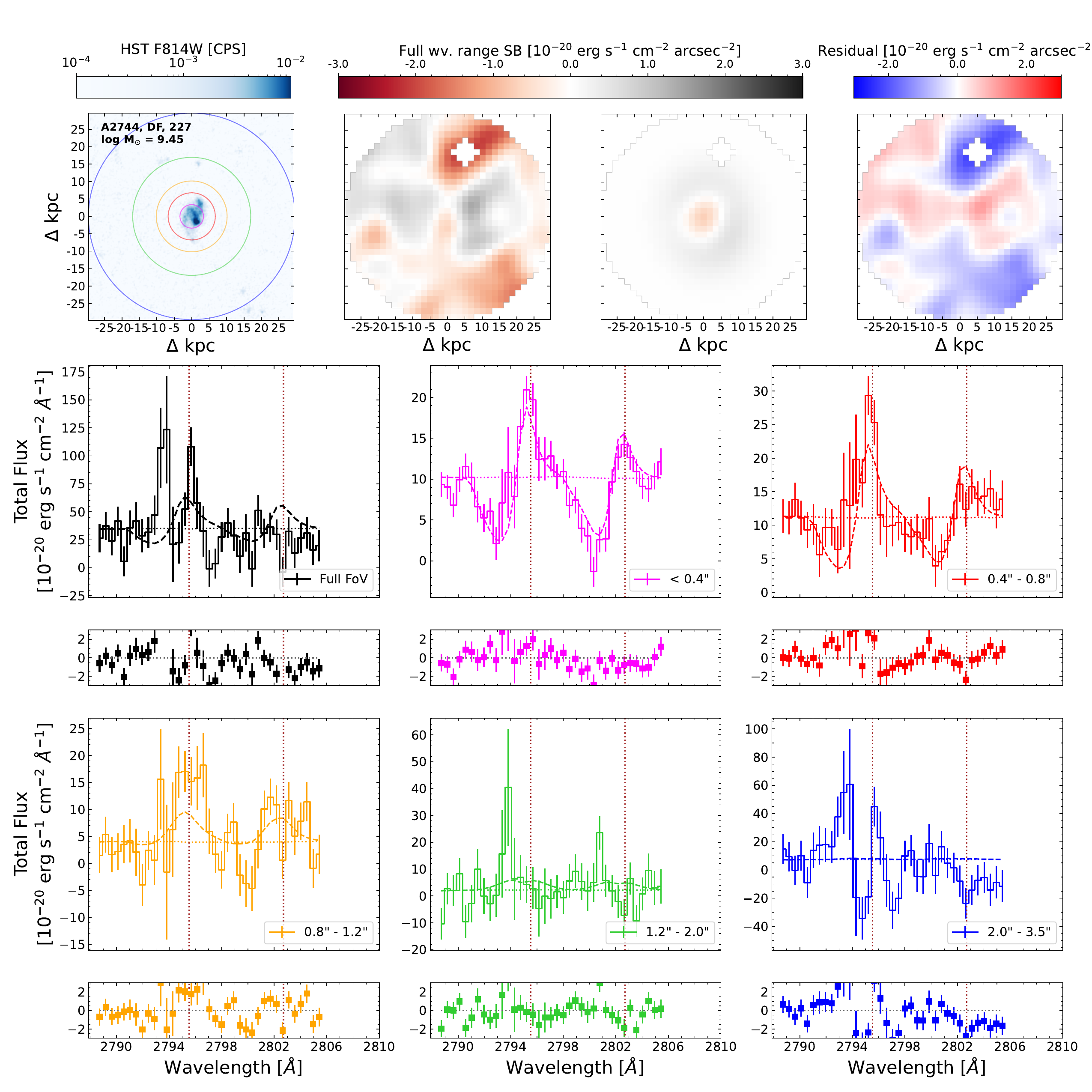}
    \end{minipage}%

    \begin{minipage}{0.49\textwidth}
        \centering
        \includegraphics[width=0.99\columnwidth]{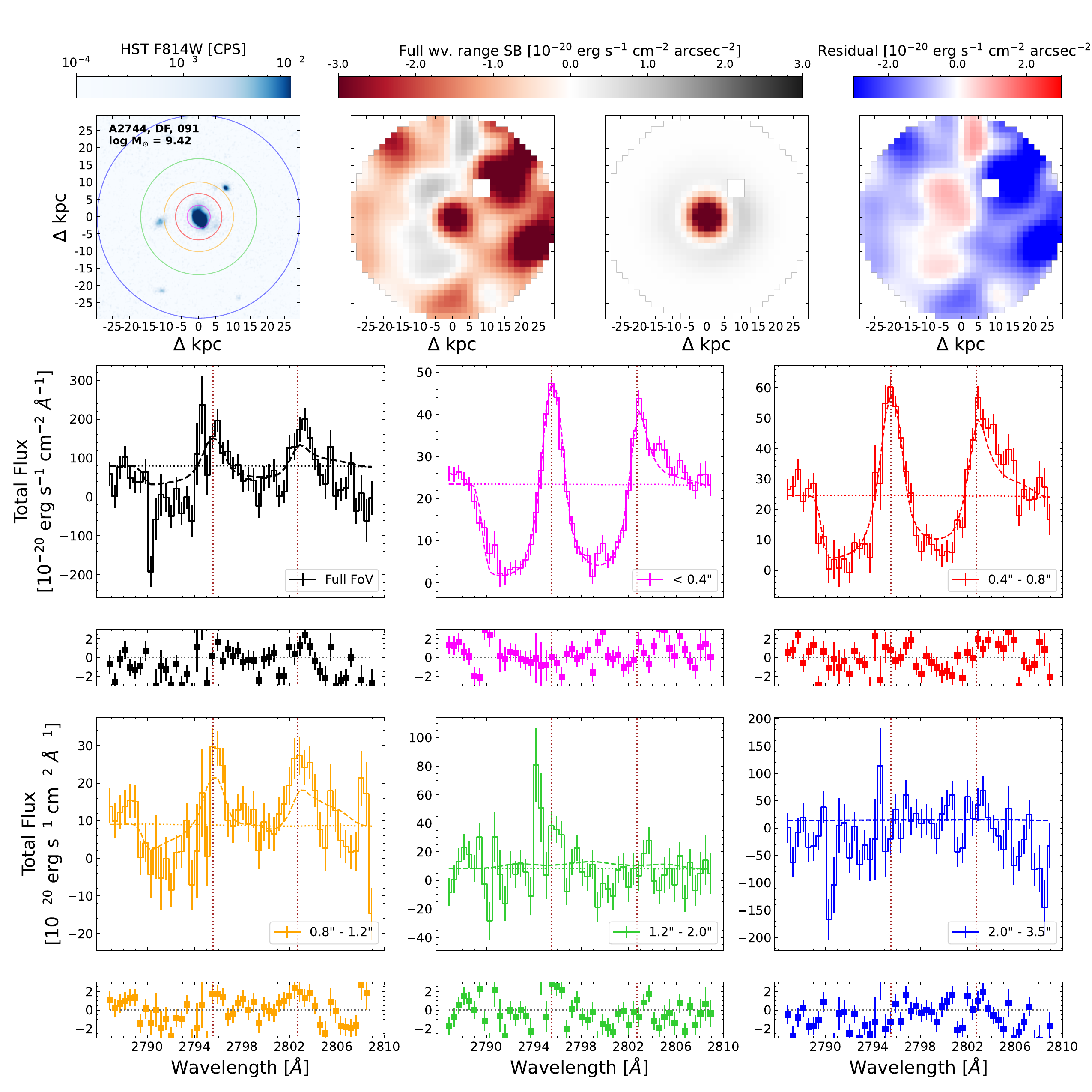}
    \end{minipage}
        \begin{minipage}{.49\textwidth}
        \centering
        \includegraphics[width=0.99\columnwidth]{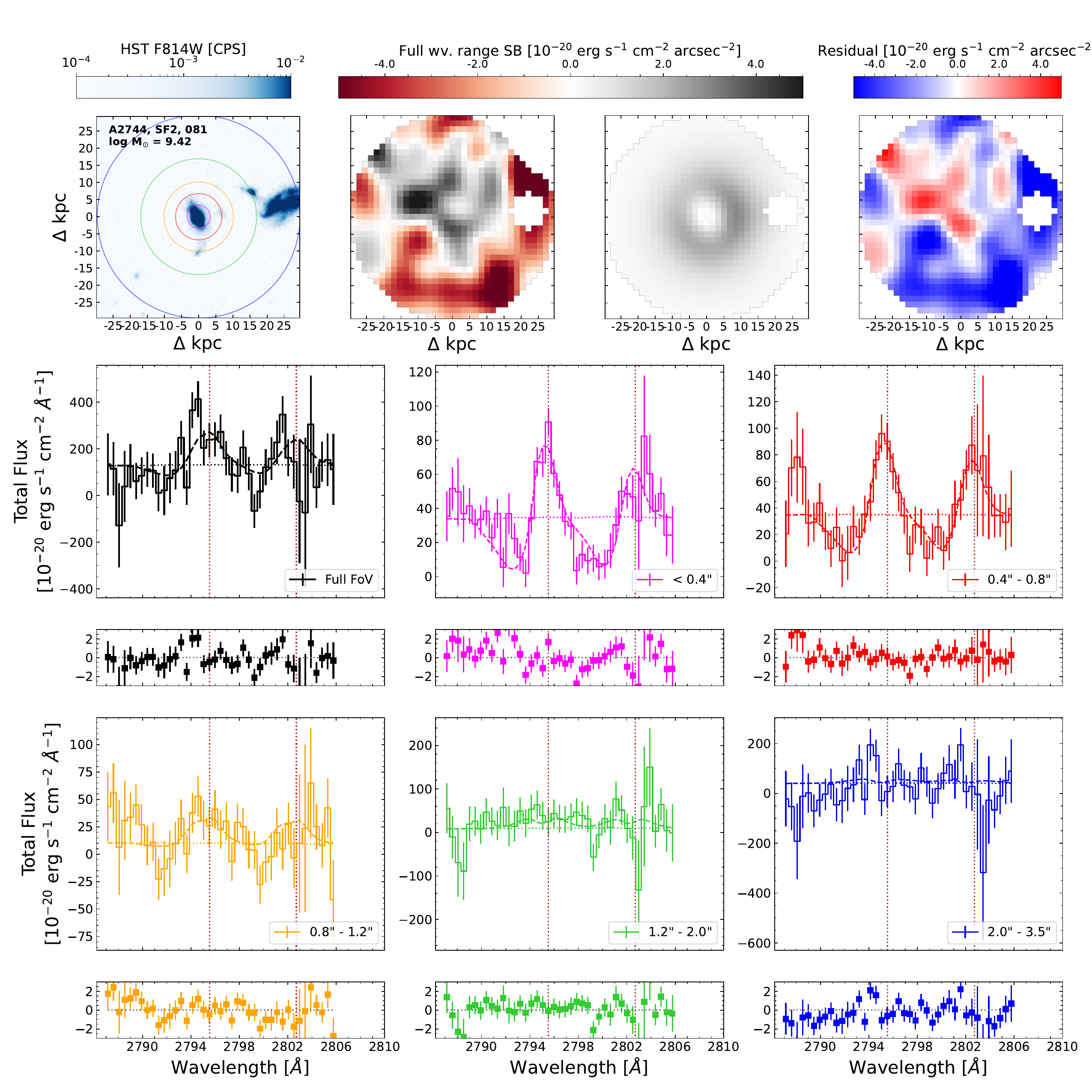}
    \end{minipage}%

    \caption{Same as Fig.~\ref{fig:master_1}, for other four galaxies in our sample.}
    \label{fig:master_5}
\end{figure*}

\begin{figure*}[!htb]
    \centering
    \begin{minipage}{0.49\textwidth}
        \centering
        \includegraphics[width=0.99\columnwidth]{Master_figures/M0416_SF07_012_Master_figure.pdf}
    \end{minipage} 
    \begin{minipage}{.49\textwidth}
        \centering
        \includegraphics[width=0.99\columnwidth]{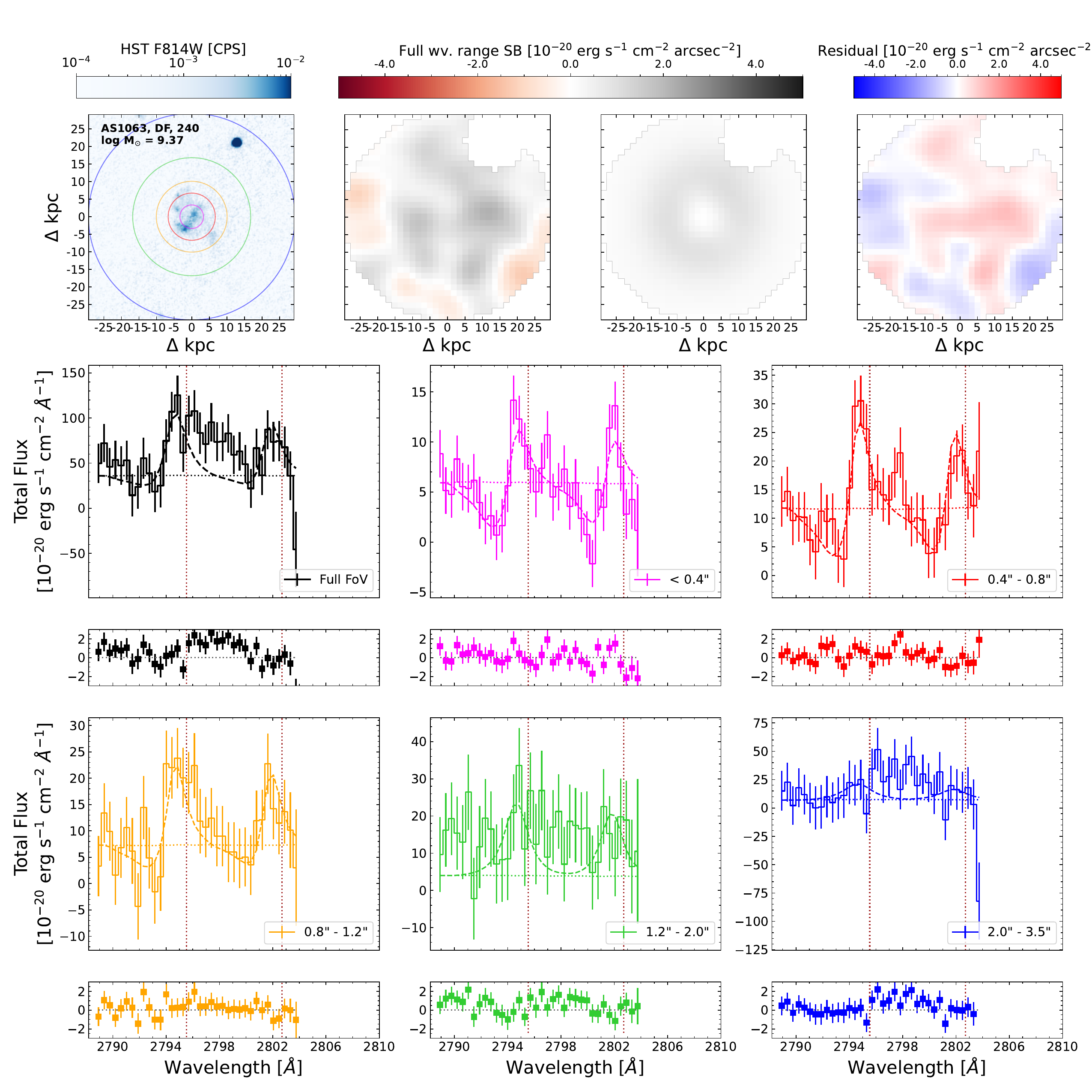}
    \end{minipage}%

    \begin{minipage}{0.49\textwidth}
        \centering
        \includegraphics[width=0.99\columnwidth]{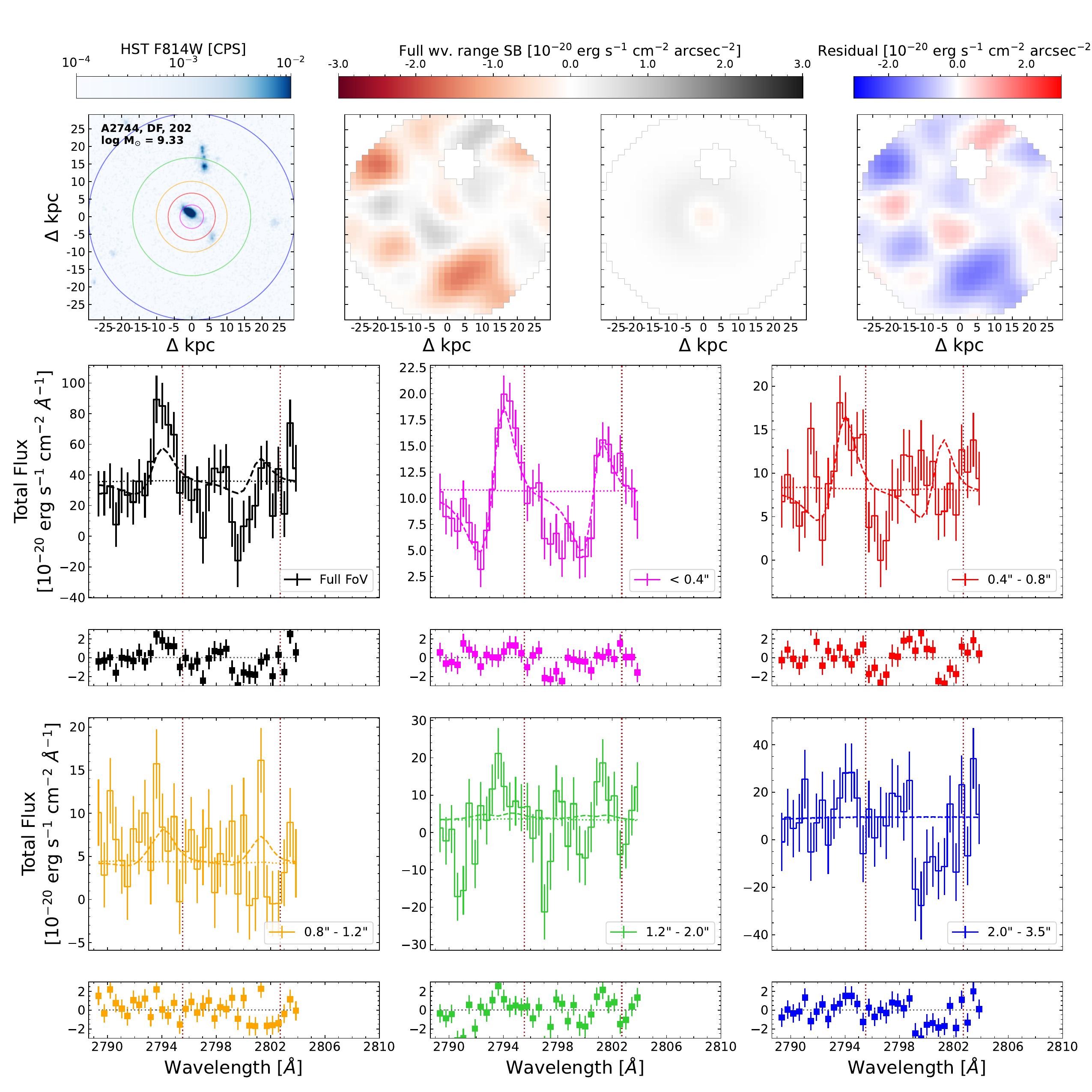}
    \end{minipage}
    \begin{minipage}{.49\textwidth}
        \centering
        \includegraphics[width=0.99\columnwidth]{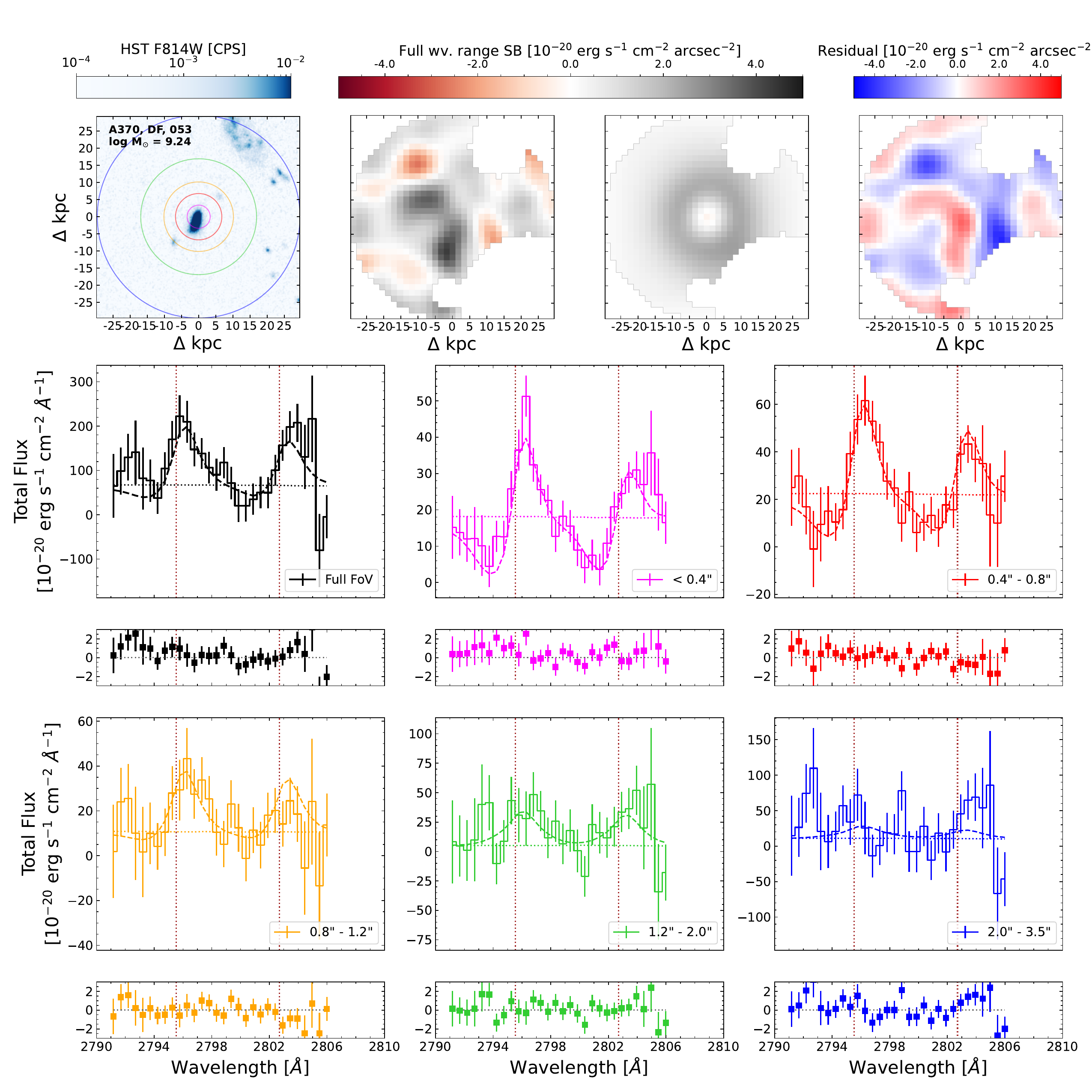}
    \end{minipage}%
    \caption{Same as Fig.~\ref{fig:master_1}, for other four galaxies in our sample.}
    \label{fig:master_6}
\end{figure*}    
\begin{figure*}[!htb]
    \centering
    
    \begin{minipage}{0.49\textwidth}
        \centering
        \includegraphics[width=0.99\columnwidth]{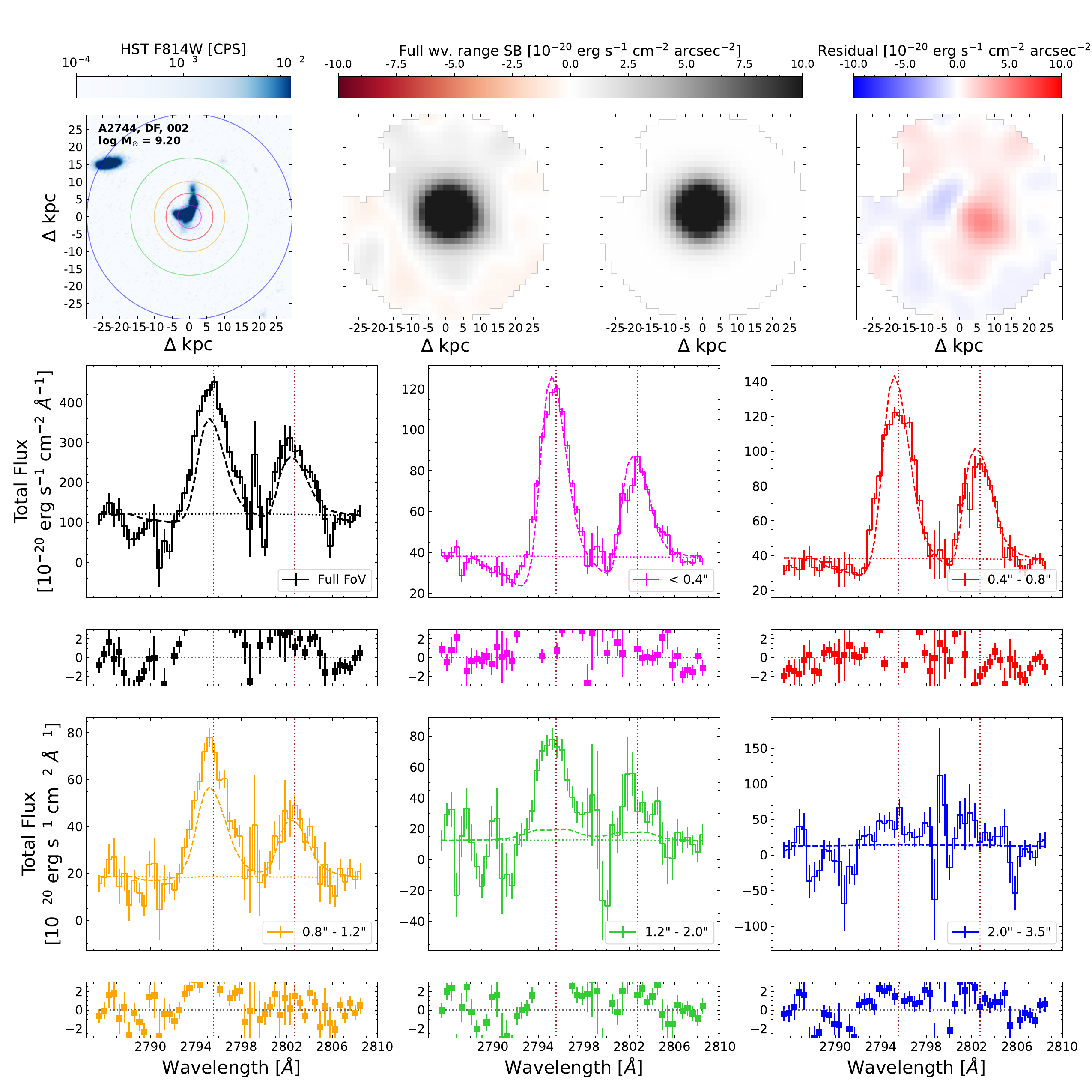}
    \end{minipage}
    \begin{minipage}{.49\textwidth}
        \centering
        \includegraphics[width=0.99\columnwidth]{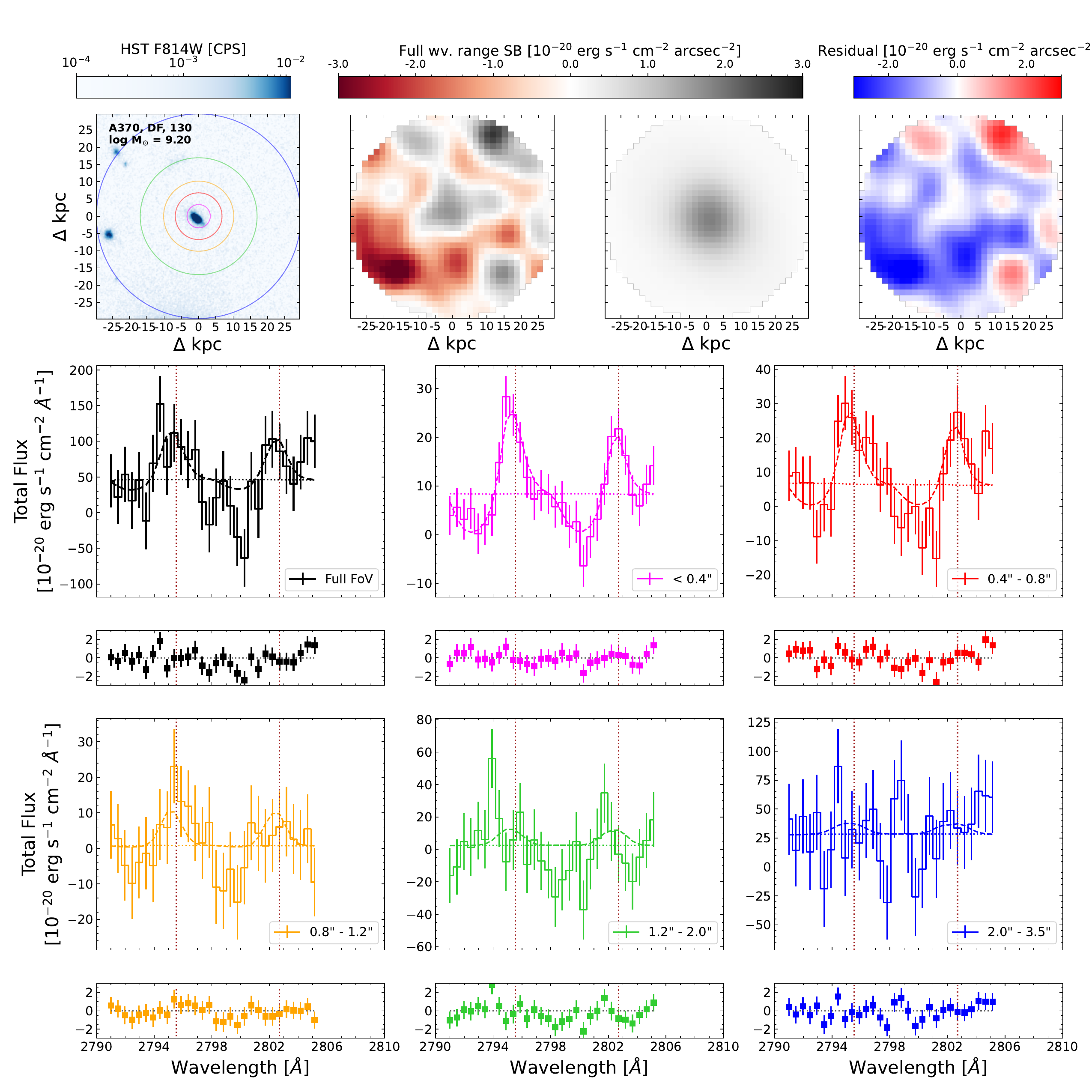}
    \end{minipage}%

    \begin{minipage}{0.49\textwidth}
        \centering
        \includegraphics[width=0.99\columnwidth]{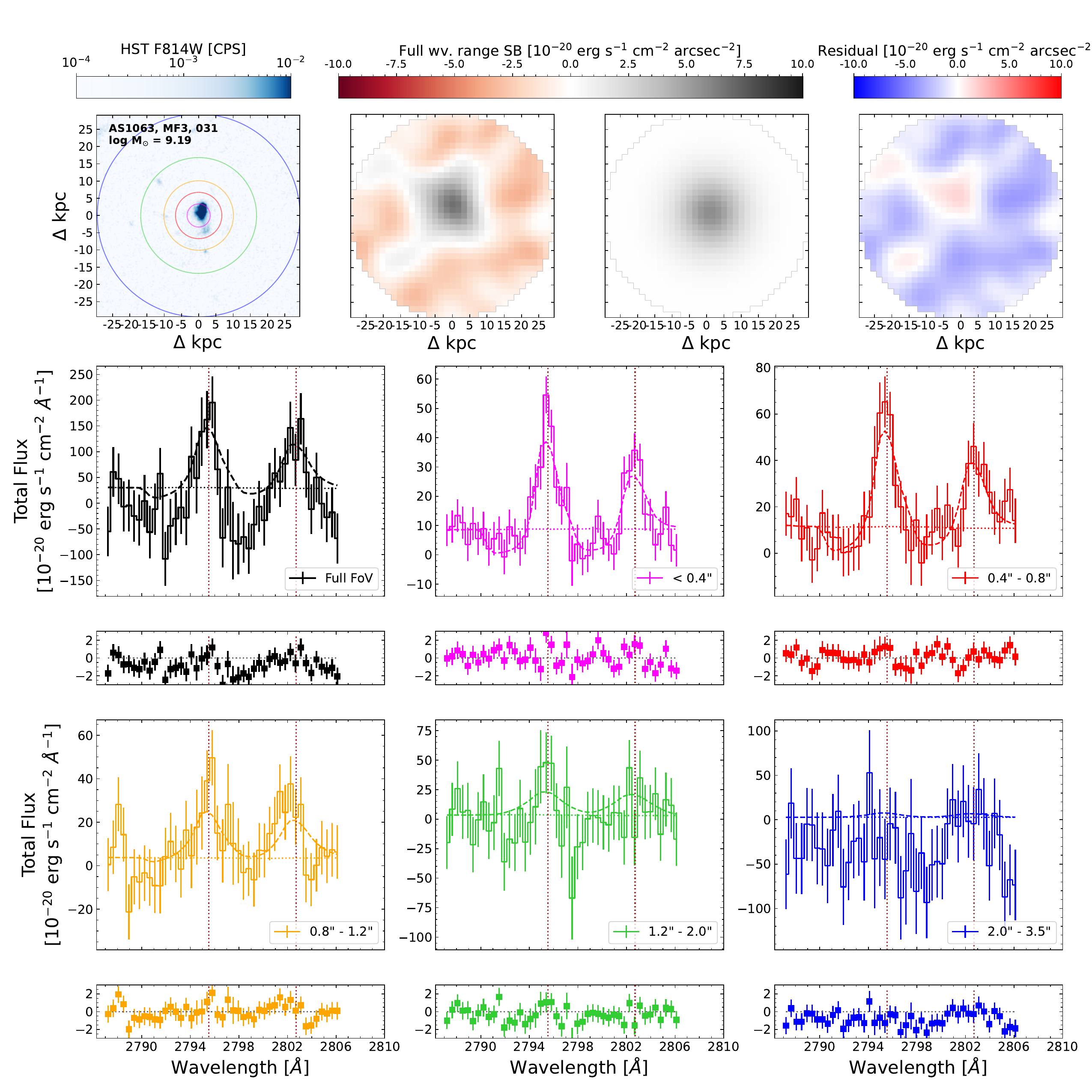}
    \end{minipage}
    \begin{minipage}{.49\textwidth}
        \centering
        \includegraphics[width=0.99\columnwidth]{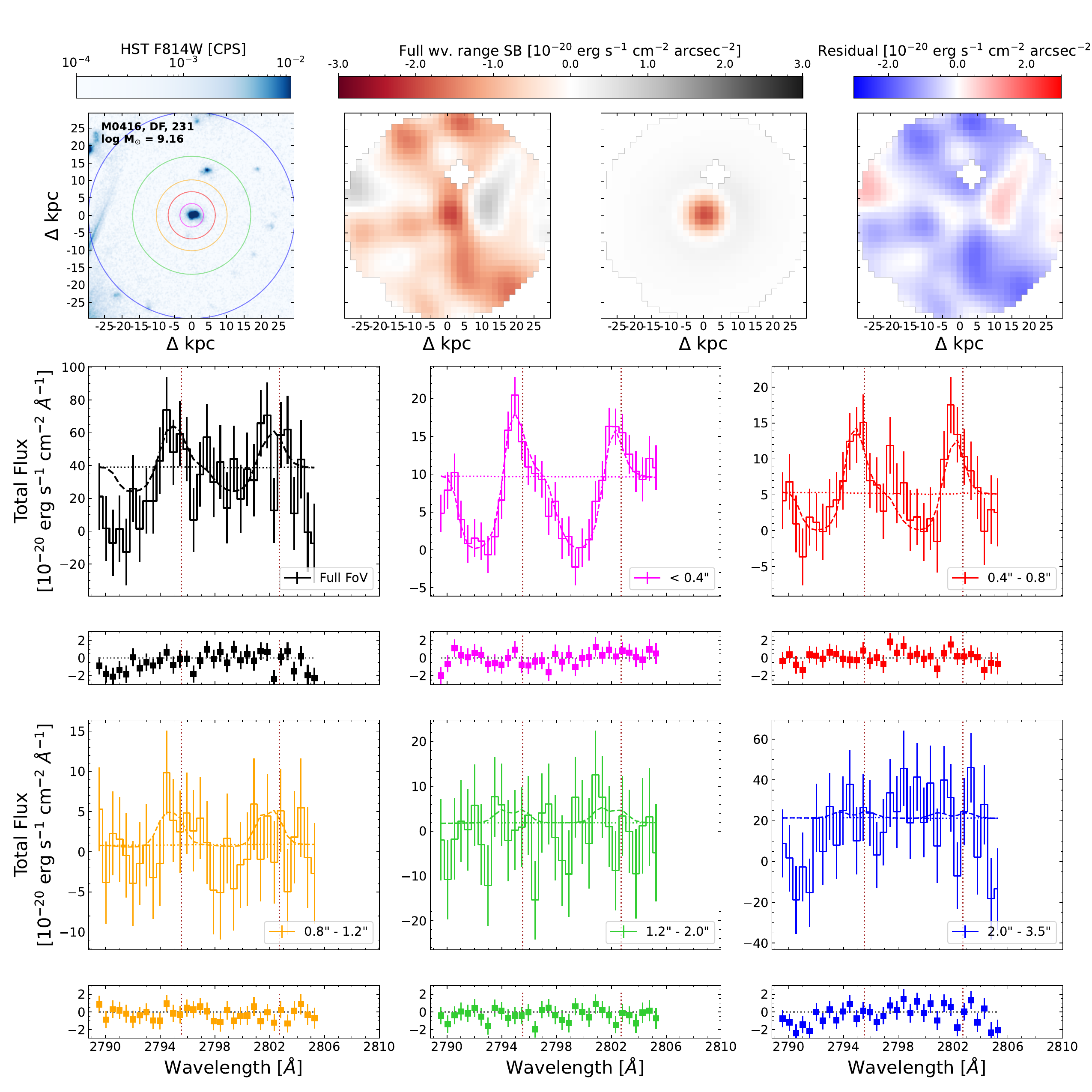}
    \end{minipage}%
    \caption{Same as Fig.~\ref{fig:master_1}, for other four galaxies in our sample.}
    \label{fig:master_7}
\end{figure*}    
\begin{figure*}[!htb]
    \centering
    
    \begin{minipage}{0.49\textwidth}
        \centering
        \includegraphics[width=0.99\columnwidth]{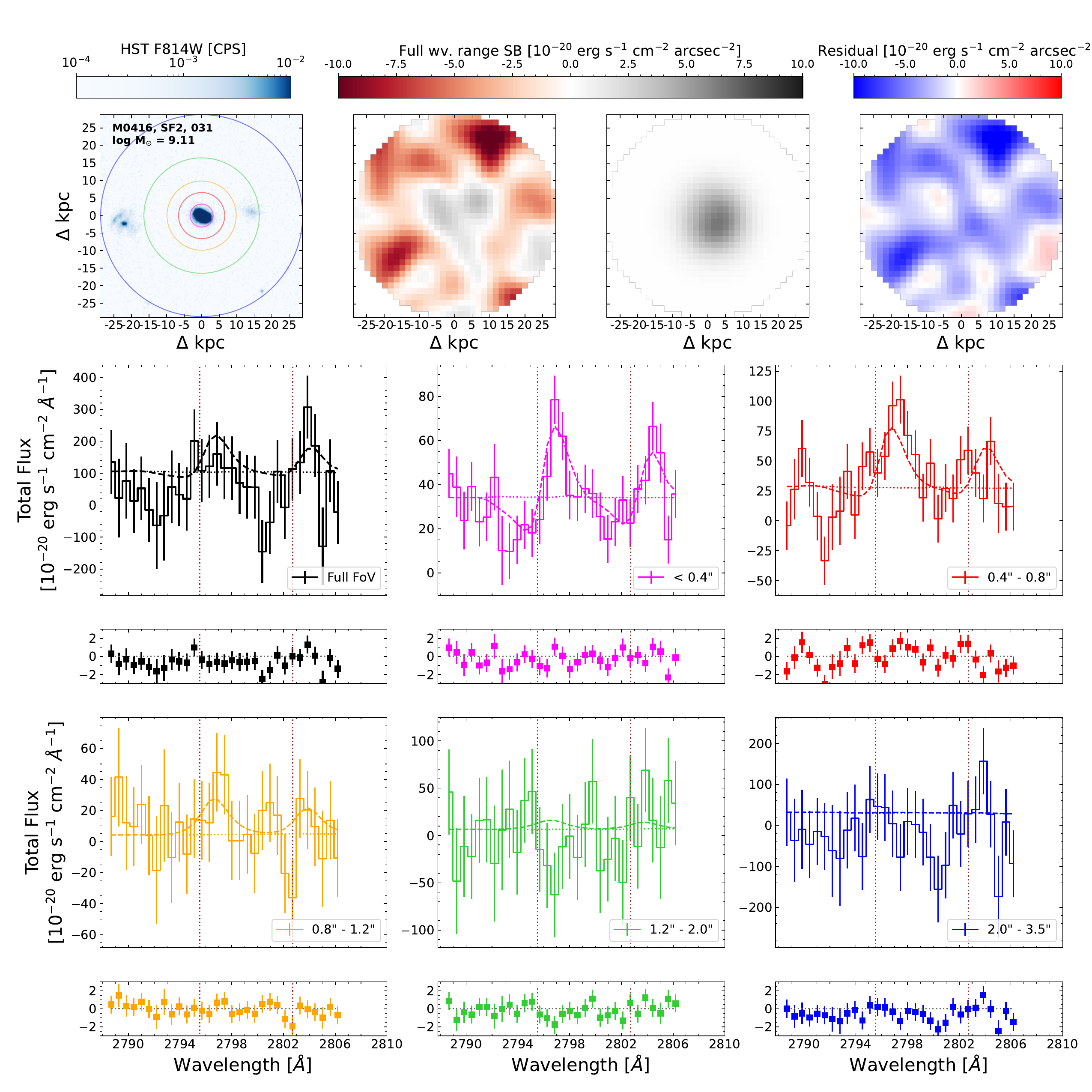}
    \end{minipage}    
    \begin{minipage}{.49\textwidth}
        \centering
        \includegraphics[width=0.99\columnwidth]{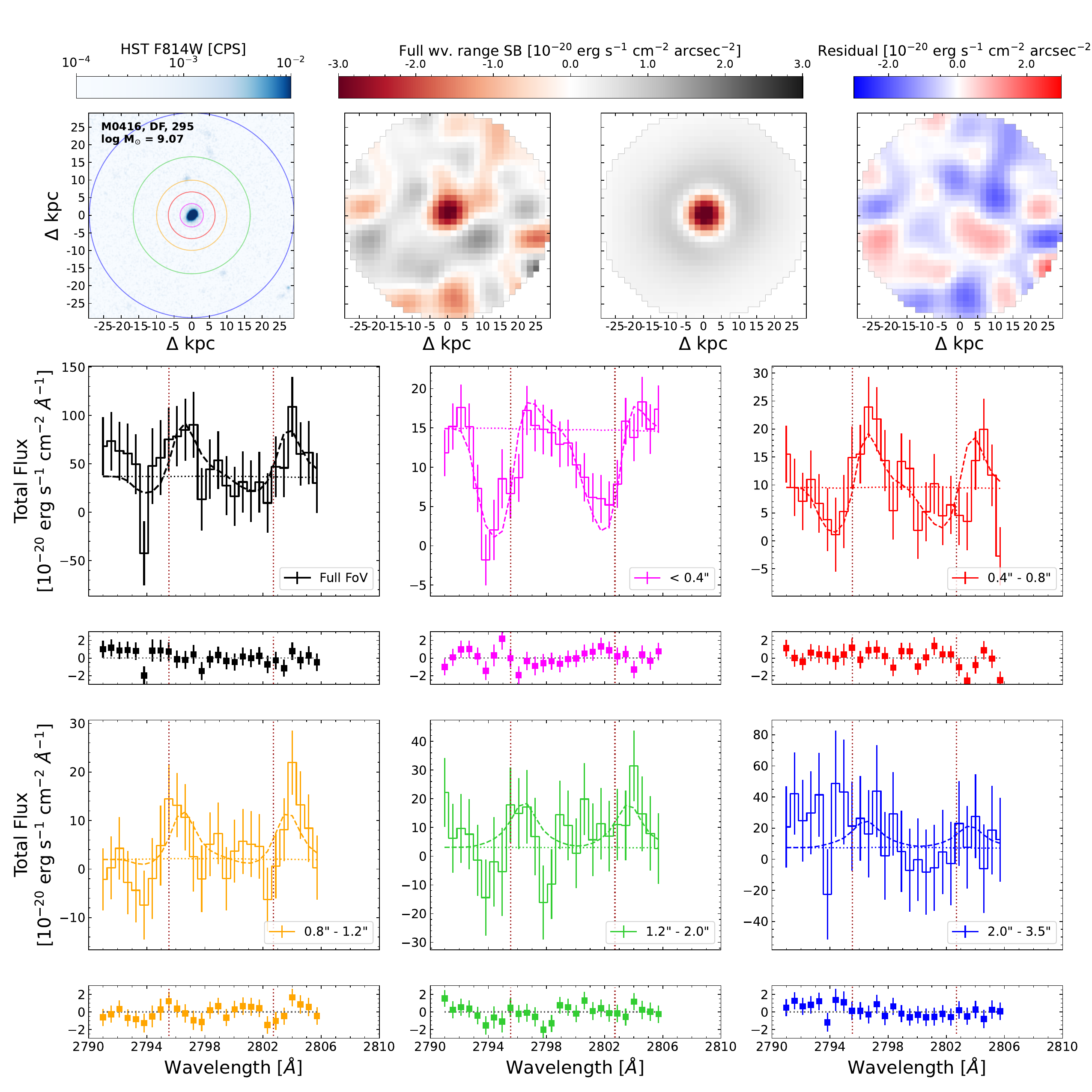}
    \end{minipage}%
 
    \begin{minipage}{0.49\textwidth}
        \centering
        \includegraphics[width=0.99\columnwidth]{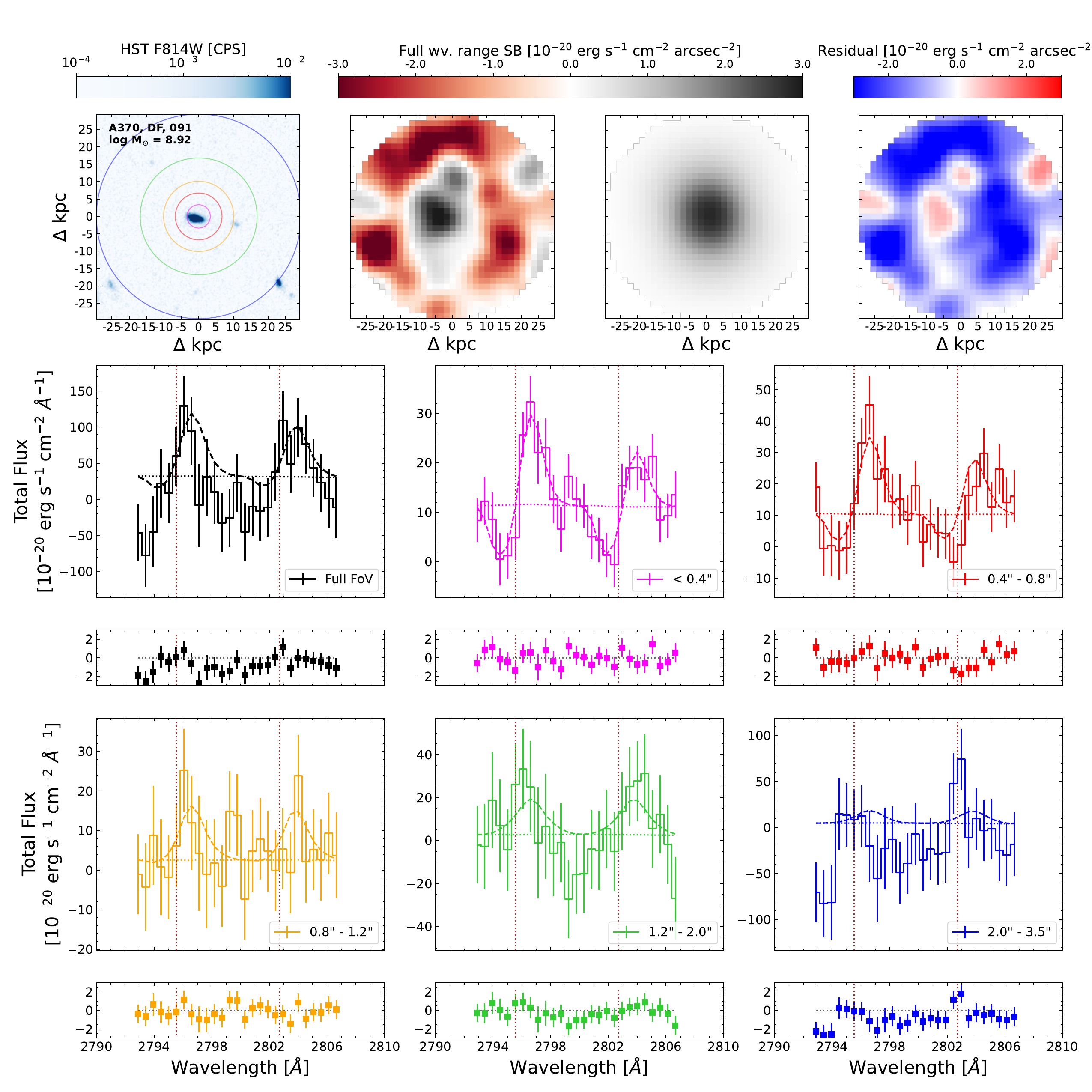}
    \end{minipage}
    \begin{minipage}{.49\textwidth}
        \centering
        \includegraphics[width=0.99\columnwidth]{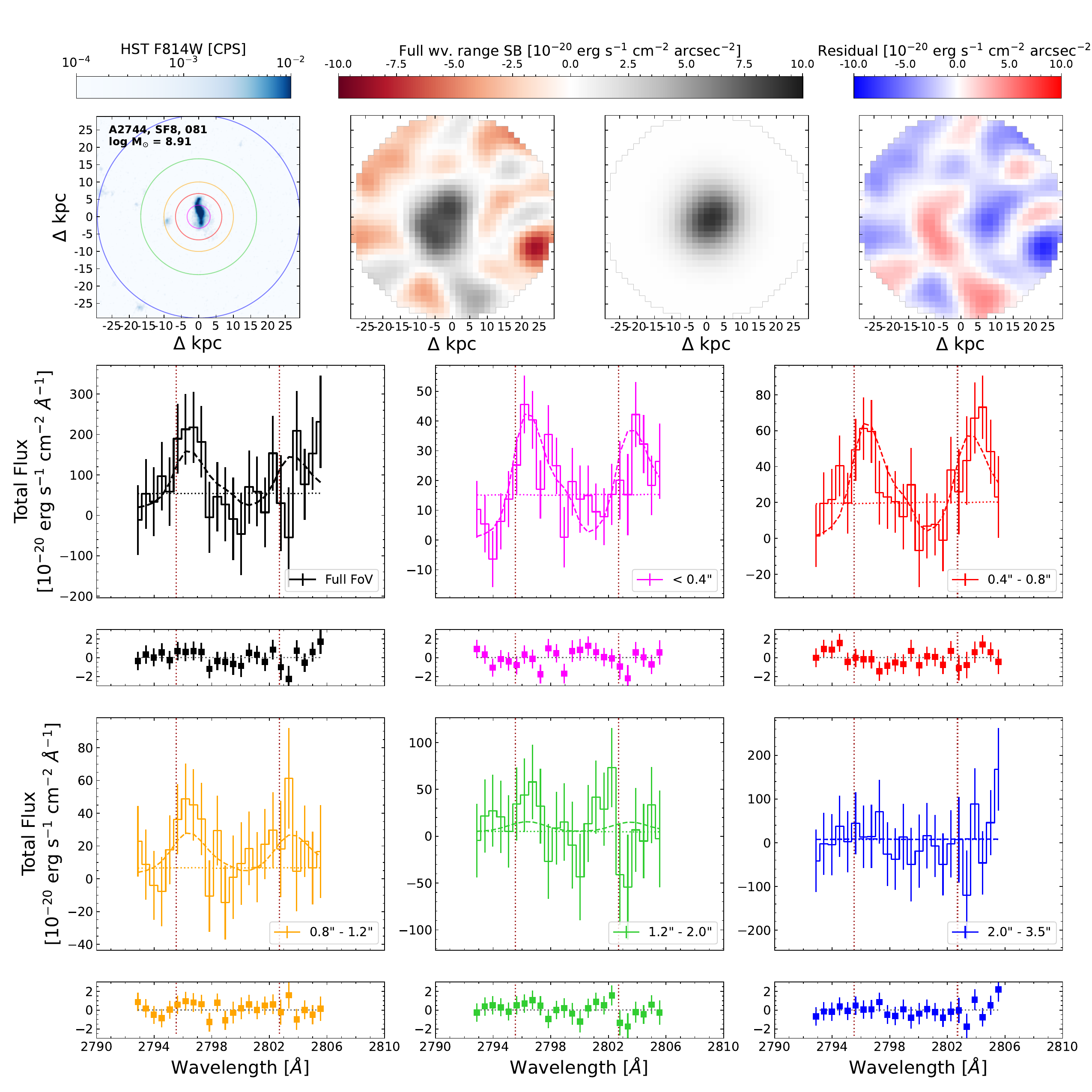}
    \end{minipage}%
    \caption{Same as Fig.~\ref{fig:master_1}, for other four galaxies in our sample.}
    \label{fig:master_8}
\end{figure*}    
\begin{figure*}[!htb]
    \centering
    
    \begin{minipage}{0.49\textwidth}
        \centering
        \includegraphics[width=0.99\columnwidth]{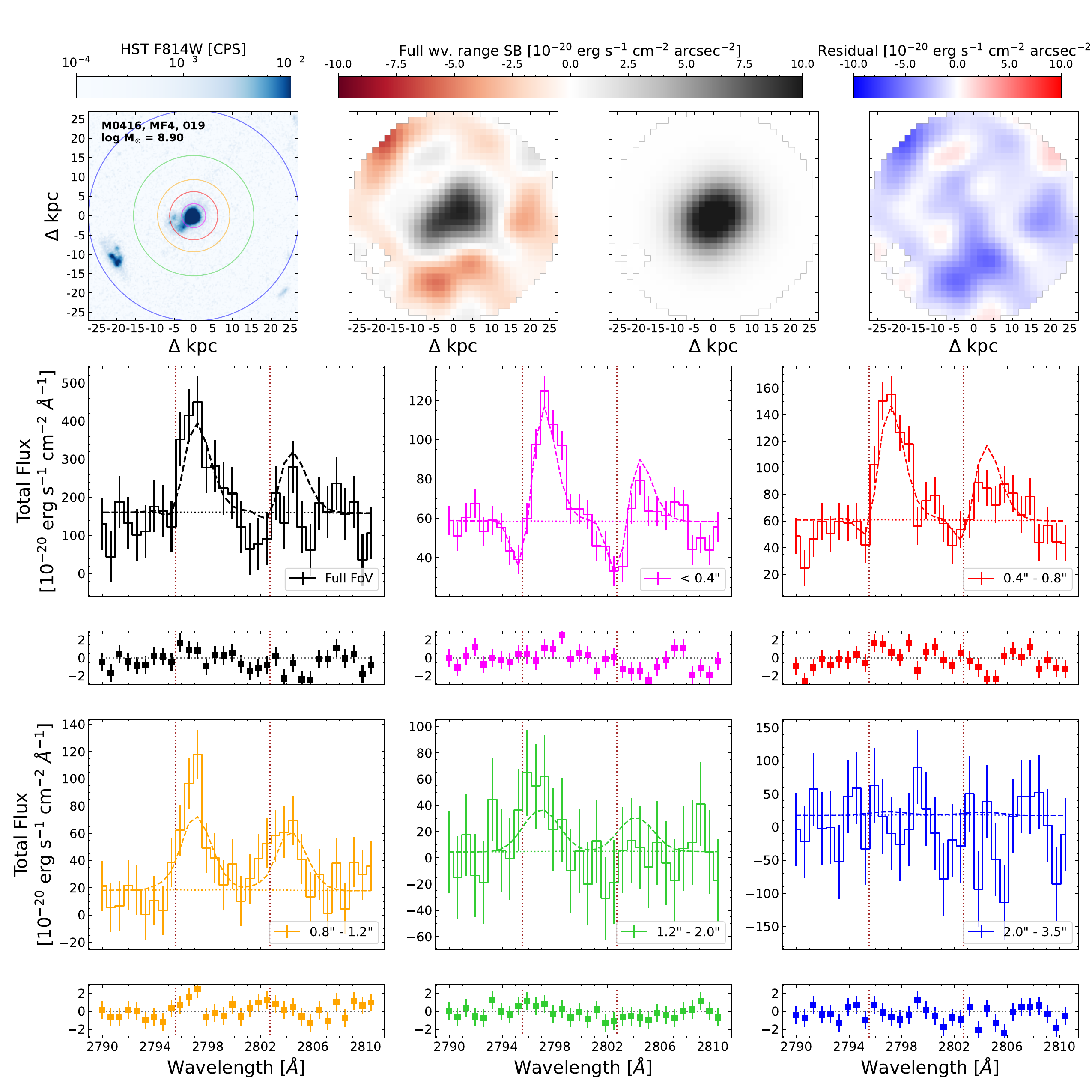}
    \end{minipage}
    \begin{minipage}{.49\textwidth}
        \centering
        \includegraphics[width=0.99\columnwidth]{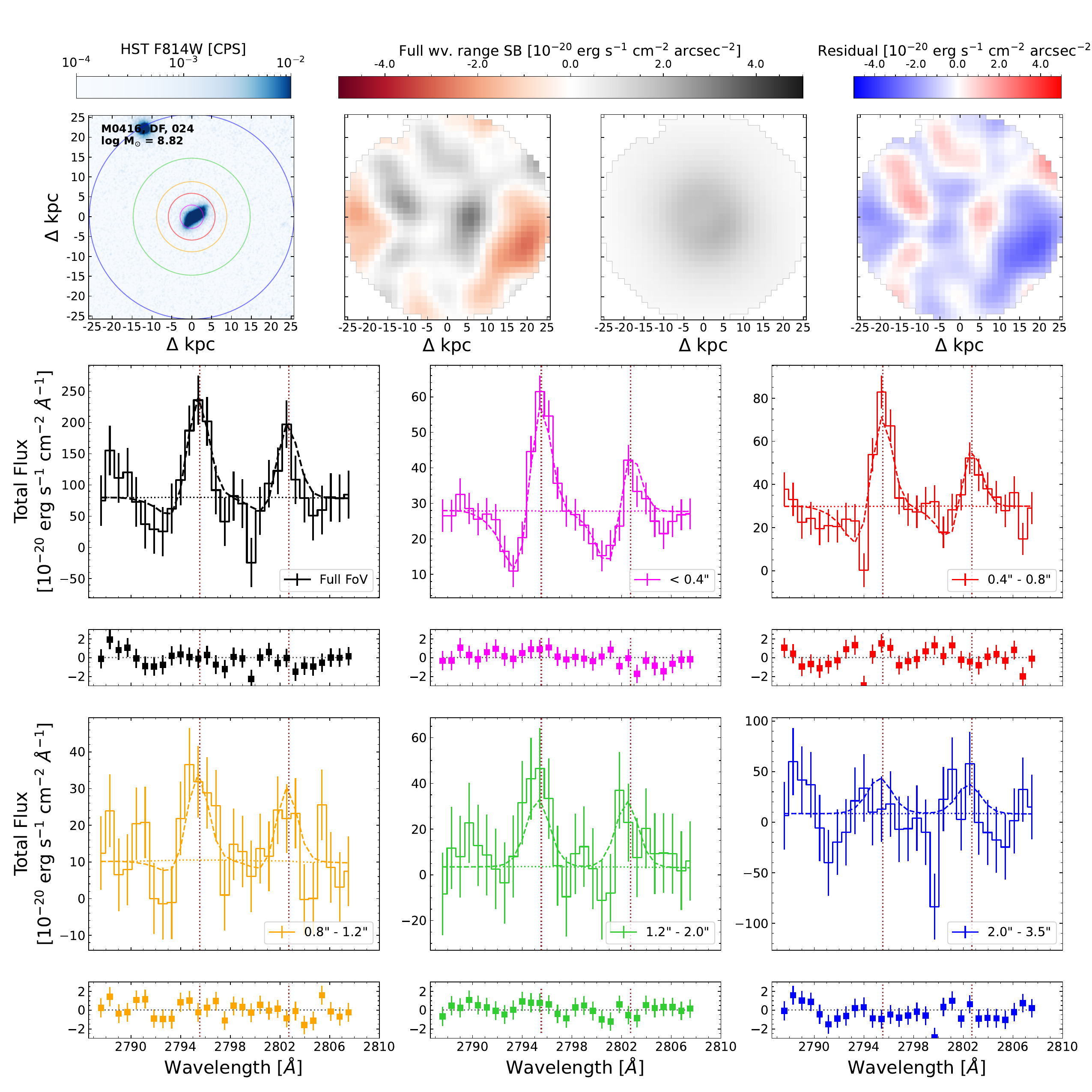}
    \end{minipage}%

    \begin{minipage}{0.49\textwidth}
        \centering
        \includegraphics[width=0.99\columnwidth]{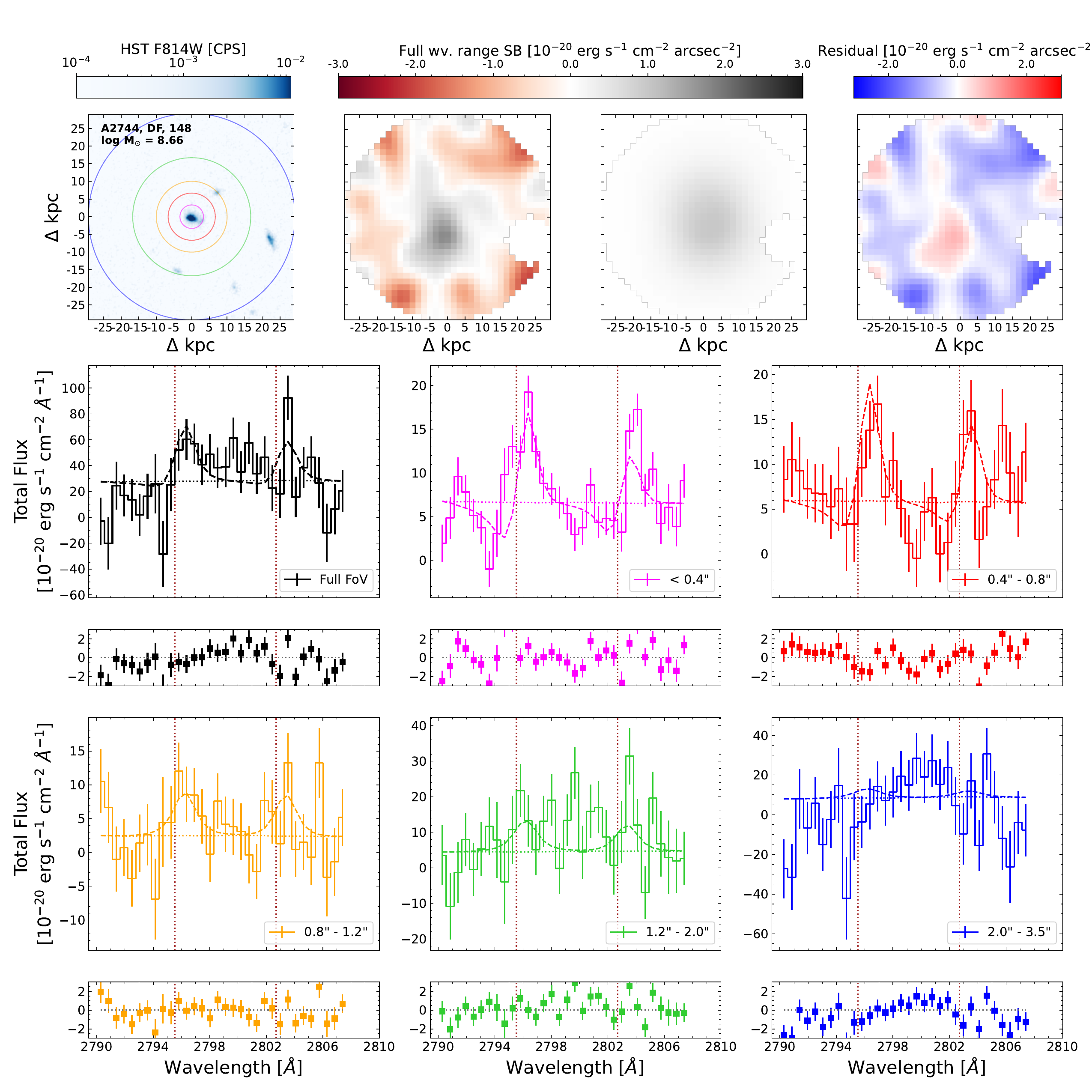}
    \end{minipage}
    \begin{minipage}{.49\textwidth}
        \centering
        \includegraphics[width=0.99\columnwidth]{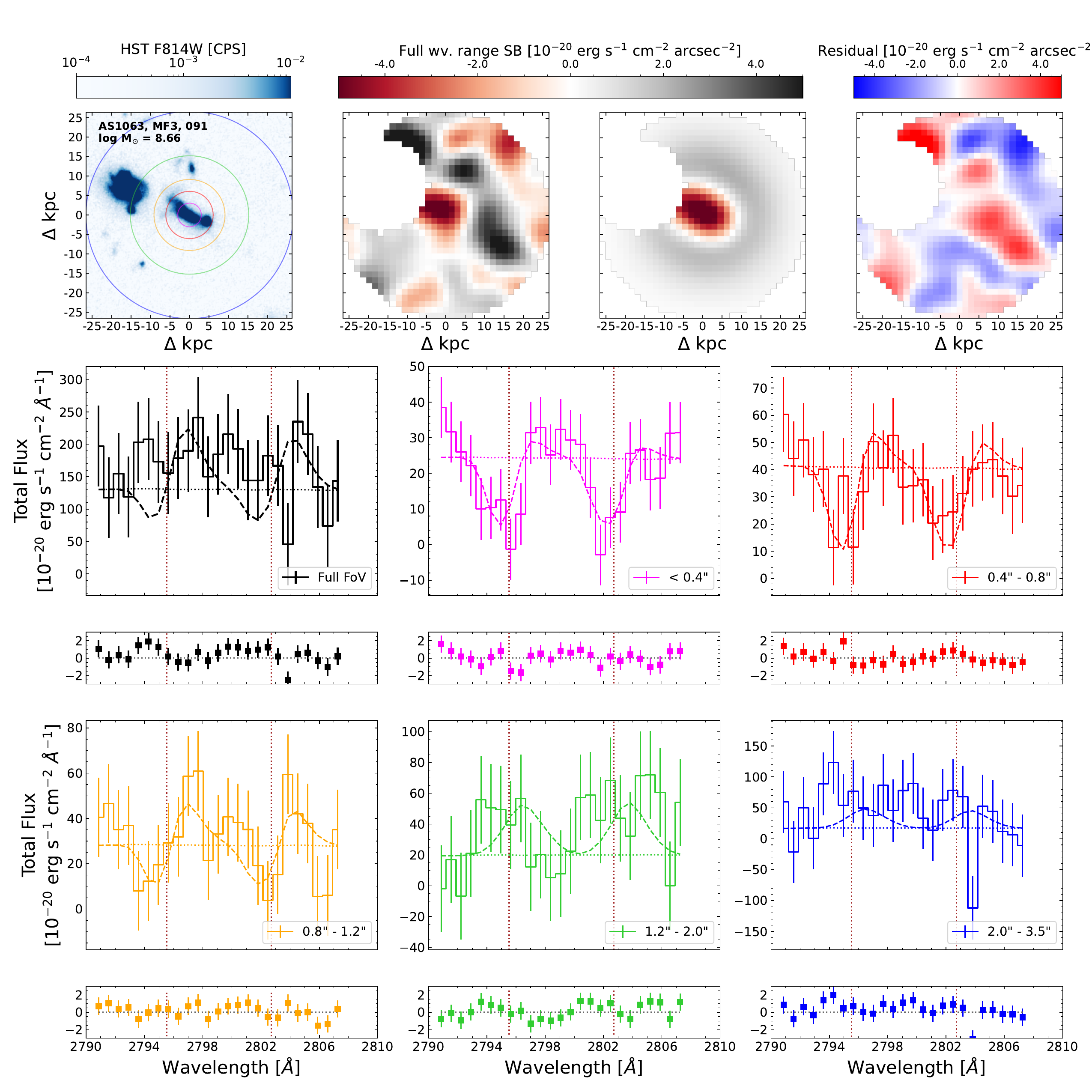}
    \end{minipage}%

    \caption{Same as Fig.~\ref{fig:master_1}, for other four galaxies in our sample.}
    \label{fig:master_9}
\end{figure*}

\begin{figure*}[!htb]
    \centering
    
    \begin{minipage}{0.49\textwidth}
        \centering
        \includegraphics[width=0.99\columnwidth]{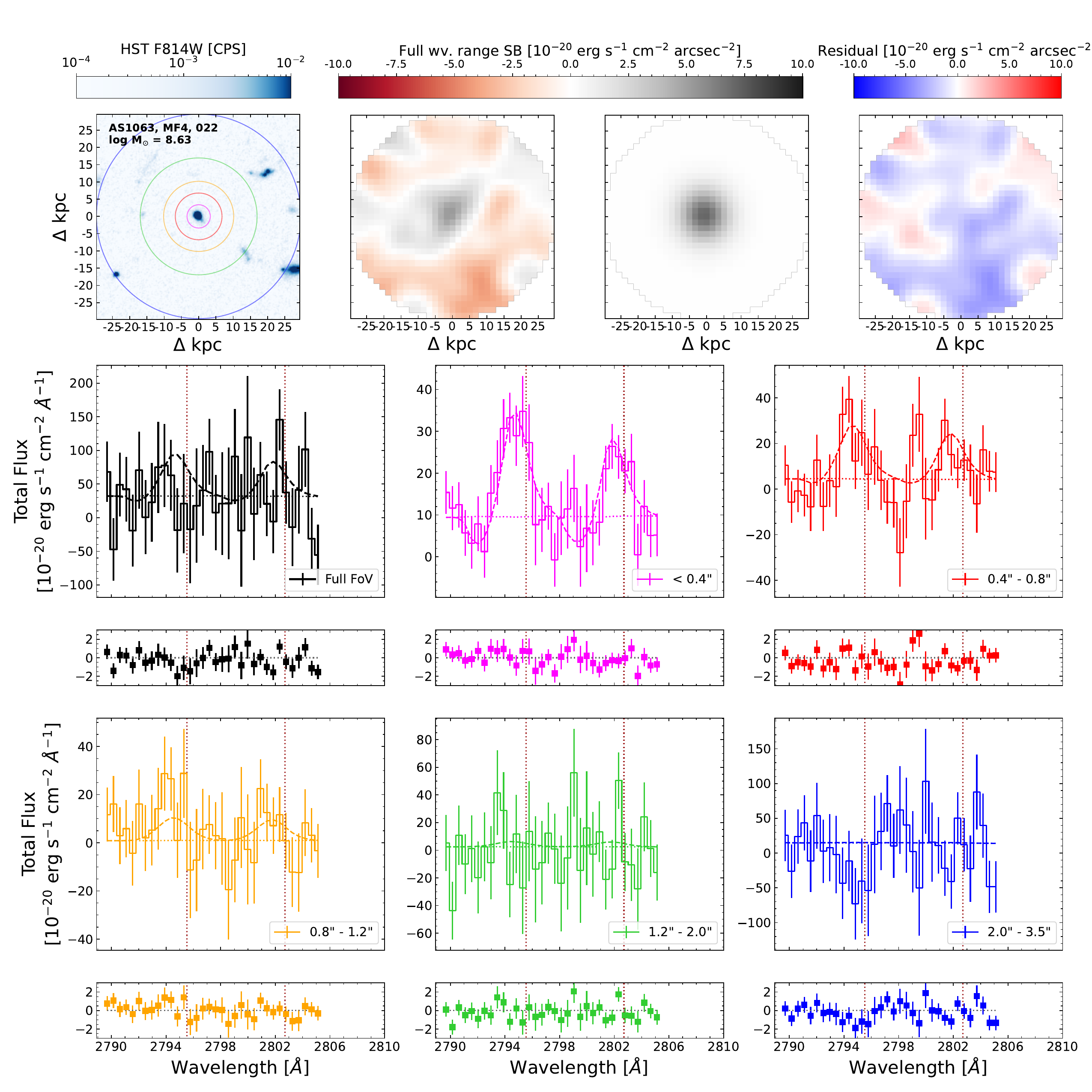}
    \end{minipage}
    \begin{minipage}{.49\textwidth}
        \centering
        \includegraphics[width=0.99\columnwidth]{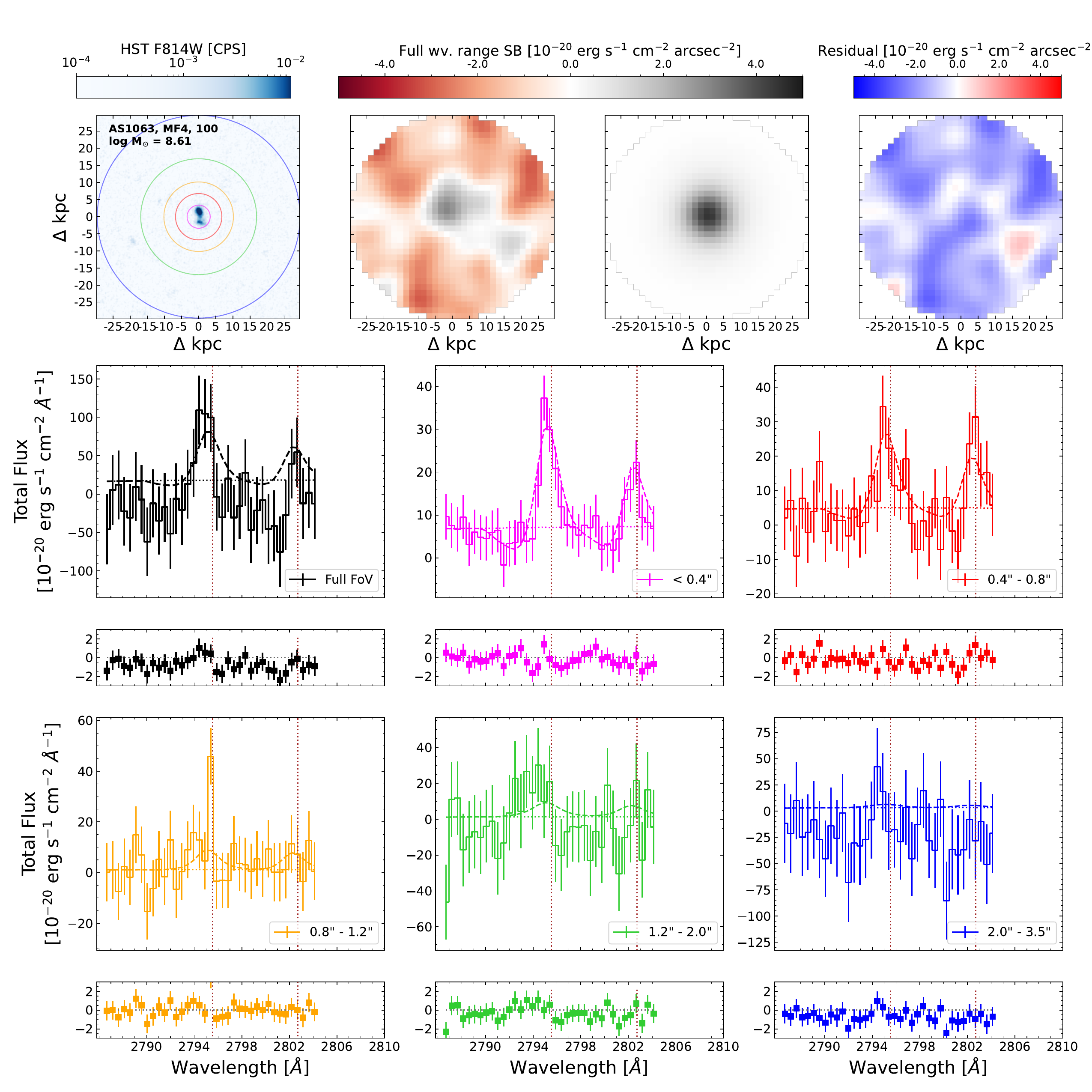}
    \end{minipage}%

    \begin{minipage}{0.49\textwidth}
        \centering
        \includegraphics[width=0.99\columnwidth]{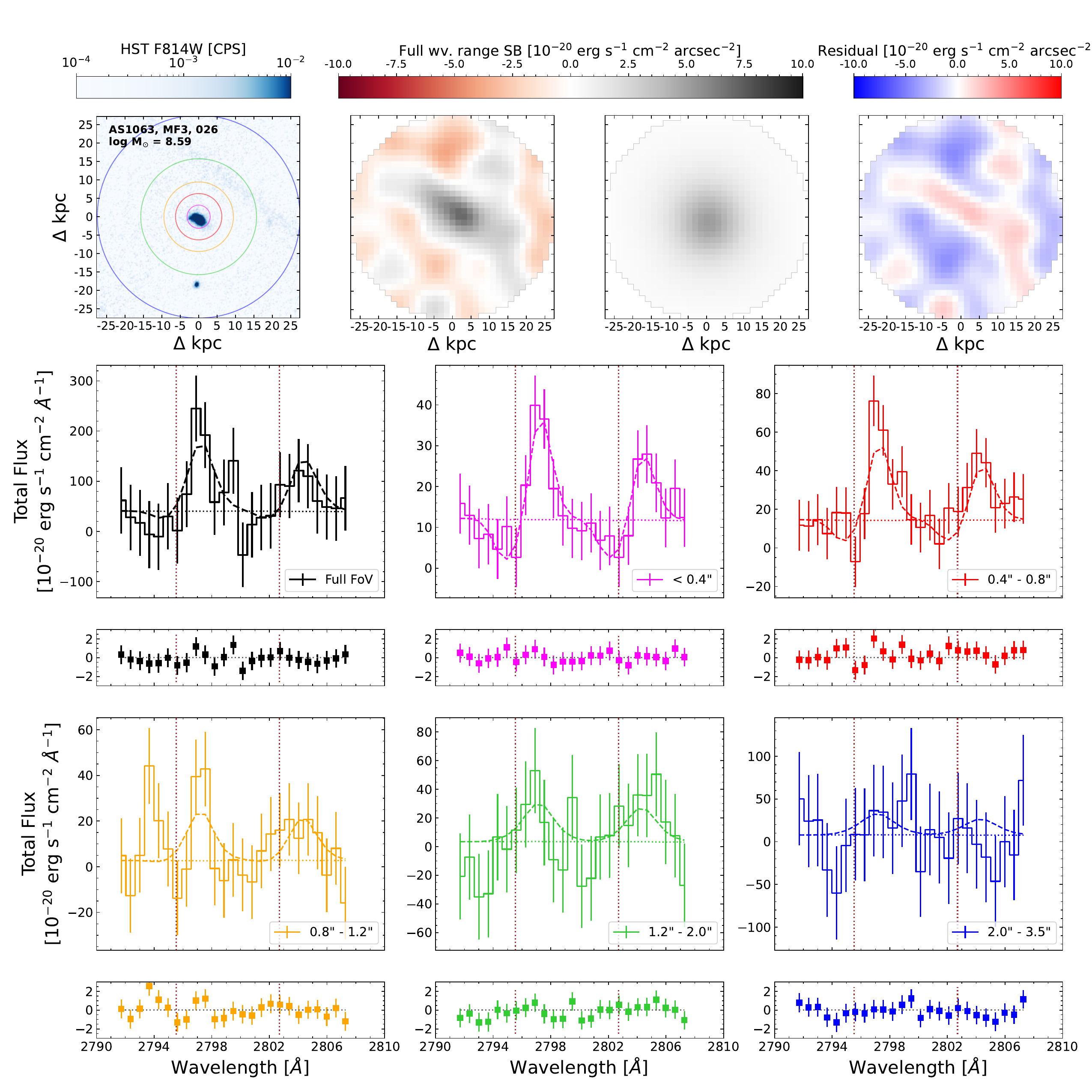}
    \end{minipage}
    \begin{minipage}{.49\textwidth}
        \centering
        \includegraphics[width=0.99\columnwidth]{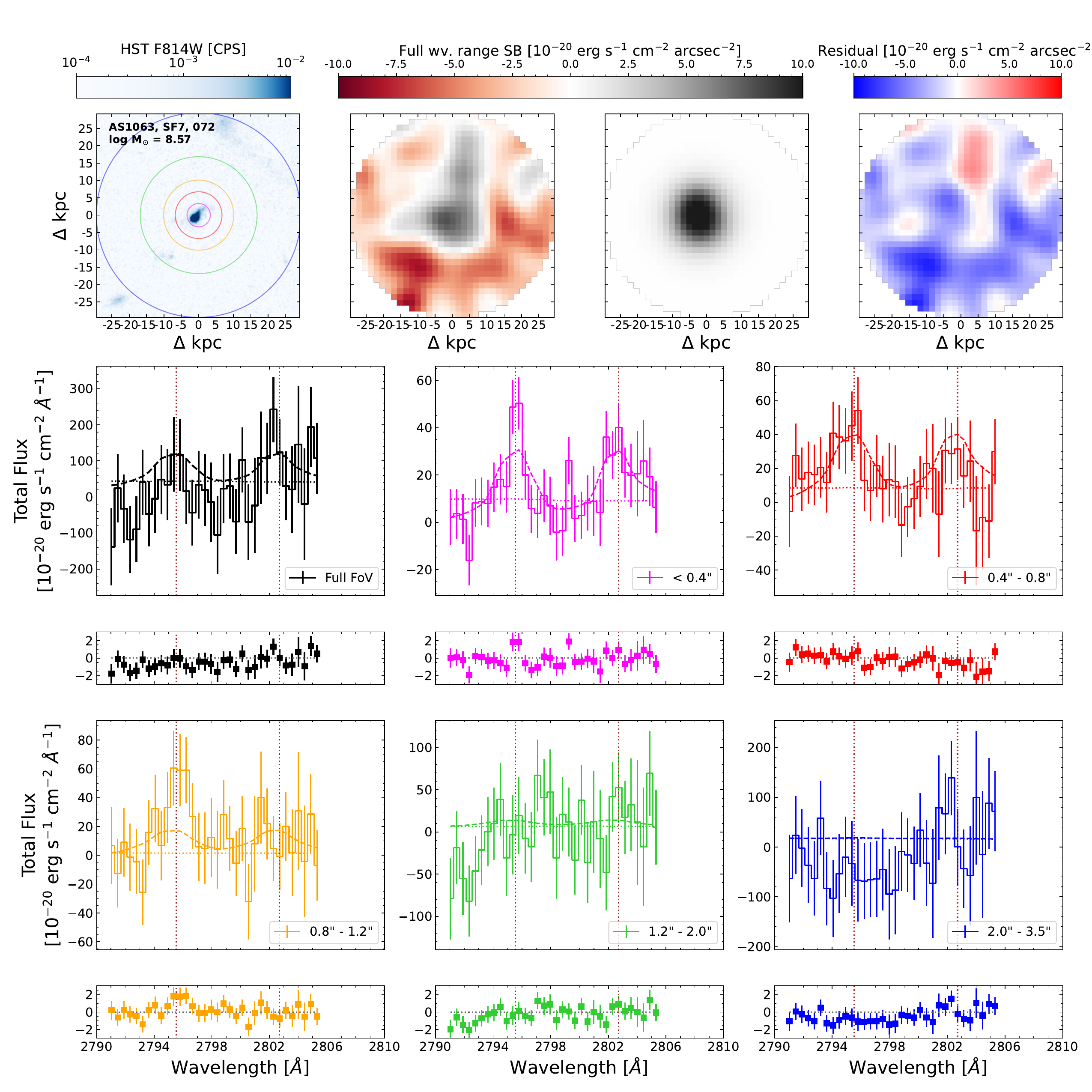}
    \end{minipage}%

    \caption{Same as Fig.~\ref{fig:master_1}, for other four galaxies in our sample.}
    \label{fig:master_10}
\end{figure*}

\begin{figure*}[!htb]
    \centering
    
    \begin{minipage}{0.49\textwidth}
        \centering
        \includegraphics[width=0.99\columnwidth]{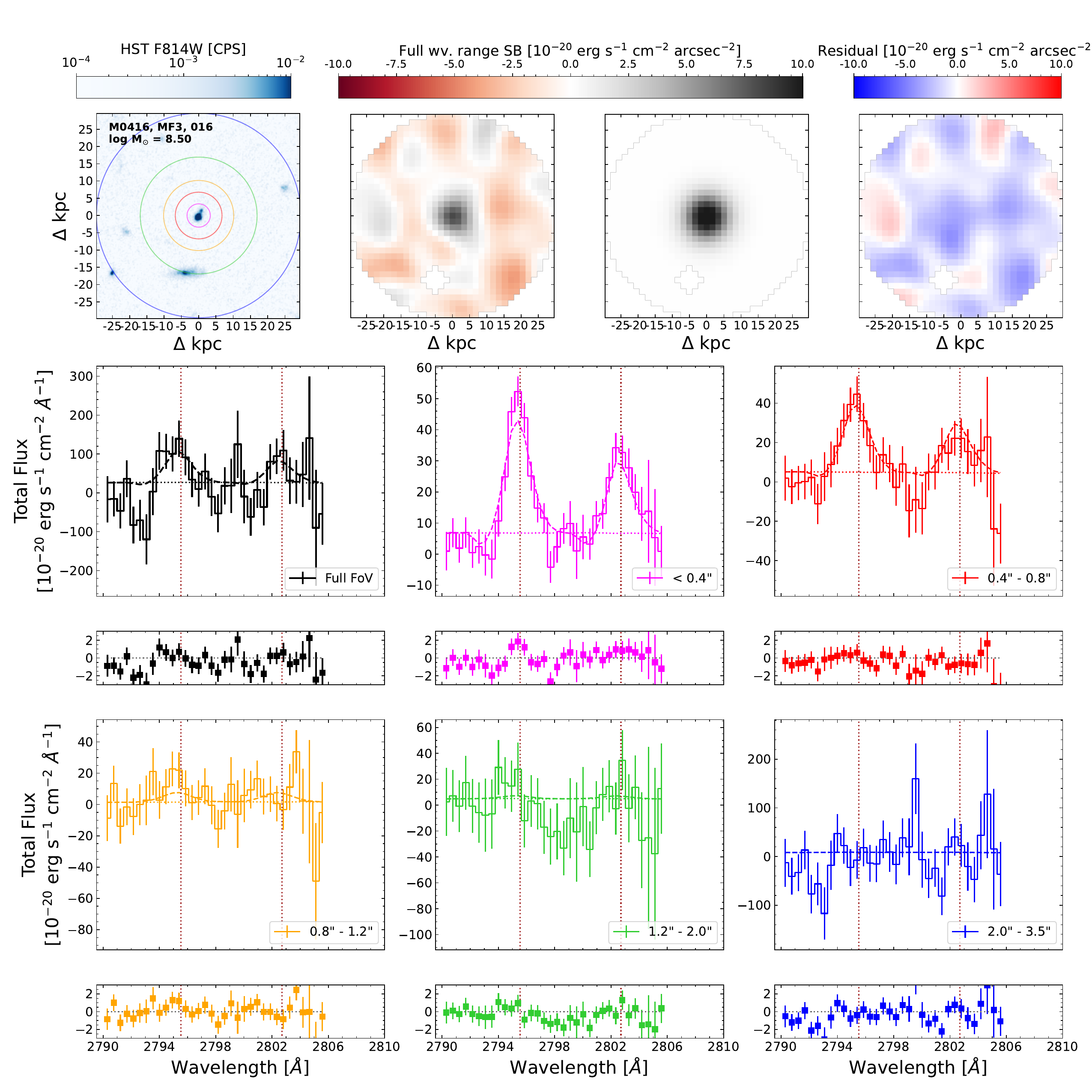}
    \end{minipage}
    \begin{minipage}{.49\textwidth}
        \centering
        \includegraphics[width=0.99\columnwidth]{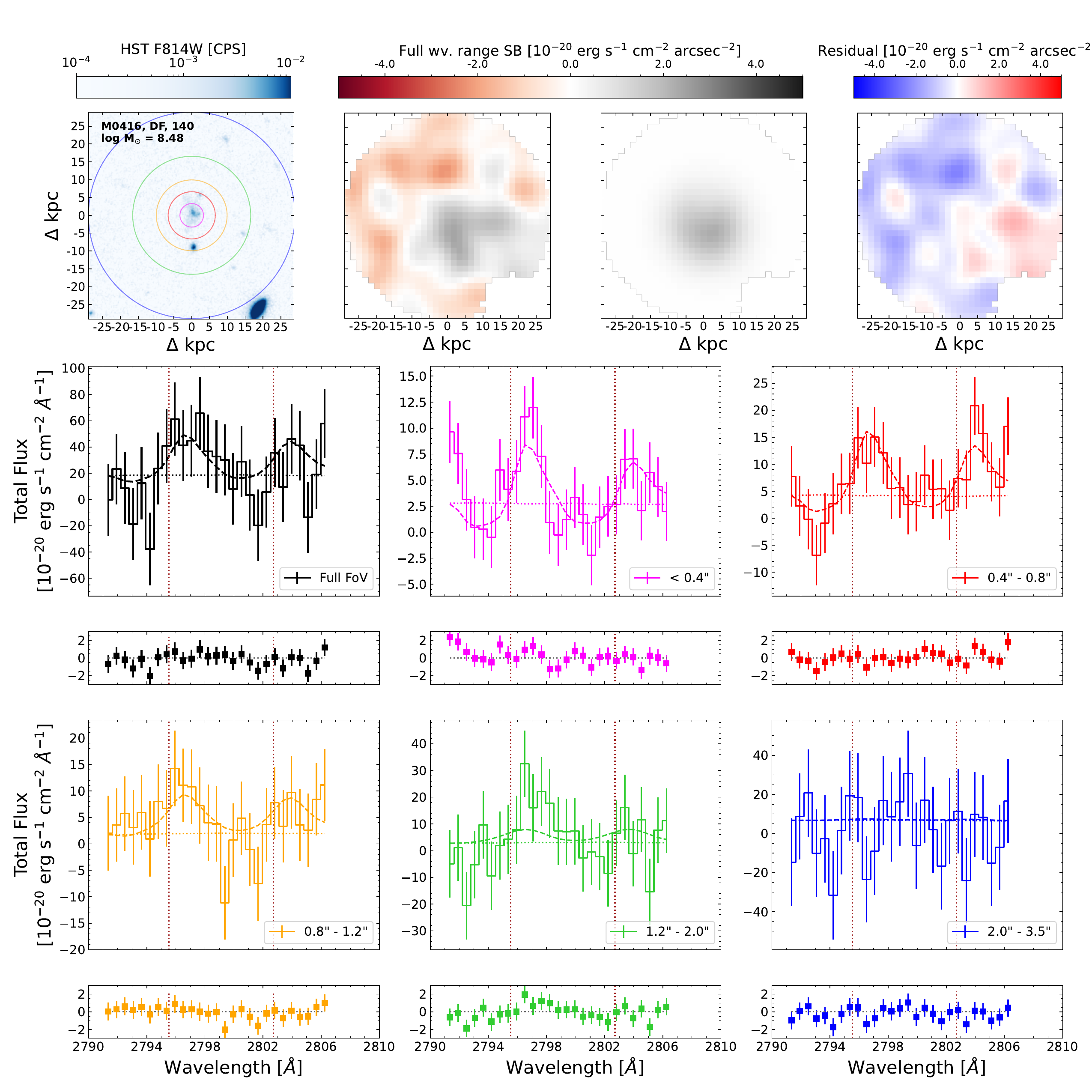}
    \end{minipage}%

    \begin{minipage}{0.49\textwidth}
        \centering
        \includegraphics[width=0.99\columnwidth]{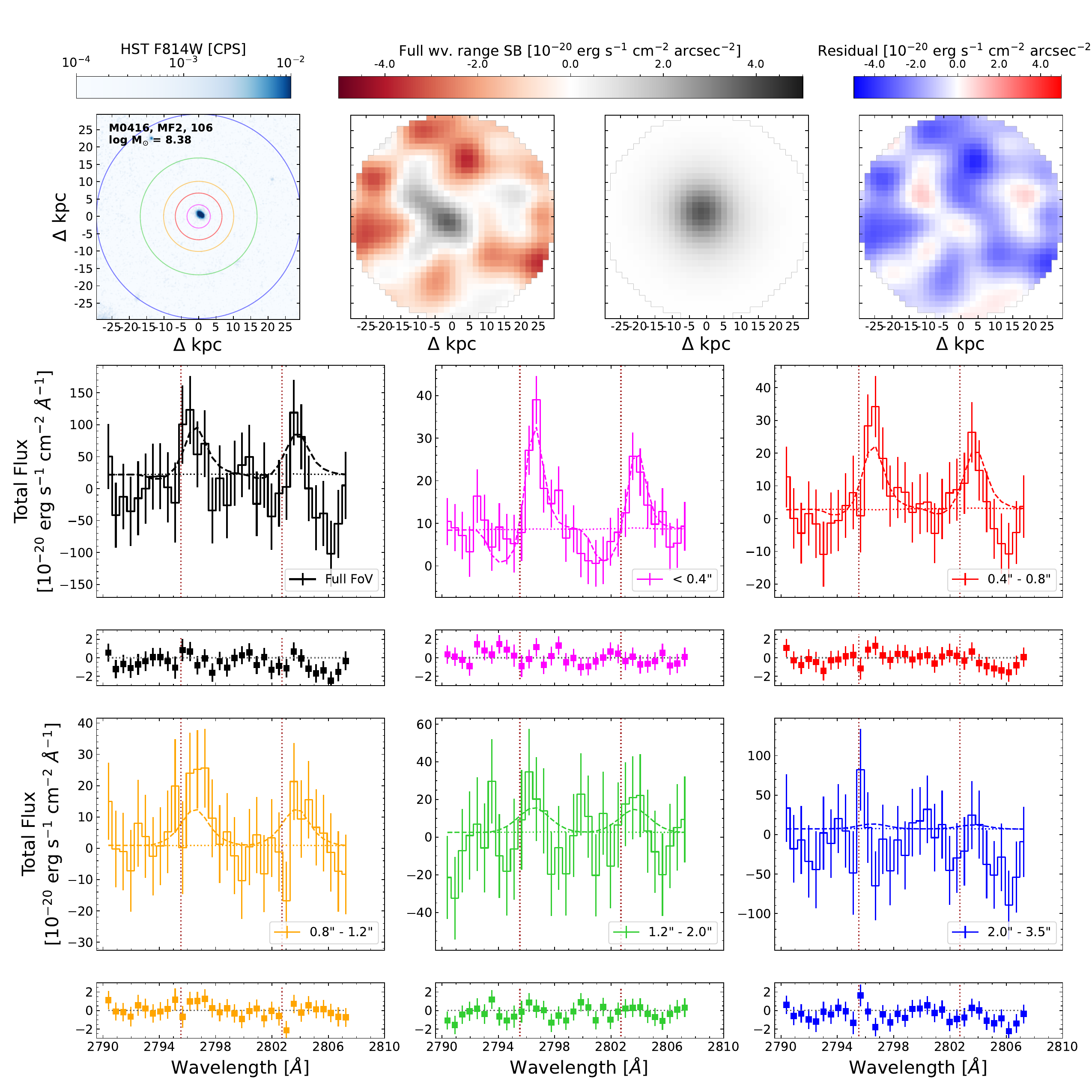}
    \end{minipage}
    \begin{minipage}{.49\textwidth}
        \centering
        \includegraphics[width=0.99\columnwidth]{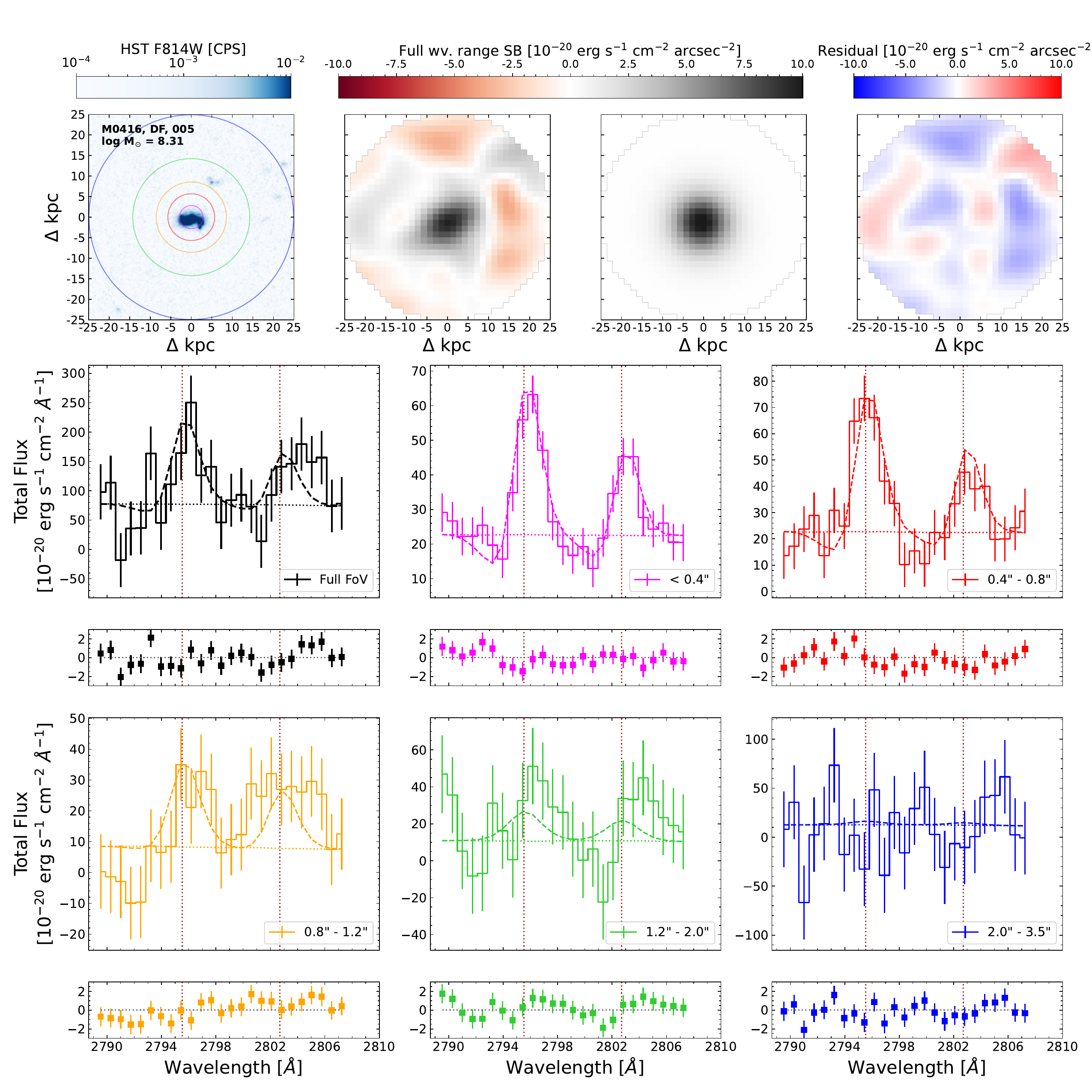}
    \end{minipage}%

    \caption{Same as Fig.~\ref{fig:master_1}, for other four galaxies in our sample.}
    \label{fig:master_11}
\end{figure*}

\begin{figure*}[!htb]
    \centering
    
    \begin{minipage}{0.49\textwidth}
        \centering
        \includegraphics[width=0.99\columnwidth]{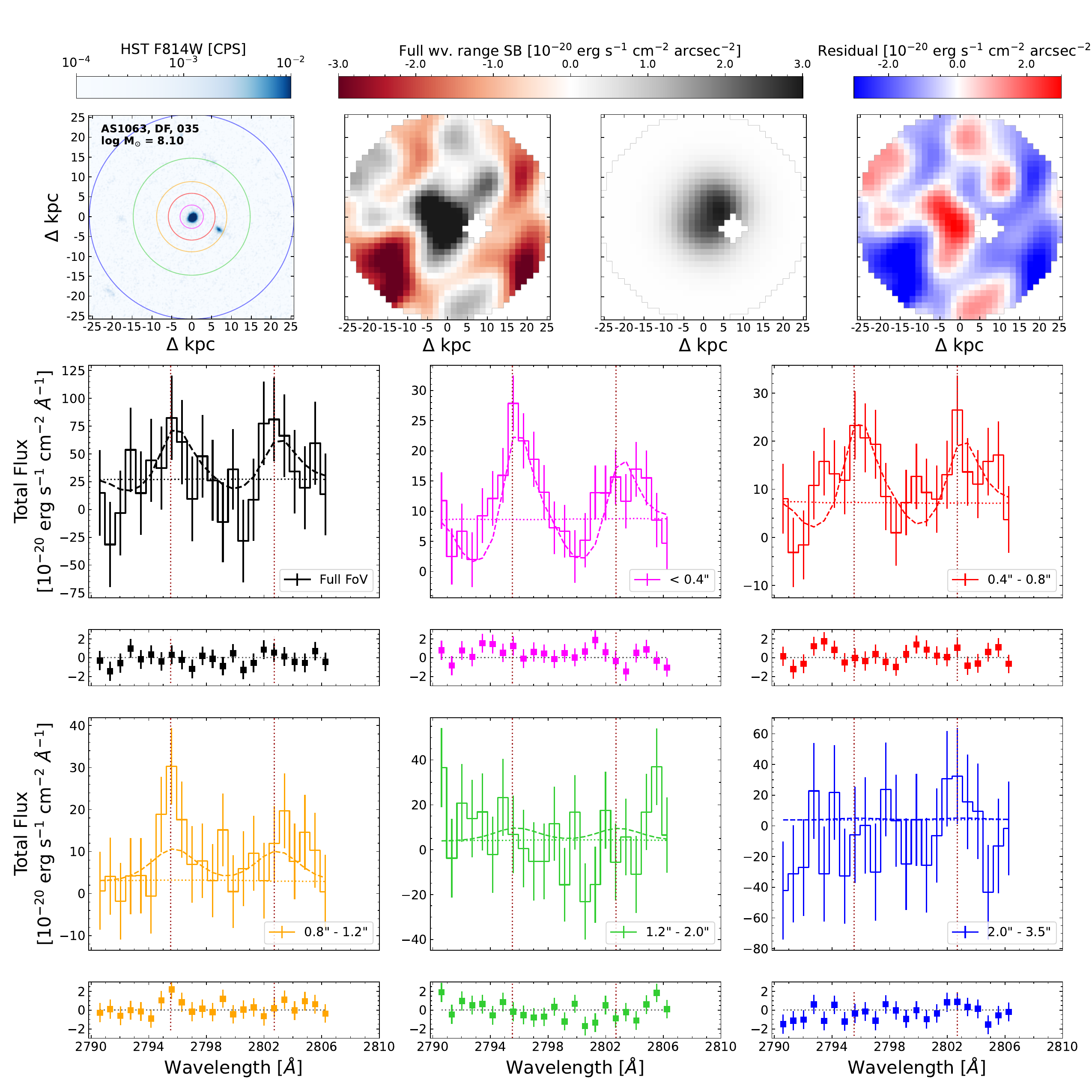}
    \end{minipage}
    \begin{minipage}{.49\textwidth}
        \centering
        \includegraphics[width=0.99\columnwidth]{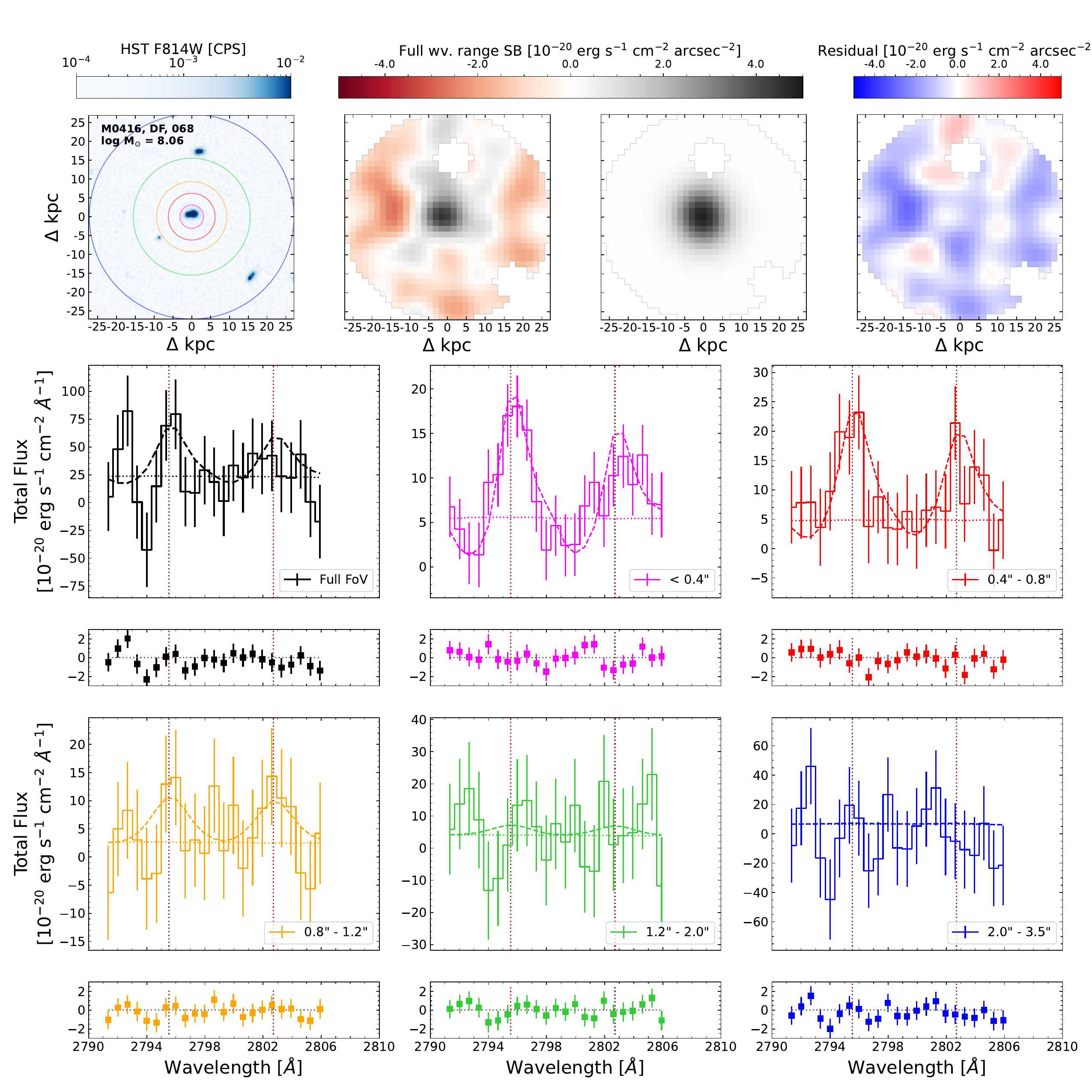}        
    \end{minipage}%

    \begin{minipage}{0.49\textwidth}
        \centering
        \includegraphics[width=0.99\columnwidth]{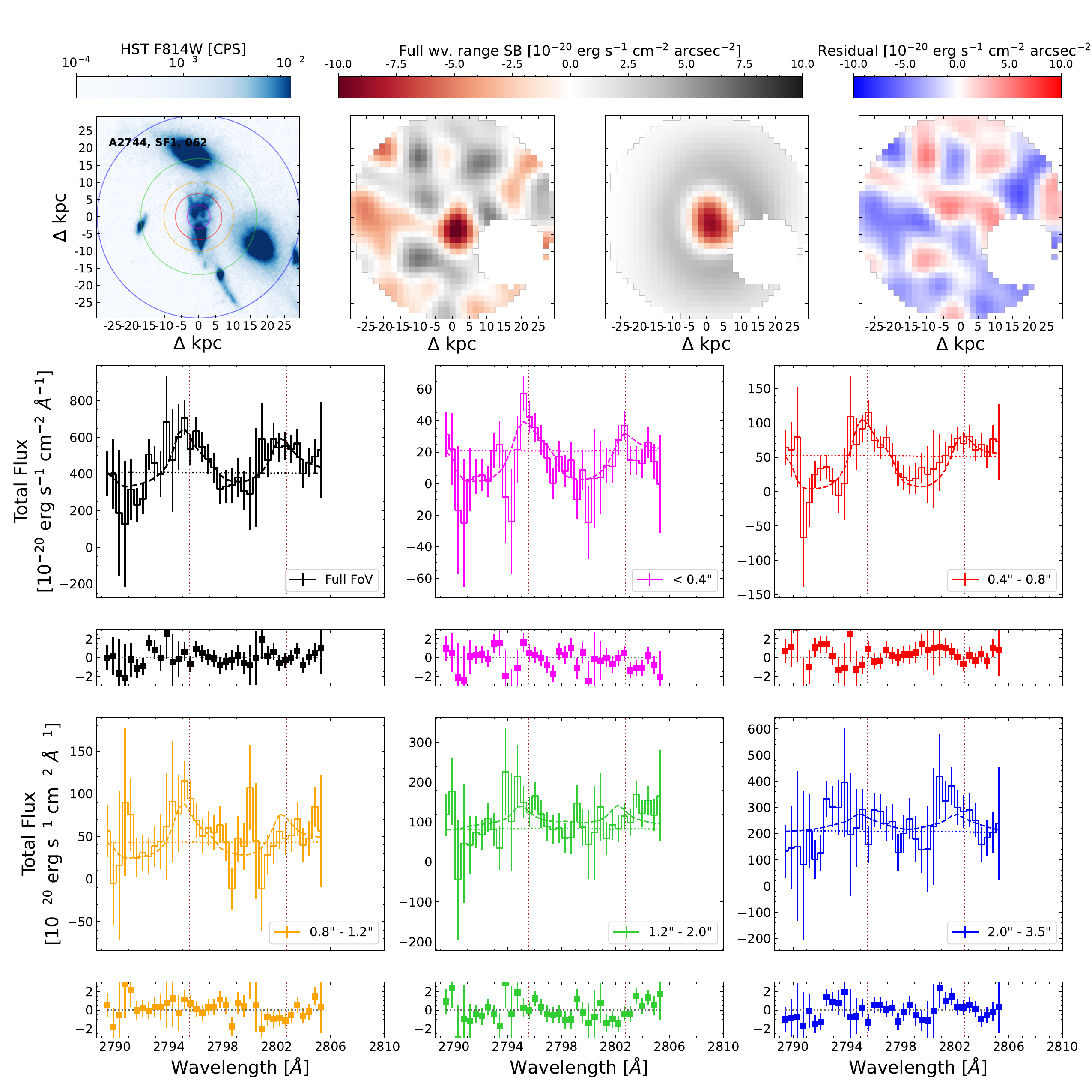}
    \end{minipage}
    \begin{minipage}{.49\textwidth}
        \centering
    \end{minipage}%

    \caption{Same as Fig.~\ref{fig:master_1}, for other three galaxies in our sample.}
    \label{fig:master_12}
\end{figure*}

\twocolumn
\FloatBarrier

\onecolumn
\section{Discussion on particular cases}
\label{sec:particular_cases}
In this appendix, we discuss two particular cases that could potentially deviate from a simple outflow model, in a broader sense than exhibiting more complex \ion{Mg}{II} velocity components within their \ion{Mg}{II} halo.

\subsection{A potentially interacting system}
\label{sec:interacting_system}
Two of our sample galaxies (AS1063-MF3-004 and AS1063-MF3-091)  are actually physical neighbors, with a projected separation of $\sim18$ kpc, and a velocity difference of $\sim160$ km s$^{-1}$. Figure~\ref{fig:binary_system} shows the HST, continuum-subtracted \ion{Mg}{II} pseudo-narrowband images, and the best-fitting model for these two galaxies.

The pseudo-narrowband images show that there is indeed a significant detection of \ion{Mg}{II} emission surrounding both of these galaxies. This `shared' \ion{Mg}{II} halo is also seen in the corresponding panels of Fig.~\ref{fig:sample_1}. We have modeled the \ion{Mg}{II} halo of each galaxy independently, neglecting any additional contribution from the neighbor galaxy (the neighbor galaxy has been masked out during each fitting). This can potentially lead to the misrepresentation of the \ion{Mg}{II} emission by the model seen in Fig.~\ref{fig:binary_system}, especially in the outer apertures. There is more \ion{Mg}{II} emission in the outer apertures in the data cube, compared to the best-fitting model (especially for AS1063-MF3-004), and with seemingly more complex kinematics.

Another possibility is that the \ion{Mg}{II} halo belongs to (in the sense that it is produced by) only one of the galaxies in the system. In that line, the kinematics of the \ion{Mg}{II} emission are more consistent with the more massive galaxy of the pair AS1063-MF3-004 (offset with respect to systemic velocity $\Delta v = 4$ km s$^{-1}$) than with the less massive one AS1063-MF3-091 ($\Delta v = 120$ km s$^{-1}$). However, an offset of 120 km s$^{-1}$ is not unusual in our sample, and it is consistent with the rest of the sample, composed of essentially isolated galaxies. On the other hand, the morphology of the halo does not suggest that the halo is produced by just one of the galaxies of the pair. If this were the case, it would imply an extremely asymmetric and elongated halo. One could speculate that the gravitational interaction could, in principle, produce such morphology, driving the outflowing gas toward the second galaxy. Nevertheless, there is not yet strong evidence that this is the case, and modeling both galaxies independently produces a reasonable output. While using a more sophisticated model is outside the scope of this paper, systems like this can provide observational evidence of how the physical properties of cool gas halos are affected by nearby galaxies, and how they can impact the baryon cycle.

\FloatBarrier
\begin{figure*}
\centering
    \begin{minipage}{.45\textwidth}
        \centering
        \includegraphics[width=\columnwidth]{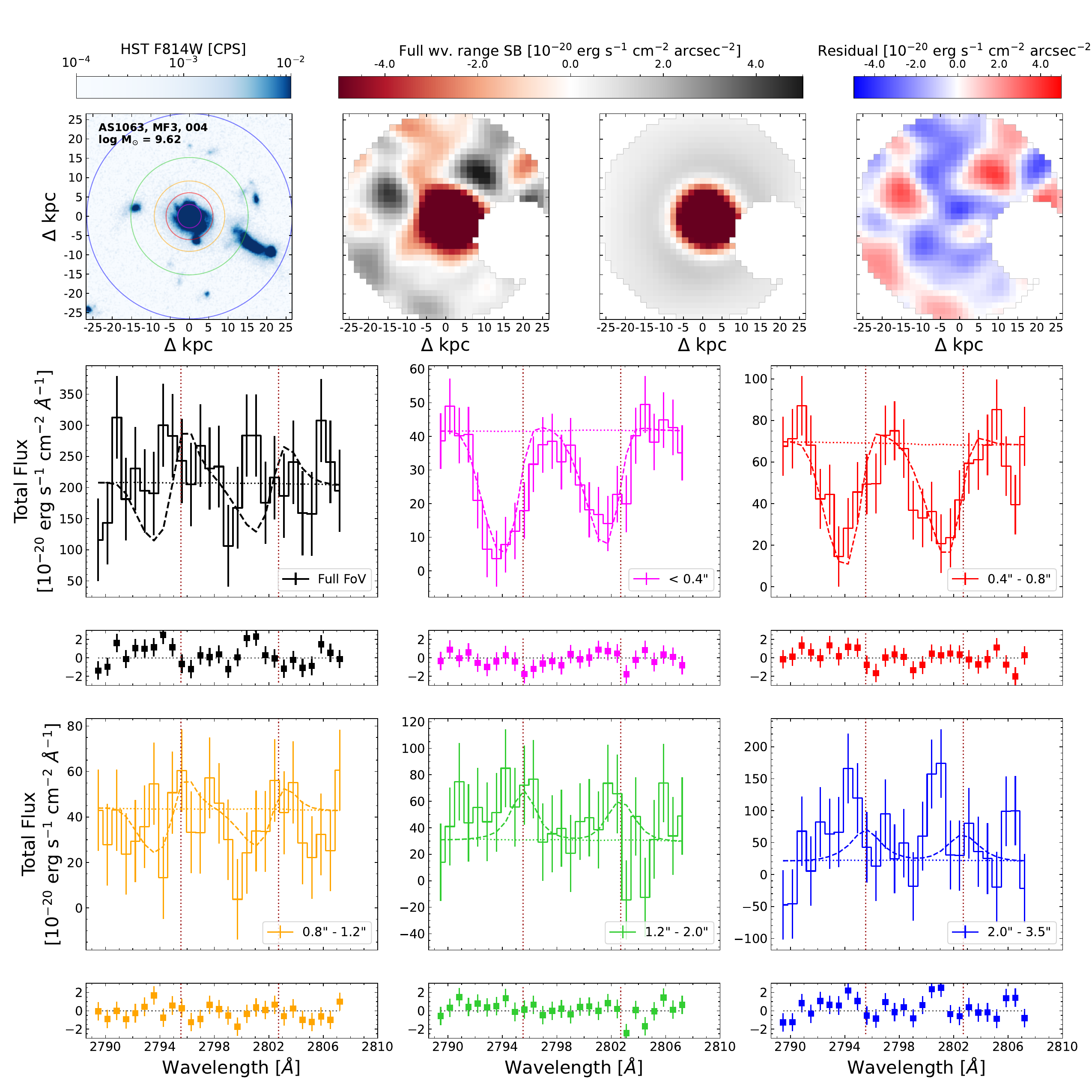}
    \end{minipage}%
    \begin{minipage}{0.45\textwidth}
        \centering
        \includegraphics[width=\columnwidth]{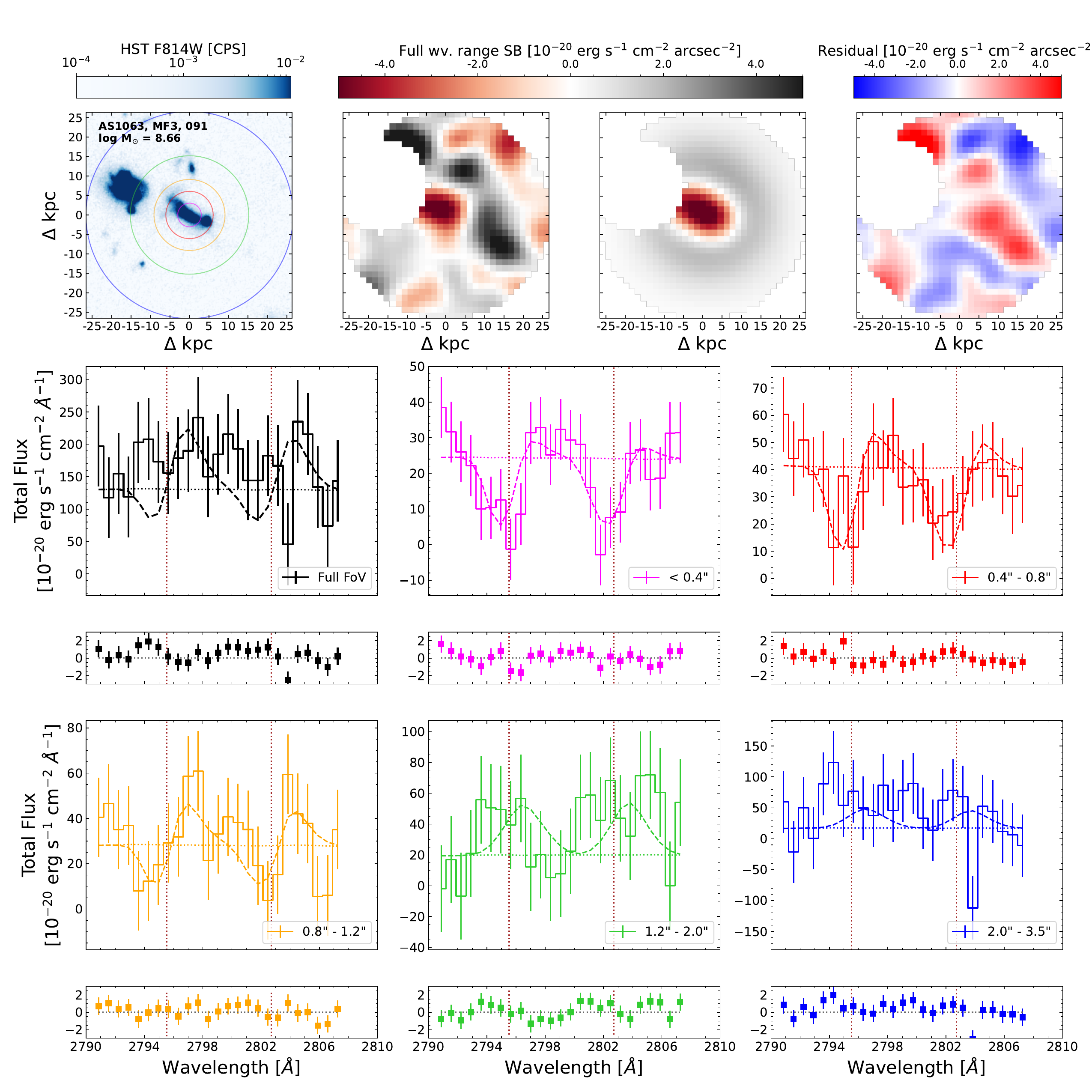}
    \end{minipage}
    
\caption{Comparison between data and model cube for two galaxies in our sample  (AS1063-MF3-004 and AS1063-MF3-091) that are physical neighbors, with a projected separation of $\sim18$ kpc, and a velocity difference of $\sim160$ km s$^{-1}$, whose \ion{Mg}{II} halo could potentially be interacting, and thus, deviating from a simple outflow model. The set of panels on the left side of the figure shows the data and best-fitting model comparison for AS1063-MF3-004, and the right set of panels shows the same for AS1063-MF3-091. The layout of each set of panels is the same as in Fig.~\ref{fig:AS1063_MF02_011_Master_figure}, comparing the continuum-subtracted \ion{Mg}{II} pseudo-narrowband images and the spectra extracted from annular apertures from the data and best-fitting model cube. The nearby galaxies have been masked during the fitting of each individual galaxy.}
\label{fig:binary_system}
\end{figure*}

\FloatBarrier
\subsection{Possible inflow signature}
\label{sec:inflow_galaxy}
While there is evidence that suggests that cool accretion flows could become the dominant mechanisms shaping the CGM at large impact parameters  \citep[$b \gtrsim 50$ kpc, see, e.g.,][]{Turner2017,Chung2019}, absorption line studies find that extended accretion disks aligned with the semi-major axis of galaxies can be present at smaller impact parameters \citep[$b\sim10$ kpc][]{Zabl:2019ija}. \citet{Figueroa2025} also find evidence that suggests that ultra-strong \ion{Mg}{II} absorbers in the sightlines of background QSOs at $b \lesssim 30$ kpc, and down the barrel, could be probing cool-gas inflows. This additional inflowing component could certainly leave an imprint in the \ion{Mg}{II} emission halo that our outflow model might not be able to capture. In this line, one of our sample galaxies (A2744-DF-013) exhibits a morphology and \ion{Mg}{II} spectral profile that seems to be incompatible with our outflow model. 

The left section of Fig.~\ref{fig:inflow_galaxy} shows the data and best-fitting model for this galaxy, using our fiducial outflow model. It is clear that the data present characteristics that are too complex for the model to reproduce. The inner apertures are dominated by a strong and broad absorption, with likely more than one absorbing kinematic component, and the outer apertures exhibit strong \ion{Mg}{II} emission. In between, the spectral profile of the line becomes extremely complex, with even a seemingly inverted P-Cygni profile, which could be indicative of inflowing gas. Interestingly, the full aperture-integrated spectrum does not show any clear \ion{Mg}{II} P-Cygni profile, and only the \ion{Mg}{II} 2796 line shows up cleanly in emission. The high S/N of the data implies that this complex spectral profile is real, and not a noise-driven feature. Additionally, both the absorption and the emission features are well centered at the systemic velocity. 

While the model attempts to capture these features, it clearly lacks the complexity to reproduce the data. Specifically, the model suggests a nearly face-on outflow (R.A. = 71$^{\circ}$) that would indeed exhibit only absorption at low impact parameters, and only emission at larger galactocentric distances. However, the predicted level of emission falls short with respect to the data in the outer apertures by a significant margin. The overall shape and velocity of the \ion{Mg}{II} spectrum are also not well captured by the model.

In order to improve the model and provide a more accurate representation of the data, we have included an additional inflowing absorption contribution. The motivation is that inflowing gas absorbing light from the central galaxy in the innermost line of sights could produce a signal similar to what is seen in the data, such as strong absorption at nearly the same velocity as the \ion{Mg}{II} emission, that could totally or partially cancel out in the integrated spectrum, as well as the inverted P-Cygni profile seen in the yellow aperture.

To parametrize the gas inflow, we have employed a similar formalism used for the outflowing gas, following the implementation from \citet{Carr2022}, where the velocity of the inflowing gas also varies radially as:

\begin{equation}
\label{eq:vel_field_inflow}
\begin{aligned}
    &v = v_{\infty} - v_{0,\mathrm{in}} \, \biggl(\frac{r}{R_{0}} \biggr)^{\gamma_{\mathrm{in}}} &\mathrm{\;for\;} r < R_{\rm max,in} \\
    &v =0   &\mathrm{\;for\;} r \geq R_{\rm max,in}\\
\end{aligned}
\end{equation}

where $v = v_{\infty} - v_{0,\mathrm{in}}$ would represent the maximum velocity of the inflowing gas, reached at $r = R_0$. The accretion begins from rest at $r = R_{\rm max,in}$, and then the gas is radially accelerated during its infall up to the maximum velocity, at a rate that depends on $v_{0,\mathrm{in}}$ and $\gamma_{\mathrm{in}}$. This simple scenario is consistent with gas that is ejected from the galaxy and eventually reaccreted \citep{Barbani2025}.

This different assumed velocity law implies that instead of solving Eq.~\ref{eq:radius_velocity} to determine the bijective relation between observed velocity and radius, we have to solve instead:

\begin{equation}
    \label{eq:radius_velocity_inflow}
    r^{2} - R_{0}^{2} \, \biggl(1 - \biggl[\frac{\epsilon}{r}\biggr]^{2} \biggr)^{\frac{\gamma_{\mathrm{in}} - 1}{\gamma_{\mathrm{in}}}} \biggl(\frac{v_{\infty} (1 - \frac{\epsilon^2}{r^2} )^{0.5} - v_{\mathrm{obs}}}{v_{0,\mathrm{in}}} \biggr)^{2/\gamma_{\mathrm{in}}} - \epsilon^{2}= 0
\end{equation}

Finally, to compute the inflow optical depth $\tau_{\mathrm{in}}$ at a given radius, we use the same Eq.~\ref{eq:optical_depth} as for the outflow, but using a different parameter $\tau_\mathrm{0,in}$ to describe the optical depth at $r = R_0$ for the inflowing gas, and $\gamma_{\mathrm{in}}$ for the acceleration rate. A total of four additional parameters are required to add this additional contribution into the model ($ v_{\infty}$, $v_{0,\mathrm{in}}$, $\gamma_{\mathrm{in}}$, and $\tau_\mathrm{0,in}$).

In \cite{Carr2022}, the authors show that using this parametrization, the isovelocity curves (i.e., surfaces of constant observed velocity, see also Appendix~\ref{sec:model}) bend back into the central source, meaning that for impact parameters larger than the source size $R_0$, any given sightline will intersect the same isovelocity curve more than once \citep[as opposed from the outflow case, where the isovelocity curves do not bend, and are intersected always only once by any given sightline, see, e.g., ][]{Pessa2024}. Essentially, this means that for large impact parameters, Eq.~\ref{eq:radius_velocity_inflow} will have more than one real solution, and a single observed velocity value would be probing more than one location inside the inflow. However, the absorption contribution of the inflows occurs, by definition, at impact parameters enclosed by the source size; thus, this ambiguity is not a problem when accounting only for the absorption contribution of the inflowing gas.

The model obtained after adding this additional absorbing contribution is shown in the right section of Fig.~\ref{fig:inflow_galaxy}. With this additional contribution, the data is now described as a nearly edge-on outflow (R.A. = $10^{\circ}$), and a wide opening angle (although not isotropic) of O.A = $75^{\circ}$. In this scenario, both the outflowing and the inflowing gas contribute to the absorption profile along the lines of sight at low impact parameters. With this additional complexity, the model is able to reproduce better (or at least get closer to reproducing) several of the features exhibited by the data, such as the anisotropic morphology, more complex profiles, kinematic centers of the emission and absorbing components (and thus, partial cancellation of emission and absorption in the integrated spectrum), and the overall strength of the emission.

Despite this improvement, we stress that our goal here is not to provide a robust inflow plus outflow modeling for this particular galaxy. In fact, our implementation of the inflowing gas is extremely simplistic. We neglect that the inflowing gas could also produce additional extended \ion{Mg}{II} emission. \citet{Zabl:2019ija} and \citet{Kacprzak2025} find that gas accretion is aligned with the semi-major axis of galaxies, corotating with the stellar disk. Thus, not accounting for the emission of accreting gas could potentially lead to broader opening angles. The reasons to not include it in our model are, firstly, that as described earlier, the relation between observed velocity and radius inside the inflow is no longer bijective toward large impact parameters \citep[because the isovelocity curves bend back into the central source, see ][for a detailed explanation]{Carr2022}, and secondly, the geometry of the emission produced by the inflowing gas is uncertain. It could be, in principle, modeled like an accretion disk, following findings from absorption line studies that are consistent with the biconical outflow plus accretion disk CGM model \citep[see, e.g.,][]{Bouche:2012ho, Tumlinson2017, Zabl:2019ija, Schroetter:2019es, Zabl:2020js}. However, this would still require the modeling of the disk geometry and its orientation with respect to the line of sight, adding several degrees of freedom to the model, and possibly resulting in degeneracies between the outflow emission and inflow emission.

Nevertheless, even with our simplified approach to incorporating inflows into the model, we obtain promising results. This may be the first object in which the fountain CGM model can be directly observed in \ion{Mg}{II} emission. These findings encourage the exploration of more sophisticated models that account for both outflows and inflows, as well as additional phenomena such as underlying stellar and ISM absorption (which we do not consider here). Furthermore, the disturbed morphology of this object in the HST image suggests it may be a merging system, making the interpretation of the emission line profiles much more complex. In this context, modern radiative transfer simulations \citep[e.g.,][]{Chang2024, Chang2025} can serve as a powerful tool for interpreting observations. This will help to (i) determine the mechanisms truly driving the observed features in the \ion{Mg}{II} emission and (ii) place constraints on the gas properties.

\begin{figure*}
\centering
    \begin{minipage}{.45\textwidth}
        \centering
        \includegraphics[width=\columnwidth]{Master_figures/A2744_013_Master_figure.pdf}
    \end{minipage}%
    \begin{minipage}{0.45\textwidth}
        \centering
        \includegraphics[width=\columnwidth]{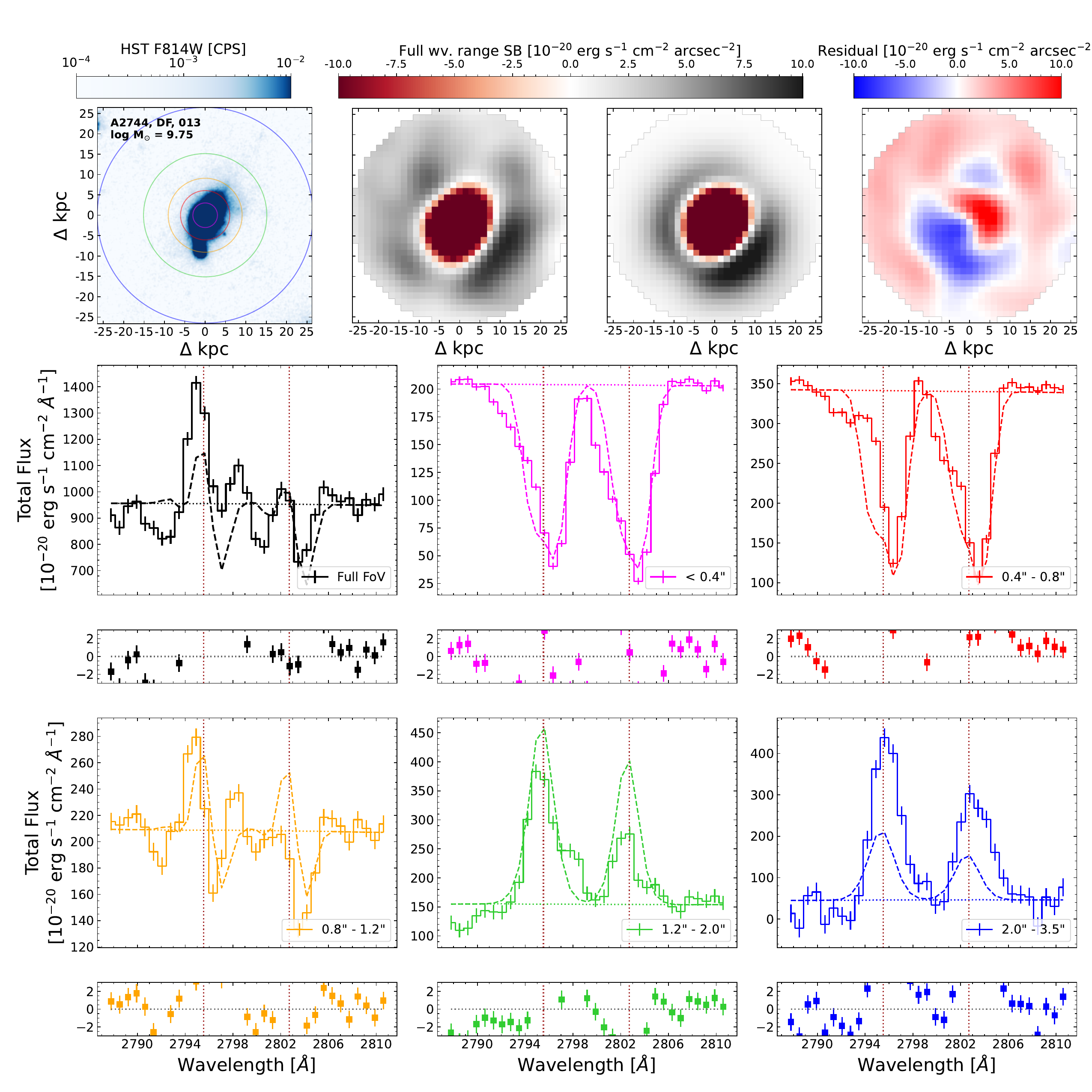}
    \end{minipage}
    
\caption{Comparison between data and model cube for one of our sample galaxies (A2744-DF-013) whose \ion{Mg}{II} emission halo exhibits characteristics that are inconsistent with our purely-outflow emission model. By including an additional inflow-absorbing contribution to the model, we are able to reproduce the data more closely. The parameterization of the inflow component is detailed in the main text (see Sec~\ref{sec:inflow_galaxy}). The set of panels on the left side of the figure shows the data and best-fitting model comparison for A2744-DF-013, using our fiducial outflow-only model described in Appendix~\ref{sec:model}. The right set of panels shows the same for our outflow+inflow model. The layout of each set of panels is the same as in Fig.~\ref{fig:AS1063_MF02_011_Master_figure}, comparing the continuum-subtracted \ion{Mg}{II} pseudo-narrowband images and the spectra extracted from annular apertures from the data and best-fitting model cube.}
\label{fig:inflow_galaxy}
\end{figure*}

\twocolumn
\FloatBarrier
\subsection{Blueshifting of the \ion{Mg}{II} emission toward larger galactocentric distances}

In this section, we discuss, in particular, two galaxies from our sample whose \ion{Mg}{II} emission exhibits a clear blueshifting toward larger radii with respect to the emission closer to the galaxy, and also with respect to the expectations from a simple outflow model.

 Figure~\ref{fig:AS1063_MF02_011_Master_figure} shows the datacube and best-fitting cube comparison for the first object. The continuum-subtracted pseudo-narrowband images show a close-to-isotropic morphology, dominated by \ion{Mg}{II} in emission. The P-Cygni profiles of both \ion{Mg}{II} lines shown in the inner apertures are well reproduced by the model. Remarkably, a second \ion{Mg}{II} kinematic component starts to show up in emission at distances greater than 0.8' (yellow spectrum). This second component is blueshifted relative to the main component and becomes progressively stronger toward larger galactocentric distances, becoming dominant in the farthest aperture (blue spectrum), where it is clear that the observed spectrum is significantly blueshifted with respect to the model expectation. As a result, the spectrum integrated in the full modeled aperture (black spectrum) displays \ion{Mg}{II} in emission only, since the \ion{Mg}{II} emission of this second component overcomes the central absorption at nearly the same velocity. 

Similarly, Fig.~\ref{fig:AS1063_109_Master_figure} shows the datacube and best-fitting cube comparison for the first object for the second object discussed in this section. It exhibits strong extended \ion{Mg}{II} emission and central absorption. There is significant \ion{Mg}{II} emission even in the farthest aperture, and the pseudo-narrowband images show that it extends up to the edges of the modeled region in some directions. Furthermore, the emission in the outer aperture appears blueshifted relative to the \ion{Mg}{II} emission closer to the galaxy, and to the systemic velocity. However, unlike Fig.~\ref{fig:AS1063_MF02_011_Master_figure}, there are no indications of a secondary \ion{Mg}{II} kinematic component that could be driving this blueshifting. Instead, the blueshift appears to be due to a change in the gas kinematics that is not captured by the single power law used by the model. At face value, it could indicate that the assumption of a velocity that increases with radius is no longer a good approximation at large galactocentric distances, where the gas could potentially stop accelerating, and or even begin an infall \citep[see, e.g., hydrodynamical N-body simulations from][]{Barbani2025}. \citet{Guo2023} and Kozlova et al. (in prep) also report a similar blueshifting of the Ly$\alpha$ emission peak toward larger galactocentric distances, possibly due to the presence of cool-gas inflows. Nevertheless, a more detailed model that allows for more complex kinematics (potentially involving radiative transfer modeling) would be required to thoroughly investigate the nature of this change in the gas kinematics.

 While a more complex modeling of the extended \ion{Mg}{II} emission is beyond the scope of this paper, these residuals show how our data captures the complexity of the CGM, and have the potential to serve as benchmark to provide constraints on other mechanisms shaping the galactic environment, such as inflows from the intergalactic medium into the galaxy halo, or even past outflow events. 

Similar to what we reported in a previous work \citep{Pessa2024}, we generally find a good agreement with the data in the inner regions, where the outflow emission likely dominates, and relatively higher residuals at larger galactocentric distances where other mechanisms, such as inflows or satellites, could become relevant (or dominant) in shaping the CGM of galaxies \citep{Dylan2019b}.

Finally, while here we discuss only these particular objects as representative cases of this phenomenon, there are more galaxies in our sample that could show a similar behavior (i.e., a blueshifting of the \ion{Mg}{II} emission toward larger radii), although they are less clear due to their lower S/N  (e.g., M0416-DF-032,  M0416-DF-147).

\begin{figure}
\includegraphics[width=\columnwidth]{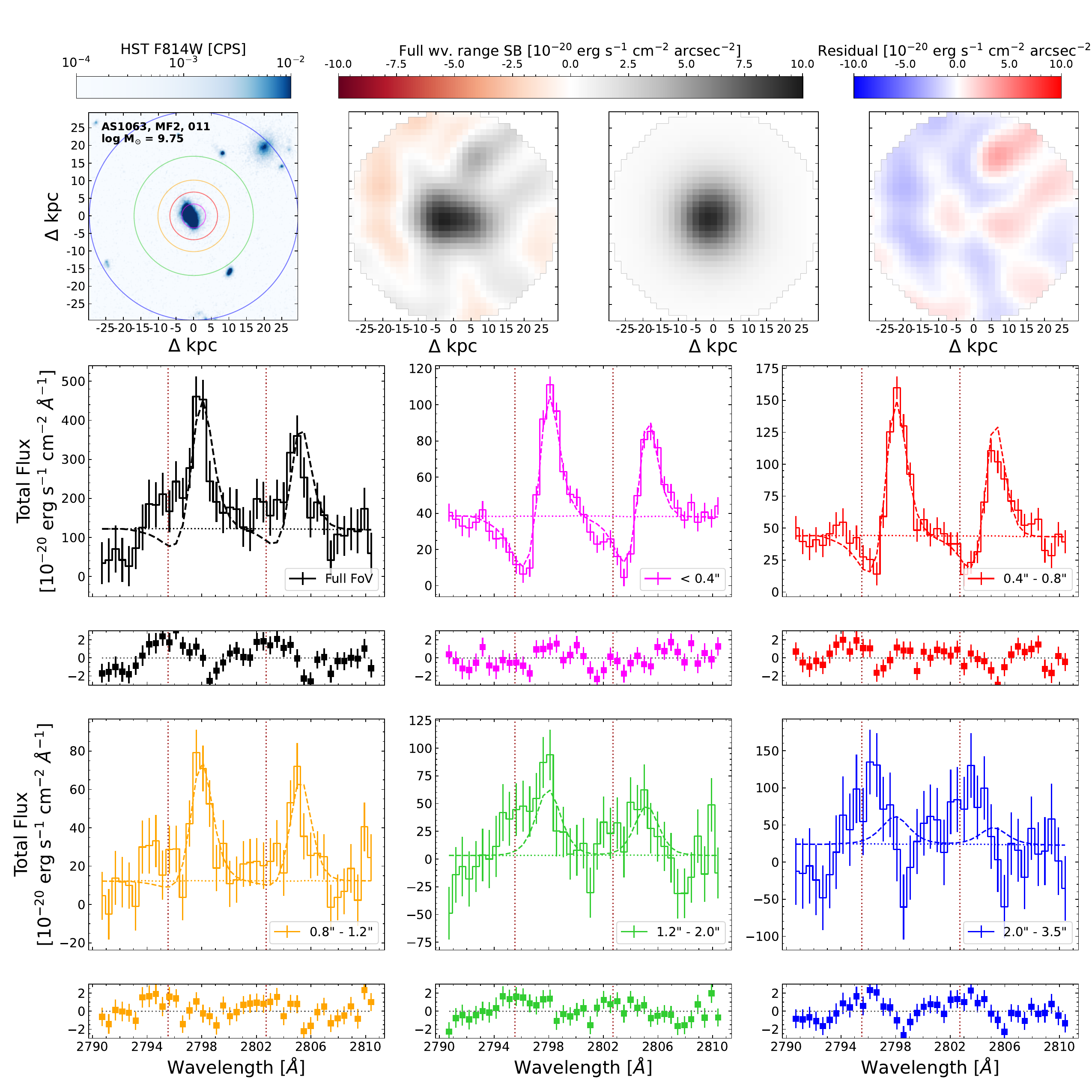}
\caption{Same as Fig.~\ref{fig:M0416_SF07_012_Master_figure}, for another galaxy in our sample (AS1063-MF2-011).}
\label{fig:AS1063_MF02_011_Master_figure}
\end{figure}

\begin{figure}
\includegraphics[width=\columnwidth]{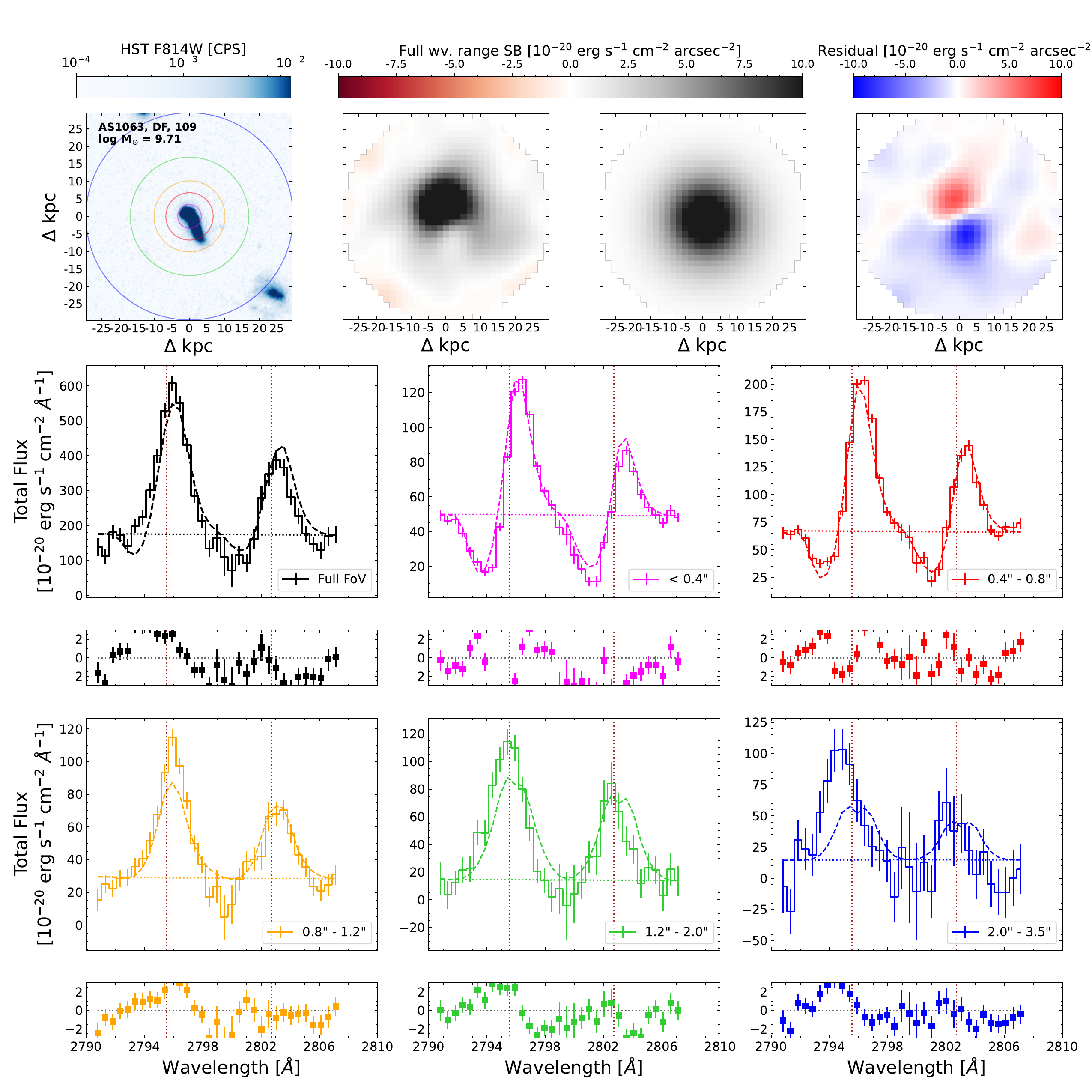}
\caption{Same as Fig.~\ref{fig:M0416_SF07_012_Master_figure}, for another galaxy in our sample (AS1063-DF-109).}
\label{fig:AS1063_109_Master_figure}
\end{figure}

\section{Methods}
\label{sec:methods}

\subsection{Outflow model}
\label{sec:model}

The outflow model utilized to interpret our MUSE data was introduced and thoroughly described in \citet{Pessa2024}. For a comprehensive description of the model, we refer the reader to that work. Here, we summarize its most relevant aspects.

The model employs the Sobolev approximation \citep{Ambartsumian1958, Sobolev1960, Grinin2001} to characterize the interactions between scattered continuum photons and ions in the outflowing material, rather than performing full radiative transfer calculations. It consists of a spherically symmetric source with radius $R_{0}$ that emits continuum radiation isotropically, encased by a spherical (or biconical) gaseous envelope (wind) that expands from $R_{0}$ to a terminal radius $R_{\rm max}$. The expansion velocity of the envelope ($v$) increases with radius ($r$) following a power-law profile with exponent $\gamma$, expressed as:

\begin{equation}
\label{eq:vel_field}
\begin{aligned}
    &v = v_{0} \, \biggl(\frac{r}{R_{0}} \biggr)^{\gamma} &\mathrm{\;for\;} r < R_{\rm max} \\
    &v =v_{\rm max}   &\mathrm{\;for\;} r \geq R_{\rm max}\\
\end{aligned}
\end{equation}

\noindent where $v_{0}$ corresponds to the initial velocity of the wind (i.e., at $R_{0}$), and $v_{\mathrm{max}}$ is the terminal velocity of the wind at $R_{\mathrm{max}}$. Under the assumption of mass conservation through the outflow, the density profile corresponding to Eq.~\ref{eq:vel_field} is:

\begin{equation}
\label{eq:density}
    n(r) = n_{0} \, \biggl( \frac{R_{0}}{r}  \biggr)^{\gamma + 2}
\end{equation}

\noindent where $n_{0}$ is the gas density at $r = R_{0}$. Provided the velocity gradient in Eq.~\ref{eq:vel_field} is large, the photons produced by the central source will interact with the outflowing material only at the specific radius where the absorbing ions are at resonance (due to their Doppler shift, i.e., Sobolev approximation).

Due to projection effects, the observed line-of-sight velocity $v_{\rm obs}$ of a given spherical shell at radius $r$ will differ from its true underlying velocity. This observed velocity depends on the projected distance ($\epsilon$) of the line of sight from the central source, commonly referred to as the impact parameter.

For a line of sight at an arbitrary impact parameter $\epsilon$, the corresponding radial location $r$ within the envelope that produces an observed velocity $v_{\rm obs}$ is given by the expression:

\begin{equation}
    \label{eq:radius_velocity}
    r^{2} - R_{0}^{2} \, \biggl( \frac{v_{\rm obs}}{v_{0}} \biggr)^{2/\gamma} \biggl(1 - \biggl[\frac{\epsilon}{r}\biggr]^{2} \biggr)^{\frac{\gamma - 1}{\gamma}} - \epsilon^{2}= 0
\end{equation}

The optical depth ($\tau$) can be evaluated on the corresponding interaction surface, as a function of the physical properties of the wind and atomic constants, following \citet{Castor1970}:

\begin{equation}
    \label{eq:definition_tau}
    \tau (r) = \frac{\pi e^{2}}{mc} \, f_{lu} \, \lambda_{lu} \, n_{l}(r) \, \biggl[ 1 - \frac{n_{u} \, g_{l}}{n_{l} \, g_{u}} \biggr] \frac{r/v}{1 + \sigma \cos^{2} \phi }
\end{equation}

\noindent where $\phi$ is the angle between the velocity and the trajectory of the photon, $f_{ul}$ and $\lambda_{ul}$ correspond to the oscillator strength and wavelength for the $ul$ transition, respectively, and $\sigma = \frac{d \ln (v)}{d \ln (r)} - 1$. Assuming the velocity law in Eq.~\ref{eq:vel_field} and neglecting stimulated emission (i.e., $\biggl[ 1 - \frac{n_{u} \, g_{l}}{n_{l} \, g_{u}} \biggr] = 1$), we get that the optical depth can be expressed as:
\begin{equation}
 \tau (r) = \frac{\pi e^{2}}{mc} \, f_{lu} \, \lambda_{lu} \, n_{0} \biggl( \frac{R_{0}}{r}  \biggr)^{\gamma + 2} \frac{r/v}{1 + (\gamma - 1) \cos^{2} \phi }
\end{equation}

\noindent Finally, defining:

\begin{equation}
\label{eq:tau_0_n_0}
    \tau_{0} = \frac{\pi e^{2}}{mc} \, f_{lu} \, \lambda_{lu} \, n_{0} \, \frac{R_{0}}{v_{0}}
\end{equation}
we can write the optical depth as:
\begin{equation}
    \label{eq:optical_depth}
    \tau (r) = \frac{ \tau_{0}}{1 + (\gamma - 1) \, \cos^{2} \phi } \biggl( \frac{R_{0}}{r}  \biggr)^{2\gamma + 1}
\end{equation}

Altogether, Eqs.~\ref{eq:radius_velocity} and ~\ref{eq:optical_depth} provide a bijective relation between observed velocity and radius of resonance, that is, the radius where continuum photons will interact with the outflowing ions, given their Doppler shift (for a given sightline, defined by its impact parameter), and the value of optical depth at any radius within the outflow, respectively. Combining these two equations, it is possible to calculate $\tau$ as a function of $v_{\rm obs}$ (and, therefore, wavelength $\lambda$).

While Eq.~\ref{eq:radius_velocity} establishes a bijective relationship between the observed velocity and the resonance radius, Eq.~\ref{eq:optical_depth} defines the optical depth at any given radius within the outflow. By combining them, one can express $\tau$ as a function of $v_{\rm obs}$, and consequently, as a function of wavelength $\lambda$.

\subsubsection{Modeling integral field spectroscopic data}

Each MUSE spaxel is considered, in principle, as an individual sightline, characterized by a single impact parameter value. For a given MUSE spaxel, the fraction of the continuum flux absorbed at each wavelength by the spherical envelope located at resonance with the photons emitted by the central source at wavelength $\lambda$ is given by:

\begin{equation}
    I_{\rm abs}(\lambda) = 1 - e^{-\tau(\lambda)}
\end{equation}

On the other hand, the profile of the emission for any given MUSE spaxel, as a function of wavelength, can then be expressed in units of the continuum radiation as:

\begin{equation}
\label{eq:emission_part}
    I_{\rm ems}(\lambda) = \frac{1 - e^{-\tau (\lambda)}}{4 \pi r_{\lambda}^{2}}
\end{equation}

\noindent where $r_{\lambda}$ is the radius of resonance for the photons emitted at wavelength $\lambda$, calculated following Eq.~\ref{eq:radius_velocity}. 

However, note that while the absorbing component is scaled by the level of continuum in any specific sightline, the emission contribution must be scaled to the total continuum radiation emitted by the central source. This is because while the absorption is only meaningful relative to a continuum, the emission component does not depend on the presence of a continuum in the same sightline, but it is powered by the total production of continuum photons by the central source. 

By construction, the absorption component is confined to wavelengths that correspond to $v_{\rm obs} < 0$, since only the part of the spherical envelope moving toward the observer, between the observer and the central source, is able to absorb continuum radiation in the line of sight. On the other hand, both the blueshifted and redshifted parts of any given spherical shell are able to scatter photons into the line of sight. However, note that since $R_{0}$ represents the launching radius of the wind, wavelengths that correspond to a radius smaller than $R_{0}$ do not absorb or emit scattered radiation.

The resulting P-Cygni profile of the outflow can be computed for any specific sightline as:

\begin{equation}
    I(\lambda) = I_{\rm cont, local} (1 - I_{\rm abs}) + I_{\rm cont, total}\;I_{\rm ems}
\end{equation}
where $I_{\rm cont, local}$ and $I_{\rm cont, total}$ correspond to the local level of continuum in the sightline, and the total continuum of the galaxy in the corresponding wavelengths, respectively, with $R_0$, $\tau_{0}$, $\gamma$, $v_{0}$, and $v_{\rm max}$ being free parameters of the model.

\subsubsection{Nebular emission contribution}

Besides the \ion{Mg}{II} emission produced by resonant scattering of continuum photons produced by the central galaxy with ions in the CGM, photoionization models of \ion{H}{II} regions also predict the production of nebular \ion{Mg}{II} emission \citep[see, e.g.,][]{Henry:2018gd}.

We model this possible extra source of \ion{Mg}{II} photons in a given sightline as a Gaussian (centered at $0$ km s$^{-1}$) whose amplitude is proportional to the local stellar continuum, through a free parameter $f_{C}$. Specifically, we define:

\begin{equation}
    I_{\rm Neb}(\lambda) = (f_{C}\times I_{\rm cont, local}) e^{-\frac{1}{2}\frac{(\lambda - \lambda_{0})^{2}}{\sigma_{\rm Neb}^{2}}}
\end{equation}

\noindent where the standard deviation of the nebular emission $\sigma_{\rm Neb}$ is fixed to the value of velocity dispersion measured for the [\ion{O}{II}] line if available. Otherwise, we use the mean of the distribution computed for those galaxies where [\ion{O}{II}] is available ($80\pm30$ km s$^{-1}$). Lastly, the intrinsic emissivity ratio of nebular \ion{Mg}{II} $\lambda$2796 to \ion{Mg}{II} $\lambda$2803 emission is set by atomic physics to be two \citep[see, e.g.,][]{Chisholm2020}, so we fix this ratio in our model by using half of the amplitude for the \ion{Mg}{II} $\lambda$2803 line.

\subsubsection{Biconical geometry}
\label{sec:biconic_geometry}
Observational evidence suggests that the circumgalactic medium is not isotropically distributed around a central galaxy \citep[e.g.,][]{Schroetter:2019es}. Moreover, the data are consistent with a picture of \ion{Mg}{II} being predominantly present in outflow cones and extended disc-like structures.  \citep[e.g.,][]{Bouche:2012ho,Schroetter:2019es,Zabl:2019ija,FernandezFigueroa2022, Guo2023b}. Thus, a strictly spherical geometry represents an oversimplification of the problem.

The biconical geometry can be fully described through three additional parameters; an opening angle (O.A.), a position angle (P.A.), and a rotation angle (R.A.). Figure~\ref{fig:bicone} is the same as \citet{Pessa2024}, and provides a schematic representation of the angles that describe this parametrization. The opening angle represents the angle between the central axis and the surface of each cone. The position angle corresponds to the angle of the outflow with respect to the horizontal axis, in the plane of the sky (i.e., perpendicular to the line of sight, see the left panel of Fig.~\ref{fig:bicone}), and the rotation angle measures the rotation of the outflow in the plane defined by $x = 0$, with respect to the $y$-axis. That is, a rotation perpendicular to the plane of the sky, toward the observer. Hence, for a given set of O.A., P.A., and R.A., we set the scattered emission outside the biconical volume to zero, whereas inside the outflow it is given by Eq.~\ref{eq:emission_part}

\begin{figure}
\includegraphics[width=\columnwidth]{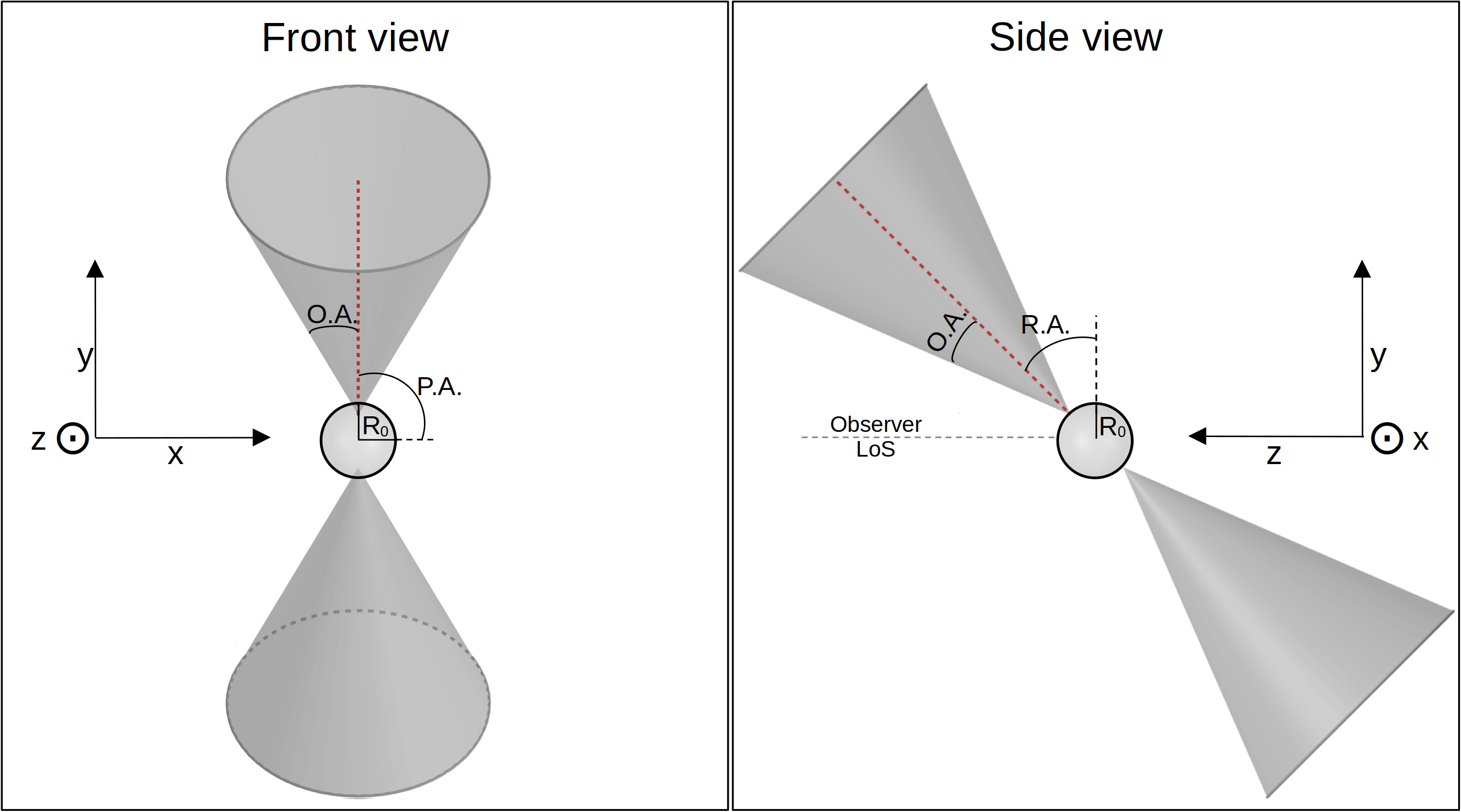}
\caption{Schematic representation of the outflow geometry parametrization described in Appendix~\ref{sec:biconic_geometry}, taken from \citet{Pessa2024}. The figure illustrates the front \textit{(left)} and side \textit{(right)} view of an outflow characterized by an arbitrary O.A.,  P.A. $= 90^{\circ}$ and R.A. $= 45^{\circ}$. The opening angle represents the angle between the axis and the surface of each outflow cone, the position angle corresponds to the angle of the outflow with respect to the horizontal axis, in the plane of the sky, and the rotation angle measures the rotation of the outflow in the plane defined by $x = 0$, with respect to the $y$-axis. The red line shows the cone axis. The observer is located at a large positive $z$ value.}
\label{fig:bicone}
\end{figure}

\subsubsection{PSF and LSF convolution}
\label{sec:psf_lsf}

Accounting for instrumental LSF and PSF is essential for a meaningful comparison between our model and the observed data. The LSF describes the spectral broadening of a given spectral line due to instrumental effects, and the PSF characterizes the spatial broadening of the light distribution of a point-like source in the detector.

The MUSE LSF is modeled as a Gaussian function, whose width was measured by fitting a Gaussian to a set of sky lines in \citet{Bacon:2017hn}. The FWHM of the LSF varies smoothly with wavelength, as a second-degree polynomial:

\begin{equation}
\mathrm{FWHM}(\lambda) = 5.835 \times 10^{-8} \lambda^{2} - 9.080 \times 10^{-4} \lambda + 5.983
\end{equation}

\noindent where $\lambda$ and FWHM are both in $\AA$.

On the other hand, the MUSE PSF is accurately described by a circular Moffat function with a fixed $\beta$ and a FWHM that varies linearly with wavelength. For MUSCATEL, the FWHM and $\beta$ values were determined based on available stars in the field.

The PSF of the MUSCATEL data is well characterized by a Moffat function with $\beta = 2.6$ and a FWHM that decreases linearly with wavelength, ranging from a median value of $0.71\pm0.13$ arcseconds on the bluest side of the MUSE wavelength range to $0.50\pm0.11$ arcseconds at the reddest end of the wavelength range. For each galaxy, we adopt the corresponding FWHM based on its MUSCATEL field and the wavelength at which the \ion{Mg}{II} transition is observed.

\subsection{Fitting method}
\label{sec:fitting_method}
\subsubsection{Modeled signal}
We use the model described in the previous section to fit the \ion{Mg}{II} emission of our sample galaxies. We consider the wavelength interval around the \ion{Mg}{II} transition, approximately the 2790 - 2809 $\AA$ range  (rest-frame), modeling both \ion{Mg}{II} transitions simultaneously. For some galaxies, the modeled wavelength interval is slightly larger or smaller, if they exhibit, for instance, a broader spectral profile, contamination due to sky emission lines, or a nearby galaxy (in projected distance) that would require truncating the wavelength range. We fit the \ion{Mg}{II} emission profile in a circular field of view with a radius of $3\farcs5$, which corresponds to $26-30$ kpc at redshifts $0.7-2.2$, respectively.

We model the \ion{Mg}{II} $\lambda \lambda 2796$, $2803$ transitions at their corresponding rest-frame velocities, using the same set of parameters to model both transitions, except for $\tau_{0}$. This is because  the oscillator strength of \ion{Mg}{II} $\lambda 2796$ is essentially two times larger than that of \ion{Mg}{II} $\lambda 2803$ \citep{Kelleher2008}, thus, we use an optical depth $\tau = \tau_{0}/2$ for \ion{Mg}{II} $\lambda 2803$, as the optical depth depends linearly on the oscillator strength

\subsubsection{Description of the fitting}
\label{sec:fitting_desc}

To find the best fitting parameters for the MUSE data of our sample galaxies, we used a Markov Chain Monte Carlo (MCMC) analysis implemented in the {\sc emcee} python package \citep{ForemanMackey:2013io}.  For each MCMC step, we compute a model cube and the respective likelihood of the used set of parameters.

We used broad, uninformative priors for most of the outflow parameters described in Appendix~\ref{sec:model}:
\begin{gather}
\tau_0\,\,\,\,\, \sim \mathcal{L}\mathcal{U}(0,5) \\
\gamma \,\,\,\,\,\,\, \sim \mathcal{U}(0.2,2.0) \\
v_0 \,\,\,\,\, \sim \mathcal{U}(10,200) \\
v_{\mathrm{max}} \sim \mathcal{U}(200,750) \\
f_{C} \,\,\, \sim \mathcal{L}\mathcal{U}(-3.0,1.0)
\end{gather}

where  $\mathcal{N}(\mu,\sigma)$ stands for a Normal distribution with mean $\mu$ and standard deviation $\sigma$. $\mathcal{L}\mathcal{U}(a,b)$ and $\mathcal{U}(a,b)$ stand for Log-Uniform and Uniform distributions, within the $(a,b)$ intervals, respectively. This is our standard setup of priors, however, for some galaxies we modified these priors in subsequent runs, if for instance, a posterior distribution converges toward the edge of a prior.

For $R_{0}$, we chose a prior distribution tailored to each individual galaxy. As described in previous sections, $R_{0}$ corresponds to the launching radius of the galactic wind, and should be on the same order as the size of the stellar component of each galaxy. Thus, we used \textsc{GALFIT} \citep{Galfit_a, Galfit_b} to fit a S\'ersic model to the HST F814W image of each galaxy, which should be dominated by stellar continuum emission, to obtain a first estimation of the galaxy size, parametrized by the half-light radius of the S\'ersic profile. Then, we chose a normal prior for $R_{0}$ centered on this half-light radius, with standard deviation of $1$ kpc, namely:

\begin{equation}
    R_{0}\,\,\,\, \sim \mathcal{N}(\mathrm{HLR}_{\mathrm{GALFIT}},1) \\
\end{equation}

Thus, we exclude scenarios where $R_0$ is much larger or much smaller than the actual size of a given galaxy. For galaxies with more irregular geometries, where the size is less constrained by a simple S\'ersic model, we use a broader prior, with a standard deviation of $2$ kpc instead. The continuum HLR measured for each galaxy is reported in Table~\ref{tab:sample_table}.

An additional parameter $\Delta v$ accounts for any kinematic offset between the systemic velocity and the P-Cygni emission. Its prior is given by:
\begin{equation}
\Delta v \sim \mathcal{U}(-200,200) \\
\end{equation}
and it has units of km s$^{-1}$. For the geometric parameters, we also use completely agnostic priors, namely:

\begin{gather}
\mathrm{O.A.} \sim \mathcal{U}(0,90) \\
\mathrm{P.A.} \sim \mathcal{U}(0,180) \\
\mathrm{R.A.} \sim \mathcal{U}(0,90)
\end{gather}
 
\noindent in units of degrees. Finally, we included an additional term $f$ to account for the existence of additional variance in the data. The extra variance ($\sigma$) is proportional to the model, such that:

\begin{equation}
    \sigma^{2} =  f^{2} M^{2}
\end{equation}
where $M$ is the value of the model, evaluated in the independent variable, given a set of parameters. The prior on $f$ is given by:
\begin{equation}
f \sim \mathcal{L}\mathcal{U}(-5,-1)    
\end{equation}

We use 22 Markov Chain Monte Carlo (MCMC) chains to sample the posterior distributions, each with 6000 iterations. We discard 500 burn-in iterations before convergence for each chain. The convergence of the MCMC chains is ensured by using the $\hat{R} \approx 1$ criterion \citep{Gelman1992}. 

\onecolumn
\section{Fraction of galaxies in our sample for different exposure times}
\label{sec:frac_depth_levels}
In Sec.~\ref{sec:paren_sample_comparison} we show how the presence of a \ion{Mg}{II} traced galactic-scale outflow correlated with host-galaxy properties, for our full sample. Here, we investigate the role of the depth of the field in the prevalence of the outflows. Figure~\ref{fig:SFR_vs_depth} shows the SFR distribution of our parent and final sample of galaxies, and the ratio between the number of galaxies in both samples as a function of the galaxy SFR. Same as the top-right panel of Fig.~\ref{fig:parent_sample_comparison}, but now separated by depth of the MUSCATEL fields (e.g., SF, MF, and DF). While the trend with SFR is qualitatively the same for the different depth levels (i.e., higher prevalence of galactic-scale outflows in higher SFR galaxies), the absolute fraction of galaxies that exhibit an outflow is significantly higher for the deeper fields, compared to the shallow field.
\FloatBarrier
\begin{figure*}[ht!]
\centering
\includegraphics[width=0.95\textwidth]{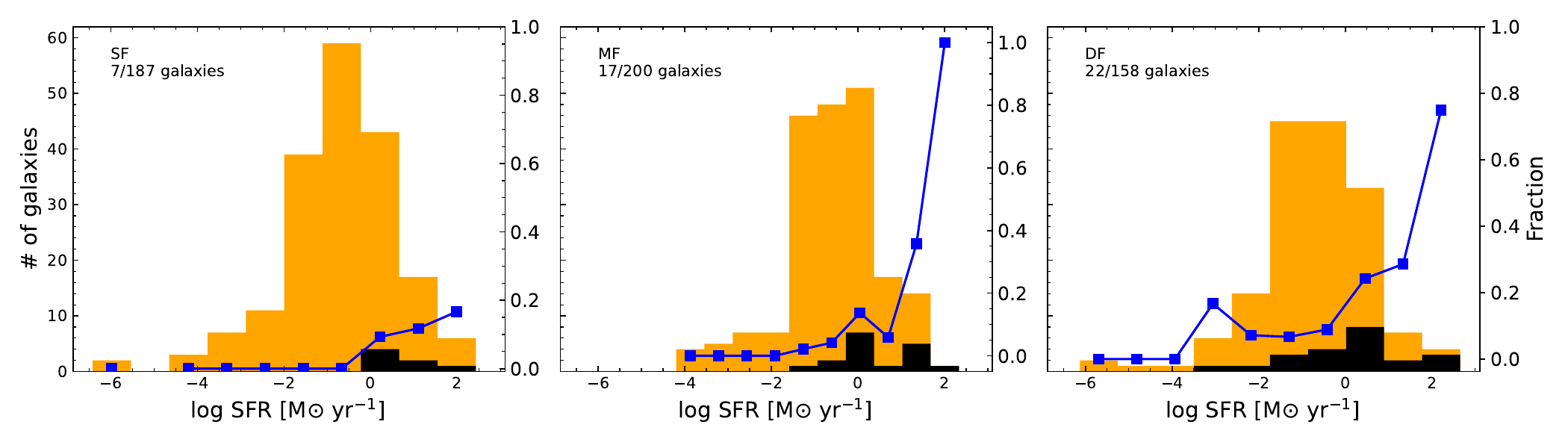}
\caption{Comparison of the SFR distributions for our final sample of galaxies (black) and our parent sample (orange), separated by depth of the field (SF, MF, and DF, see MUSCATEL description in Sec.~\ref{sec:muscatel}). The blue line shows the fraction of galaxies from the parent sample in our final sample, in each SFR bin. The right-side $y$-axis indicates the fraction shown by the blue line. In the top left corner of each panel is indicated the field depth, and the number of galaxies in the parent/final sample.}
\label{fig:SFR_vs_depth}
\end{figure*}

\twocolumn

\end{appendix}

\end{document}